\documentclass[12pt]{article} 
\usepackage{amssymb}          
\usepackage{latexsym}         

\font\twelveeufm=eufm10 scaled\magstep1    
\newcommand{\goth}[1]{\mbox{\twelveeufm #1}} 
\newcommand{\slt}{\goth{sl}_2}

\newcommand{\slth}{\widehat{\goth{sl}}_2}
\newcommand{\slnh}{\widehat{\goth{sl}}_n}
\newcommand{\slNh}{\widehat{\goth{sl}}_N}
\newcommand{\g}{{\goth{g}}} 
\newcommand{\gothh}{{\goth{h}}} 
\newcommand{\Z}{{\mathbb Z}} 
\newcommand{\C}{{\mathbb C}} 

\newcommand{\be}{\begin{equation}}
\newcommand{\ee}{\end{equation}}
\newcommand{\ba}{\begin{eqnarray}}
\newcommand{\ea}{\end{eqnarray}}
\newcommand{\ban}{\begin{eqnarray*}}
\newcommand{\ean}{\end{eqnarray*}}
\newcommand{\n}{\nonumber \\}
\newcommand{\eq}[1]{(\ref{#1})}
\newcommand{\sfrac}[2]{{\textstyle \frac{#1}{#2}}}
\newcommand{\ds}{\displaystyle}
\newcommand{\scr}{\scriptstyle}

\newcommand{\F}{{\cal F}}
\newcommand{\T}{{\cal T}}
\newcommand{\cH}{{\cal H}}
\newcommand{\cR}{{\cal R}}
\newcommand{\cL}{{\cal L}}
\newcommand{\Aqp}{{\cal A}_{q,p}}
\newcommand{\Bqla}{{{\cal B}_{q,\lambda}}}
\newcommand{\B}{{\cal B}}
\newcommand{\ve}{\varepsilon}
\newcommand{\la}{\lambda}
\newcommand{\La}{\Lambda}
\newcommand{\Wb}{\overline{W}}
\newcommand{\dz}{\underline{dz}}
\newcommand{\dw}{\underline{dw}}
\newcommand{\dbr}[1]{{\langle\!\langle #1 \rangle\!\rangle}} 
\newcommand{\bra}[1]{\langle #1 |}        
\newcommand{\ket}[1]{{| #1 \rangle}}      
\newcommand{\ketb}[1]{{| #1 \rangle_{\!B}}}  
\newcommand{\ketbs}[1]{{| #1 \rangle_{\!B}^*}} 
\newcommand{\rs}{{r^*}}
\newcommand{\BW}[5]{\Bigl({#1\atop#3}\ {#2\atop#4}\Bigl|#5\Bigr)}
\newcommand{\az}{\alpha_0}
\newcommand{\id}{{\rm id}}
\newcommand{\tr}{{\rm tr}}
\newcommand{\qed}{\hfill \fbox{}\medskip}
\newcommand{\no}{{\textstyle{\circ\atop\circ}}}
\newcommand{\verylongrightarrow}{\relbar\joinrel\relbar\joinrel
  \relbar\joinrel\rightarrow}
\newcommand{\kuru}{\curvearrowleft} 

\newcommand{\ignore}[1]{}

\newcommand{\hep}[1]{(#1)}

\newcommand{\maprightu}[2]
  {\smash{\mathop{\hbox to #1{\rightarrowfill}}\limits^{#2}}}
\newcommand{\maprightd}[2]
  {\smash{\mathop{\hbox to #1{\rightarrowfill}}\limits_{#2}}}
\newcommand{\mapleftu}[2]
  {\smash{\mathop{\hbox to #1{\leftarrowfill}}\limits^{#2}}}
\newcommand{\mapleftd}[2]
  {\smash{\mathop{\hbox to #1{\leftarrowfill}}\limits_{#2}}}
\newcommand{\mapdownl}[1]
  {\Bigg\downarrow\llap{$\vcenter{\hbox{$\scriptstyle#1\;\;\;$}}$}}
\newcommand{\mapdownr}[1]
  {\Bigg\downarrow\rlap{$\vcenter{\hbox{$\scriptstyle#1$}}$}}
\newcommand{\mapupl}[1]
  {\Bigg\uparrow\llap{$\vcenter{\hbox{$\scriptstyle#1\;\;\;$}}$}}
\newcommand{\mapupr}[1]
  {\Bigg\uparrow\rlap{$\vcenter{\hbox{$\scriptstyle#1$}}$}}

\begin{document}
\font\csc=cmcsc10 scaled\magstep1

{\baselineskip=14pt
 \rightline{
 \vbox{October 1999\hfill
       \hbox{DPSU-99-5}
}}}

\vskip 11mm
\begin{center}
{\Large\bf Beyond CFT : \\\vskip.2in
Deformed Virasoro and Elliptic Algebras\footnote{
Lecture given at 1999 CRM Summer School
``Theoretical Physics at the End of the XXth Century'',
June 27 - July 10, Banff (Alberta), Canada.
To appear in {\sl CRM Series in Mathematical Physics}, Springer Verlag.
}}
\end{center}
\vskip11mm
\begin{center}
{\csc Satoru Odake}
\\ 
{\baselineskip=15pt
\it\vskip.35in 
\setcounter{footnote}{0}\renewcommand{\thefootnote}{\arabic{footnote}}
%
Department of Physics, Faculty of Science \\
Shinshu University, Matsumoto 390-8621, Japan\\
odake@azusa.shinshu-u.ac.jp
\vskip.1in 
}
\end{center}

\vskip5mm

\begin{abstract}
In this lecture we discuss `beyond CFT' from symmetry point of view.
After reviewing the Virasoro algebra, we introduce deformed Virasoro algebras
and elliptic algebras. These algebras appear in solvable lattice models and
we study them by free field approach. 
\end{abstract}

\vskip20mm
hep-th/9910226

\newpage
\tableofcontents
\newpage

\setcounter{footnote}{0}
\renewcommand{\thefootnote}{\arabic{footnote})}

\setcounter{section}{0}
\setcounter{equation}{0}
\section{Introduction}

The conformal field theory (CFT) is a theory which is invariant
under the conformal transformation. 
CFT in 2 dimensions \cite{BPZ} can be applied to the string theory as a 
worldsheet theory and statistical critical phenomena in 2 dimensional space.
This theory has made a remarkable progress contacting with various branches
of mathematics \cite{CFT,string}. 
The main reason is that in two dimensional space 
(or 1+1 dimensional spacetime), the group of conformal transformations 
is infinite dimensional. Its algebra is known as the Virasoro algebra 
in the field theory realization, and this symmetry is very powerful. 
By using its detailed representation theory one can determine the spectrum 
and even calculate correlation functions. 
Statistical critical phenomena in 2 dimensional space are understood 
systematically by CFT and the list of critical exponents is obtained by
the representation theory of the Virasoro algebra.
In the first superstring revolution (middle 80's), the string theory was 
regarded as CFT on a worldsheet, and the knowledge obtained in the study
of CFT developed string theories very much.  
Recent progress in string theory (the second superstring revolution 
(middle 90's)) is based on spacetime symmetry consideration,
e.g., duality, D-brane, and also AdS/CFT correspondence, 
but importance of worldsheet symmetries remains unchanged.

Quantum field theory and critical phenomena are the systems with
infinite degrees of freedom. Consequently they are difficult to treat.
However we can sometimes solve some models, so-called solvable models.
Here we loosely use the word `solvable' if some physical quantities of 
that model can be calculated exactly, for example, 2D Ising model, 
XYZ spin chain, solvable lattice model, CFT, 4D super Yang-Mills theory, etc.
Although solvable models themselves are interesting for us, a main purpose
for physicists to study solvable models is to create new idea and concept
and to develop them through the study of solvable models.
Physics should explain the real world. The matter in the real world is 
very complicated and we need some approximations to study it. 
An approximation is not a bad thing if it grasps the essence of the problem.
Physicists have developed a variety of concepts, approximation methods, 
calculation techniques, etc. through the study of solvable models, 
in order to apply them to the real world.
 
Symmetry is one of the main idea of modern physics.
If the model has some symmetry, 
its analysis becomes much simpler 
by using the representation theory of the symmetry algebra.
By reversing this direction,
symmetry is also used for a model building. 
For example, the general relativity and the gauge theory are constructed 
by imposing the invariance under the general coordinate transformation 
and the gauge transformation respectively.
{}From symmetry point of view, `solvable' in the system with infinite 
degrees of freedom is stated as the following `equation':
\ba
  &&\frac{\mbox{System of {\bf infinite} degrees of freedom}}
  {\mbox{{\bf Infinite} dimensional symmetry}}\\
  &=&
  \mbox{System can be described by {\bf finite} degrees of freedom.}
  \nonumber
\ea
This is the reason why we are interested in infinite dimensional
symmetries.

Common feature of many solvable models is the factorization of the 
scattering $S$ matrix \cite{Zam79}, 
in other words the Yang-Baxter equation \cite{J,Bax,J89}.
Solutions of the Yang-Baxter equation are related to the Lie algebras
\cite{JMO1,DJKMO,JMO3}, 
and three types of solutions are known; rational, trigonometric and elliptic. 
Associated for each type of solution ($R$ matrix), 
algebras are defined \cite{D},
$$
  \begin{array}{lcl}
  \mbox{rational}&\rightarrow&\mbox{Yangian},\\
  \mbox{trigonometric}&\rightarrow&\mbox{quantum group (quantum algebra)},\\
  \mbox{elliptic}&\rightarrow&\mbox{elliptic quantum group (elliptic algebra)}.
  \end{array}
$$
This elliptic algebra is one of the topics in this lecture.

Since CFT is invariant under the scale transformation, 
CFT has no scale, in other words, it is a massless theory.
If we add to CFT the perturbation which breaks the conformal symmetry,
then the theory becomes massive. General massive theories are very difficult.
So we restrict ourselves to its subset, massive integrable models (MIM).
If we perturb CFT in a `good' manner (for example $(1,3)$ or $(1,2)$ 
perturbation \cite{Zam87,EY,Zam89,Smi91}), 
infinitely many conserved quantities survive. 
In the terminology of statistical mechanics, CFT corresponds
to on-critical theory, and perturbation corresponds to off-critical
procedure, and a lattice analogue of MIM is a solvable lattice model.
CFT is controlled by the Virasoro symmetry, but MIM is massive, therefore
there is no Virasoro symmetry.
$$
\begin{tabular}{cccccc}
  CFT&~~~~~~~&on-critical&&massless&Virasoro\\
  $\mapdownr{\mbox{good perturbation}}$&&$\mapdownr{\mbox{off-critical}}$&&
  $\mapdownr{}$&$\mapdownr{}$\\
  massive&&solvable&&massive&no\\
  integrable model&&lattice model&&&Virasoro
\end{tabular}
$$
\noindent
A natural question arises:
\begin{quote}
  What symmetry ensures the integrability or infinitely many conserved 
  quantities of MIM or solvable lattice model ?
\end{quote}
We would like to answer this question. 
This is our main motivation for recent study.

In some cases the Yangian or the quantum group symmetry plays 
an important role.
Kyoto group investigated the XXZ spin chain and clarified its symmetry,
the quantum affine Lie algebra $U_q(\slth)$ \cite{DFJMN}.
They studied XXZ spin chain from the representation theory point of view and
developed vertex operator calculation technique.
But naively one expected some deformation of the Virasoro algebra.
Such algebras, deformed Virasoro algebra (DVA) and deformed $W_N$ 
algebras (DWA), were constructed in different points of view 
\cite{SKAO,AKOS95,FR95,FF95},
using a correspondence of singular vectors and multivariable orthogonal 
symmetric polynomials or using the Wakimoto realization at the critical level.
Later it was shown that this deformed Virasoro algebra appears
in the Andrews-Baxter-Forester (ABF) model as a symmetry \cite{LP96}.
This DVA corresponds to $A_1^{(1)}$ algebra.
DVA corresponding to $A_2^{(2)}$ was obtained in \cite{BL97}.

Another possibility is elliptic quantum groups (elliptic algebras).
Corresponding to the two types of elliptic solutions of the Yang-Baxter 
equation, there are two type of elliptic quantum groups \cite{FIJKMY,Fel95}.
These two elliptic quantum groups have a common structure \cite{Fron};
They are quasi-Hopf algebras \cite{D}. 
Along this line, explicit formulas for the twistors were presented and 
the vertex type algebra $\Aqp(\slnh)$ and the face type 
algebra $\Bqla(\g)$ were defined in \cite{JKOS1}.

In this lecture we would like to 
(i) introduce the deformed Virasoro algebras and elliptic algebras, 
(ii) present free field approach and vertex operator calculation technique
for the solvable lattice models.
Contents of this lecture is presented in previous pages.
In section 2 we review the Virasoro algebra which is needed to understand
the deformed case. The deformed Virasoro algebra of type $A_1^{(1)}$ is
defined and its properties are given in section 3.
In section 4 we review solvable lattice models and introduce the elliptic
quantum groups. Section 5 is devoted to an application of these idea
to the ABF model. ABF model in regime III corresponds to the 
$(1,3)$-perturbation of the minimal unitary CFT.
Vertex operators are bosonized and local height 
probabilities (LHP's) are calculated. Another deformed Virasoro algebra
DVA($A_2^{(2)}$) is defined in section 6 and the dilute $A_L$ models are 
studied by free field approach. 
Dilute $A_L$ model in regime $2^+$ corresponds to the 
$(1,2)$-perturbation of the minimal unitary CFT.
In section 7 we mention other topics 
that are not treated in this lecture. Appendix A is a summary of notations 
and formulas used throughout this lecture.

\setcounter{section}{1}
\setcounter{equation}{0}
\section{Conformal Field Theory and Virasoro Algebra}

\subsection{Conformal field theory}

The conformal field theory (CFT) is a theory which is invariant
under the conformal transformation. 
In two dimensional space (or 1+1 dimensional spacetime)
the conformal transformation is any holomorphic map
$z\mapsto w=w(z)$ where $z$ is a complex coordinate of the space. 
Therefore the conformal group is infinite dimensional.

An important property of the CFT (bulk theory) is the factorization into
a holomorphic part ($z$, left mover) and an antiholomorphic part ($\bar{z}$, 
right mover).
We can treat them independently. Usually we treat $z$ part (chiral part)
only. To get final physical quantities, however, we have to glue chiral and 
antichiral parts with appropriate physical conditions.

As a quantum field theory, infinitesimal conformal transformation,
\be
  l_n\;:\;z\mapsto z+\epsilon z^{n+1},
\ee
is generated by the chiral part of the energy momentum tensor, $L(z)$.
The algebra generated by this $L(z)$ is called the Virasoro algebra.
Invariance under the conformal transformation imposes that correlation 
functions satisfy the conformal Ward identity. 
Correlation functions are severely controlled by this infinite dimensional
Virasoro symmetry.

General review of the CFT is not the aim of this lecture. 
Since many good review and books are available now,
for various topics of the CFT, see \cite{BPZ,CFT,string}.
  
\subsection{Virasoro algebra}\label{sec:2.2}

In this and next subsections we review some properties of the Virasoro 
algebra in order to compare them with the deformed one in the next section.

\subsubsection{definition and consistency}\label{sec:2.2.1}

\noindent{\bf Definition}\quad
The Virasoro algebra is a Lie algebra over $\C$ generated by $L_n$ ($n\in\Z$)
and $c$, and their relation is
\be
  [L_n,L_m]=(n-m)L_{n+m}+\frac{c}{12}(n^3-n)\delta_{n+m,0},\qquad
  [L_n,c]=0.
  \label{Vir}
\ee
In terms of the Virasoro current $\ds L(z)=\sum_{n\in\Z}L_nz^{-n-2}$, 
this relation is equivalent to the following operator product expansion (OPE),
\be
  L(z)L(w)=\frac{c}{2(z-w)^4}+\frac{2L(w)}{(z-w)^2}+\frac{\partial L(w)}{z-w}
  +\mbox{reg.}
\ee
As a formal power series this can be written as
\be
  [L(z),L(w)]=\frac{1}{z^4}\delta'''\Bigl(\frac{w}{z}\Bigr)\frac{c}{12}
  +\frac{1}{z^2}\delta'\Bigl(\frac{w}{z}\Bigr)2L(w)
  +\frac{1}{z^1}\delta\Bigl(\frac{w}{z}\Bigr)\partial L(w),
\ee
or 
\be
  [z^2L(z),w^2L(w)]=
  \Bigl(\frac{w}{z}\Bigr)^2\delta'''\Bigl(\frac{w}{z}\Bigr)
  \frac{c}{12}+
  \frac{w}{z}\delta'\Bigl(\frac{w}{z}\Bigr)\Bigl(z^2L(z)+w^2L(w)\Bigr),
  \label{Vir2}
\ee
where $\ds\delta(z)=\sum_{n\in\Z}z^n$ (see appendix \ref{app:a.2}).
Meaning of \eq{Vir2} is that the coefficient of $z^nw^m$ in \eq{Vir2} 
gives \eq{Vir}.

\medskip

\noindent{\bf Consistency}\quad
Mathematically the Virasoro algebra is a one dimensional central
extension of diff($S^1$), Lie algebra of diffeomorphism group of
a circle $S^1$. Vector fields on $S^1$ form a Lie algebra,
\be
  [l_n,l_m]=(n-m)l_{n+m},
\ee
where $l_n=-z^{n+1}\frac{d}{dz}=ie^{in\theta}\frac{d}{d\theta}$
($z=e^{i\theta}$ is a coordinate of $S^1$).
Let us consider its central extension,
\be
  [L_n,L_m]=(n-m)L_{n+m}+cf(n,m),
  \label{c.ext}
\ee
where $c$ is a central element and $f(n,m)$ is a number.
{}From antisymmetry of bracket $[~,~]$ and the Jacobi identity, 
$f(n,m)$ should satisfy 
\ba
  &&f(n,m)=-f(m,n),
  \label{anti}\\
  &&(n-m)f(l,n+m)+(m-l)f(n,m+l)+(l-n)f(m,l+n)=0.
  \label{Jacobi}
\ea
Its nontrivial solution has the following form, 
\be
  f(n,m)=\mbox{const}\cdot n^3\delta_{n+m,0}.
\ee
{\it Proof.}\quad
{}From \eq{Jacobi} with $l=0$, we have
$f(n,m)=\frac{n-m}{n+m}f(n+m,0)$ for $n+m\neq 0$, which satisfies
eqs.(\ref{anti},\ref{Jacobi}) with $n+m+l\neq 0$.
By setting $L'_n=L_n+\frac{c}{n}f(n,0)$ ($n\neq 0$) and $L'_0=L_0$, 
\eq{c.ext} becomes
$$
  [L'_n,L'_m]=(n-m)L'_{n+m}+c\delta_{n+m,0}f(n,-n).
$$
Hence we are enough to determine $a_n=f(n,-n)$. 
Eq.(\ref{Jacobi}) with $l=-n-m$ implies
$$
  -(n-m)a_{n+m}+(2m+n)a_n-(2n+m)a_m=0.
$$
For $m=1$, we obtain 
$$
  a_n=\sfrac16(n^3-n)a_2+(n-\sfrac16(n^3-n))a_1,
$$
and this solution satisfies eqs.(\ref{anti}, \ref{Jacobi}) for all cases.
The term proportional to $n$ in $a_n$, say $bn$, can be deleted by 
$L^{\prime\prime}_n=L'_n+\frac{c}{2}b\delta_{n,0}$.\qed

The central term in Virasoro algebra is chosen so that it vanishes
for $n=1,0,-1$.

\subsubsection{representation theory}\label{sec:2.2.2}

We consider the highest representation of the Virasoro algebra.
The highest weight state $\ket{h}$ ($h\in\C$) is characterized by 
\be
  L_n\ket{h}=0\quad(n>0),\qquad L_0\ket{h}=h\ket{h},
\ee
and the Verma module is 
\be
  M=\bigoplus_{l\geq 0}\bigoplus_{n_1\geq\cdots\geq n_l>0}\C
  L_{-n_1}\cdots L_{-n_l}\ket{h}.
\ee
Descendant $L_{-n_1}\cdots L_{-n_l}\ket{h}$ is also an eigenstate of 
$L_0$ with eigenvalue $\ds h+\sum_{i=1}^ln_i$. $\ds\sum_{i=1}^ln_i$ is called 
a level of the state.
Since $L_0$ corresponds to energy (exactly speaking Hamiltonian is
$L_0+\bar{L}_0$), this representation has energy bounded below.
The highest weight state $\ket{h}$ is created by a primary field with
conformal weight $h$, $\phi_h(z)$, from the vacuum $\ket{\bf 0}$ 
($L_n\ket{\bf 0}=0$ for $n\geq -1$); 
$\ds\ket{h}=\lim_{z\rightarrow 0}\phi_h(z)\ket{\bf 0}$.

At level $N$ there are $p(N)$ independent 
states, $L_{-n_1}\cdots L_{-n_l}\ket{\lambda}$ 
($n_1\geq\cdots\geq n_l>0$, $\ds\sum_{i=1}^ln_i=N$).
Here $p(N)$ is the number of partition and its
generating function is given by
\be
  \sum_{N=0}^{\infty}p(N)y^N=\prod_{n=1}^{\infty}\frac{1}{1-y^n}.
  \label{p(N)}
\ee
Let us number these states by the reverse lexicographic ordering for 
$(n_1,\cdots,n_l)$, i.e., 
$\ket{h;N,1}=L_{-N}\ket{\lambda}$, 
$\ket{h;N,2}=L_{-N+1}L_{-1}\ket{\lambda}$, $\cdots$, 
$\ket{h;N,p(N)}=L_{-1}^N\ket{\lambda}$.

The first problem of representation theory is whether $M$ is 
irreducible or not, namely $M$ has Virasoro invariant subspaces or not.
If $M$ has a singular vector at level $N$, $\ket{\chi}$, 
which is defined by
\be
  L_n\ket{\chi}=0\quad(n>0),\qquad L_0\ket{\chi}=(h+N)\ket{\chi},
\ee
then $\ket{\chi}$ generates an invariant subspace and we have to 
quotient out it from $M$ in order to get an irreducible module $\cL$.
Existence of singular vectors can be detected by considering an `inner 
product' (bilinear form). Let introduce dual module $M^*$ on which the 
Virasoro algebra act as $L_n^{\dagger}=L_{-n}$.
$M^*$ is generated by $\bra{h}$ which satisfies 
$\bra{h}L_n=0$ ($n<0$), $\bra{h}L_0=h\bra{h}$ and $\bra{h}h\rangle=1$.
At level $N$ there are $p(N)$ states
$\bra{h;N,1}=\bra{h}L_N$, 
$\bra{h;N,2}=\bra{h}L_1L_{N-1}$, $\cdots$,  
$\bra{h;N,p(N)}=\bra{h}L_1^N$.  
A state $\ket{\psi}\in M$ is called a null state if it is orthogonal to
all states, i.e., $\bra{\psi'}\psi\rangle=0$ for $\forall\bra{\psi'}\in M^*$.
Descendants of $\ket{\chi}$ are null state, because 
(i) $\bra{h}\chi\rangle=0$ (consider $\bra{h}L_0\ket{\chi}$) and
(ii) $\bra{h}\cdots L_{m_2}L_{m_1}\cdot L_{-n_1}L_{-n_2}\cdots\ket{\chi}
=\sum(\cdots)\bra{h}\chi\rangle=0$.
Physical meaning of quotienting out invariant subspaces is projecting out 
null states which decouple from all the states.

We give an example of inner product at level $1$, $2$, $3$.
\ban
  &&\bra{h}L_1L_{-1}\ket{h}=2h,\\
  &&\left(\begin{array}{cc}
  \bra{h}L_2L_{-2}\ket{h}&\bra{h}L_2L_{-1}^2\ket{h}\\
  \bra{h}L_{1}^2L_{-2}\ket{h}&\bra{h}L_1^2L_{-1}^2\ket{h}
  \end{array}\right)=
  \left(\begin{array}{cc}
  4h+\sfrac12c&6h\\
  6h&4h(2h+1)
  \end{array}\right),\\
  &&\left(\begin{array}{ccc}
  \bra{h}L_3L_{-3}\ket{h}&\bra{h}L_3L_{-2}L_{-1}\ket{h}&
  \bra{h}L_3L_{-1}^3\ket{h}\\
  \bra{h}L_1L_2L_{-3}\ket{h}&\bra{h}L_1L_2L_{-2}L_{-1}\ket{h}&
  \bra{h}L_1L_2L_{-1}^3\ket{h}\\
  \bra{h}L_1^3L_{-3}\ket{h}&\bra{h}L_1^3L_{-2}L_{-1}\ket{h}&
  \bra{h}L_1^3L_{-1}^3\ket{h}
  \end{array}\right)\\
  &&\qquad=
  \left(\begin{array}{ccc}
  6h+2c&10h&24h\\
  10h&h(8h+8+c)&12h(3h+1)\\
  24h&12h(3h+1)&24h(h+1)(2h+1)
  \end{array}\right).
\ean
A determinant of this matrix is called the Kac determinant and 
its zeros indicate existence of null states.
The Kac determinant at level 2 is $2h(16h^2+2(c-5)h+c)$
 which vanishes for $h=0$ and $h=h_{\pm}=
\frac{1}{16}(5-c\pm\sqrt{(1-c)(c-25)})$.
For $h=0$, $L_{-1}^2\ket{h}$ is a null state which is a descendant
of the singular vector at level 1 $L_{-1}\ket{h}$.
For $h=h_{\pm}$, there exists a new singular vector,
\be
  \ket{\chi_{\pm}}=\Bigl(L_{-2}-\frac{3}{2(2h_{\pm}+1)}L_{-1}^2\Bigr)
  \ket{h_{\pm}}.
  \label{chipm}
\ee

At general level $N$, the Kac determinant is given by \cite{Kac79}
\be
  \det\Bigl(\langle h;N,i|h;N,j\rangle\Bigr)_{1\leq i,j\leq p(N)}
  =\prod_{l,k\geq 1\atop lk\leq N}\Bigl(2lk(h-h_{l,k})\Bigr)^{p(N-lk)},
  \label{Kacdet}
\ee
Here we have parametrized the central charge $c$ and conformal 
weight $h_{l,k}$ by a parameter $\beta$,
\ba
  c&\!\!=\!\!&1-6\az^2,
  \label{cbeta}\\
  h_{l,k}&\!\!=\!\!&
  \frac14\Biggl(\Bigl(\sqrt{\beta}l-\frac{1}{\sqrt{\beta}}k\Bigr)^2
  -\az^2\Biggr),  
  \label{hlkbeta}
\ea
where $\az$ is
\be
  \az=\sqrt{\beta}-\frac{1}{\sqrt{\beta}}.
  \label{az}
\ee
These $\beta$ and $\az$ will be used throughout this lecture.
Remark that $c$ and $h_{l,k}$ are invariant under 
$\beta\rightarrow\beta^{-1}$.

Detailed study of the Kac determinant shows that representations of 
the Virasoro algebra are classified into several classes. 
The most interesting class is so-called minimal series \cite{BPZ}.
In minimal series $\beta$ is a rational number (we take $\beta>1$.),
\be
  \beta=\frac{p''}{p'},\quad p',p''\in\Z_{>0},\quad p''>p',\quad (p',p'')=1,
  \label{minimalbeta}
\ee
and in this case eqs.(\ref{cbeta}, \ref{hlkbeta}) become
\ba
  c&\!\!=\!\!&1-6\frac{(p''-p')^2}{p'p''},\\
  h_{l,k}&\!\!=\!\!&\frac{(p''l-p'k)^2-(p''-p')^2}{4p'p''}.
\ea
An operator algebra of $\phi_h(z)$ closes for the following finite number
of $h=h_{l,k}$,
\be
  1\leq l\leq p'-1,\quad 1\leq k\leq p''-1,\quad p''l-p'k>0.
  \label{lkrange}
\ee
If one prefers $0<\beta<1$, it is achieved by $p'\leftrightarrow p''$ 
($\beta\leftrightarrow\beta^{-1}$), and then $h_{l,k}\leftrightarrow h_{k,l}$.

Let denote the Verma module with $h=h_{l,k}$ as $M_{l,k}$.
{}From the Kac determinant and the property of $h_{l,k}$,
\ba
  &&h_{l,k}=h_{-l,-k}=h_{l+p'n,k+p''n}\quad(n\in\Z),
  \label{hprop}\\
  &&h_{l,-k}=h_{l,k}+lk,
\ea
there are two basic singular vectors in $M_{l,k}$. One is at level $lk$ 
($h_{l,-k}=h_{l,k}+lk$, $h_{l,-k}=h_{-l,k}$) and the other is at 
level $(p'-l)(p''-k)$ 
($h_{l,k}=h_{p'-l,p''-k}$, $h_{-p'+l,p''-k}=h_{l,k}+(p'-l)(p''-k)$,  
$h_{-p'+l,p''-k}=h_{-l+2p',k}$).
Therefore we have
\be
  M_{l,k}\supset(M_{-l+2p',k}+M_{-l,k}).
\ee
To get an irreducible module we have to factor out invariant subspace,
$M_{l,k}/(M_{-l+2p',k}+M_{-l,k})$.
But the story has not ended yet because $M_{-l+2p',k}+M_{-l,k}$
does not coincide with $M_{-l+2p',k}\oplus M_{-l,k}$.

{}From the Kac determinant, $M_{-l,k}$ also has two singular vectors,
\ban
  M_{-l,k}&\!\!=\!\!&M_{p'-l,p''+k}\supset M_{-p'+l,p''+k}=M_{l-2p',k}\\
  &=\!\!&M_{p'+l,p''-k}\supset M_{-p'-l,p''-k}=M_{l+2p',k}.
\ean
Here indices $(p'-l,p''+k)$ and $(p'+l,p''-k)$ are positive and minimal
with respect to translation of $(p',p'')$ (see \eq{hprop}). 
Similarly 
\ban
  M_{-l+2p',k}&\!\!=\!\!&M_{2p'-l,k}\supset M_{-2p'+l,k}=M_{l-2p',k}\\
  &=\!\!&M_{l,2p''-k}\supset M_{-l,2p''-k}=M_{l+2p',k}.
\ean
These submodules of $M_{-l+2p',k}$ and $M_{-l,k}$ coincide, because if 
they do not coincide it implies more zeros of the Kac determinant.
Therefore $M_{-l+2p',k}$ and $M_{-l,k}$ share two invariant submodules
and we have
$$
  M_{-l,k}+M_{-l+2p',k}=(M_{-l,k}\oplus M_{-l+2p',k})/
  (M_{l-2p',k}+M_{l+2p',k}).
$$
General embedding pattern is illustrated in the following figure
\be
\setlength{\unitlength}{0.8mm}
\begin{picture}(140,35)(0,4)
\put(9,19){$\bullet$}
\put(19,29){$\bullet$}
\put(19,9){$\bullet$}
\put(39,29){$\bullet$}
\put(39,9){$\bullet$}
\put(59,29){$\bullet$}
\put(59,9){$\bullet$}
\put(79,29){$\bullet$}
\put(79,9){$\bullet$}
\put(5,19){$s_0$}
\put(18,33){$s_1$}
\put(18,5){$s'_1$}
\put(38,33){$s_2$}
\put(38,5){$s'_2$}
\put(58,33){$s_3$}
\put(58,5){$s'_3$}
\put(78,33){$s_4$}
\put(78,5){$s'_4$}
\put(11.5,21.5){\vector(1,1){7}}
\put(11.5,18.5){\vector(1,-1){7}}
\put(21.5,30){\vector(1,0){17}}
\put(21.5,28.5){\vector(1,-1){17}}
\put(21.5,11.5){\vector(1,1){17}}
\put(21.5,10){\vector(1,0){17}}
\put(41.5,30){\vector(1,0){17}}
\put(41.5,28.5){\vector(1,-1){17}}
\put(41.5,11.5){\vector(1,1){17}}
\put(41.5,10){\vector(1,0){17}}
\put(61.5,30){\vector(1,0){17}}
\put(61.5,28.5){\vector(1,-1){17}}
\put(61.5,11.5){\vector(1,1){17}}
\put(61.5,10){\vector(1,0){17}}
\put(81.5,30){\vector(1,0){17}}
\put(81.5,28.5){\vector(1,-1){17}}
\put(81.5,11.5){\vector(1,1){17}}
\put(81.5,10){\vector(1,0){17}}
\put(105,19){$\cdots\cdots$}
\put(125,10){.}
\end{picture}
\label{embed}
\ee
Conformal weights of singular vectors are given by
\ba
  s_0&:&L_0=h_{l,k}=A(0)+\sfrac{c-1}{24},\\
  s'_{2m-1}&:&L_0=h_{-l+2mp',k}=B(-m)+\sfrac{c-1}{24} \quad(m\geq 1),\\
  s_{2m+1}&:&L_0=h_{-l-2mp',k}=B(m)+\sfrac{c-1}{24} \quad(m\geq 0),\\
  s'_{2m}&:&L_0=h_{l-2mp',k}=A(-m)+\sfrac{c-1}{24} \quad(m\geq 1),\\
  s_{2m}&:&L_0=h_{l+2mp',k}=A(m)+\sfrac{c-1}{24} \quad(m\geq 1),
\ea
where $A(m)$ and $B(m)$ are given by
\be
  A(m)=\frac{(p''l-p'k+2mp'p'')^2}{4p'p''},\quad
  B(m)=\frac{(p''l+p'k+2mp'p'')^2}{4p'p''}.
  \label{AB}
\ee
Irreducible module $\cL_{l,k}$ is 
\be
  \cL_{l,k}=M_{l,k}-(M_{-l+2p',k}\oplus M_{-l,k})
  +(M_{l-2p',k}\oplus M_{l+2p',k})-\cdots.
  \label{Llk}
\ee
This information is encoded into the character.

The character of a representation is a tool that counts the number of
states at each level, 
\be
  ch=\tr~q^{L_0-\frac{c}{24}}=\sum_{N=0}^{\infty}
  \dim\{\mbox{level $N$ states}\}\cdot q^{h+N-\frac{c}{24}},
\ee
where $\frac{c}{24}$ is included for the modular transformation property.
For the Verma module $M$ with $h$ we have
\be
  \tr_{M}q^{L_0-\frac{c}{24}}=q^{h-\frac{c}{24}}
  \prod_{n=1}^{\infty}\frac{1}{1-q^n}.
\ee
{}From this and \eq{Llk}, the character of the irreducible Virasoro 
module $\cL_{l,k}$ is \cite{R-C}
\be
  \chi_{l,k}(q)=\tr_{\cL_{l,k}}q^{L_0-\frac{c}{24}}
  =\frac{1}{\eta(\tau)}\sum_{m\in\Z}\Bigl(q^{A(m)}-q^{B(m)}\Bigr),
  \label{Virch}
\ee
where the Dedekind eta function $\eta(\tau)$ is
\be
  \eta(\tau)=q^{\frac{1}{24}}\prod_{n=1}^{\infty}(1-q^n),\quad
  q=e^{2\pi i\tau}.
\ee

The representation is called unitary if the inner product is positive
definite. The unitary minimal series is \cite{FQS} 
\be
  \beta=\frac{m}{m-1}\mbox{ or }\frac{m-1}{m},\quad
  c=1-\frac{6}{m(m-1)}\quad (m=3,4,\cdots).  
\ee
This is shown by drawing the vanishing line of the Kac determinant 
in ($c,h$) plane. Use
\ba
  \beta&\!\!=\!\!&\frac{1}{12}\Bigl(13-c\pm\sqrt{(1-c)(25-c)}\Bigr)
  \raisebox{-0.5ex}{ $\stackrel{\displaystyle>}{<}$ } 1,\\
  h_{l,k}&\!\!=\!\!&\frac{1-c}{24}(lk-1)+\frac{13-c}{48}\pm
  \frac{l^2-k^2}{48}\sqrt{(1-c)(25-c)}.
\ea

\subsubsection{primary fields and correlation functions}\label{sec:2.2.3}

Primary fields are fundamental fields in CFT. Under the conformal 
transformation, a primary field with the conformal weight $h$, 
$\phi_h(z)$, transforms as a rank $h$ form ($\phi_h(z)dz^h$). We have the 
OPE,
\be
  L(z)\phi_h(w)=\frac{h\phi_h(w)}{(z-w)^2}+\frac{\partial\phi_h(w)}{z-w}
  +\mbox{reg.},
\ee
or in mode
\be
  [L_n,\phi_{h,m}]=\Bigl((h-1)n-m\Bigr)\phi_{h,n+m},
  \label{primary}
\ee
where $\ds\phi_h(z)=\sum_{n\in\Z-h}\phi_{h,n}z^{-n-h}$.
($\phi_h(z)$ is called a quasiprimary field if \eq{primary} 
holds for $n=1,0,-1$.)
We remark that the zero mode of a $h=1$ field commutes with the Virasoro
algebra, $[L_n,\phi_{1,0}]=0$. 

States are created by fields from the vacuum, 
$\ds\ket{\psi}=\lim_{z\rightarrow 0}\psi(z)\ket{\bf 0}$, 
and they are one-to-one correspondence. 
The highest weight state corresponds to the primary field and 
descendant states correspond to secondary fields. For example, 
\ban
  \ket{h}&\leftrightarrow&\phi_h(z),\\
  L_{-1}\ket{h}&\leftrightarrow&\partial\phi_h(z),\\
  L_{-3}L_{-2}\ket{h}&\leftrightarrow&\hat{L}_{-3}\hat{L}_{-2}\phi_h(z),
\ean
where $\hat{L}_n$ is 
\be
  \hat{L}_{-n}\psi(z)=\oint_z\frac{dy}{2\pi i}(y-z)^{1-n}L(y)\psi(z).
\ee

In quantum field theory physical information is obtained from the Green
functions (correlation functions). 
In CFT conformal symmetry imposes that correlation functions obey
the conformal Ward identity \cite{BPZ}, 
which enables us to reduce calculation of 
correlation function of secondary fields to that of primary fields.
Moreover if null states exist, the correlation functions satisfy
differential equations \cite{BPZ}. We illustrate this by taking an example. 
For $h=h_{1,2}$ or $h_{2,1}$, from \eq{chipm}, we have a null field
\be
  \chi(z)=\Bigl(\hat{L}_{-2}-\frac{3}{2(2h+1)}\hat{L}_{-1}^2\Bigr)\phi_h(z).
\ee
Since a null field decouples from all the fields, correlation functions
including it vanish,
\ba
  0&\!\!=\!\!&\langle\chi(z)X\rangle\n
  &=\!\!&\Bigl(\cL_{-2}(z)-\frac{3}{2(2h+1)}\cL_{-1}(z)^2\Bigr)
  \langle\phi_h(z)X\rangle\\
  &=\!\!&\left(\sum_{i=1}^N\Bigl(\frac{h_i}{(z-z_i)^2}
  +\frac{1}{z-z_i}\partial_{z_i}\Bigr)
  -\frac{3}{2(2h+1)}\partial_z^2\right)\langle\phi_h(z)X\rangle,
  \nonumber
\ea
where $X=\phi_{h_1}(z)\cdots\phi_{h_N}(z)$.
Here we have used 
\ba
  \langle\hat{L}_{-n}\phi_h(z)X\rangle&\!\!=\!\!&
  \oint_z\frac{dy}{2\pi i}(y-z)^{1-n}\langle L(y)\phi_h(z)X\rangle
  \qquad(n\geq 1)\n
  &=\!\!&-\sum_{i=1}^N\oint_{z_i}\frac{dy}{2\pi i}(y-z)^{1-n}
  \Bigl(\frac{h_i}{(y-z_i)^2}\langle\phi_h(z)X\rangle
  +\frac{1}{y-z_i}\partial_{z_i}\langle\phi_h(z)X\rangle\Bigr)\n
  &=\!\!&\cL_{-n}(z)\langle\phi_h(z)X\rangle,\\
  \cL_{-n}(z)&\!\!=\!\!&
  \sum_{i=1}^N\Bigl(\frac{(n-1)h_i}{(z_i-z)^n}-\frac{1}{(z_i-z)^{n-1}}
  \partial_{z_i}\Bigr).
\ea
Due to translational invariance, we have $\cL_{-1}(z)=\partial_z$.
By solving this differential equation the correlation function can be
calculated. Another method for calculation of correlation functions is 
free field realization, see \cite{DF,CFT}.

\subsection{Free field realization}\label{sec:2.3}

In the previous subsection the Virasoro algebra is treated in abstract way.
In this subsection we treat it more explicitly by using our familiar
free boson. 

\subsubsection{free field realization}\label{sec:2.3.1}

Let us introduce free boson oscillator $a_n$ ($n\in\Z_{\neq 0})$ and
zero mode $a_0$, $Q$,
\be
  [a_n,a_m]=2n\delta_{n+m},\qquad [a_n,Q]=2\delta_{n,0}.
  \label{an}
\ee
Here $2$ is the Cartan matrix of $A_1$ Lie algebra.
The Fock space with momentum $\alpha$, $\F_{\alpha}$, is defined by
\be
  \F_{\alpha}=\bigoplus_{l\geq 0}\bigoplus_{n_1\geq\cdots\geq n_l>0}\C
  a_{-n_1}\cdots a_{-n_l}\ketb{\alpha},
\ee
where $\ketb{\alpha}$ is characterized by
\be
  a_n\ketb{\alpha}=0\quad(n>0),\qquad 
  a_0\ketb{\alpha}=\alpha\ketb{\alpha}.
  \label{ketalpha}
\ee
$\ketb{\alpha}$ is obtained from $\ketb{0}$ 
($a_n\ketb{0}=0$ for $n\geq 0$), 
\be
  \ketb{\alpha}=e^{\frac12\alpha Q}\ketb{0}.
\ee
A boson field $\phi(z)$ is 
\be
  \phi(z)=Q+a_0\log z-\sum_{n\neq 0}\frac{1}{n}a_nz^{-n},\qquad
  \partial\phi(z)=\sum_{n\in\Z}a_nz^{-n-1}.
\ee
($\phi(z)$ here and usual one used in string theory is related as 
$\phi(z)=i\sqrt{2}\phi^{string}(z)$. $Q$ here is antihermitian.)
Normal ordering prescription $:~:$ is defined by 
\be
  \left\{\begin{array}{lll}
  \mbox{move}&\mbox{$a_n (n>0)$ and $a_0$}&\mbox{to right},\\
  \mbox{move}&\mbox{$a_n (n<0)$ and $Q$}&\mbox{to left}.
  \end{array}\right.
\ee

The Virasoro algebra is realized by a free boson in the following way,
\be
  L(z)=\frac14:\partial\phi(z)\partial\phi(z):+\frac12\az\partial^2\phi(z),
  \label{L-phi}
\ee
where the back ground charge $\az$ is given in \eq{az}.
In mode it is
\be
  L_n=\frac14:\sum_ma_{n-m}a_m:-\frac12\az(n+1)a_n,
  \label{Ln}
\ee
or more explicitly,
\ba
  L_n&\!\!=\!\!&\frac14\sum_{m\neq 0,n}a_{n-m}a_m+\frac12a_na_0
  -\frac12\az(n+1)a_n\quad(n\neq 0),\\
  L_0&\!\!=\!\!&\frac12\sum_{m>0}a_{-m}a_m
  +\frac14\Bigl((a_0-\az)^2-\az^2\Bigr).
\ea
This Virasoro current realizes the Virasoro algebra with the central charge
\eq{cbeta}. $\ketb{\alpha}$ is the highest weight state of the Virasoro
algebra with the conformal weight $h=h(\alpha)$,
\ba
  &&\ketb{\alpha}=\ket{h(\alpha)},\\
  &&h(\alpha)=\frac14\Bigl((\alpha-\az)^2-\az^2\Bigr),
  \label{halpha}
\ea
and the corresponding primary field is
\be
  V_{\alpha}(z)=\,:e^{\frac12\alpha\phi(z)}:.
  \label{Valpha}
\ee
Due to the background charge $\az$, $\partial\phi(z)$ is not a primary field,
\be
  [L_n,a_m]=-ma_{n+m}-\az n(n+1)\delta_{n+m,0}.
\ee

The dual Fock space $\F_{\alpha}^*$ also has the Virasoro module structure
by the pairing $<~,~>$ $:$ $\F_{\alpha}^*\times\F_{\alpha}\rightarrow\C$,
\be
  <{}^{t\!}Af,v>\;=\;<f,Av>,\quad f\in\F_{\alpha}^*,\;v\in\F_{\alpha},\quad
  A:\mbox{ operator}.
  \label{pairing}
\ee
Here ${}^tL_n$ and ${}^ta_n$ are given by
\ba
  {}^tL_n&\!\!=\!\!&L_{-n},
  \label{tLn}\\
  {}^ta_n&\!\!=\!\!&-a_{-n}\quad(n\neq 0),
  \label{tan}\\
  {}^ta_0&\!\!=\!\!&-a_0+2\az,
  \label{ta0}
\ea
and 
\be
  <\ketbs{\alpha},\ketb{\alpha}>=1.
  \label{<|>=1}
\ee
We write the following pairings as in the last lines,
\ba
  &&<{}^tL_{m_1}\cdots{}^tL_{m_k}\ketbs{\alpha},
  L_{-n_1}\cdots L_{-n_l}\ketb{\alpha}>\n
  &=\!\!&
  <\ketb{\alpha},L_{m_k}\cdots L_{m_1}L_{-n_1}\cdots L_{-n_l}
  \ketb{\alpha}>\n
  &=:\!\!&
  \!\!\!{\phantom{\rangle}}_{B\!\!}\!\!\!{\phantom{\rangle}}^{*\!}\bra{\alpha}
  L_{m_k}\cdots L_{m_1}L_{-n_1}\cdots L_{-n_l}
  \ketb{\alpha},\\
  &&<{}^ta_{m_1}\cdots{}^ta_{m_k}\ketbs{\alpha},
  a_{-n_1}\cdots a_{-n_l}\ketb{\alpha}>\n
  &=\!\!&
  <\ketb{\alpha},a_{m_k}\cdots a_{m_1}a_{-n_1}\cdots a_{-n_l}
  \ketb{\alpha}>\n
  &=:\!\!&
  \!\!\!{\phantom{\rangle}}_{B\!\!}\!\!\!{\phantom{\rangle}}^{*\!}\bra{\alpha}
  a_{m_k}\cdots a_{m_1}a_{-n_1}\cdots a_{-n_l}
  \ketb{\alpha}.
\ea
By \eq{tan} and \eq{ta0}, $\F_{\alpha}^*$ is isomorphic 
to $\F_{2\az-\alpha}$ as a Virasoro module,
\ba
  \F_{\alpha}^*&\!\!\cong\!\!&\F_{2\az-\alpha}\quad(\mbox{Vir. module}).\\
  \ketbs{\alpha}&\!\!\leftrightarrow\!\!&\ketb{2\az-\alpha}.
\ea
Of course $h(\alpha)=h(2\az-\alpha)$ by \eq{halpha}.

For later use we define $a'_0$ by,
\be
  a'_0=a_0-\az,\quad a'_0\ketb{\alpha}=(\alpha-\az)\ketb{\alpha},\quad
  {}^ta'_0=-a'_0.
  \label{a'0}
\ee
The Virasoro current \eq{L-phi} can be rewritten as
\be
  z^2L(z)=\frac14:D\phi'(z)D\phi'(z):+\frac12\az D^2\phi'(z),
  \label{L-phi2}
\ee
where $\phi'(z)$ and $D$ are
\be
  \phi'(z)=\phi(z)\Bigl|_{a_0\rightarrow a'_0},\qquad
  \partial\phi(z)=\partial\phi'(z)+\az z^{-1},
\ee
\be
  D=z\frac{\partial}{\partial z}.
  \label{D}
\ee

\subsubsection{singular vectors and Kac determinant}\label{sec:2.3.2}

In the case of $h=h_{l,k}$ there are singular vectors. 
In the free boson realization they can be expressed by using
the screening currents. 
Screening currents $S_{\pm}(z)$ are primary fields with $h=1$,
\be
  S_{\pm}(z)=V_{2\alpha_{\pm}}=\,:e^{\alpha_{\pm}\phi(z)}:,\quad
  \alpha_+=\sqrt{\beta},\quad\alpha_-=\frac{-1}{\sqrt{\beta}}\,.
  \label{S}
\ee
Hence their zero modes (screening charge) commute with the Virasoro algebra,
\be
  \Bigl[L_n,\oint_0\frac{dz}{2\pi i}S_{\pm}(z)\Bigr]=0.
\ee
We will use $S_+(z)$ in what follows. $S_-(z)$ can be treated similarly.
We follow the method in \cite{KM}.

The representation with $h=h_{l,k}$ is realized on $\F_{\alpha_{l,k}}$,
\ba
  h_{l,k}&\!\!=\!\!&h(\alpha_{l,k}),\\
  \alpha_{l,k}&\!\!=\!\!&\sqrt{\beta}(1-l)-\frac{1}{\sqrt{\beta}}(1-k).
  \label{alphalk}
\ea
In the following we write $\F_{\alpha_{l,k}}=\F_{l,k}$.
To contact with our previous paper \cite{AMOS}($\alpha_{r,s}$ there is 
$\propto\alpha_{-r,-s}$ here.), 
we consider dual space $\F_{l,k}^*=\F_{-l,-k}$.
In the Verma module with $h_{-l,-k}=h_{l,k}$, 
the singular vector at level $lk$ is expressed as
\ba
  \!\!\!\!\!&&\ket{\chi_{-l,-k}}\n
  \!\!\!\!\!&=\!\!&\int\prod_{j=1}^l\frac{dz_j}{2\pi i}\cdot
  S_+(z_1)\cdots S_+(z_l)\ketb{\alpha_{l,-k}}
  \label{sv}\\
  \!\!\!\!\!&=\!\!&\int\prod_{j=1}^l\dz_j\cdot
  \prod_{1\leq i<j\leq l}(z_i-z_j)^{2\beta}\cdot
  \prod_{i=1}^lz_i^{\beta(1-l)-k}\cdot\prod_{j=1}^l
  \exp\Bigl(\sqrt{\beta}\sum_{n>0}\frac{1}{n}a_{-n}z_j^n\Bigr)\cdot
  \ketb{\alpha_{-l,-k}}.
  \nonumber
\ea
Here $\dz_j$ is given in \eq{dz} and
the integration contour is summarized in appendix\ref{app:a.4} 
(for example take $I_{BMP}$). 
Since this integral of $S_+$'s, which is a map from $\F_{l,-k}$ 
to $\F_{-l,-k}$, commutes with $L_n$, $\ket{\chi_{-l,-k}}$ is annihilated
by $L_n$ ($n>0$). This state $\ket{\chi_{-l,-k}}$ is non zero because of
\eq{exI'DF}.

To prove the Kac determinant formula, we first study the relation between 
states in the Verma module and Fock space.
Let us introduce following notation,
\ba
  &&I=\{n_1,\cdots,n_l\},\quad n_1\geq\cdots\geq n_l>0,\qquad \ell(I)=l,\\
  &&L_I=L_{n_l}\cdots L_{n_1},\quad L_{-I}=L_{-n_1}\cdots L_{-n_l},\\
  &&a_I=a_{n_l}\cdots a_{n_1},\quad a_{-I}=a_{-n_1}\cdots a_{-n_l},
\ea
and $I$ is ordered by the reverse lexicographic ordering.
States at level $N$ in the Verma module are linear combinations of 
those of Fock space,
\be
  L_{-I}\ketb{\alpha}=\sum_{J}C(N,\alpha)_{I,J}a_{-J}\ketb{\alpha}.
\ee
For example,
\ban
  L_{-1}\ketb{\alpha}&\!\!=\!\!&\sfrac12\alpha a_{-1}\ketb{\alpha},\\
  \left(\begin{array}{c}
  L_{-2}\ketb{\alpha}\\cL_{-1}^2\ketb{\alpha}
  \end{array}\right)
  &\!\!=\!\!&\left(\begin{array}{cc}
  \sfrac12(\alpha+\az)&\sfrac14 a_{-1}^2\\
  \sfrac12\alpha&\sfrac14\alpha^2
  \end{array}\right)
  \left(\begin{array}{c}
  a_{-2}\ketb{\alpha}\\a_{-1}^2\ketb{\alpha}
  \end{array}\right),\\
  \left(\begin{array}{c}
  L_{-3}\ketb{\alpha}\\cL_{-2}L_{-1}\ketb{\alpha}\\cL_{-1}^3\ketb{\alpha}
  \end{array}\right)&\!\!=\!\!&
  \left(\begin{array}{ccc}
  \sfrac12\alpha+\az&\sfrac12&0\\
  \sfrac12\alpha&\sfrac14\alpha(\alpha+\az)&\sfrac18\alpha\\
  \alpha&\sfrac34\alpha^2&\sfrac18\alpha^3
  \end{array}\right)
  \left(\begin{array}{c}
  a_{-3}\ketb{\alpha}\\a_{-2}a_{-1}\ketb{\alpha}\\a_{-1}^3\ketb{\alpha}
  \end{array}\right).
\ean
The determinant of matrix $C(N,\alpha)$ is given by
\be
  \det C(N,\alpha)_{I,J}=\prod_{l,k\geq 1\atop lk\leq N}
  \Bigl(\sfrac12(\alpha-\alpha_{l,k})\Bigr)^{p(N-lk)}.
  \label{detC}
\ee
{\it Proof.}\quad
Since $L_{-n}=\frac12a_{-n}a_0+\cdots$, leading term of $\alpha$ in 
$\det C(N,\alpha)$ is 
$$
  \prod_{\{k_i\}\atop{\scriptstyle\Sigma}_iik_i=N}\prod_i
  \Bigl(\frac{\alpha}{2}\Bigr)^{k_i}=
  \prod_{l,k\geq 1\atop lk\leq N}\Bigl(\frac{\alpha}{2}\Bigr)^{p(N-lk)}.
$$
Here we have used \eq{Fk}.
In the last paragraph we have constructed singular vectors as same
number as this order of $\alpha$.\qed

Next we consider a dual $\F_{\alpha}^*$. With the notation,
\ba
  &&({}^tL)_I={}^t(L_I)={}^tL_{n_1}\cdots{}^tL_{n_l},\\
  &&({}^ta)_I={}^t(a_I)={}^ta_{n_1}\cdots{}^ta_{n_l},
\ea
we define a matrix $C'(N,\alpha)$,
\be
  ({}^tL)_I\ketbs{\alpha}=\sum_JC'(N,\alpha)_{I,J}({}^ta)_J\ketbs{\alpha}.
\ee
By \eq{tLn} and \eq{tan}, the left and right hand sides of this equation are 
\ba
  &&\mbox{LHS}=L_{-I}\ketbs{\alpha}=
  \sum_KC(N,2\az-\alpha)_{I,K}a_{-K}\ketbs{\alpha},\\
  &&\mbox{RHS}=\sum_{J,K}C'(N,\alpha)_{I,J}D_{J,K}a_{-K}\ketbs{\alpha},
\ea
where matrix $D$ is
\be
  D_{J,K}=\delta_{J,K}(-1)^{\ell(J)}.
  \label{DJK}
\ee
Hence $C'$ can be expressed by $C$,
\be
  C'(N,\alpha)_{I,J}=\sum_KC(N,2\az-\alpha)_{I,K}D_{K,J},
  \label{C'CD} 
\ee
and its determinant is 
\ba
  \det C'(N,\alpha)_{I,J}&\!\!=\!\!&
  \det C(N,2\az-\alpha)\cdot\det D\n
  &=\!\!&\prod_{l,k\geq 1\atop lk\leq N}
  \Bigl(\sfrac12(\alpha-\alpha_{-l,-k})\Bigr)^{p(N-lk)}.
\ea

By gathering these results, the inner product of two states in the Verma 
module becomes
\ba
  \bra{h}L_IL_{-J}\ket{h}&\!\!=\!\!&
  <({}^tL)_I\ketbs{\alpha},L_{-J}\ketb{\alpha}>\n
  &=\!\!&<\sum_KC'(N,\alpha)_{I,K}({}^ta)_K\ketbs{\alpha},
  \sum_LC(N,\alpha)_{J,L}a_{-L}\ketb{\alpha}>\n
  &=\!\!&\sum_{K,L}C'(N,\alpha)_{I,K}G_{K,L}C(N,\alpha)_{J,L}.
  \label{LILJ}
\ea
Here $G_{K,L}$ is
\ba
  G_{K,L}&\!\!=\!\!&<({}^ta)_K\ketbs{\alpha},a_{-L}\ketb{\alpha}>\n
  &=\!\!&
  \!\!\!{\phantom{\rangle}}_{B\!\!}\!\!\!{\phantom{\rangle}}^{*\!}\bra{\alpha}
  a_{m_k}\cdots a_{m_1}a_{-n_1}\cdots a_{-n_l}\ketb{\alpha}\n
  &=\!\!&\delta_{K,L}
  {\phantom{\rangle}}_{B\!\!}\!\!\!{\phantom{\rangle}}^{*\!}\bra{\alpha}
  \cdots a_2^{k_2}a_1^{k_1}a_{-1}^{k_1}a_{-2}^{k_2}\cdots\ketb{\alpha}\n
  &=\!\!&\delta_{K,L}\prod_i(2i)^{k_i}k_i!,
\ea
where $(m_1,\cdots,m_k)=1^{k_1}2^{k_2}\cdots$. 
By using \eq{k_i!} and \eq{Fk}, its determinant becomes
\be
  \det G_{K,L}=
  \prod_{\{k_i\}\atop{\scriptstyle\Sigma}_iik_i=N}
  \prod_i(2i)^{k_i}k_i!
  =\prod_{l,k\geq 1\atop lk\leq N}\Bigl(2lk\Bigr)^{p(N-lk)}.
\ee

Therefore we obtain the Kac determinant \eq{Kacdet},
\ba
  \det\bra{h}L_IL_{-J}\ket{h}&\!\!=\!\!&
  \det C'(N,\alpha)\cdot\det G\cdot\det{}^tC(N,\alpha)\n
  &=\!\!&\prod_{l,k\geq 1\atop lk\leq N}
  \Bigl(2lk(h-h_{l,k})\Bigr)^{p(N-lk)}.
\ea
Here we have used 
$(\alpha-\alpha_{l,k})(\alpha-\alpha_{-l,-k})=4(h-h(\alpha_{l,k}))$.

\subsubsection{Felder complex}\label{sec:2.3.3}

Irreducible Virasoro module $\cL_{l,k}$ is realized on the Fock space 
$\F_{l,k}$, but this $\F_{l,k}$ is bigger than $\cL_{l,k}$.
So we have to discuss how $\cL_{l,k}$ is obtained from $\F_{l,k}$.
The Virasoro structure of $\F_{l,k}$ was investigated by 
Feigin-Fuchs \cite{FeFu} and Felder \cite{Fel89}.
Here we review it following \cite{Fel89}.
We consider minimal series eqs.(\ref{minimalbeta})-(\ref{lkrange}). 

The Becchi-Rouet-Stora-Tyupin (BRST) charge $Q_m$ is defined by using 
the screening current $S_+(z)$,
\ba
  Q_m&\!\!=\!\!&
  \int\prod_{j=1}^m\frac{dz_j}{2\pi i}\cdot S_+(z_1)\cdots S_+(z_m),\\
  &=\!\!&\int\prod_{j=1}^m\frac{dz_j}{2\pi i}\cdot 
  \prod_{1\leq i<j\leq m}(z_i-z_j)^{2\beta}\cdot
  :S_+(z_1)\cdots S_+(z_m):,\nonumber
\ea
where the integration contour is taken as $I_{BMP}$ in appendix\ref{app:a.4}.
On $\F_{l',k'}$ conditions \eq{condi} and \eq{condii} become
\be
  \mbox{(i) }m\equiv 0\!\!\pmod{p'},\quad
  \mbox{(ii) }m\equiv l'\!\!\pmod{p'}.
\ee
For example $Q_l$ is well defined on $\F_{l',k}$ 
($l'\equiv l$ ($\bmod p'$)), 
and the contour can be deformed to $I_F$ type by \eq{IBIF}.
Importance property of the BRST charge is commutativity with 
the Virasoro algebra,
\ba
  &&Q_l\,:\,\F_{l',k}\rightarrow\F_{l'-2l,k},\quad 
  l'\equiv l\!\!\pmod{p'},\quad [L_n,Q_l]=0,\\
  &&Q_{p'-l}\,:\,\F_{l',k}\rightarrow\F_{l'-2(p'-l),k},\quad 
  l'\equiv -l\!\!\pmod{p'},\quad [L_n,Q_{p'-l}]=0,
\ea
and the nilpotency,
\be
  Q_lQ_{p'-l}=Q_{p'-l}Q_l=Q_{p'}=0.
\ee
Here $Q_lQ_{p'-l}=Q_{p'}$ and $Q_{p'-l}Q_l=Q_{p'}$ are shown by $I_{BMP}$ 
type contour, and $Q_{p'}=0$ is shown by \eq{IBIF}.
The singular vector \eq{sv} can be written as 
$\ket{\chi_{-l,-k}}=Q_l\ketb{\alpha_{l,-k}}$.

Let us consider the Felder complex $C_{l,k}$,
\be
  \cdots ~{\buildrel X_{-3} \over \longrightarrow }~
  C_{-2} ~{\buildrel X_{-2} \over \longrightarrow }~
  C_{-1} ~{\buildrel X_{-1} \over \longrightarrow }~
  C_0 ~{\buildrel X_0 \over \longrightarrow }~
  C_1 ~{\buildrel X_1 \over \longrightarrow }~
  C_2 ~{\buildrel X_2 \over \longrightarrow }~ \cdots, 
\ee
where $C_j$ and $X_j\;:\;C_j\rightarrow C_{j+1}$ ($j\in\Z$) are
\ba
  &&C_{2j}=\F_{l-2p'j,k},\quad C_{2j+1}=\F_{-l-2p'j,k},\\
  &&X_{2j}=Q_l,\quad X_{2j+1}=Q_{p'-l}.
\ea
$X$ satisfies the BRST property,
\be
  X_jX_{j-1}=0.
\ee

Structure of the Felder complex is illustrated in the following figure:
\be
\setlength{\unitlength}{1mm}
\begin{picture}(140,65)(-5,0)
\put(5,15){\makebox(0,0){$\cdots$}}
\put(12,10){\vector(1,0){5.5}}\put(12,9.5){\line(0,1){1}}
\put(12,20){\vector(1,0){5.5}}\put(12,19.5){\line(0,1){1}}
\put(20,10){\makebox(0,0){$\scr v_{-1}$}}
\put(30,10){\makebox(0,0){$\scr v_1$}}
\put(20,20){\makebox(0,0){$\scr u_1$}}
\put(30,20){\makebox(0,0){$\scr w_0$}}
\put(30,30){\makebox(0,0){$\scr v_0$}}
\put(25,5){\makebox(0,0){$\vdots$}}
\put(20,12){\vector(0,1){6}}
\put(30,18){\vector(0,-1){6}}
\put(28,12){\vector(-1,1){6}}
\put(28,18){\vector(-1,-1){6}}
\put(30,22){\vector(0,1){6}}
\put(28,28){\vector(-1,-1){6}}
\put(32,10){\vector(1,0){5.5}}\put(32,9.5){\line(0,1){1}}
\put(32,20){\vector(1,0){5.5}}\put(32,19.5){\line(0,1){1}}
\put(32,30){\vector(1,0){5.5}}\put(32,29.5){\line(0,1){1}}
\put(40,10){\makebox(0,0){$\scr u_2$}}
\put(50,10){\makebox(0,0){$\scr w_1$}}
\put(40,20){\makebox(0,0){$\scr v_{-1}$}}
\put(50,20){\makebox(0,0){$\scr v_1$}}
\put(40,30){\makebox(0,0){$\scr u_1$}}
\put(50,30){\makebox(0,0){$\scr w_0$}}
\put(50,40){\makebox(0,0){$\scr v_0$}}
\put(45,5){\makebox(0,0){$\vdots$}}
\put(40,18){\vector(0,-1){6}}
\put(50,12){\vector(0,1){6}}
\put(48,12){\vector(-1,1){6}}
\put(48,18){\vector(-1,-1){6}}
\put(40,22){\vector(0,1){6}}
\put(50,28){\vector(0,-1){6}}
\put(48,22){\vector(-1,1){6}}
\put(48,28){\vector(-1,-1){6}}
\put(50,32){\vector(0,1){6}}
\put(48,38){\vector(-1,-1){6}}
\put(52,10){\vector(1,0){5.5}}\put(52,9.5){\line(0,1){1}}
\put(52,20){\vector(1,0){5.5}}\put(52,19.5){\line(0,1){1}}
\put(52,30){\vector(1,0){5.5}}\put(52,29.5){\line(0,1){1}}
\put(52,40){\vector(1,0){5.5}}\put(52,39.5){\line(0,1){1}}
\put(60,10){\makebox(0,0){$\scr v_{-2}$}}
\put(70,10){\makebox(0,0){$\scr v_2$}}
\put(60,20){\makebox(0,0){$\scr u_2$}}
\put(70,20){\makebox(0,0){$\scr w_1$}}
\put(60,30){\makebox(0,0){$\scr v_{-1}$}}
\put(70,30){\makebox(0,0){$\scr v_1$}}
\put(60,40){\makebox(0,0){$\scr u_1$}}
\put(70,40){\makebox(0,0){$\scr w_0$}}
\put(60,50){\makebox(0,0){$\scr v_0$}}
\put(65,5){\makebox(0,0){$\vdots$}}
\put(60,12){\vector(0,1){6}}
\put(70,18){\vector(0,-1){6}}
\put(68,12){\vector(-1,1){6}}
\put(68,18){\vector(-1,-1){6}}
\put(60,28){\vector(0,-1){6}}
\put(70,22){\vector(0,1){6}}
\put(68,22){\vector(-1,1){6}}
\put(68,28){\vector(-1,-1){6}}
\put(60,32){\vector(0,1){6}}
\put(70,38){\vector(0,-1){6}}
\put(68,32){\vector(-1,1){6}}
\put(68,38){\vector(-1,-1){6}}
\put(60,48){\vector(0,-1){6}}
\put(68,42){\vector(-1,1){6}}
\put(72,10){\vector(1,0){5.5}}\put(72,9.5){\line(0,1){1}}
\put(72,20){\vector(1,0){5.5}}\put(72,19.5){\line(0,1){1}}
\put(72,30){\vector(1,0){5.5}}\put(72,29.5){\line(0,1){1}}
\put(72,40){\vector(1,0){5.5}}\put(72,39.5){\line(0,1){1}}
\put(80,10){\makebox(0,0){$\scr u_2$}}
\put(90,10){\makebox(0,0){$\scr w_1$}}
\put(80,20){\makebox(0,0){$\scr v_{-1}$}}
\put(90,20){\makebox(0,0){$\scr v_1$}}
\put(80,30){\makebox(0,0){$\scr u_1$}}
\put(90,30){\makebox(0,0){$\scr w_0$}}
\put(80,40){\makebox(0,0){$\scr v_0$}}
\put(85,5){\makebox(0,0){$\vdots$}}
\put(80,18){\vector(0,-1){6}}
\put(90,12){\vector(0,1){6}}
\put(88,12){\vector(-1,1){6}}
\put(88,18){\vector(-1,-1){6}}
\put(80,22){\vector(0,1){6}}
\put(90,28){\vector(0,-1){6}}
\put(88,22){\vector(-1,1){6}}
\put(88,28){\vector(-1,-1){6}}
\put(80,38){\vector(0,-1){6}}
\put(88,32){\vector(-1,1){6}}
\put(92,10){\vector(1,0){5.5}}\put(92,9.5){\line(0,1){1}}
\put(92,20){\vector(1,0){5.5}}\put(92,19.5){\line(0,1){1}}
\put(92,30){\vector(1,0){5.5}}\put(92,29.5){\line(0,1){1}}
\put(100,10){\makebox(0,0){$\scr v_{-1}$}}
\put(110,10){\makebox(0,0){$\scr v_1$}}
\put(100,20){\makebox(0,0){$\scr u_1$}}
\put(110,20){\makebox(0,0){$\scr w_0$}}
\put(100,30){\makebox(0,0){$\scr v_0$}}
\put(105,5){\makebox(0,0){$\vdots$}}
\put(100,12){\vector(0,1){6}}
\put(110,18){\vector(0,-1){6}}
\put(108,12){\vector(-1,1){6}}
\put(108,18){\vector(-1,-1){6}}
\put(100,28){\vector(0,-1){6}}
\put(108,22){\vector(-1,1){6}}
\put(112,10){\vector(1,0){5.5}}\put(112,9.5){\line(0,1){1}}
\put(112,20){\vector(1,0){5.5}}\put(112,19.5){\line(0,1){1}}
\put(125,15){\makebox(0,0){$\cdots$}}
\put(5,58){\makebox(0,0){$\cdots$}}
\put(25,58){\makebox(0,0){$C_{-2}$}}
\put(45,58){\makebox(0,0){$C_{-1}$}}
\put(65,58){\makebox(0,0){$C_0$}}
\put(85,58){\makebox(0,0){$C_1$}}
\put(105,58){\makebox(0,0){$C_2$}}
\put(125,58){\makebox(0,0){$\cdots$}}
\put(9,58){\vector(1,0){12}}
\put(29,58){\vector(1,0){12}}
\put(49,58){\vector(1,0){12}}
\put(69,58){\vector(1,0){12}}
\put(89,58){\vector(1,0){12}}
\put(109,58){\vector(1,0){12}}
\put(35,61){\makebox(0,0){$\scr X_{-2}$}}
\put(55,61){\makebox(0,0){$\scr X_{-1}$}}
\put(75,61){\makebox(0,0){$\scr X_0$}}
\put(95,61){\makebox(0,0){$\scr X_1$}}
\put(115,61){\makebox(0,0){$\scr X_2$}}
\end{picture}
\ee
Each Fock space has the structure
\ba
\setlength{\unitlength}{0.8mm}
\begin{picture}(140,35)(0,4)
\put(9,19){$\bullet$}
\put(19,29){$\bullet$}
\put(19,9){$\bullet$}
\put(39,29){$\bullet$}
\put(39,9){$\bullet$}
\put(59,29){$\bullet$}
\put(59,9){$\bullet$}
\put(79,29){$\bullet$}
\put(79,9){$\bullet$}
\put(5,19){$v_0$}
\put(18,33){$w_0$}
\put(18,5){$u_1$}
\put(38,33){$v_1$}
\put(38,5){$v_{-1}$}
\put(58,33){$w_1$}
\put(58,5){$u_2$}
\put(78,33){$v_2$}
\put(78,5){$v_{-2}$}
\put(18.5,28.5){\vector(-1,-1){7}}
\put(11.5,18.5){\vector(1,-1){7}}
\put(21.5,30){\vector(1,0){17}}
\put(21.5,28.5){\vector(1,-1){17}}
\put(38.5,28.5){\vector(-1,-1){17}}
\put(38.5,10){\vector(-1,0){17}}
\put(58.5,30){\vector(-1,0){17}}
\put(41.5,28.5){\vector(1,-1){17}}
\put(58.5,28.5){\vector(-1,-1){17}}
\put(41.5,10){\vector(1,0){17}}
\put(61.5,30){\vector(1,0){17}}
\put(61.5,28.5){\vector(1,-1){17}}
\put(78.5,28.5){\vector(-1,-1){17}}
\put(78.5,10){\vector(-1,0){17}}
\put(98.5,30){\vector(-1,0){17}}
\put(81.5,28.5){\vector(1,-1){17}}
\put(98.5,28.5){\vector(-1,-1){17}}
\put(81.5,10){\vector(1,0){17}}
\put(105,19){$\cdots\cdots$}
\put(125,10){,}
\end{picture}
\ea
which is obtained by combining the information from $C(N,\alpha)$
\ba
\setlength{\unitlength}{0.8mm}
\begin{picture}(140,35)(0,4)
\put(9,19){$\bullet$}
\put(19,29){$\bullet$}
\put(19,9){$\bullet$}
\put(39,29){$\bullet$}
\put(39,9){$\bullet$}
\put(59,29){$\bullet$}
\put(59,9){$\bullet$}
\put(79,29){$\bullet$}
\put(79,9){$\bullet$}
\put(99,9){$\bullet$}
\put(5,19){$0$}
\put(19,33){$1$}
\put(19,5){$0$}
\put(39,33){$1$}
\put(39,5){$1$}
\put(59,33){$2$}
\put(59,5){$1$}
\put(79,33){$2$}
\put(79,5){$2$}
\put(99,5){$2$}
\put(11.5,21.5){\vector(1,1){7}}
\put(11.5,18.5){\vector(1,-1){7}}
\put(21.5,30){\vector(1,0){17}}
\put(21.5,28.5){\vector(1,-1){17}}
\put(21.5,11.5){\vector(1,1){17}}
\put(21.5,10){\vector(1,0){17}}
\put(41.5,30){\vector(1,0){17}}
\put(41.5,28.5){\vector(1,-1){17}}
\put(41.5,11.5){\vector(1,1){17}}
\put(41.5,10){\vector(1,0){17}}
\put(61.5,30){\vector(1,0){17}}
\put(61.5,28.5){\vector(1,-1){17}}
\put(61.5,11.5){\vector(1,1){17}}
\put(61.5,10){\vector(1,0){17}}
\put(81.5,30){\vector(1,0){17}}
\put(81.5,28.5){\vector(1,-1){17}}
\put(81.5,11.5){\vector(1,1){17}}
\put(81.5,10){\vector(1,0){17}}
\put(105,19){$\cdots\cdots$}
\put(125,10){,}
\end{picture}
\nonumber
\ea
and the information from $C'(N,\alpha)$
\ba
\setlength{\unitlength}{0.8mm}
\begin{picture}(140,35)(0,4)
\put(9,19){$\bullet$}
\put(19,29){$\bullet$}
\put(19,9){$\bullet$}
\put(39,29){$\bullet$}
\put(39,9){$\bullet$}
\put(59,29){$\bullet$}
\put(59,9){$\bullet$}
\put(79,29){$\bullet$}
\put(79,9){$\bullet$}
\put(99,29){$\bullet$}
\put(5,19){$0$}
\put(19,33){$0$}
\put(19,5){$1$}
\put(39,33){$1$}
\put(39,5){$1$}
\put(59,33){$1$}
\put(59,5){$2$}
\put(79,33){$2$}
\put(79,5){$2$}
\put(99,33){$2$}
\put(18.5,28.5){\vector(-1,-1){7}}
\put(18.5,11.5){\vector(-1,1){7}}
\put(38.5,30){\vector(-1,0){17}}
\put(38.5,11.5){\vector(-1,1){17}}
\put(38.5,28.5){\vector(-1,-1){17}}
\put(38.5,10){\vector(-1,0){17}}
\put(58.5,30){\vector(-1,0){17}}
\put(58.5,11.5){\vector(-1,1){17}}
\put(58.5,28.5){\vector(-1,-1){17}}
\put(58.5,10){\vector(-1,0){17}}
\put(78.5,30){\vector(-1,0){17}}
\put(78.5,11.5){\vector(-1,1){17}}
\put(78.5,28.5){\vector(-1,-1){17}}
\put(78.5,10){\vector(-1,0){17}}
\put(98.5,30){\vector(-1,0){17}}
\put(98.5,11.5){\vector(-1,1){17}}
\put(98.5,28.5){\vector(-1,-1){17}}
\put(98.5,10){\vector(-1,0){17}}
\put(105,19){$\cdots\cdots$}
\put(125,10){.}
\end{picture}
\nonumber
\ea
$v_0$, $u_m$, $w_{m-1}$, $v_{-m}$, $v_m$ ($m\geq 1$) have the same 
conformal weight as $s_0$, $s'_{2m-1}$, $s_{2m-1}$, $s'_{2m}$, $s_{2m}$ in
\eq{embed} respectively. 
$s_{2m-1}$ is $u_m$ but $s'_{2m-1}$ vanishes on the Fock space.
At the level of $s'_{2m-1}$, there is a state which does not belong the Verma
module. That state is $w_{m-1}$.  

We write states in $C_j$ with superfix $(j)$. Conformal weights of $v,u,w$
are following: 
\ba
  v^{(2j)}_0&:&L_0=A(-j)+\sfrac{c-1}{24},\\
  v^{(2j+1)}_0&:&L_0=B(j)+\sfrac{c-1}{24},\\
  u^{(2j)}_m&:&L_0=B(-|j|-m)+\sfrac{c-1}{24} \quad(m\geq 1),\\
  u^{(2j+1)}_m&:&L_0=A(\sfrac{|2j+1|-1}{2}+m)+\sfrac{c-1}{24} \quad(m\geq 1),\\
  v^{(2j)}_{-m}&:&L_0=A(-|j|-m)+\sfrac{c-1}{24} \quad(m\geq 1),\\
  v^{(2j+1)}_{-m}&:&L_0=B(\sfrac{|2j+1|-1}{2}+m)+\sfrac{c-1}{24}
  \quad(m\geq 1),\\
  v^{(2j)}_m&:&L_0=A(|j|+m)+\sfrac{c-1}{24} \quad(m\geq 1),\\
  v^{(2j+1)}_m&:&L_0=B(-\sfrac{|2j+1|+1}{2}-m)+\sfrac{c-1}{24} 
  \quad(m\geq 1),\\
  w^{(2j)}_m&:&L_0=B(|j|+m)+\sfrac{c-1}{24} \quad(m\geq 0),\\
  w^{(2j+1)}_m&:&L_0=A(-\sfrac{|2j+1|+1}{2}-m)+\sfrac{c-1}{24} \quad(m\geq 0),
\ea
where $A$ and $B$ are given in \eq{AB}.
BRST charge $X_j$ maps $u,v,w$ in the following way:
\ba
  j<0&:&X_jw^{(j)}_m=v^{(j+1)}_{-m-1}\quad(m\geq 0),\quad
  X_jv^{(j)}_m=u^{(j+1)}_{m+1}\quad(m\geq 0),\n
  &&X_ju^{(j)}_m=0\quad(m\geq 1),\quad X_jv^{(j)}_{-m}=0\quad(m\geq 1),\\
  j\geq 0&:&X_jw^{(j)}_m=v^{(j+1)}_{-m}\quad(m\geq 0),\quad
  X_jv^{(j)}_m=u^{(j+1)}_m\quad(m\geq 1),\n
  &&X_ju^{(j)}_m=0\quad(m\geq 1),\quad X_jv^{(j)}_{-m}=0\quad(m\geq 0).
\ea

The cohomology groups of the complex $C_{l,k}$ are \cite{Fel89}
\be
  H^j(C_{l,k})=\mbox{Ker }X_j/\mbox{Im }X_{j-1}=
  \left\{\begin{array}{ll}
  0&j\neq 0,\\
  \cL_{l,k}&j=0.
  \end{array}\right.
\ee
Using this fact, the trace of operator ${\cal O}$ over the irreducible 
Virasoro module can be converted to the alternated sum of those over 
the Fock spaces. If we can draw the commutative diagram
\be
  \cdots
  \begin{array}{ccccccc}
  \maprightu{13mm}{}&C_{-1}&\maprightu{13mm}{X_{-1}}&C_0&
  \maprightu{13mm}{X_0}&C_1&\maprightu{13mm}{X_1}\\
  &\mapdownr{{\cal O}^{(-1)}}&&\mapdownr{{\cal O}^{(0)}}&&
  \mapdownr{{\cal O}^{(1)}}&\\
  \maprightu{13mm}{}&C_{-1}&\maprightu{13mm}{X_{-1}}&C_0&
  \maprightu{13mm}{X_0}&C_1&\maprightu{13mm}{X_1}
  \end{array}
  \cdots, 
  \label{comdiag}
\ee
where ${\cal O}^{(j)}$ is an operator ${\cal O}$ realized on $C_j$, then 
by Euler-Poincar\'e principle, we have 
\be
  \tr_{\cL_{l,k}}{\cal O}=\tr_{H^0(C_{l,k})}{\cal O}^{(0)}=
  \tr_{H^*(C_{l,k})}{\cal O}
  =\sum_{j\in\Z}(-1)^j\tr_{C_j}{\cal O}^{(j)}.
\ee
For example let us calculate the character. 
Since the trace over the Fock space is
$$
  \tr_{C_{2m}}q^{L_0-\frac{c}{24}}=q^{-\frac{1}{24}+A(-m)}
  \prod_{n>0}\frac{1}{1-q^n},\quad
  \tr_{C_{2m+1}}q^{L_0-\frac{c}{24}}=q^{-\frac{1}{24}+B(m)}
  \prod_{n>0}\frac{1}{1-q^n},
$$
we obtain the Virasoro character \eq{Virch},
\be
  \tr_{\cL_{l,k}}q^{L_0-\frac{c}{24}}=
  \sum_{j\in\Z}(-1)^j\tr_{C_j}q^{L_0-\frac{c}{24}}
  =\frac{1}{\eta(\tau)}\sum_{m\in\Z}\Bigl(q^{A(-m)}-q^{B(m)}\Bigr).
\ee

If we use $S_-(z)$ instead of $S_+(z)$, another complex in which $k'$ 
of $\F_{l,k'}$ changes is obtained. 

\medskip

\noindent{\it Example:}\quad
$c=\frac12$ (Ising model)\\
Parameters and conformal weights are
\ban
  &c=\frac12,\quad p'=3,\quad p''=4,\quad\beta=\frac43,\quad
  \az=\frac{1}{2\sqrt{3}},
  \quad\alpha_+=\frac{2}{\sqrt{3}},\quad\alpha_-=-\frac{\sqrt{3}}{2},\\
  &h_{1,1}=0,\quad h_{2,1}=\frac12,\quad h_{2,2}=\frac{1}{16},\quad
  \alpha_{1,1}=0,\quad \alpha_{2,1}=-\frac{2}{\sqrt{3}},\quad 
  \alpha_{2,2}=-\frac{1}{2\sqrt{3}}.
\ean
$h=0,\frac12,\frac{1}{16}$ correspond to identity operator,
energy operator, spin operator respectively.
Remark that 
\ban
  &h_{2,3}=0,\quad h_{1,3}=\frac12,\quad h_{1,2}=\frac{1}{16},\quad
  \alpha_{2,3}=\frac{1}{\sqrt{3}},\quad \alpha_{1,3}=\sqrt{3},\quad 
  \alpha_{1,2}=\frac{\sqrt{3}}{2}.
\ean
Basic two singular vectors are
\ba
  h=0&:&
  \ket{\chi}=L_{-1}\ket{0},\\
  &&\ket{\chi'}=(L_{-6}+\sfrac{22}{9}L_{-4}L_{-2}-\sfrac{31}{36}L_{-3}^2
  -\sfrac{16}{27}L_{-2}^3)\ket{0}\n
  &&\qquad\quad+(b_1L_{-5}+b_2L_{-4}L_{-1}+b_3L_{-3}L_{-2}\\
  &&\qquad\qquad
  +b_4L_{-3}L_{-1}^2+b_5L_{-2}^2L_{-1}+b_6L_{-2}L_{-1}^3
  +b_7L_{-1}^5)\ket{\chi},\n
%
  h=\frac12&:&
  \ket{\chi}=(L_{-2}-\sfrac34L_{-1}^2)\ket{\sfrac12},\\
  &&\ket{\chi'}=(L_{-3}-\sfrac13L_{-1}^3)\ket{\sfrac12}
  +aL_{-1}\ket{\chi},\\
  h=\frac{1}{16}&:&
  \ket{\chi}=(L_{-2}-\sfrac43L_{-1}^2)\ket{\sfrac{1}{16}},\\
  &&\ket{\chi'}=(L_{-4}+\sfrac{40}{31}L_{-3}L_{-1}
  -\sfrac{256}{465}L_{-1}^4)\ket{\sfrac{1}{16}}
  +(bL_{-2}+b'L_{-1}^2)\ket{\chi},
\ea
where 
\ban
  &&b_1=-\sfrac{41}{225},\quad b_2=-\sfrac{78}{25},\quad
  b_3=\sfrac{14}{75},\quad b_4=\sfrac{34}{25},\quad
  b_5=\sfrac{172}{75},\quad b_6=-\sfrac85,\quad b_7=\sfrac{4}{25},\\
  &&a=-\sfrac45,\quad b=\sfrac{147}{3100},\quad b'=-\sfrac{401}{775}, 
\ean
are determined by the requirement that $\ket{\chi'}$ is orthogonal 
(with double zero) to descendants of $\ket{\chi}$. 

In the free field realization these singular vectors become as follows.
For $h=0$,
\ba
  \ket{0}=\ketb{\alpha_{1,1}}&:&
  \ket{\chi}=0,\\
  &&\ket{\chi'}=-\sfrac{1}{216\sqrt{3}}(a_{-6}+6\sqrt{3}a_{-5}a_{-1}
  -32\sqrt{3}a_{-4}a_{-2}-51a_{-4}a_{-1}^2\n
  &&\qquad\quad
  +\sfrac{51\sqrt{3}}{2}a_{-3}^2+75a_{-3}a_{-2}a_{-1}
  -18\sqrt{3}a_{-3}a_{-1}^3-\sfrac{97}{3}a_{-2}^3\\
  &&\qquad\quad 
  +\sfrac{31\sqrt{3}}{2}a_{-2}^2a_{-1}^2+6a_{-2}a_{-1}^4
  +2\sqrt{3}a_{-1}^6)\ketb{\alpha_{1,1}},\n
  \ket{0}=\ketb{\alpha_{2,3}}&:&
  \ket{\chi}=\sfrac{1}{2\sqrt{3}}a_{-1}\ketb{\alpha_{2,3}},\\
  &&\ket{\chi'}=0,
\ea
for $h=\frac12$,
\ba
  \ket{\sfrac12}=\ketb{\alpha_{2,1}}&:&
  \ket{\chi}=0,\\
  &&\ket{\chi'}=\sfrac{1}{6\sqrt{3}}(a_{-3}+\sqrt{3}a_{-2}a_{-1}
  +\sfrac23a_{-1}^3)\ketb{\alpha_{2,1}},\\
  \ket{\sfrac12}=\ketb{\alpha_{1,3}}&:&
  \ket{\chi}=\sfrac{5}{8\sqrt{3}}(a_{-2}-\sfrac{\sqrt{3}}{2}a_{-1}^2
  )\ketb{\alpha_{1,3}},\\
  &&\ket{\chi'}=0,
\ea
and for $h=\frac{1}{16}$,
\ba
  \ket{\sfrac{1}{16}}=\ketb{\alpha_{2,2}}&:&
  \ket{\chi}=\sfrac{1}{3\sqrt{3}}(a_{-2}+\sfrac{2}{\sqrt{3}}a_{-1}^2
  )\ketb{\alpha_{2,2}},\\
  &&\ket{\chi'}=0,\\
  \ket{\sfrac{1}{16}}=\ketb{\alpha_{1,2}}&:&
  \ket{\chi}=0,\\
  &&\ket{\chi'}=-\sfrac{\sqrt{3}}{310}(
  a_{-4}-8\sqrt{3}a_{-3}a_{-1}+\sfrac{37}{2\sqrt{3}}a_{-2}^2-2a_{-2}a_{-1}^2
  \\
  &&\qquad\quad
  +2\sqrt{3}a_{-1}^4)\ketb{\alpha_{1,3}}.\nonumber
\ea
On $\F_{\alpha}$, BRST charges are
\ba
  Q_1&\!\!=\!\!&\int_{C_{KM}}\frac{dz_1}{2\pi i}S_+(z_1)\n
  &=\!\!&\oint_0\frac{dz}{2\pi i}
  \exp\Bigl(\frac{2}{\sqrt{3}}\sum_{n>0}\frac{1}{n}a_{-n}z^n\Bigr)
  \exp\Bigl(-\frac{2}{\sqrt{3}}\sum_{n>0}\frac{1}{n}a_nz^{-n}\Bigr)
  e^{\frac{2}{\sqrt{3}}Q}z^{\frac{2}{\sqrt{3}}a_0},\\
  Q_2&\!\!=\!\!&\frac12(e^{2\pi i\alpha_+^2}+1)(e^{2\pi i\alpha_+\alpha}-1)
  \int_{C_{KM}}\frac{dz_1}{2\pi i}\frac{dz_2}{2\pi i}S_+(z_1)S_+(z_2)\n
  &=\!\!&\frac12(e^{2\pi i\frac13}+1)(e^{2\pi i\frac{2}{\sqrt{3}}\alpha}-1)
  \frac{1}{2\pi i}\oint_0\frac{dz}{2\pi iz}\int_0^1du\,
  z^{\frac{14}{3}}(1-u)^{\frac83}
  e^{\frac{4}{\sqrt{3}}Q}z^{\frac{4}{\sqrt{3}}a_0}
  u^{\frac{2}{\sqrt{3}}a_0},\n
  &&\quad\times
  \exp\Bigl(\frac{2}{\sqrt{3}}\sum_{n>0}\frac{1}{n}a_{-n}z^n(1+u^n)\Bigr)
  \exp\Bigl(-\frac{2}{\sqrt{3}}\sum_{n>0}\frac{1}{n}a_nz^{-n}(1+u^{-n})\Bigr),
\ea
where we have taken $I_{KM}$ contour (see appendix \ref{app:a.4}) and 
changed integration variables $z_1=z$, $z_2=zu$.
We give lower level examples. 
We perform $\oint_0dz$ first, and next use 
$\int_0^1duu^{a-1}(1-u)^{b-1}=B(a,b)$ for `arbitrary' values $a,b$ 
(analytic continuation).
For $h=\frac12$ and $\ket{\frac12}=\ketb{\alpha_{2,1}}$,
\ba
  &&L_{-2}\ketb{\alpha_{2,1}}=\sfrac34L_{-1}^2\ketb{\alpha_{2,1}}=
  -\sfrac{\sqrt{3}}{4}(a_{-2}-\sfrac{1}{\sqrt{3}}a_{-1}^2)
  \ketb{\alpha_{2,1}},\\
  &&Q_1\ketb{\alpha_{4,1}}=
  \sfrac{2}{3\sqrt{3}}(a_{-3}+\sqrt{3}a_{-2}a_{-1}
  +\sfrac23a_{-1}^3)\ketb{\alpha_{2,1}},\\
  &&Q_2(Aa_{-2}+Ba_{-1}^2)\ketb{\alpha_{2,1}}=-\sfrac{120}{7\pi}
  \sfrac{\Gamma(\frac23)^2}{\Gamma(\frac13)}
  (A+\sqrt{3}B)\ketb{\alpha_{-2,1}},\\
  &&Q_2(A'a_{-3}+B'a_{-2}a_{-1}+C'a_{-1}^3)\ketb{\alpha_{2,1}}\n
  &&\quad=-\sfrac{90\sqrt{3}}{7\pi}\sfrac{\Gamma(\frac23)^2}{\Gamma(\frac13)}
  (A'+\sfrac{1}{\sqrt{3}}B'-3C')a_{-1}\ketb{\alpha_{-2,1}},
\ea
therefore we have
\be
  Q_1\ketb{\alpha_{4,1}}=4\ket{\chi'},\quad
  Q_2L_{-2}\ketb{\alpha_{2,1}}=Q_2L_{-1}^2\ketb{\alpha_{2,1}}=0,\quad
  Q_2Q_1\ketb{\alpha_{4,1}}=0.
\ee
For the complex $C_{2,1}$, we have
\ba
  &&v_0^{(-1)}=\ketb{\alpha_{4,1}},\quad v_0^{(0)}=\ketb{\alpha_{2,1}},\quad
  v_0^{(1)}=\ketb{\alpha_{-2,1}},\\
  &&u_1^{(0)}=Q_1v_0^{(-1)}=\sfrac{2}{3\sqrt{3}}
  (a_{-3}+\sqrt{3}a_{-2}a_{-1}+\sfrac23a_{-1}^3)
  \ketb{\alpha_{2,1}},\\
  &&Q_2w_0^{(0)}=v_0^{(1)},\quad
  w_0^{(0)}=-\sfrac{7\pi}{300}\sfrac{\Gamma(\frac13)}{\Gamma(\frac23)^2}
  (a_{-2}+\sfrac{\sqrt{3}}{2}a_{-1}^2)\ketb{\alpha_{2,1}}.
\ea
Remark that the last equation is the dual of the following state,
\be
  Q_2\ketb{\alpha_{2,-1}}=\sfrac{30}{7\pi}
  \sfrac{\Gamma(\frac23)^2}{\Gamma(\frac13)}
  (a_{-2}-\sfrac{\sqrt{3}}{2}a_{-1}^2)\ketb{\alpha_{-2,-1}}.
\ee

For more examples and explicit expressions of the Virasoro singular vectors,
see \cite{CFT,KY92,CDK92}.

\subsubsection{Calogero-Sutherland model and Jack symmetric polynomial}
\label{sec:2.3.4}

In this subsection we review the relation between the Virasoro singular
vector and the Calogero-Sutherland model (CSM) \cite{MY,AMOS}. 
Calogero-Sutherland model is a many body quantum mechanical system on 
a circle with length $L$ under the $1/r^2$ potential. 
Its Hamiltonian and momentum are
\ba
  H_{CS}&\!\!=\!\!&\sum_{j=1}^{N_0}\frac{1}{2m}\hat{p}_j^2
  +\frac{\bar{\beta}(\bar{\beta}-\hbar)}{m}\Bigl(\frac{\pi}{L}\Bigr)^2
  \sum_{1\leq i<j\leq N_0}\frac{1}{\sin^2\frac{\pi}{L}(q_i-q_j)},
  \label{CS}\\
  P_{CS}&\!\!=\!\!&\sum_{j=1}^{N_0}\hat{p}_j,\quad  
  \hat{p}_j=\frac{\hbar}{i}\frac{\partial}{\partial q_j},
\ea
where $q_j$ is a coordinate of $j$-th particle, $p_j$ is its momentum,
$N_0$ is a number of particles,  
$m$ is a mass of particles, $\bar{\beta}$ is a (dimensionful) 
coupling constant and $\hbar$ is a Planck constant.
Since this Hamiltonian can be rewritten as
\ba
  H_{CS}&\!\!=\!\!&\sum_j\frac{1}{2m}\Pi_j^{\dagger}\Pi_j
  +\frac{\bar{\beta}^2}{m}\Bigl(\frac{\pi}{L}\Bigr)^2\frac{N_0^3-N_0}{6},\\
  \Pi_j&\!\!=\!\!&\hat{p}_j+i\bar{\beta}\frac{\pi}{L}
  \sum_{k\neq j}\cot\frac{\pi}{L}(q_j-q_k),
\ea
the ground state $\Psi_0$ is determined by $\Pi_j\Psi_0=0$,
\be
  \Psi_0={\cal N}\Biggl(\prod_{i<j}\sin\frac{\pi}{L}(q_i-q_j)\Biggr)^{\beta}
  \propto\Biggl(\prod_{i<j}\sqrt{\frac{x_i}{x_j}}\Bigl(1-\frac{x_j}{x_i}\Bigr)
  \Biggr)^{\beta},
\ee
where ${\cal N}$ is a normalization constant, and $\beta$ and $x_j$ are
dimensionless quantities,
\be
  \beta=\frac{\bar{\beta}}{\hbar},\quad x_j=e^{2\pi i\frac{q_j}{L}}.
  \label{betaxj}
\ee
Under the interchange of particle, this ground state gives a phase 
$(-1)^{\beta}$ which means that this system obeys fractional statistics.
Excited states that we seek for have the form $\Psi=\psi\Psi_0$, where 
$\psi$ is a symmetric function of $x_j$ because $\Psi$ has same statistical
property as $\Psi_0$.
By removing the contribution from the ground state,
\ba
  \Psi_0^{-1}\circ H_{CS}\circ\Psi_0&\!\!=\!\!&
  \frac{1}{2m}\Bigl(\frac{2\pi\hbar}{L}\Bigr)^2\Bigl(
  H_{\beta}+\sfrac{1}{12}\beta^2(N_0^3-N_0)\Bigr),\\
  \Psi_0^{-1}\circ P_{CS}\circ\Psi_0&\!\!=\!\!&
  \frac{2\pi\hbar}{L}P,
\ea
Hamiltonian and momentum that act directly $\psi$ are
\ba
  H_{\beta}&\!\!=\!\!&\sum_{i=1}^{N_0}D_i^2+\beta\sum_{1\leq i<j\leq N_0}
  \frac{x_i+x_j}{x_i-x_j}(D_i-D_j),
  \label{Hbeta}\\
  P&\!\!=\!\!&\sum_{i=1}^{N_0}D_i,\quad D_i=x_i\frac{\partial}{\partial x_i}.
\ea

Here we recall basic definitions of symmetric polynomials \cite{Mac}:
\ba
  \!\!\!\!\!\!\!\!\!\!\!\!\!\!\!\!\!\!\!\!&&\mbox{partition: }
  \lambda=(\lambda_1,\lambda_2,\cdots)=1^{m_1}2^{m_2}\cdots,\quad
  \lambda_1\geq\lambda_2\geq\cdots\geq 0,\quad m_i\geq 0,\n
  \!\!\!\!\!\!\!\!\!\!\!\!\!\!\!\!\!\!\!\!&&\qquad\qquad
  |\lambda|=\sum_i\lambda_i=\sum_iim_i,\quad
  \ell(\lambda)=\mathop{\max}_i\{\lambda_i>0\},\\
  \!\!\!\!\!\!\!\!\!\!\!\!\!\!\!\!\!\!\!\!&&
  \mbox{dominance (partial)ordering: }
  \lambda\geq\mu\n
  \!\!\!\!\!\!\!\!\!\!\!\!\!\!\!\!\!\!\!\!&&\qquad\qquad
  \Leftrightarrow|\lambda|=|\mu|,\quad
  \lambda_1+\cdots+\lambda_i=\mu_1+\cdots+\mu_i\quad(\forall i),\\
  \!\!\!\!\!\!\!\!\!\!\!\!\!\!\!\!\!\!\!\!&&
  \mbox{monomial symmetric function: }m_{\lambda}=\sum_{\alpha}x^{\alpha},
  \quad x^{\alpha}=x_1^{\alpha_1}x_2^{\alpha_2}\cdots,\n
  \!\!\!\!\!\!\!\!\!\!\!\!\!\!\!\!\!\!\!\!&&\qquad\qquad
  \alpha:\mbox{all distinct permutation of $\lambda=(\lambda_1,\lambda_2,
  \cdots)$},\\
  \!\!\!\!\!\!\!\!\!\!\!\!\!\!\!\!\!\!\!\!&&
  \mbox{power sum symmetric function: }p_{\lambda}=p_{\lambda_1}p_{\lambda_2}
  \cdots p_{\lambda_{\ell(\lambda)}},\quad p_n=\sum_ix_i^n,\\
  \!\!\!\!\!\!\!\!\!\!\!\!\!\!\!\!\!\!\!\!&&
  \mbox{inner product: }\langle p_{\lambda},p_{\mu}\rangle_{\beta}=
  \delta_{\lambda,\mu}\prod_ii^{m_i}m_i!\cdot\beta^{-\sum_im_i},
\ea
The Jack symmetric polynomial $J_{\lambda}=J_{\lambda}(x;\beta)$ is
uniquely determined by the following two conditions \cite{St,Mac},
\ba
  &\mbox{(i)}&J_{\lambda}(x)=\sum_{\mu\leq\lambda}u_{\lambda,\mu}
  m_{\mu}(x),\quad u_{\lambda,\lambda}=1,\\
  &\mbox{(ii)}&\langle J_{\lambda},J_{\mu}\rangle_{\beta}=0\quad
  \mbox{if $\lambda\neq\mu$}.
\ea
The condition (ii) can be replaced by (ii)',
\be
  \mbox{(ii)'}\quad H_{\beta}J_{\lambda}=\ve_{\beta,\lambda}J_{\lambda},\quad
  \ve_{\beta,\lambda}=\sum_{i=1}^{N_0}\Bigl(\lambda_i^2
  +\beta(N_0+1-2i)\lambda_i\Bigr).
\ee
Therefore excited states of Calogero-Sutherland model are described by
the Jack symmetric polynomials. Using properties of the Jack symmetric
polynomial, some dynamical correlation functions were calculated \cite{Ha}.

There exists another inner product,
\ba
  &&\langle f,g\rangle'_{l,\beta}=
  \frac{1}{l!}\oint\prod_{j=1}^l\underline{dx}_j\cdot
  \bar{\Delta}(x)f(\bar{x})g(x),\\
  &&\bar{\Delta}(x)=\prod_{i\neq j}\Bigl(1-\frac{x_j}{x_i}\Bigr)^{\beta},
  \quad f(\bar{x})=f\Bigl(\frac{1}{x_1},\frac{1}{x_2},\cdots\Bigr),\quad
  \underline{dx}_j=\frac{dx_j}{2\pi ix_j}.
\ea
This is a usual inner product in quantum mechanics. These two inner products
are proportional,
\be
  \langle~,~\rangle'_{l,\beta}\propto\langle~,~\rangle_{\beta}.
\ee
Let us introduce two transformations for symmetric polynomials,
\ba
  {\cal G}_k&:&\Bigl({\cal G}_kf\Bigr)(x_1,\cdots,x_l)
  =\prod_{i=1}^lx^k\cdot f(x_1,\cdots,x_l),\\
  {\cal N}_{l',l}&:&\Bigl({\cal N}_{l',l}f\Bigr)(x'_1,\cdots,x'_{l'})
  =\oint\prod_{j=1}^l\underline{dx}_j\cdot\bar{\Pi}(x',\bar{x})
  \bar{\Delta}(x)f(x_1,\cdots,x_l),\\
  &&\bar{\Pi}(x,y)=\prod_i\prod_j(1-x_iy_j)^{-\beta}.
\ea
Then the Jack symmetric polynomial satisfies the following two properties,
\ba
  &&J_{(k^l)+\lambda}={\cal G}_kJ_{\lambda}\quad
  (\mbox{$l$ variables}),\\
  &&J_{\lambda}=\frac{\langle J_{\lambda},J_{\lambda}\rangle_{\beta}}
  {l!\langle J_{\lambda},J_{\lambda}\rangle'_{l,\beta}}
  {\cal N}_{l',l}J_{\lambda}\quad
  {\mbox{($J_{\lambda}$ in LHS : $l'$ variables)} \atop
  \mbox{($J_{\lambda}$ in RHS : $l$ variables)}}.
\ea
By successive action of these two transformations to $J_{\phi}(x)=1$,
an integral representation of the Jack symmetric polynomial can
be obtained \cite{MY,AMOS},
\be
  J_{\lambda}(x)\propto{\cal N}_{l,l_1}{\cal G}_{k_1}
  {\cal N}_{l_1,l_2}{\cal G}_{k_2}\cdots
  {\cal N}_{l_{N-2},l_{N-1}}{\cal G}_{k_{N-1}}{\cal N}_{l_{N-1},0}\cdot 1,
\ee
where the partition $\lambda$ is 
$\lambda'=((l_1)^{k_1},(l_2)^{k_2},\cdots,(l_{N-1})^{k_{N-1}})$ ,
namely corresponds to the following Young diagram,
\def\generalYoung{
\vskip.25cm
\noindent
\makebox[  4cm]{ }
\makebox[  2cm]{$k_1$}\hskip-.4pt
\makebox[1.7cm]{$k_2$}
\makebox[1.4cm]{$\!\cdots\cdot$}
\makebox[1.4cm]{$\!\!k_{N-2}$}\hskip-.5pt
\makebox[1.3cm]{$\!\!k_{N-1}$}
\hfill\break
 \makebox[  4cm][r]{$\hfill\lambda\:=\;$}
\framebox[  2cm][l]{\rule[  -1cm]{0cm}{  2cm}
                    \raisebox{.2cm}{$\!l_1$}}\hskip-.4pt
\framebox[1.7cm][l]{\rule[-0.7cm]{0cm}{1.7cm}
                    \raisebox{.2cm}{$\!l_2$}}
\hskip-0.15cm\rule[1.105cm]{1.6cm}{0.4pt}\hskip-1.55cm
 \makebox[1.4cm]   {\raisebox{.2cm}{$\,\,\cdots\cdot$}}
\framebox[1.4cm][l]{\rule[-0.3cm]{0cm}{1.3cm}
                    \raisebox{.2cm}{$\!l_{N-2}$}}\hskip-.4pt
\framebox[1.3cm][l]{\rule[0.0cm]{0cm}{1.0cm}
                    \raisebox{.2cm}{$\!l_{N-1}$}}
\makebox[0.2cm][r]{.}
\vskip.4cm
}
\generalYoung

\noindent
For example, the Jack symmetric polynomial with a rectangular Young 
diagram $(k^l)$ is
\be
  J_{(k^l)}(x)\propto
  \oint\prod_{j=1}^l\dz_j\cdot
  \prod_{j=1}^l\exp\Bigl(\beta\sum_{n>0}\frac{1}{n}z_j^n\sum_ix_i^n\Bigr)\cdot
  \prod_{i\neq j}\Bigl(1-\frac{z_j}{z_i}\Bigr)^{\beta}\cdot
  \prod_{j=1}^lz_j^{-k}.
  \label{Jrec}
\ee

We will show the relation between the Jack symmetric polynomials and
the Virasoro singular vectors \cite{AMOS}. 
States in the Fock space and symmetric polynomials have one-to-one 
correspondence,
\ba
  \F_{\alpha}&\longrightarrow&\{\mbox{symmetric function}\}\n
  \ket{f}&\mapsto&f(x)
  =\bra{\alpha}\exp\Bigl(\frac12\sqrt{\beta}
  \sum_{n>0}\frac{1}{n}a_np_n\Bigr)\ket{f}. 
  \label{Focktosym}
\ea
Namely oscillator $a_n$ and power sum $p_n$ correspond as, 
\be
  \frac{a_{-n}}{\sqrt{\beta}}\leftrightarrow p_n,\quad
  \frac{a_n}{\sqrt{\beta}}\leftrightarrow\frac{2n}{\beta}
  \frac{\partial}{\partial p_n}.
\ee
Using this correspondence, Hamiltonian $H_{\beta}$ \eq{Hbeta}, which is a
differential operator with respect to $x_j$ (or $p_n$), is bosonized,
\be
  H_{\beta}\bra{\alpha}\exp\Bigl(\frac12\sqrt{\beta}
  \sum_{n>0}\frac{1}{n}a_np_n\Bigr)
  =\bra{\alpha}\exp\Bigl(\frac12\sqrt{\beta}
  \sum_{n>0}\frac{1}{n}a_np_n\Bigr)\hat{H}_{\beta}.
\ee
Bosonized Hamiltonian $\hat{H}_{\beta}$ is an operator on $\F_{\alpha}$.
It is cubic in $a_n$ and can be rewritten by using the Virasoro generator 
$L_n$ \eq{Ln},
\ba
  \hat{H}_{\beta}&\!\!=\!\!&\sfrac14\sqrt{\beta}
  \sum_{n,m>0}(a_{-n-m}a_na_m+2a_{-n}a_{-m}a_{n+m})
  +\sfrac12\sum_{n>0}a_{-n}a_n\Bigl((1-\beta)n+N_0\beta\Bigr)\n
  &=\!\!&\sqrt{\beta}\sum_{n>0}a_{-n}L_n+\sfrac12\sum_{n>0}a_{-n}a_n
  \Bigl(N_0\beta+\beta-1-\sqrt{\beta}a_0\Bigr).
\ea
When $\hat{H}_{\beta}$ acts on the singular vector \eq{sv}, 
the first term of $\hat{H}_{\beta}$ vanishes because of the property of 
singular vector, and the second term is already diagonal,
\be
  \hat{H}_{\beta}\ket{\chi_{-l,-k}}
  =lk\Bigl(N_0\beta+\beta-1-\sqrt{\beta}\alpha_{-l,-k}\Bigr)\ket{\chi_{-l,-k}}
  =\ve_{\beta,(k^l)}\ket{\chi_{-l,-k}}.
\ee
Therefore $\ket{\chi_{-l,-k}}$ is an eigenstate of $\hat{H}_{\beta}$, 
i.e., it gives the Jack symmetric polynomial by the map \eq{Focktosym}.
In fact the polynomial obtained from \eq{sv} by \eq{Focktosym} agrees
with the integral representation of the Jack symmetric polynomial with
the partition $(k^l)$, \eq{Jrec}.

The Jack symmetric polynomial with general partition whose Young diagram
is composed of $N-1$ rectangles is related to the singular vector of 
$W_N$ algebra \cite{AMOS}.

\setcounter{section}{2}
\setcounter{equation}{0}
\section{Deformed Virasoro Algebra ($A_1^{(1)}$ type)} \label{sec:3}

\subsection{Definition and consistency}\label{sec:3.1}

\noindent{\bf Definition}\quad
Deformed Virasoro algebra(DVA) ($A_1^{(1)}$ type) is an associative algebra 
over $\C$ generated by $T_n$ ($n\in\Z$) with two parameters $x$ and $r$, 
and their relation is \cite{SKAO} (see \eq{qtpxr} for correspondence of 
parameters)
\be
  [T_n,T_m]=-\sum_{\ell=1}^{\infty}f_{\ell}
  (T_{n-\ell}T_{m+\ell}-T_{m-\ell}T_{n+\ell})
  -(x-x^{-1})^2[r]_x[r-1]_x[2n]_x\delta_{n+m,0},
  \label{DVA}
\ee
where the structure constants $f_{\ell}$ is given by
\ba
  f(z)&\!\!=\!\!&\sum_{\ell=0}^{\infty}f_{\ell}z^{\ell}
  =\exp\Biggl(-\sum_{n>0}\frac{z^n}{n}
  \frac{(x^{rn}-x^{-rn})(x^{(r-1)n}-x^{-(r-1)n})}{x^n+x^{-n}}\Biggr)
  \label{f}\\
  &=\!\!&\frac{1}{1-z}\frac{(x^{2r}z,x^{-2(r-1)}z;x^4)_{\infty}}
  {(x^{2+2r}z,x^{2-2(r-1)}z;x^4)_{\infty}}.\nonumber
\ea
Here we have used the notation \eq{[n]_x}, \eq{pinf}. 
By introducing DVA current $\ds T(z)=\sum_{n\in\Z}T_nz^{-n}$, 
the above relation can be written as a formal power series,
\be
  f(\sfrac{w}{z})T(z)T(w)-T(w)T(z)f(\sfrac{z}{w})=
  -(x-x^{-1})[r]_x[r-1]_x
  \Bigl(\delta(x^2\sfrac{w}{z})-\delta(x^{-2}\sfrac{w}{z})\Bigr).
  \label{DVAfTT}
\ee
For later use we add a grading operator $d$,
\be
  [d,T_n]=-nT_n.
\ee

The above relation \eq{DVA} is invariant under 
\be
  T_n\mapsto-T_n.
  \label{T-T}
\ee
It is also invariant under the following two transformations:
\ba
  \mbox{(i)}&\theta:&x\mapsto x^{-1},\quad r\mapsto r,
  \label{theta}\\
  \mbox{(ii)}&\omega:&x\mapsto x,\quad r\mapsto 1-r.
\ea
In the case of (i) $f(z)$ is understood as the first line of \eq{f}. 
Let us introduce $\beta$ as
\be
  \beta=\frac{r}{r-1},
  \label{betar}
\ee
then 
\be
  \theta\cdot\beta=\beta,\quad\omega\cdot\beta=\beta^{-1}.
\ee
$\az$ in \eq{az} is
\be
  \az=\frac{1}{\sqrt{r(r-1)}}.
\ee

\medskip

\noindent{\bf Consistency}\quad
In subsection \ref{sec:2.2.1} the central term of the Virasoro algebra
is determined by the Jacobi identity. Here we will show that the structure
function $f(z)$ is determined by associativity \cite{AKOS3}.

Let us consider the following relation,
\be
  f(\sfrac{w}{z})T(z)T(w)-T(w)T(z)f(\sfrac{z}{w})=
  c_0\Bigl(\delta(x^2\sfrac{w}{z})-\delta(x^{-2}\sfrac{w}{z})\Bigr),
\ee
where $c_0$ is a normalization constant and $x$ is a parameter, and
$f(z)$ is an unknown Taylor series 
$\ds f(z)=\sum_{\ell=0}^{\infty}f_{\ell}z^{\ell}$.
By using this relation, 
$f(\frac{z_2}{z_1})f(\frac{z_3}{z_1})f(\frac{z_3}{z_2})T(z_1)T(z_2)T(z_3)$ 
is related to 
$f(\frac{z_1}{z_2})f(\frac{z_1}{z_3})f(\frac{z_2}{z_3})T(z_3)T(z_2)T(z_1)$ 
in two ways,
$$
  \begin{array}{ccccc}
  (123)&\rightarrow&(132)&\rightarrow&(312)\\
  \downarrow&&&&\downarrow\\
  (213)&\rightarrow&(231)&\rightarrow&(321)
  \end{array}.
$$
These two results should agree. 
So we obtain an equation containing delta functions,
\be
  c_0T(z_1)
  \Bigl(\delta(x^2\sfrac{z_3}{z_2})-\delta(x^{-2}\sfrac{z_3}{z_2})\Bigr)
  \Bigl(f(\sfrac{z_2}{z_1})f(\sfrac{z_3}{z_1})
  -f(\sfrac{z_1}{z_2})f(\sfrac{z_1}{z_3})\Bigr)+\mbox{cyclic}=0.
\ee
This is equivalent to
\be
  c_0T(z_1)
  \Bigl(\delta(x^2\sfrac{z_3}{z_2})g(\sfrac{z_2}{z_1})
  -\delta(x^{-2}\sfrac{z_3}{z_2})g(\sfrac{z_3}{z_1})\Bigr)
  +\mbox{cyclic}=0,
\ee
where $g(z)$ is 
\be
  g(z)=f(z)f(x^{-2}z)-f(z^{-1})f(x^2z^{-1})=-g(x^2z^{-1}).
  \label{defg}
\ee
In mode expansion $\ds g(z)=\sum_{n\in\Z}g_nz^n$,
this equation becomes
\ba
  &&(x^{2n}-x^{2m})g_{n+m}+(x^{2m}-x^{2l})g_{m+l}+(x^{2l}-x^{2n})g_{l+n}=0,
  \label{(x-x)g}\\
  &&g_{-n}=-x^{2n}g_n.
  \label{g-n}
\ea
{}From \eq{g-n}, we have $g_0=0$. 
{}From \eq{(x-x)g} with $(m,l)=(n-1,1-n)$, $(m,l)=(n-2,2-n)$, we have
$g_{2n-1}=\frac{1-x^{-2(2n-1)}}{1-x^{-2}}g_1$ and 
$g_{2n-2}=\frac{1-x^{-2(2n-2)}}{1-x^{-4}}g_2$.
And from \eq{(x-x)g} with $(m,l)=(n-1,n+1)$ we have $g_2=(1+x^{-2})g_1$.
Combining these we get $g_n=(1-x^{-2n})g_1$ and this satisfies both 
\eq{(x-x)g} and \eq{g-n}.
Therefore the solution is 
\be
  g_n=c'_0(1-x^{-2n}),\quad
  g(z)=c'_0\Bigl(\delta(z)-\delta(x^{-2}z)\Bigr),
  \label{solg}
\ee
where $c'_0$ is a constant.
By setting $F(z)=f(z)f(x^2z)$, \eq{defg} and \eq{solg} become
\be
  F(z)-F(x^{-2}z^{-1})=-c'_0\sum_{n\in\Z}(1-x^{2n})z^n.
\ee
Since $F(z)$ is also a Taylor series, this equation implies
\be
  F(z)=-c'_0\Bigl(\alpha+\sum_{n>0}(1-x^{2n})z^n\Bigr),
\ee
where a new parameter $\alpha$ has appeared. 
If we express this $\alpha$ by a new parameter $r$ as
\be
  \alpha=\frac{x-x^{-1}}{(x^r-x^{-r})(x^{r-1}-x^{-(r-1)})},
\ee
then $F(z)$ becomes
\be
  F(z)=-c'_0\alpha\frac{(1-x^{2r}z)(1-x^{-2(r-1)})}{(1-z)(1-x^2z)}.
\ee
{}From this equation $f(z)$ is calculated as ($|x|<1$)
\ba
  f(z)&\!\!=\!\!&\frac{F(z)}{F(x^2z)}f(x^4z)=\frac{F(z)}{F(x^2z)}
  \frac{F(x^4z)}{F(x^6z)}\cdots\times f(0)\n
  &=\!\!&f(0)\exp\Bigl(\sum_{n>0}\frac{z^n}{n}
  \frac{(1-x^{2rn})(1-x^{-2(r-1)n})}{1+x^{2n}}\Bigr).
\ea
We choose the normalization of $f(z)$ as $f(0)=1$ ($c'_0=-\alpha^{-1}$),
then this $f(z)$ is just \eq{f}.
We fix the normalization of $T(z)$ by $c_0=-\alpha^{-1}$.

\subsection{Conformal limit}\label{sec:3.2}

Quantum group(algebra) $U_q(\g)$ is a deformation of $U(\g)$ and
it reduces to undeformed one in the $q\rightarrow 1$ limit,
$$
  U_q(\g)\stackrel{q\rightarrow 1}{\verylongrightarrow}U(\g).
$$
The deformed Virasoro algebra defined in the previous subsection
is a deformation of the Virasoro algebra as expected from its name. 
Then in what limit it reduces to the usual Virasoro algebra?
$$
  \mbox{DVA}\stackrel{??}{\verylongrightarrow}\mbox{Vir}.
$$
Since DVA contains two parameters $x$ and $r$
(or $\beta$ \eq{betar}), it admits various limits\cite{AKOS3}.
In this subsection we will show that DVA reduces to the Virasoro 
algebra in the conformal limit.

The conformal limit is
\be
  x\rightarrow 1,\quad \mbox{$r$ (or $\beta$) : fixed}.
  \label{conflim}
\ee
To study this limit we write $x$ as
\be 
  x=e^{-\frac12\az\hbar},
\ee
where $\az$ is given in \eq{az} and $\hbar$ is a fictitious Plank constant,
and take $\hbar\rightarrow 0$.
$f(z)$ has the expansion,
\be
  f(z)=1+\hbar^2f^{[2]}(z)+\hbar^4f^{[4]}(z)+\cdots.
\ee
Taking into account the invariance under \eq{theta}, we assume the expansion
\be
  T(z)=2+\hbar^2T^{[2]}(z)+\hbar^4T^{[4]}(z)+\hbar^6T^{[6]}(z)
  +\cdots,
  \label{Texp}
\ee
(Remark that $-T(z)$ is also a solution.) and we set 
\be
  T^{[2]}(z)=z^2L(z)+\sfrac14\az^2.
  \label{T2}
\ee
Then LHS of \eq{DVAfTT} is
\ba
  &&\hbar^24\Bigr(f^{[2]}(\zeta)-f^{[2]}(\zeta^{-1})\Bigr)
  +\hbar^4\Biggl(\Bigl[T^{[2]}(z),T^{[2]}(w)\Bigr]\n
  &&\quad
  +2\Bigl(f^{[2]}(\zeta)-f^{[2]}(\zeta^{-1})\Bigr)
  \Bigl(T^{[2]}(z)+T^{[2]}(w)\Bigr)
  +4\Bigl(f^{[4]}(\zeta)-f^{[4]}(\zeta^{-1})\Bigr)\Biggr)+\cdots\n
  &=\!\!&\hbar^2\Bigr(-2\zeta\delta'(\zeta)\Bigr)
  +\hbar^4\Biggl(\Bigl[z^2L(z),w^2L(w)\Bigr]\n
  &&\quad
  -\zeta\delta'(\zeta)\Bigl(z^2L(z)+w^2L(w)\Bigr)
  -\zeta\delta'(\zeta)\frac{1+2\az^2}{6}
  -\zeta^2\delta'''(\zeta)\frac{1-2\az^2}{12}\Biggr)+\cdots,
\ea
where $\zeta=\frac{w}{z}$. 
On the other hand RHS of \eq{DVAfTT} is
\be
  \hbar^2\Bigr(-2\zeta\delta'(\zeta)\Bigr)
  +\hbar^4\Biggl(-\zeta\delta'(\zeta)\frac{1+2\az^2}{6}
  -\zeta^2\delta'''(\zeta)\frac{\az^2}{3}\Biggr)+\cdots.
\ee
Comparing these equations and \eq{Vir2} shows that $L(z)$ in \eq{T2} is 
the Virasoro current with the central charge \eq{cbeta}.

Therefore the Virasoro current $L(z)$ is found in $\hbar^2$ term of the DVA 
current $T(z)$. And also (conformal)spin $4$ current $T^{[4]}(z)$, 
spin $6$ current $T^{[6]}(z)$, $\cdots$ exist in the DVA current.
In massless case only one Virasoro current $L(z)$ controls 
(chiral part of)CFT. In massive case, however, $L(z)$ and infinitely many
higher spin currents $T^{[2n]}(z)$ are needed to control massive theory, 
and they gather and form the DVA current $T(z)$. 
DVA current is a `dressed' Virasoro current.

\subsection{Representation theory}\label{sec:3.3}

Let us consider the highest weight representation. The highest weight
state $\ket{\lambda}$ ($\lambda\in\C$) is characterized by
\be
  T_n\ket{\lambda}=0\quad(n>0),\qquad T_0\ket{\lambda}=\lambda\ket{\lambda},
\ee
and the Verma module is 
\be
  M=\bigoplus_{l\geq 0}\bigoplus_{n_1\geq\cdots\geq n_l>0}\C
  T_{-n_1}\cdots T_{-n_l}\ket{\lambda}.
\ee
$M$ is a graded module with grading $d$,
$$
  d\ket{\lambda}=d_{\lambda}\ket{\lambda}\; (d_{\lambda}\in\C),\quad
  d\cdot T_{-n_1}\cdots T_{-n_l}\ket{\lambda}
  =\Bigl(d_{\lambda}+\sum_{i=1}^ln_l\Bigr)
  T_{-n_1}\cdots T_{-n_l}\ket{\lambda},
$$ 
and we call $\sum_{i=1}^ln_l$ as a level.

To obtain the irreducible module from the Verma module we have to quotient
out invariant submodules, which are generated by singular vectors.  
Singular vector at level $N$, $\ket{\chi}$, is 
\be
  T_n\ket{\chi}=0\quad(n>0),\quad T_0\ket{\chi}=\lambda_{\chi}\ket{\chi},
  \quad d\ket{\chi}=(d_{\lambda}+N)\ket{\chi},
\ee
where $\lambda_{\chi}$ is some eigenvalue.
Existence of singular vectors is detected by zeros of the 
Kac determinant. To define the Kac determinant let us introduce the dual
module $M^*$ on which the DVA act as $T_n^{\dagger}=T_{-n}$. 
$M^*$ is generated by $\bra{\lambda}$ which satisfies 
$\bra{\lambda}T_n=0$ ($n<0$), $\bra{\lambda}T_0=\lambda\bra{\lambda}$ 
and $\bra{\lambda}\lambda\rangle=1$.
At level $N$ there are $p(N)$ states
$\ket{\lambda;N,1}=T_{-N}\ket{\lambda}$, 
$\ket{\lambda;N,2}=T_{-N+1}T_{-1}\ket{\lambda}$, $\cdots$, 
$\ket{\lambda;N,p(N)}=T_{-1}^N\ket{\lambda}$ in $M$, and
$\bra{\lambda;N,1}=\bra{\lambda}T_N$, 
$\bra{\lambda;N,2}=\bra{\lambda}T_1T_{N-1}$, $\cdots$,  
$\bra{\lambda;N,p(N)}=\bra{\lambda}T_1^N$ in $M^*$.  

The Kac determinant at level $N$ is given by \cite{SKAO,BP97}
\be
  \det\langle\lambda;N,i|\lambda;N,j\rangle
  =\prod_{l,k\geq 1\atop lk\leq N}\Biggl(
  \frac{(x^{rl}-x^{-rl})(x^{(r-1)l}-x^{-(r-1)l})}{x^l+x^{-l}}
  (\lambda^2-\lambda_{l,k}^2)\Biggr)^{p(N-lk)},
  \label{DVAKacdet}
\ee
where $\lambda_{l,k}$ is
\be
  \lambda_{l,k}=x^{rl-(r-1)k}+x^{-rl+(r-1)k}.
\ee
$\lambda$ dependence appears through $\lambda^2$ because of the symmetry 
\eq{T-T}.
Conformal limit \eq{conflim} of this determinant is
\ba
  &&\prod_{l,k\geq 1\atop lk\leq N}\Bigl(
  \frac{\hbar^2l^2}{2}\cdot\hbar^2(h-h_{l,k})\cdot 4\Bigr)^{p(N-lk)}
  +\cdots,\n
  &=\!\!&\hbar^A
  \prod_{l,k\geq 1\atop lk\leq N}\Bigl(2lk(h-h_{l,k})\Bigr)^{p(N-lk)}
  +\cdots,\quad A=4\sum_{l,k\geq 1\atop lk\leq N}p(N-lk). 
\ea
This is just the Kac determinant of the Virasoro algebra \eq{Kacdet},
and the order of $\hbar$ is consistent with \eq{Texp} (use \eq{Fk}).

For generic values of $x$ and $r$, this Kac determinant \eq{DVAKacdet} 
has essentially the same structure as the Virasoro one \eq{Kacdet}. 
Therefore embedding pattern is same as \eq{embed}, and the character 
which counts the degeneracy at each level ($\tr\,q^d$) is also the same.
For special value of $x$ and $r$, for example $x^r$ is a root of unity, 
the Kac determinant has more zeros. In this case we need special study,
see \cite{BP97}.

\subsection{Free field realization}\label{sec:3.4}

\subsubsection{free field realization}\label{sec:3.4.1}

Let us introduce free boson oscillator $h_n$ ($n\in\Z_{\neq 0}$),
\be
  [h_n,h_m]=(x-x^{-1})^2\frac{1}{n}
  \frac{[n]_x[rn]_x[(r-1)n]_x}{[2n]_x}\delta_{n+m,0},
  \label{hn}
\ee
and use zero mode $a_0$ and $Q$ defined in \eq{an} (or $a'_0$ in \eq{a'0}).
The Fock space $\F_{\alpha}$ is defined by
\be
  \F_{\alpha}=\bigoplus_{l\geq 0}\bigoplus_{n_1\geq\cdots\geq n_l>0}\C
  h_{-n_1}\cdots h_{-n_l}\ketb{\alpha},
\ee
where $\ketb{\alpha}$ is given by \eq{ketalpha} with replacing $a_n$ by $h_n$.

The DVA current $T(z)$ is realized as follows:
\ba
  &&T(z)=\Lambda_+(z)+\Lambda_-(z),\n
  &&\Lambda_{\pm}(z)=\,
  :\exp\Bigl(\pm\sum_{n\neq 0}h_n(x^{\pm 1}z)^{-n}\Bigr):
  \times x^{\pm\sqrt{r(r-1)}a'_0}.
  \label{Tfr}
\ea
To prove this we need the OPE formula,
\ba
  f(\sfrac{w}{z})\Lambda_{\pm}(z)\Lambda_{\pm}(w)&\!\!=\!\!&
  :\Lambda_{\pm}(z)\Lambda_{\pm}(w):,\n
  f(\sfrac{w}{z})\Lambda_{\pm}(z)\Lambda_{\mp}(w)&\!\!=\!\!&
  :\Lambda_{\pm}(z)\Lambda_{\mp}(w):\gamma(x^{\mp 1}\sfrac{w}{z}),
\ea
and the relation of $\Lambda_{\pm}$, 
\be
  :\Lambda_+(x^{-1}z)\Lambda_-(xz):\,=1.
\ee
Here $\gamma(z)$ is
\be
  \gamma(z)=\frac{(1-x^{2r-1}z)(1-x^{-(2r-1)}z)}{(1-xz)(1-x^{-1}z)},
  \label{gamma}
\ee
and from \eq{deltaformulaex} we have
\be
  \gamma(z)-\gamma(z^{-1})=-(x-x^{-1})[r]_x[r-1]_x\
  \Bigl(\delta(xz)-\delta(x^{-1}z)\Bigr).
\ee
The grading operator $d$ is realized by
\ba
  &&d=d^{\rm osc}+d^{\rm zero},\n
  &&d^{\rm osc}=\sum_{n>0}\frac{n^2[2n]_x}{(x-x^{-1})^2[n]_x[rn]_x[(r-1)n]_x}
  h_{-n}h_n,\quad d^{\rm zero}=\frac14a_0^{\prime\,2}-\frac{1}{24},
  \label{d}
\ea
which satisfies 
\be
  [d,h_n]=-nh_n,\quad [d,Q]=a'_0,\quad
  d\ketb{\alpha_{l,k}}=(h_{l,k}-\sfrac{c}{24})\ketb{\alpha_{l,k}},
\ee 
where $c$ and $h_{l,k}$ are
given by \eq{cbeta} and \eq{hlkbeta} respectively.

$\ketb{\alpha}$ is the highest weight state of DVA with 
$\lambda=\lambda(\alpha)$,
\ba
  &&\ketb{\alpha}=\ket{\lambda(\alpha)},\\
  &&\lambda(\alpha)
  =x^{\sqrt{r(r-1)}(\alpha-\az)}+x^{-\sqrt{r(r-1)}(\alpha-\az)}.
  \label{lambdaalpha}
\ea
The dual space $\F_{\alpha}^*$ becomes a DVA module by \eq{pairing} with
\ba
  {}^tT_n&\!\!=\!\!&T_{-n},\\
  {}^th_n&\!\!=\!\!&-h_{-n},\quad  \mbox{eq.}\eq{ta0},
  \label{thn}
\ea
and \eq{<|>=1}.
By \eq{thn}, $\F_{\alpha}^*$ is isomorphic to $\F_{2\az-\alpha}$ as 
DVA module,
\ba
  \F_{\alpha}^*&\!\!\cong\!\!&\F_{2\az-\alpha}\quad(\mbox{DVA module}).\\
  \ketbs{\alpha}&\!\!\leftrightarrow\!\!&\ketb{2\az-\alpha}.
\ea
Remark that $\lambda(\alpha)=\lambda(2\az-\alpha)$ by \eq{lambdaalpha}.

In the conformal limit \eq{conflim}, oscillator $h_n$ is expressed by
$a_n$ in \eq{an} as follows:
\be
  h_n=\frac{\hbar}{2}\sqrt{\frac{2}{x^n+x^{-n}}
  \frac{x^{rn}-x^{-rn}}{rn\az\hbar}  
  \frac{x^{(r-1)n}-x^{-(r-1)n}}{(r-1)n\az\hbar}}a_n.
\ee
Substituting this expression into \eq{Tfr} and expanding in $\hbar$, 
we get \eq{Texp} with $L(z)$ in \eq{L-phi} (or \eq{L-phi2}).

\subsubsection{singular vectors and Kac determinant}\label{sec:3.4.2}

In the case of $\lambda=\lambda_{l,k}$ there are singular vectors. 
In the free boson realization they can be expressed by using 
the screening currents. 
For later convenience we denote screening currents as $x_{\pm}(z)$.
$x_{\pm}(z)$ is defined by
\ba
  x_+(z)&\!\!=\!\!&
  :\exp\Bigl(-\sum_{n\neq 0}\frac{\alpha_n}{[n]_x}z^{-n}\Bigr):\times 
  e^{\sqrt{\frac{r}{r-1}}Q}z^{\sqrt{\frac{r}{r-1}}a'_0+\frac{r}{r-1}},
  \label{x+first}\\
  x_-(z)&\!\!=\!\!&
  :\exp\Bigl(\sum_{n\neq 0}\frac{\alpha'_n}{[n]_x}z^{-n}\Bigr):\times
  e^{-\sqrt{\frac{r-1}{r}}Q}z^{-\sqrt{\frac{r-1}{r}}a'_0+\frac{r-1}{r}},
  \label{x-first}
\ea
where oscillators $\alpha_n$, $\alpha'_n$ ($n\in\Z_{\neq 0}$) are related
to $h_n$ as
\be
  h_n=(x-x^{-1})(-1)^n\frac{[(r-1)n]_x}{[2n]_x}\alpha_n
  =(x-x^{-1})\frac{[rn]_x}{[2n]_x}\alpha'_n.
  \label{haa}
\ee
Conformal limit \eq{conflim} of $x_-(z)$ is $zS_-(z)$ in \eq{S}
and that of $x_+(z)$ is $zS_+(-z)$ (up to phase) due to $(-1)^n$ factor 
in \eq{haa}.

Commutation relation of DVA generator and screening currents is
a total difference
\ba
  \!\!\!\!\!
  \lbrack T_n,x_+(w)]&\!\!=\!\!&(x^r-x^{-r})\Bigl((-x^{r-1}w)^nA_+(x^{r-1}w)
  -(-x^{-(r-1)}w)^nA_+(x^{-(r-1)}w)\Bigr),
  \label{Tnx+}\\
  \!\!\!\!\!
  \lbrack T_n,x_-(w)]&\!\!=\!\!&(x^{r-1}-x^{-(r-1)})\Bigl((x^rw)^nA_-(x^rw)
  -(x^{-r}w)^nA_-(x^{-r}w)\Bigr),
\ea
where $A_{\pm}(w)$ is
\ba
  A_+(w)&\!\!=\!\!&x^{\pm r}:\Lambda_{\pm}(-w)x_+(x^{\mp(r-1)}w):,\\
  A_-(w)&\!\!=\!\!&x^{\pm(r-1)}:\Lambda_{\mp}(w)x_-(x^{\mp r}w):.
\ea
Hence their zero modes (screening charge) commute with DVA,
\be
  \Bigl[T_n,\oint_0\frac{dz}{2\pi iz}x_{\pm}(z)\Bigr]=0.
\ee
(Exactly speaking we have to specify the Fock space on which they act.
See subsection \ref{sec:3.4.3}.)
We will use $x_+(z)$ in what follows. $x_-(z)$ can be treated similarly.

The representation with $\lambda=\lambda_{l,k}$ is realized on 
$\F_{l,k}=\F_{\alpha_{l,k}}$,
\be
  \lambda_{l,k}=\lambda(\alpha_{l,k}),
  \label{lambdalk}
\ee
where $\alpha_{l,k}$ is given in \eq{alphalk}.
To contact with our previous paper \cite{SKAO,AKOS95}
($\alpha_{r,s}$ there is $\propto\alpha_{-r,-s}$ here.), 
we consider dual space $\F_{l,k}^*=\F_{-l,-k}$.
Like as the Virasoro case \eq{sv}, singular vectors of DVA are expressed 
by product of screening charges. Moreover this product becomes more clear
for the deformed case, if we include Lukyanov's zero mode factor (see 
subsection \ref{sec:3.4.3}). 
In the Verma module with $\lambda_{-l,-k}$, the singular vector at 
level $lk$ is expressed as
\ba
  \!\!\!\!\!&&\ket{\chi_{-l,-k}}=Q_l\ketb{\alpha_{l,-k}}
  \label{DVAsv}\\
  \!\!\!\!\!&=\!\!&\int\prod_{j=1}^l\dz_j\cdot
  \prod_{i,j=1\atop i\neq j}^l\frac{(\frac{z_i}{z_j};x^{2(r-1)})_{\infty}}
  {(x^{2r}\frac{z_i}{z_j};x^{2(r-1)})_{\infty}}\cdot
  C(z)\prod_{i=1}^lz_i^{-k}\cdot\prod_{j=1}^l
  e^{\sum_{n>0}\frac{1}{[n]_x}\alpha_{-n}z_j^n}\ketb{\alpha_{-l,-k}},
  \nonumber
\ea
where $\dz_j$ is given in \eq{dz}, the BRST charge $Q_l$ will be given 
in \eq{Q_m}, and $C(z)$ is 
\ba
  C(z)&\!\!=\!\!&\prod_{1\leq i<j\leq l}
  \frac{(x^{2r}\frac{z_i}{z_j};x^{2(r-1)})_{\infty}}
  {(\frac{z_i}{z_j};x^{2(r-1)})_{\infty}}
  \frac{(x^{-2}\frac{z_j}{z_i};x^{2(r-1)})_{\infty}}
  {(x^{2(r-1)}\frac{z_j}{z_i};x^{2(r-1)})_{\infty}}\cdot
  \prod_{i=1}^lz_i^{\frac{r}{r-1}(l+1-2i)}\n
  &&\times\prod_{i=1}^l\frac{[-u_i+\frac12-(2i-l)]^*}{[-u_i-\frac12]^*}.
  \label{C(z)}
\ea
Here we have used the notation \eq{[u]^*} with $\rs=r-1$ and $z_i=x^{2u_i}$.
Since the BRST charge $Q_l$, which is a map from $\F_{l,-k}$ 
to $\F_{-l,-k}$, commutes with $T_n$, $\ket{\chi_{-l,-k}}$ is annihilated
by $T_n$ ($n>0$). This state $\ket{\chi_{-l,-k}}$ is non zero because its 
conformal limit gives non zero state \eq{sv}.

Like as in subsection \ref{sec:2.3.2} let us introduce matrices 
$C(N,\alpha)$ and $C'(N,\alpha)$,
\ba
  T_{-I}\ketb{\alpha}&\!\!=\!\!&
  \sum_{J}C(N,\alpha)_{I,J}h_{-J}\ketb{\alpha},
  \label{CDVA}\\
  ({}^tT)_I\ketbs{\alpha}&\!\!=\!\!&
  \sum_JC'(N,\alpha)_{I,J}({}^th)_J\ketbs{\alpha}.
  \label{C'DVA}
\ea
$C'$ can be expressed by $C$ as \eq{C'CD} with \eq{DJK}.
Their determinants are given by
\ba
  \det C(N,\alpha)_{I,J}&\!\!=\!\!&\prod_{l,k\geq 1\atop lk\leq N}
  \Bigl(x^{\sqrt{r(r-1)}(\alpha-\alpha_{l,k})}
  -x^{-\sqrt{r(r-1)}(\alpha-\alpha_{l,k})}\Bigr)^{p(N-lk)},
  \label{DVAdetC}\\
  \det C'(N,\alpha)_{I,J}&\!\!=\!\!&\det C(N,2\az-\alpha)\cdot\det D\n
  &=\!\!&\prod_{l,k\geq 1\atop lk\leq N}
  \Bigl(x^{\sqrt{r(r-1)}(\alpha+\alpha_{l,k}-2\az)}
  -x^{-\sqrt{r(r-1)}(\alpha+\alpha_{l,k}-2\az)}\Bigr)^{p(N-lk)}.
  \label{DVAdetC'}
\ea
The inner product of two states in the Verma module becomes
(see \eq{LILJ})
\ba
  \bra{\lambda}T_IT_{-J}\ket{\lambda}&\!\!=\!\!&
  <({}^tT)_I\ketbs{\alpha},T_{-J}\ketb{\alpha}>\n
  &=\!\!&\sum_{K,L}C'(N,\alpha)_{I,K}G_{K,L}C(N,\alpha)_{J,L}.
  \label{hTTh}
\ea
Here $G_{K,L}$ is
\ba
  G_{K,L}&\!\!=\!\!&<({}^th)_K\ketbs{\alpha},h_{-L}\ketb{\alpha}>\n
  &=\!\!&\delta_{K,L}\prod_i\Biggl(
  \frac{1}{i}\frac{(x^{ri}-x^{-ri})(x^{(r-1)i}-x^{-(r-1)i})}{x^i+x^{-i}}
  \Biggr)^{k_i}k_i!,
\ea
and its determinant is (use \eq{k_i!} and \eq{Fk}) 
\ba
  \det G_{K,L}&\!\!=\!\!&
  \prod_{\{k_i\}\atop{\scriptstyle\Sigma}_iik_i=N}
  \prod_i\Biggl(
  \frac{1}{i}\frac{(x^{ri}-x^{-ri})(x^{(r-1)i}-x^{-(r-1)i})}{x^i+x^{-i}}
  \Biggr)^{k_i}k_i!\n
  &=\!\!&\prod_{l,k\geq 1\atop lk\leq N}\Biggl(
  \frac{(x^{rl}-x^{-rl})(x^{(r-1)l}-x^{-(r-1)l})}{x^l+x^{-l}}
  \Biggr)^{p(N-lk)}.
\ea
Therefore we obtain the Kac determinant \eq{DVAKacdet} \cite{SKAO,BP97},
\ba
  \det\bra{\lambda}T_IT_{-J}\ket{\lambda}&\!\!=\!\!&
  \det C'(N,\alpha)\cdot\det G\cdot\det{}^tC(N,\alpha)\n
  &=\!\!&\prod_{l,k\geq 1\atop lk\leq N}
  \Biggl(
  \frac{(x^{rl}-x^{-rl})(x^{(r-1)l}-x^{-(r-1)l})}{x^l+x^{-l}}
  (\lambda^2-\lambda_{l,k}^2)\Biggr)^{p(N-lk)}.
\ea

\subsubsection{Felder complex}\label{sec:3.4.3}

We consider the representation of $\lambda=\lambda_{l,k}$ in \eq{lambdalk}
with \eq{minimalbeta} and \eq{lkrange}, i.e.
\be
  r=\frac{p''}{p''-p'}. 
\ee
We set 
$$
  \rs=r-1.
$$
Let us define the screening operator $X(z)$ as the integral of $x_+(z')$
plus extra zero mode factor \cite{JLMP},
\be
  X(z)=\oint_{C}\dz'x_+(z')
  \frac{[u-u'+\frac12-\hat{l}\,]^*}{[u-u'-\frac12]^*},
  \label{X}
\ee
where $z=x^{2u}$, $z'=x^{2u'}$, $\dz'=\frac{dz'}{2\pi iz'}$, 
and the integration contour $C$ is
a simple closed curve that encircles $z'=x^{-1+2(r-1)n}z$ but
not $z'=x^{-1-2(r-1)(n+1)}z$ ($n=0,1,2,\cdots$).
$\hat{l}$ is $\hat{l}=l\times\id$ on $\F_{l,k}$ (see subsection \ref{sec:5.2}).
$z$ is an arbitrary point, for example we take $z=1$.

This $X(z)$ is well defined on $\F_{l',k'}$ ($\forall l',k'$)
\be
  X(z):\F_{l',k'}\mapsto\F_{l'-2,k'}.
\ee
Product of $X(z)$'s is \cite{JLMP}
\ba
  &&X(z)^m\;\;:\;\F_{l',k'}\mapsto\F_{l'-2m,k'}\n
  &\!\!=\!\!&\oint\prod_{j=1}^m\dz_j\cdot
  x_+(z_1)\cdots x_+(z_m)
  \prod_{i=1}^m\frac{[u-u_i+\frac12-(\hat{l}-2(m-i))]^*}
  {[u-u_i-\frac12]^*}\n
  &\!\!=\!\!&\oint\prod_{j=1}^m\dz_j\cdot
  x_+(z_1)\cdots x_+(z_m)\n
  &&\quad\times
  \frac{1}{m!}\sum_{\sigma\in S_m}\prod_{i=1}^m
  \frac{[u-u_{\sigma(i)}+\frac12-(\hat{l}-2(m-i))]^*}
  {[u-u_{\sigma(i)}-\frac12]^*}\cdot
  \prod_{1\leq i<j\leq m\atop\sigma(i)>\sigma(j)}
  h^*(u_{\sigma(i)}-u_{\sigma(j)})\n
  &\!\!=\!\!&\oint\prod_{j=1}^m\dz_j\cdot
  x_+(z_1)\cdots x_+(z_m)\n
  &&\quad\times
  \frac{1}{m!}\prod_{i=1}^m\frac{[i]^*}{[1]^*}\cdot
  \prod_{1\leq i<j\leq m}\frac{[u_i-u_j]^*}{[u_i-u_j+1]^*}\cdot
  \prod_{i=1}^m\frac{[u-u_i-\frac12+m-\hat{l}\,]^*}
  {[u-u_i-\frac12]^*},
  \label{X^m}
\ea
where $z=x^{2u}$, $z_i=x^{2u_i}$ and $\dz_j$ is given in \eq{dz}.
Here we have used $x_+(z_1)x_+(z_2)=h^*(u_1-u_2)x_+(z_2)x_+(z_1)$ (as a
meromorphic function) with $h^*(u)=\frac{[u+1]^*}{[u-1]^*}$ (see subsection
\ref{sec:5.2}), and the formula \eq{JLMPlem4} with $r\rightarrow r-1$.
{}From the factor $[i]^*$ in the last line of \eq{X^m}, 
we have (use \eq{prop[u]})
\be
  X(z)^{p'}=0.
\ee
Moreover $X(z)$ has the property,
\be
  [T_n,X(z)^l]=0 \quad\mbox{on $\F_{l',k}$}\quad l'\equiv l\!\!\pmod{p'}.
\ee
This can be proved by \eq{Tnx+} and careful analysis of location of 
poles \cite{JLMP}.
We define the BRST charge $Q_m$ as
\be
  Q_m=X(1)^m.
  \label{Q_m}
\ee

Let us consider the Felder complex $C_{l,k}$,
\be
  \cdots ~{\buildrel X_{-3} \over \longrightarrow }~
  C_{-2} ~{\buildrel X_{-2} \over \longrightarrow }~
  C_{-1} ~{\buildrel X_{-1} \over \longrightarrow }~
  C_0 ~{\buildrel X_0 \over \longrightarrow }~
  C_1 ~{\buildrel X_1 \over \longrightarrow }~
  C_2 ~{\buildrel X_2 \over \longrightarrow }~ \cdots, 
\ee
where $C_j$ and $X_j\;:\;C_j\rightarrow C_{j+1}$ ($j\in\Z$) are
\ba
  &&C_{2j}=\F_{l-2p'j,k},\quad C_{2j+1}=\F_{-l-2p'j,k},
  \label{Cj} \\
  &&X_{2j}=Q_l,\quad X_{2j+1}=Q_{p'-l}.
  \label{Xj}
\ea
$X$ satisfies the BRST property,
\be
  X_jX_{j-1}=0.
\ee
We assume that this Felder complex has the same structure as the Virasoro
case because it formally tends to Virasoro one in the conformal limit
($x\rightarrow 1$ and $r$ and $z=x^{2u}$ fixed kept).
Then the cohomology groups of the complex $C_{l,k}$ are
\be
  H^j(C_{l,k})=\mbox{Ker }X_j/\mbox{Im }X_{j-1}=
  \left\{\begin{array}{ll}
  0&j\neq 0,\\
  \cL_{l,k}&j=0,
  \end{array}\right.
  \label{cohomologyDVA}
\ee
where $\cL_{l,k}$ is the irreducible DVA module of 
$\lambda=\lambda_{l,k}$.
The trace of operator ${\cal O}$ over $\cL_{l,k}$ can be written as
(see \eq{comdiag})
\be
  \tr_{\cL_{l,k}}{\cal O}=\tr_{H^0(C_{l,k})}{\cal O}^{(0)}=
  \tr_{H^*(C_{l,k})}{\cal O}
  =\sum_{j\in\Z}(-1)^j\tr_{C_j}{\cal O}^{(j)},
  \label{trO}
\ee
where ${\cal O}^{(j)}$ is an operator ${\cal O}$ realized on $C_j$.

If we use $x_-(z)$ instead of $x_+(z)$, another complex $C'_{l,k}$ 
in which $k'$ of $\F_{l,k'}$ changes is obtained. 
The corresponding screening operator is
\be
  X'(z)=\oint_{C'}\dz'x_-(z')
  \frac{[u-u'-\frac12+\hat{k}]}{[u-u'+\frac12]},
  \label{X'}
\ee
where $[u]$ in \eq{[u]}, $z=x^{2u}$, $z'=x^{2u'}$, $\dz'=\frac{dz'}{2\pi iz'}$,
and the integration contour $C'$ is
a simple closed curve that encircles $z'=x^{1+2rn}z$ but
not $z'=x^{1-2r(n+1)}z$ ($n=0,1,2,\cdots$).
$\hat{k}$ is $\hat{k}=k\times\id$ on $\F_{l,k}$ (see subsection \ref{sec:5.2}).
$X'(z)$ is well defined on $\F_{l',k'}$ ($\forall l',k'$), 
$X'(z):\F_{l',k'}\mapsto\F_{l',k'-2}$,
and $X'(z)$ satisfies
\ba
  &&X'(z)^{p''}=0,\\
  &&[T_n,X'(z)^k]=0 \quad
  \mbox{on $\F_{l,k'}$}\quad k'\equiv k\!\!\pmod{p''}.
\ea
By setting the BRST charge as $Q'_m=X'(1)^m$, another Felder complex 
$C'_{l,k}=\{C'_j, X'_j:C'_j\rightarrow C'_{j+1}\}$ is
$C'_{2j}=\F_{l,k-2p''j}$, $C'_{2j+1}=\F_{l,-k-2p''j}$,
$X'_{2j}=Q'_k$, $X'_{2j+1}=Q'_{p''-k}$,
and the cohomology is 
$H^j(C'_{l,k})=\mbox{Ker }X'_j/\mbox{Im }X'_{j-1}=\delta_{j,0}\cL_{l,k}$.

\subsubsection{Trigonometric Ruijsenaars-Schneider model and 
Macdonald symmetric polynomial}\label{sec:3.4.4}

The trigonometric Ruijsenaars-Schneider model (tRSM) is a 
relativistic version of the Calogero-Sutherland model.
The Ruijsenaars-Schneider model is a many body system with 
sufficiently enough conserved quantities $H_{\pm k}$ ($k=1,\cdots,N_0$) 
\cite{RS} (see also \cite{Ko96}),
\be
  H_{\pm k}=\sum_{I\subset\{1,\cdots,N_0\}\atop|I|=k}
  \prod_{i\in I\atop j\not\in I}
  h(\pm(q_i-q_j))^{\frac12}\cdot e^{\mp\sum_{i\in I}\theta_i}\cdot
  \prod_{i\in I\atop j\not\in I}h(\mp(q_i-q_j))^{\frac12},
\ee
where $\theta_i$ ($i=1,\cdots,N_0$) is a rapidity and $\bar{q}_i$ is 
its conjugate variable and $\ds q_i=\frac{\bar{q}_i}{mc}$ is a coordinate.
Their dimensions are $[\theta_i]=1$, $[\bar{q}_i]=\hbar$ and $[q_i]=L$.
(Exactly speaking $\theta_i$ and $\bar{q}_i$ are related to 
the coordinate $q_i$ and momentum $p_i$ by 
$\bar{q}_i=mcq_i\cosh\theta_i$ and $p_i=mc\sinh\theta_i$.
But we skip this procedure. See \cite{BrSa}.)
$\theta_j$ is 
\be
  \theta_j=\frac{\hbar}{i}\frac{\partial}{\partial\bar{q}_j}=
  \frac{1}{i}\frac{\hbar}{mc}\frac{\partial}{\partial q_j},
\ee
and $h(q)$ is
\be
  h(q)=\frac{\sigma(\bar{q}+i\bar{\beta})}{\sigma(\bar{q})},
\ee
where $\bar{\beta}$ is a coupling constant and 
$\sigma(z)$ is the Weierstrass $\sigma$ function.
These conserved quantities commute mutually
\be
  [H_k,H_l]=0\quad(k,l=-N_0,\cdots,N_0).
\ee
This model is a `relativistic' model because it has Poincar\'{e}
invariance,
\be
  \begin{array}{lll}
  H=mc^2\sfrac12(H_{-1}+H_1),&&[H,P]=0,\\
  P=mc(H_{-1}-H_1),&&[H,B]=i\hbar P,\\
  B=-\frac{1}{c}\sum_{i=1}^{N_0}\bar{q}_i,&&[P,B]=i\hbar\frac{1}{c^2}H.
  \end{array}
\ee
In the `non-relativistic' limit (the speed of light $c\rightarrow\infty$),
the Hamiltonian $H$ becomes
\be
  \lim_{c\rightarrow\infty}(H-N_0mc^2)=\sum_{j=1}^{N_0}
  \frac{1}{2m}\Bigl(\frac{\hbar}{i}\frac{\partial}{\partial q_j}\Bigr)^2
  +\frac{1}{m}\bar{\beta}(\bar{\beta}-\hbar)
  \sum_{1\leq i<j\leq N_0}\wp(q_i-q_j),
\ee
where $\wp(z)$ is the Weierstrass $\wp$ function 
and we have rescaled the periods of $\sigma(z)$.

tRSM is a trigonometric case of RSM, 
\be
  h(q)=\frac{\sin\frac{\pi}{mcL}(\bar{q}+i\bar{\beta})}
  {\sin\frac{\pi}{mcL}\bar{q}}.
\ee
Its `non-relativistic' limit is
\be
  \lim_{c\rightarrow\infty}(H_{tRS}-N_0mc^2)=
  H_{CS},
\ee
where $H_{CS}$ is given in \eq{CS}.
Let us introducing dimensionless quantities $x_j$ and $\beta$ in 
\eq{betaxj} and
\be
  q=e^{-2\pi\frac{\hbar}{mcL}},\quad t=q^{\beta}.
  \label{qt}
\ee
Removing the contribution from the ground state, we obtain
\be
  \Delta^{-\frac12}H_{\pm k}\Delta^{\frac12}=t^{\mp\frac12k(N_0-1)}
  D_k(q^{\pm 1},t^{\pm 1}).
\ee
Here $\Delta$ is 
\be
  \Delta=\Delta(x)=\Delta(x;q,t)=\prod_{i,j=1\atop i\neq j}^{N_0}
  \frac{(\frac{x_i}{x_j};q)_{\infty}}{(t\frac{x_i}{x_j};q)_{\infty}},
\ee
and $D_k(q,t)$ ($k=1,\cdots,N_0$) is the Macdonald operator
\be
  D_k(q,t)=t^{\frac12k(k-1)}\sum_{I\subset\{1,\cdots,N_0\}\atop|I|=k}
  \prod_{i\in I\atop j\not\in I}\frac{tx_i-x_j}{x_i-x_j}\cdot
  \prod_{i\in I}q^{D_i},
\ee 
where $D_i=x_i\frac{\partial}{\partial x_i}$. $q^{D_i}$ is a $q$-shift,
$q^{D_i}f(x_1,\cdots,x_i,\cdots,x_{N_0})=f(x_1,\cdots,qx_i,\cdots,x_{N_0})$.

The Macdonald symmetric polynomial $P_{\lambda}=P_{\lambda}(x;q,t)$ is 
a multivariable orthogonal polynomial with two parameters $q$ and $t$,
which is determined uniquely by the following conditions \cite{Mac},
\ba
  &\mbox{(i)}&P_{\lambda}=\sum_{\mu\leq\lambda}u_{\lambda,\mu}
  m_{\mu}(x),\quad u_{\lambda,\lambda}=1,\\
  &\mbox{(ii)}&\langle P_{\lambda},P_{\mu}\rangle_{q,t}=0\quad
  \mbox{if $\lambda\neq\mu$},
\ea
where the inner product is 
\be
  \langle p_{\lambda},p_{\mu}\rangle_{q,t}=
  \delta_{\lambda,\mu}\prod_ii^{m_i}m_i!\cdot
  \prod_{i=1}^{\ell(\lambda)}\frac{1-q^{\lambda_i}}{1-t^{\lambda_i}}.
  \label{ipqt}
\ee
The condition (ii) can be replaced by (ii)',
\be
  \mbox{(ii)'}\quad D_1(q,t)P_{\lambda}=\sum_{i=1}^{N_0}
  t^{N_0-i}q^{\lambda_i}\cdot P_{\lambda}.
\ee
Moreover the Macdonald symmetric polynomial is the simultaneous eigenfunction
of the Macdonald operators,
\be
  \sum_{k=0}^{N_0}(-u)^kD_k(q,t)P_{\lambda}(x;q,t)=
  \prod_{i=1}^{N_0}(1-ut^{N_0-i}q^{\lambda_i})\cdot P_{\lambda}(x;q,t),
\ee
or
\be
  D_k(q^{\pm 1},t^{\pm 1})P_{\lambda}(x;q,t)=
  \sum_{1\leq i_1<\cdots<i_k\leq N_0}\prod_{l=1}^kt^{N_0-i_l}q^{\lambda_{i_l}}
  \cdot P_{\lambda}(x;q,t).
\ee
Remark that $P_{\lambda}(x;q,t)=P_{\lambda}(x;q^{-1},t^{-1})$ and 
$\langle f,g\rangle_{q^{-1},t^{-1}}=(qt^{-1})^{N_0}\langle f,g\rangle_{q,t}$.
In the `conformal limit',
\be
  t=q^{\beta},\quad q\rightarrow 1,\quad\mbox{$\beta$ : fixed},
\ee
which corresponds to the non-relativistic limit ($c\rightarrow\infty$) 
(see \eq{qt}), the Macdonald polynomial reduces to the Jack polynomial
\be
  \lim_{q\rightarrow 1\atop t=q^{\beta}}P_{\lambda}(x;q,t)
  =J_{\lambda}(x;\beta).
\ee
$\beta\rightarrow 1$ limit of the Jack polynomial is the Schur 
polynomial $\ds s_{\lambda}(x)=\lim_{\beta\rightarrow 1}J_{\lambda}(x;\beta)$.
$t\rightarrow q$ limit of the Macdonald polynomial is also
$\ds s_{\lambda}(x)=\lim_{t\rightarrow q}P_{\lambda}(x;q,t)$.

There exists another inner product,
\be
  \langle f,g\rangle'_{l;q,t}=
  \frac{1}{l!}\oint\prod_{j=1}^l\underline{dx}_j\cdot
  \Delta(x;q,t)f(\bar{x})g(x),
  \quad \underline{dx}_j=\frac{dx_j}{2\pi ix_j},
\ee
which satisfies
\be
  \langle~,~\rangle'_{l;q,t}\propto\langle~,~\rangle_{q,t}.
\ee
Like as the Jack symmetric polynomial, the Macdonald symmetric polynomial
has an integral representation obtained by the following two transformations,
\ba
  {\cal G}_k&:&\Bigl({\cal G}_kf\Bigr)(x_1,\cdots,x_l)
  =\prod_{i=1}^lx^k\cdot f(x_1,\cdots,x_l),\\
  {\cal N}_{l',l}&:&\Bigl({\cal N}_{l',l}f\Bigr)(x'_1,\cdots,x'_{l'})
  =\oint\prod_{j=1}^l\underline{dx}_j\cdot\Pi(x',\bar{x})
  \Delta(x)f(x_1,\cdots,x_l),\\
  &&\Pi(x,y)=\Pi(x,y;q,t)=\prod_i\prod_j
  \frac{(tx_iy_j;q)_{\infty}}{(x_iy_j;q)_{\infty}}.
\ea
The Macdonald symmetric polynomial with $\lambda$ 
($\lambda'=((l_1)^{k_1},(l_2)^{k_2},\cdots,(l_{N-1})^{k_{N-1}})$) is 
expressed as \cite{AOS}
\be
  P_{\lambda}(x)\propto{\cal N}_{l,l_1}{\cal G}_{k_1}
  {\cal N}_{l_1,l_2}{\cal G}_{k_2}\cdots
  {\cal N}_{l_{N-2},l_{N-1}}{\cal G}_{k_{N-1}}{\cal N}_{l_{N-1},0}\cdot 1,
\ee
because the Macdonald symmetric polynomial has two properties 
\ba
  &&P_{(k^l)+\lambda}={\cal G}_kP_{\lambda}\quad
  (\mbox{$l$ variables}),\\
  &&P_{\lambda}=\frac{\langle P_{\lambda},P_{\lambda}\rangle_{q,t}}
  {l!\langle P_{\lambda},P_{\lambda}\rangle'_{l;q,t}}
  {\cal N}_{l',l}P_{\lambda}\quad
  {\mbox{($P_{\lambda}$ in LHS : $l'$ variables)} \atop
  \mbox{($P_{\lambda}$ in RHS : $l$ variables)}}.
\ea
This second property is valid if ${\cal N}_{l',l}$ is modified in the 
following way \cite{AKOS95},
\ba
  &&P_{\lambda}\propto{\cal N}'_{l',l}P_{\lambda},\n
  &&\Bigl({\cal N}'_{l',l}f\Bigr)(x'_1,\cdots,x'_{l'})
  =\oint\prod_{j=1}^l\underline{dx}_j\cdot\Pi(x',\bar{x})
  \Delta(x)C(x)f(x_1,\cdots,x_l),
\ea
where $C(x)=C(x_1,\cdots,x_l)$ is an arbitrary pseudoconstant with
respect to the $q$-shift, $q^{D_i}C(x)=C(x)$ ($\forall i$).
(Remark that $\langle D_1(q,t)f,g\rangle'_{l;q,t}=
\langle f,D_1(q,t)g\rangle'_{l;q,t}$.)

\noindent
For example, the Macdonald symmetric polynomial with a rectangular Young 
diagram $(k^l)$ is
\be
  P_{(k^l)}(x)\propto
  \oint\prod_{j=1}^l\dz_j\cdot
  \prod_{j=1}^l\exp\Bigl(\sum_{n>0}\frac{1}{n}\frac{1-t^n}{1-q^n}z_j^n
  \sum_ix_i^n\Bigr)\cdot
  \prod_{i,j=1\atop i\neq j}^l\frac{(\frac{z_i}{z_j};q)_{\infty}}
  {(t\frac{z_i}{z_j};q)_{\infty}}\cdot
  \prod_{j=1}^lz_j^{-k}.
  \label{Prec}
\ee

Next let us consider bosonization of the Macdonald operator 
${\cal D}=D_1(q,t)$ \cite{SKAO,AKOS95}. 
${\cal D}$ can be expressed by power sum polynomials,
\ba
  {\cal D}&\!\!=\!\!&\sum_{i=1}^{N_0}\prod_{j\neq i}
  \frac{tx_i-x_j}{x_i-x_j}\cdot q^{D_i}\\
  &=\!\!&\frac{1-t^{N_0}}{1-t}+\frac{t^{N_0}}{t-1}\Biggl(
  \oint_0\dz
  \exp\Bigl(\sum_{n>0}\frac{1}{n}(1-t^{-n})p_nz^n\Bigr)
  \exp\Bigl(\sum_{n>0}(q^n-1)\frac{\partial}{\partial p_n}z^{-n}\Bigr)
  -1\Biggr).\nonumber
\ea
{\it Proof.}\quad
Since $q^{D_i}p_n=p_n+x_i^n(q^n-1)$, $q^{D_i}$ can be written as
\be
  q^{D_i}=\exp\Bigl(\sum_{n>0}x_i^n(q^n-1)\frac{\partial}{\partial p_n}\Bigr)
  =\sum_{n\geq 0}x_i^n\tilde{q}_n(-m\frac{\partial}{\partial p_n};q),
\ee
where $\tilde{q}_n(p_m;t)=q_n(x_m;t)$ is (see \cite{Mac} p.209)
\be
  q_n(x;t)=(1-t)\sum_{i=1}^{N_0}x_i^n\prod_{j\neq i}
  \frac{x_i-tx_j}{x_i-x_j},\quad
  \sum_{n\geq 0}q_n(x;t)z^n
  =\exp\Bigl(\sum_{n>0}\frac{1}{n}(1-t^n)p_nz^n\Bigr).
  \label{qn}
\ee
Then we have
\ba
  {\cal D}&\!\!=\!\!&\sum_{i=1}^{N_0}\prod_{j\neq i}
  \frac{tx_i-x_j}{x_i-x_j}\cdot\sum_{n\geq 0}x_i^n
  \tilde{q}_n(-m\frac{\partial}{\partial p_n};q)\n
  &=\!\!&\frac{1-t^{N_0}}{1-t}+\frac{t^{N_0}}{t-1}
  \sum_{n>0}q_n(x_m;t^{-1})
  \tilde{q}_n(-m\frac{\partial}{\partial p_m};q)
\ea
and get the result by \eq{qn}. \qed

To bosonize this operator we introduce boson oscillator $\tilde{a}_n$ 
($n\in\Z_{\neq 0}$),
\be
  [\tilde{a}_n,\tilde{a}_m]=n\frac{1-q^{|n|}}{1-t^{|n|}}\delta_{n+m,0}.
\ee
A state in the Fock space is mapped to a symmetric function by
\be
  \ket{f}\mapsto f(x)=\bra{\alpha}
  \exp\Bigl(\sum_{n>0}\frac{1}{n}\frac{1-t^n}{1-q^n}\tilde{a}_np_n\Bigr)
  \ket{f},
  \label{FocktosymMac}
\ee
namely
\be
  \tilde{a}_{-n}\leftrightarrow p_n,\quad
  \tilde{a}_n\leftrightarrow
  n\frac{1-q^n}{1-t^n}\frac{\partial}{\partial p_n}.
\ee
This normalization for $\tilde{a}_n$ is chosen so that it produces
the inner product \eq{ipqt}.
Then ${\cal D}$ is realized on the Fock space :
\be
  {\cal D}\bra{\alpha}
  \exp\Bigl(\sum_{n>0}\frac{1}{n}\frac{1-t^n}{1-q^n}\tilde{a}_np_n\Bigr)
  =\bra{\alpha}
  \exp\Bigl(\sum_{n>0}\frac{1}{n}\frac{1-t^n}{1-q^n}\tilde{a}_np_n\Bigr)
  \hat{{\cal D}}.
\ee
The bosonized operator $\hat{{\cal D}}$ is 
\ba
  \hat{{\cal D}}&\!\!=\!\!&
  \frac{1-t^{N_0}}{1-t}+\frac{t^{N_0}}{t-1}\Biggl(
  \oint_0\dz
  \exp\Bigl(\sum_{n>0}\frac{1}{n}(1-t^{-n})\tilde{a}_{-n}z^n\Bigr)
  \exp\Bigl(-\sum_{n>0}(1-t^n)\tilde{a}_nz^{-n}\Bigr)
  -1\Biggr)\n
  &=\!\!&
  \frac{1-t^{N_0}}{1-t}+\frac{t^{N_0}}{t-1}\Biggl(
  \oint_0\dz\Bigl(\psi(z)T(z)
  -\exp\Bigl(-\sum_{n>0}h_nx^nz^{-n}\Bigr)x^{-2\sqrt{r(r-1)}a'_0}\Bigr)
  -1\Biggr)\n
  &=\!\!&
  \frac{1-t^{N_0}}{1-t}+\frac{t^{N_0}}{t-1}\Biggl(
  \sum_{n\geq 0}\psi_{-n}T_n-x^{-2\sqrt{r(r-1)}a'_0}
  -1\Biggr),
\ea
where $T(z)$ is the DVA current \eq{Tfr} and $\psi(z)$ is  
\be
  \psi(z)=\sum_{n\geq 0}\psi_{-n}z^n=
  \exp\Bigl(\sum_{n>0}h_{-n}x^{-n}z^n\Bigr)x^{-\sqrt{r(r-1)}a'_0}.
\ee
Here $\tilde{a}_n$ is expressed by $h_n$ in \eq{hn}
\be
  \tilde{a}_n=n\frac{1}{x^{rn}-x^{-rn}}h_n,\quad
  \tilde{a}_{-n}=nx^{-n}\frac{x^n+x^{-n}}{x^{rn}-x^{-rn}}h_{-n},\quad(n>0),
\ee
and parameters are identified as
\be
  q=x^{2(r-1)},\quad t=q^{\beta}=x^{2r},\quad p=qt^{-1}=x^{-2},\quad
  \beta=\frac{r}{r-1}.
  \label{qtpxr}
\ee

When $\hat{{\cal D}}$ acts on the singular vector \eq{DVAsv}, 
$\sum_{n>0}\psi_{-n}T_n$ term vanishes because of the property of 
singular vector and the remaining terms are already diagonal,
\be
  \hat{{\cal D}}\ket{\chi_{-l,-k}}
  =\sum_{i=1}^{N_0}t^{N_0-i}q^{\lambda_i}\ket{\chi_{-l,-k}},
\ee
where the partition $\lambda$ is $\lambda=(k^l,0^{N_0-l})$.
Therefore $\ket{\chi_{-l,-k}}$ is an eigenstate of $\hat{{\cal D}}$, 
i.e., it gives the Macdonald symmetric polynomial by the map \eq{FocktosymMac}.
In fact the polynomial obtained from \eq{DVAsv} by \eq{FocktosymMac} agrees
with the integral representation of the Macdonald symmetric polynomial with
partition $(k^l)$ in \eq{Prec} up to $x_i\rightarrow -x_i$ and 
$C(z)$ in \eq{C(z)}, 
which is irrelevant because $C(z)$ is a pseudoconstant with respect 
to the $q$-shift, $q^{D_{z_i}}C(z)=C(z)$ ($\forall i$).

The Macdonald symmetric polynomial with general partition whose Young diagram
is composed of $N-1$ rectangles is related to the singular vector of the
deformed $W_N$ algebra \cite{AKOS95}.

We have established the relation between the Macdonald symmetric polynomials
and the singular vectors of DVA in free field realization.
When the DVA was firstly formulated in \cite{SKAO}, this relation was not the 
derived property but the guiding principle to find the DVA.
At that time we knew the two facts:
\begin{enumerate}
\item
In the free field realization, the singular vectors of the Virasoro and $W_N$ 
algebras realize the Jack symmetric polynomials \cite{MY,AMOS}.
\item
The Jack symmetric polynomials have the good $q$-deformation, the Macdonald
symmetric polynomials \cite{Mac}.
\end{enumerate}
Based on these, we set up the following `natural' question:
\begin{itemize}
\item
Construct the algebras whose singular vectors in the free field realization
realize the Macdonald symmetric polynomials.
\end{itemize}
The resultant algebra are worth being called quantum deformation 
($q$-deformation) of the Virasoro and $W_N$ algebras in this sense.
This scenario is illustrated in the following figure,

\setlength{\unitlength}{1mm}
\begin{picture}(140,65)(-10,3)
\put(10,45){\framebox(20,15){\shortstack{$Vir$ \\ \\ $W_N$}}}
\put(10,10){\framebox(20,15){\shortstack{$q$-$Vir$ \\ \\ $q$-$W_N$}}}
\put(80,45){\framebox(25,15){\shortstack{Jack \\ \\ polynomial}}}
\put(80,10){\framebox(25,15){\shortstack{Macdonald \\ \\ polynomial}}}
\thicklines
\put(30,52.5){\vector(1,0){50}}
\put(30,17.5){\vector(1,0){50}}
\put(19.5,45){\vector(0,-1){20}}
\put(20.5,45){\vector(0,-1){20}}
\put(92.5,45){\vector(0,-1){20}}
\put(30,52.5){\makebox(50,7.5){free field realization}}
\put(30,45){\makebox(50,7.5){singular vector}}
\put(30,17.5){\makebox(50,7.5){free field realization}}
\put(30,10){\makebox(50,7.5){singular vector}}
\put(25,22.5){\makebox(20,25){$q$-deformation!}}
\put(97.5,22.5){\makebox(20,25){$q$-deformation}}
%
%
\put(110,52.5){\makebox(0,0)[l]{$\leftrightarrow$ CSM}}
\put(110,17.5){\makebox(0,0)[l]{$\leftrightarrow$ tRSM}}
\put(128,10){.}
\end{picture}

\noindent
See \cite{SKAO} how the DVA was found.

\subsection{Higher DVA currents}\label{sec:3.5}

In this subsection we present higher DVA currents.
We define higher DVA currents $T_{(n)}(z)$ ($n=1,2,\cdots$) by `fusion',
\ba
  T_{(1)}(z)&\!\!=\!\!&T(z),\quad T_{(0)}(z)=1,\n
  T_{(n)}(z)&\!\!=\!\!&
  f_{(1),(n-1)}(x^{rn}\sfrac{z}{z'})T_{(1)}(x^{-r(n-1)}z')T_{(n-1)}(x^rz)
  \Bigl|_{z'\rightarrow z},
  \label{Tn fusion}
\ea
where $f_{(n),(m)}(z)$ is given by
\be
  f_{(n),(m)}(z)=\prod_{i=0}^{n-1}\prod_{j=0}^{m-1}f(x^{r(2j-m+1)-r(2i-n+1)}z)
  =f_{(m),(n)}(z),
\ee
with $f(z)$ in \eq{f}.
In free boson realization \eq{Tfr}, we have
\be
  T_{(n)}(z)=\sum_{i=0}^n{}_n\!B_i\,\Lambda_{n,i}(z),
  \label{T_{(n)}}
\ee
where ${}_n\!B_i$ and $\Lambda_{n,i}(z)$ are
\ba
  &&{}_n\!B_i=\frac{b_i\,b_{n-i}}{b_n},\quad
  b_i=\prod_{i'=0}^{i-1}\frac{[ri'+1]_x}{[r(i'+1)]_x},\\
  &&\Lambda_{n,i}(z)=\,
  :\prod_{i'=0}^{i-1}\Lambda_-(x^{r(2i'-n+1)}z)\cdot
  \prod_{i'=i}^{n-1}\Lambda_+(x^{r(2i'-n+1)}z):.
\ea
$T_{(n)}$ consists of $n+1$ terms, which corresponds to the 
spin $\frac{n}{2}$ representation of $A_1$.
$T_{(n)}(z)$ satisfies the relation
\ba
  &&f_{(n),(m)}(\sfrac{z_2}{z_1})T_{(n)}(z_1)T_{(m)}(z_2)
  -T_{(m)}(z_2)T_{(n)}(z_1)f_{(m),(n)}(\sfrac{z_1}{z_2})\qquad(n\leq m)\n
  &=&-(x-x^{-1})\sum_{a=1}^n{}_n\!B_a\,{}_m\!B_a\,
  [ra]_x[ra+1]_x\prod_{b=1}^a\frac{[rb-1]_x}{[rb+1]_x}\\
  &&\qquad\times\Bigl(
  \delta(x^{r(n+m-2a)+2}\sfrac{z_2}{z_1})
  \no T_{(n-a)}(x^{ra}z_1)T_{(m-a)}(x^{-ra}z_2)\no\n
  &&\qquad\qquad
  -\delta(x^{-r(n+m-2a)-2}\sfrac{z_2}{z_1})
  \no T_{(n-a)}(x^{-ra}z_1)T_{(m-a)}(x^{ra}z_2)\no
  \Bigr).\nonumber
\ea
Here normal ordering $\no T_{(a)}(\alpha z)T_{(b)}(z)\no$ is 
\ba
  &&\no T_{(a)}(\alpha z)T_{(b)}(z)\no\n
  &\stackrel{\rm \scriptstyle def}{=}\!\!&
  \sum_{n=-\infty}^{\infty}\sum_{m=0}^{\infty}\sum_{\ell=0}^m
  f_{(a),(b),\ell}\Bigl(\alpha^{m-\ell}T_{(a),-m}T_{(b),n+m}
  +\alpha^{\ell-m-1}T_{(b),n-m-1}T_{(a),m+1}\Bigr)\cdot z^{-n}\n
  &=\!\!&
  \oint\frac{dy}{2\pi iy}\Biggl(
  \sum_{j=0}^{\infty}(\sfrac{\alpha z}{y})^j\cdot f_{(a),(b)}(\sfrac{z}{y})
  T_{(a)}(y)T_{(b)}(z)
  +\sum_{j=0}^{\infty}(\sfrac{y}{\alpha z})^{j+1}\cdot
  T_{(b)}(z)T_{(a)}(y)f_{(a),(b)}(\sfrac{y}{z})\Biggr)\n
  &=\!\!&
  \oint_{|y|>|\alpha z|}\frac{dy}{2\pi iy}
  \frac{1}{1-\frac{\alpha z}{y}}\cdot f_{(a),(b)}(\sfrac{z}{y})
  T_{(a)}(y)T_{(b)}(z)\n
  &&\qquad
  +\oint_{|y|<|\alpha z|}\frac{dy}{2\pi iy}
  \frac{\frac{y}{\alpha z}}{1-\frac{y}{\alpha z}}\cdot
  T_{(b)}(z)T_{(a)}(y)f_{(a),(b)}(\sfrac{y}{z})\n
  &=\!\!&
  \oint_{\alpha z}\frac{dy}{2\pi i}\frac{1}{y-\alpha z}\cdot
  f_{(a),(b)}(\sfrac{z}{y})T_{(a)}(y)T_{(b)}(z),
\ea
where mode expansions are
\be
  f_{(a),(b)}(z)=\sum_{\ell=0}^{\infty}f_{(a),(b),\ell}z^{\ell},\quad
  T_{(a)}(z)=\sum_{n\in\Z}T_{(a),n}z^{-n}.
\ee
This is a generalization of the normal ordering for currents used in
the CFT,
\be
  (AB)(z)=\oint_z\frac{dy}{2\pi i}\frac{1}{y-z}\cdot A(y)B(z).
\ee
We remark that $T_{(n)}(z)$ is a composite field of $T(z)$,
\be
  \no T_{(n)}(x^{-rm}z)T_{(m)}(x^{rn}z)\no=T_{(n+m)}(z).
\ee
We remark also that another higher currents are obtained 
by replacing $r$ with $1-r$.

Concerning the higher currents for the deformed $W_N$ algebra,
see appendix C in \cite{HJKOS99} where delta function terms are neglected.
(The arguments of $W_{(n)}(u)$ and $f_{(n),(m)}(u,v)$ in \cite{HJKOS99}
should be shifted in order to compare them with $T_{(n)}(z)$ and 
$f_{(n),(m)}(z)$ here.)

\setcounter{section}{3}
\setcounter{equation}{0}
\section{Solvable Lattice Models and Elliptic Algebras}\label{sec:4}

\subsection{Solvable lattice models and Yang-Baxter equation}\label{sec:4.1}

A statistical lattice model is a statistical mechanical system on a lattice
\cite{Bax}.
In this lecture we consider classical systems on two dimensional space 
square lattice.
There are two types of lattice models, vertex models and face models. 
For vertex models dynamical variables `spin' are located on edges 
and the Boltzmann weights are assigned to each vertex. 
On the other hand, for face models, 
dynamical variables `height' are located on vertices and the Boltzmann
weights are assigned to each face.
\be
\setlength{\unitlength}{1mm}
\begin{picture}(140,37)(-7,5)
\put(0,34){Vertex model :}
\put(44,30){\line(-1,0){10}}
\put(39,35){\line(0,-1){10}}
\put(40,31){$u$}
\put(31,29){$\mu$}
\put(45,29){$\mu'$}
\put(38,22){$\nu$}
\put(38,37){$\nu'$}
\put(55,30){$\ds R_{\mu'\nu',\mu\nu}(u)$,}
\put(80,30){$u=u_1-u_2$}
\put(124,30){\vector(-1,0){3}}\put(121,30){\line(-1,0){6}}
\put(114.5,30){\makebox(0,0){$\triangleleft$}}
\put(119,35){\vector(0,-1){3}}\put(119,32){\line(0,-1){6}}
\put(119,25.5){\makebox(0,0){$\scriptscriptstyle\bigtriangledown$}}
\put(118,37){$u_1$}
\put(126,29){$u_2$,}
\put(0,16){Face ~model~~:}
\put(35,10){\framebox(8,8){$u$}}
\put(32,7){$d$}
\put(44,7){$c$}
\put(32,18){$a$}
\put(44,18){$b$}
\put(55,13){$\ds W\BW{a}{b}{c}{d}{u}$,}
\put(80,13){$u=u_1-u_2$}
\multiput(125,13)(-1,0){14}{\line(-1,0){0.5}}
\put(111,13){\makebox(0,0){$\triangleleft$}}
\multiput(119,19)(0,-1){14}{\line(0,-1){0.5}}
\put(119,5){\makebox(0,0){$\scriptscriptstyle\bigtriangledown$}}
\put(118,20){$u_1$}
\put(126,12){$u_2$.}
\put(115,9){\vector(1,0){4.7}}\put(119.7,9){\line(1,0){3.3}}
\put(115,17){\vector(1,0){4.7}}\put(119.7,17){\line(1,0){3.3}}
\put(115,17){\vector(0,-1){4.7}}\put(115,12.3){\line(0,-1){3.3}}
\put(123,17){\vector(0,-1){4.7}}\put(123,12.3){\line(0,-1){3.3}}
\end{picture}
\ee
Here $\mu,\nu,\mu',\nu'$ are spins, $a,b,c,d$ are heights,
and $\ds R_{\mu'\nu',\mu\nu}(u)$ and $\ds W\BW{a}{b}{c}{d}{u}$ are 
Boltzmann weights.
We have assumed that the spectral parameters enter in the Boltzmann
weight only through their difference $u$.
Roughly speaking, a vertex model on a lattice is a face model 
on its dual lattice.
A face model is also called an interaction round a face (IRF) model or
a solid on solid (SOS) model. If the height variable takes a finite number
of states, a SOS model is called a restricted SOS (RSOS) model.

So-called solvable lattice model is a statistical lattice model whose 
Boltzmann weight satisfies the Yang-Baxter equation (YBE)
\cite{J,Bax,J89},
\be
  R^{(12)}(u-v)R^{(13)}(u)R^{(23)}(v)
  =R^{(23)}(v)R^{(13)}(u)R^{(12)}(u-v),
  \label{YBER}
\ee
for a vertex model and 
\be
  \sum_{g}W\BW{b}{c}{g}{d}{u-v}W\BW{a}{b}{f}{g}{u}W\BW{f}{g}{e}{d}{v}
  =\sum_{g}W\BW{a}{b}{g}{c}{v}W\BW{g}{c}{e}{d}{u}W\BW{a}{g}{f}{e}{u-v},
  \label{YBEW}
\ee
for a face model. 
Here $R(u)=(R_{\mu'\nu',\mu\nu}(u))$ is a matrix on $V\otimes V$ and 
\eq{YBER} is an equation on $V\otimes V\otimes V$ and the superscript
$(12),(13)$ etc. refer to the tensor components, e.g., 
$R^{(12)}(u)=R(u)\otimes\id$.
YBE \eq{YBEW} is also satisfied by the new Boltzmann weight obtained 
by the following transformation (gauge transformation),
\be
  W'\BW{a}{b}{c}{d}{u}=\frac{F(a,b)F(b,d)}{F(a,c)F(c,d)}W\BW{a}{b}{c}{d}{u},
  \label{Wgauge}
\ee
where $F(a,b)$ is an arbitrary $u$-independent function of $a$ and $b$.
The following graphical expression of the Yang-Baxter equation will 
help us.
\be
\setlength{\unitlength}{1mm}
\begin{picture}(140,32)(-7,0)
\put(10,15){\line(1,0){11}}\put(27,15){\vector(-1,0){6}}
\put(10,25){\line(1,0){11}}\put(27,25){\vector(-1,0){6}}
\put(17,8){\line(0,1){11}}\put(17,32){\vector(0,-1){13}}
\put(27,15){\line(1,1){10}}\put(37,25){\line(1,0){6}}
\put(27,25){\line(1,-1){10}}\put(37,15){\line(1,0){6}}
\put(19,17){\makebox(0,0){$v$}}
\put(19,27){\makebox(0,0){$u$}}
\put(34,20){\makebox(0,0)[l]{$u-v$}}
\put(50,20){\makebox(0,0){$=$}}
\put(70,15){\line(1,0){4}}\put(87,15){\vector(-1,0){13}}
\put(70,25){\line(1,0){4}}\put(87,25){\vector(-1,0){13}}
\put(80,8){\line(0,1){11}}\put(80,32){\vector(0,-1){13}}
\put(70,15){\line(-1,1){10}}\put(60,25){\line(-1,0){4}}
\put(70,25){\line(-1,-1){10}}\put(60,15){\line(-1,0){4}}
\put(82,27){\makebox(0,0){$v$}}
\put(82,17){\makebox(0,0){$u$}}
\put(67,20){\makebox(0,0)[l]{$u-v$}}
\put(90,15){\makebox(0,0){,}}
\end{picture}
\ee
$$
\setlength{\unitlength}{1mm}
\begin{picture}(140,26)(0,3)
\put(10,20){\vector(1,0){9}}\put(19,20){\line(1,0){6}}
\put(20,30){\vector(-1,-1){6}}\put(10,20){\line(1,1){4}}
\put(10,20){\vector(1,-1){6}}\put(20,10){\line(-1,1){4}}
\put(20,10){\vector(1,0){8}}\put(28,10){\line(1,0){7}}
\put(20,30){\vector(1,0){8}}\put(28,30){\line(1,0){7}}
\put(35,30){\vector(-1,-1){6}}\put(25,20){\line(1,1){4}}
\put(25,20){\vector(1,-1){6}}\put(35,10){\line(-1,1){4}}
\put(35,30){\vector(1,-1){6}}\put(45,20){\line(-1,1){4}}
\put(45,20){\vector(-1,-1){6}}\put(35,10){\line(1,1){4}}
\put(36,20){\makebox(0,0){$u-v$}}
\put(22.5,25){\makebox(0,0){$u$}}
\put(22.5,15){\makebox(0,0){$v$}}
\put(25,20){\makebox(0,0){$\bullet$}}
\put(28,20){\makebox(0,0){$g$}}
\put(8,20){\makebox(0,0){$f$}}
\put(47,20){\makebox(0,0){$c$}}
\put(20,32){\makebox(0,0){$a$}}
\put(35,32){\makebox(0,0){$b$}}
\put(20,8){\makebox(0,0){$e$}}
\put(35,8){\makebox(0,0){$d$}}
\put(52.5,20){\makebox(0,0){$=$}}
\put(80,20){\vector(1,0){8}}\put(88,20){\line(1,0){7}}
\put(70,30){\vector(-1,-1){6}}\put(60,20){\line(1,1){4}}
\put(60,20){\vector(1,-1){6}}\put(70,10){\line(-1,1){4}}
\put(70,10){\vector(1,0){9}}\put(79,10){\line(1,0){6}}
\put(70,30){\vector(1,0){9}}\put(79,30){\line(1,0){6}}
\put(80,20){\vector(-1,-1){6}}\put(70,10){\line(1,1){4}}
\put(70,30){\vector(1,-1){6}}\put(80,20){\line(-1,1){4}}
\put(85,30){\vector(1,-1){6}}\put(95,20){\line(-1,1){4}}
\put(95,20){\vector(-1,-1){6}}\put(85,10){\line(1,1){4}}
\put(69,20){\makebox(0,0){$u-v$}}
\put(82.5,15){\makebox(0,0){$u$}}
\put(82.5,25){\makebox(0,0){$v$}}
\put(80,20){\makebox(0,0){$\bullet$}}
\put(77,20){\makebox(0,0){$g$}}
\put(58,20){\makebox(0,0){$f$}}
\put(97,20){\makebox(0,0){$c$}}
\put(70,32){\makebox(0,0){$a$}}
\put(85,32){\makebox(0,0){$b$}}
\put(70,8){\makebox(0,0){$e$}}
\put(85,8){\makebox(0,0){$d$}}
\put(100,18){\makebox(0,0){,}}
\end{picture}
$$
where summation is taken over the height $g$ at site $\bullet$.

Solutions of the Yang-Baxter equation were studied by many people and
the relation to the Lie algebras was clarified \cite{J,JMO1,DJKMO,JMO3}.
Three types of solutions are known; rational, trigonometric and ellipic. 
Associated for each type of solution ($R$ matrix), 
algebras are defined \cite{D},
\be
  \begin{array}{lcl}
  \mbox{rational}&\rightarrow&\mbox{Yangian},\\
  \mbox{trigonometric}&\rightarrow&\mbox{quantum group (quantum algebra)},\\
  \mbox{elliptic}&\rightarrow&\mbox{elliptic quantum group (elliptic algebra)}.
  \end{array}
\ee
Elliptic quntum groups will be introduced in subsection \ref{sec:4.4}.

\noindent
{\it Example:}\quad
Vector representation of $A_n$ algebra \cite{JMO3}.
\ba
  &&W\BW{a}{a+\hat{\mu}}{a+\hat{\mu}}{a+2\hat{\mu}}{u}
  =\frac{[1+u]}{[1]},\n
  &&W\BW{a}{a+\hat{\mu}}{a+\hat{\mu}}{a+\hat{\mu}+\hat{\nu}}{u}
  =\frac{[a_{\mu}-a_{\nu}-u]}{[a_{\mu}-a_{\nu}]}
  \quad(\mu\neq\nu),
  \label{BWAn}\\
  &&W\BW{a}{a+\hat{\nu}}{a+\hat{\mu}}{a+\hat{\mu}+\hat{\nu}}{u}
  =\frac{[u]}{[1]}
  \left(\frac{[a_{\mu}-a_{\nu}+1][a_{\mu}-a_{\nu}-1]}
  {[a_{\mu}-a_{\nu}]^2}\right)^{\frac12}\quad(\mu\neq\nu).\nonumber
\ea
Notation is as follows.
$a$ is an element of the weight space of $A_n^{(1)}$.
$\hat{\mu}$ is $\hat{\mu}=\ve_{\mu}-\frac{1}{n+1}(\ve_1+\cdots+\ve_{n+1})$ 
($\mu=1,2,\cdots,n+1$), where $\ve_{\mu}$ is an orothonomal basis of 
$\C^{n+1}$. $a_{\mu}$ is $\langle a+\rho,\hat{\mu}\rangle$, where $\rho$ 
is a sum of fundamental weight of $A_1^{(1)}$. 
So $\bar{a}+\bar{\rho}=\sum_{\mu=1}^{n+1}a_{\mu}\ve_{\mu}$, 
$\sum_{\mu=1}^{n+1}a_{\mu}=0$.

The partition function of a lattice model is 
\be
  Z=\sum_{\mbox{config.}}\prod(\mbox{Boltzmann weight}),
\ee
where the summation is taken over all configurations and 
the product is taken over all verteces or faces.
This partition function can be calculated by using the transfer matrix.
The row-to-row transfer matrix $\T(u)$ is 
\be
\setlength{\unitlength}{1mm}
\begin{picture}(140,25)(0,5)
\put(10,20){\makebox(0,0)[l]{
$\ds\T(u)=\Bigl(\T(u)_{\underline{\nu}',\underline{\nu}}\Bigr)
=\sum_{\mu_1,\cdots,\mu_N}$}}
\put(64,20){\line(1,0){53}}
\put(70,25){\line(0,-1){10}}
\put(80,25){\line(0,-1){10}}
\put(110,25){\line(0,-1){10}}
\put(70,27.5){\makebox(0,0){$\nu'_1$}}
\put(80,27.5){\makebox(0,0){$\nu'_2$}}
\put(110,27.5){\makebox(0,0){$\nu'_N$}}
\put(70,13){\makebox(0,0){$\nu_1$}}
\put(80,13){\makebox(0,0){$\nu_2$}}
\put(110,13){\makebox(0,0){$\nu_N$}}
\put(72,22){\makebox(0,0){$u$}}
\put(82,22){\makebox(0,0){$u$}}
\put(95,22){\makebox(0,0){$\cdots$}}
\put(112,22){\makebox(0,0){$u$}}
\put(67,18){\makebox(0,0){$\mu_1$}}
\put(75,18){\makebox(0,0){$\mu_2$}}
\put(105,18){\makebox(0,0){$\mu_N$}}
\put(115,18){\makebox(0,0){$\mu_1$}}
\put(120,15){,}
\end{picture}
\ee
$$
\setlength{\unitlength}{1mm}
\begin{picture}(140,23)(0,6)
\put(9,20){\makebox(0,0)[l]{
$\ds\T(u)=\Bigl(\T(u)_{\underline{a},\underline{b}}\Bigr)=$}}
\put(50,25){\vector(0,-1){6}}\put(50,19){\line(0,-1){4}}
\put(60,25){\vector(0,-1){6}}\put(60,19){\line(0,-1){4}}
\put(70,25){\vector(0,-1){6}}\put(70,19){\line(0,-1){4}}
\put(100,25){\vector(0,-1){6}}\put(100,19){\line(0,-1){4}}
\put(110,25){\vector(0,-1){6}}\put(110,19){\line(0,-1){4}}
\put(50,25){\vector(1,0){6}}
\put(56,25){\vector(1,0){10}}
\put(66,25){\vector(1,0){40}}
\put(106,25){\line(1,0){4}}
\put(50,15){\vector(1,0){6}}
\put(56,15){\vector(1,0){10}}
\put(66,15){\vector(1,0){40}}
\put(106,15){\line(1,0){4}}
\put(55,20){\makebox(0,0){$u$}}
\put(65,20){\makebox(0,0){$u$}}
\put(85,20){\makebox(0,0){$\cdots$}}
\put(105,20){\makebox(0,0){$u$}}
\put(50,27){\makebox(0,0){$a_1$}}
\put(60,27){\makebox(0,0){$a_2$}}
\put(100,27){\makebox(0,0){$a_N$}}
\put(110,27){\makebox(0,0){$a_1$}}
\put(50,12.5){\makebox(0,0){$b_1$}}
\put(60,12.5){\makebox(0,0){$b_2$}}
\put(100,12.5){\makebox(0,0){$b_N$}}
\put(110,12.5){\makebox(0,0){$b_1$}}
\put(115,15){,}
\end{picture}
$$
where $\underline{\nu}=(\nu_1,\cdots,\nu_N)$, 
$\underline{\nu}'=(\nu'_1,\cdots,\nu'_N)$,
$\underline{a}=(a_1,\cdots,a_N)$ and $\underline{b}=(b_1,\cdots,b_N)$.
Here lattice size is $N$ (horizontal) by $M$ (vertical), and we assume
the periodic boundary condition.
Then the partition function is
\be
  Z=\tr\,\Bigl(\T(u)^M\Bigr).
\ee
In the thermodynamic limit ($N,M\rightarrow\infty$), only the maximal 
eigenvalue of the transfer matrix contributes to the partition function.
Once we know information about the eigenvalues and eigenvectors of the
transfer matrix, we can calculate various phyical quantities.

If the Boltzmann weight satisfies the Yang-Baxter equation, 
the transfer matrix has good properties.
For example the trasfer matrices with different spectral parameters
commute each other,
\be
  [\T(u),\T(v)]=0.
  \label{TuTv}
\ee
To show this let us introduce the monodromy matrix $\hat{\T}(u)$
\be
\setlength{\unitlength}{1mm}
\begin{picture}(140,25)(10,5)
\put(35,20){\makebox(0,0)[l]{$\ds\hat{\T}(u)=\sum_{\mu_2,\cdots,\mu_N}$}}
\put(64,20){\line(1,0){53}}
\put(70,25){\line(0,-1){10}}
\put(80,25){\line(0,-1){10}}
\put(110,25){\line(0,-1){10}}
\put(70,27.5){\makebox(0,0){$\nu'_1$}}
\put(80,27.5){\makebox(0,0){$\nu'_2$}}
\put(110,27.5){\makebox(0,0){$\nu'_N$}}
\put(70,13){\makebox(0,0){$\nu_1$}}
\put(80,13){\makebox(0,0){$\nu_2$}}
\put(110,13){\makebox(0,0){$\nu_N$}}
\put(72,22){\makebox(0,0){$u$}}
\put(82,22){\makebox(0,0){$u$}}
\put(95,22){\makebox(0,0){$\cdots$}}
\put(112,22){\makebox(0,0){$u$}}
\put(67,18){\makebox(0,0){$\mu_1$}}
\put(75,18){\makebox(0,0){$\mu_2$}}
\put(105,18){\makebox(0,0){$\mu_N$}}
\put(115,18){\makebox(0,0){$\mu_{N+1}$}}
\put(120,15){,}
\end{picture}
\ee
$$
\setlength{\unitlength}{1mm}
\begin{picture}(140,23)(10,6)
\put(32,20){\makebox(0,0)[l]{$\ds\hat{\T}(u)=$}}
\put(50,25){\vector(0,-1){6}}\put(50,19){\line(0,-1){4}}
\put(60,25){\vector(0,-1){6}}\put(60,19){\line(0,-1){4}}
\put(70,25){\vector(0,-1){6}}\put(70,19){\line(0,-1){4}}
\put(100,25){\vector(0,-1){6}}\put(100,19){\line(0,-1){4}}
\put(110,25){\vector(0,-1){6}}\put(110,19){\line(0,-1){4}}
\put(50,25){\vector(1,0){6}}
\put(56,25){\vector(1,0){10}}
\put(66,25){\vector(1,0){40}}
\put(106,25){\line(1,0){4}}
\put(50,15){\vector(1,0){6}}
\put(56,15){\vector(1,0){10}}
\put(66,15){\vector(1,0){40}}
\put(106,15){\line(1,0){4}}
\put(55,20){\makebox(0,0){$u$}}
\put(65,20){\makebox(0,0){$u$}}
\put(85,20){\makebox(0,0){$\cdots$}}
\put(105,20){\makebox(0,0){$u$}}
\put(50,27){\makebox(0,0){$a_1$}}
\put(60,27){\makebox(0,0){$a_2$}}
\put(100,27){\makebox(0,0){$a_N$}}
\put(110,27){\makebox(0,0){$a_{N+1}$}}
\put(50,12.5){\makebox(0,0){$b_1$}}
\put(60,12.5){\makebox(0,0){$b_2$}}
\put(100,12.5){\makebox(0,0){$b_N$}}
\put(110,12.5){\makebox(0,0){$b_{N+1}$}}
\put(115,15){.}
\end{picture}
$$
Then the transfer matrix is expressed as a trace of the monodromy matrix
\be
  \T(u)=\tr\,\hat{\T}(u)\qquad\Bigl(\tr =\sum_{\mu_1,\mu_{N+1}}
  \delta_{\mu_1,\mu_{N+1}}\mbox{ or }\sum_{a_{N+1}}\delta_{a_1,a_{N+1}}
  \Bigr).
\ee
The following figure (for face model see similar figure in \eq{WPhiPhi})
\be
\setlength{\unitlength}{1mm}
\begin{picture}(140,25)(5,7)
\put(10,16){\line(1,0){40}}
\put(50,16){\line(1,1){8}}
\put(58,24){\line(1,0){4}}
\put(10,24){\line(1,0){40}}
\put(50,24){\line(1,-1){8}}
\put(58,16){\line(1,0){4}}
\put(14,28){\line(0,-1){16}}
\put(22,28){\line(0,-1){16}}
\put(38,28){\line(0,-1){16}}
\put(46,28){\line(0,-1){16}}
\put(30,20){\makebox(0,0){$\cdots$}}
\put(16,26){\makebox(0,0){$u$}}
\put(24,26){\makebox(0,0){$u$}}
\put(40,26){\makebox(0,0){$u$}}
\put(48,26){\makebox(0,0){$u$}}
\put(16,18){\makebox(0,0){$v$}}
\put(24,18){\makebox(0,0){$v$}}
\put(40,18){\makebox(0,0){$v$}}
\put(48,18){\makebox(0,0){$v$}}
\put(56,20){\makebox(0,0)[l]{$w$}}
\put(66,20){\makebox(0,0){$=$}}
\put(70,16){\line(1,0){32}}
\put(102,16){\line(1,1){8}}
\put(110,24){\line(1,0){8}}
\put(70,24){\line(1,0){32}}
\put(102,24){\line(1,-1){8}}
\put(110,16){\line(1,0){8}}
\put(74,28){\line(0,-1){16}}
\put(82,28){\line(0,-1){16}}
\put(98,28){\line(0,-1){16}}
\put(114,28){\line(0,-1){16}}
\put(90,20){\makebox(0,0){$\cdots$}}
\put(76,26){\makebox(0,0){$u$}}
\put(84,26){\makebox(0,0){$u$}}
\put(100,26){\makebox(0,0){$u$}}
\put(116,26){\makebox(0,0){$v$}}
\put(76,18){\makebox(0,0){$v$}}
\put(84,18){\makebox(0,0){$v$}}
\put(100,18){\makebox(0,0){$v$}}
\put(116,18){\makebox(0,0){$u$}}
\put(108,20){\makebox(0,0)[l]{$w$}}
\put(135,20){\makebox(0,0){$(w=u-v)$}}
\end{picture}
\ee
$$
\setlength{\unitlength}{1mm}
\begin{picture}(140,20)(-12,9)
\put(45,20){\makebox(0,0){$=$}}
\put(50,16){\line(1,0){4}}
\put(54,16){\line(1,1){8}}
\put(62,24){\line(1,0){40}}
\put(50,24){\line(1,0){4}}
\put(54,24){\line(1,-1){8}}
\put(62,16){\line(1,0){40}}
\put(66,28){\line(0,-1){16}}
\put(74,28){\line(0,-1){16}}
\put(90,28){\line(0,-1){16}}
\put(98,28){\line(0,-1){16}}
\put(82,20){\makebox(0,0){$\cdots$}}
\put(68,26){\makebox(0,0){$v$}}
\put(76,26){\makebox(0,0){$v$}}
\put(92,26){\makebox(0,0){$v$}}
\put(100,26){\makebox(0,0){$v$}}
\put(68,18){\makebox(0,0){$u$}}
\put(76,18){\makebox(0,0){$u$}}
\put(92,18){\makebox(0,0){$u$}}
\put(100,18){\makebox(0,0){$u$}}
\put(60,20){\makebox(0,0)[l]{$w$}}
\end{picture}
$$
shows 
\ba
  &&R(u-v)\hat{\T}(u)\hat{\T}(v)=\hat{\T}(v)\hat{\T}(u)R(u-v)\\
  \mbox{or}&&
  W(u-v)\hat{\T}(u)\hat{\T}(v)=\hat{\T}(v)\hat{\T}(u)W(u-v).
\ea
(Here we do not write tensor product explicitly.)
Therefore we obtain \eq{TuTv} by taking a trace of 
$R(u-v)\hat{\T}(u)\hat{\T}(v)R(u-v)^{-1}=\hat{\T}(v)\hat{\T}(u)$.

If we interpret a $R$ matrix as a $L$ matrix ($L$ operator),
whose entries are operators acting on vertical vector space,
\be
  L(u)=\Bigl(L(u)_{\mu',\mu}\Bigr),\quad 
  \Bigl(L(u)_{\mu',\mu}\Bigr)_{\nu',\nu}=
  \mu\,\rule[1mm]{10mm}{0.4pt}\,\mu'
  \hspace{-9mm}\rule[-4mm]{0.4pt}{10mm}
  \hspace{-1mm}\raisebox{7mm}{$\nu'$}
  \hspace{-3mm}\raisebox{-6.5mm}{$\nu$}
  \hspace{-1mm}\raisebox{2mm}{$u$}
  \qquad,
\ee
then \eq{YBER} is rewritten as
\be
  R^{(12)}(u-v)L^{(1)}(u)L^{(2)}(v)
  =L^{(2)}(v)L^{(1)}(u)R^{(12)}(u-v).
  \label{RLL}
\ee
The monodoromy matrix is
\be
  \hat{\T}(u)=L_1(u)L_2(u)\cdots L_N(u),
\ee
where $L_i(u)$ acts  nontrivially on $i$-th component,
$L_i(u)=1\otimes\cdots\otimes\stackrel{i\atop\smile}{L(u)}
\otimes\cdots\otimes 1$.

The $R$ matrix obtained from the Boltzmann weight of the face model
satisfies the dynamical Yang-Baxter equation \cite{Fel95},
\ba
  &&R^{(12)}(u-v;\lambda+h^{(3)})R^{(13)}(u;\lambda)
  R^{(23)}(v;\lambda+h^{(1)})\n
  &&\qquad=
  R^{(23)}(v;\lambda)R^{(13)}(u;\lambda+h^{(2)})R^{(12)}(u-v;\lambda).
  \label{dYBER}
\ea
Here $h$ is an element of the Cartan subalgebra (see subsections 
\ref{sec:4.3}, \ref{sec:4.4}).
For simpliclity we consider $A_n$ vector representation case.
Associated to the face Boltzmann weight \eq{BWAn},
$R$ matrix is introduced by
\be
  R(u,\lambda)=\sum_{\mu,\nu,\mu',\nu'\atop\mu+\nu=\mu'+\nu'}
  W\BW{\lambda}{\lambda+\hat{\nu}'}{\lambda+\hat{\mu}}
  {\lambda+\hat{\mu}+\hat{\nu}}{u}
  E_{\mu',\mu}\otimes E_{\nu',\nu},
\ee
where $E_{\mu',\mu}$ is a matrix unit, 
$(E_{\mu',\mu})_{\nu',\nu}=\delta_{\mu',\mu}\delta_{\nu',\nu}$.
$h$ acts on $E_{\mu',\mu}$ as 
$hE_{\mu',\mu}=\hat{\mu}'E_{\mu',\mu}$,
$E_{\mu',\mu}h=\hat{\mu}E_{\mu',\mu}$.
Using this and \eq{YBEW} we can show \eq{dYBER}.
Similariy $L(u,\lambda)$
\ba
  &&L(u,\lambda)=\Bigl(L(u,\lambda)_{\mu',\mu}\Bigr)
  =\sum_{\mu,\mu'}L(u,\lambda)_{\mu',\mu}E_{\mu',\mu},\n
  &&L(u,\lambda)_{\mu',\mu}=\sum_{\nu,\nu'\atop\mu+\nu=\mu'+\nu'}
  W\BW{\lambda}{\lambda+\hat{\nu}'}{\lambda+\hat{\mu}}
  {\lambda+\hat{\mu}+\hat{\nu}}{u}E_{\nu',\nu},
\ea
satisfies the dynamical $RLL$-relation
\ba
  &&R^{(12)}(u-v;\lambda+h)L^{(1)}(u;\lambda)L^{(2)}(v;\lambda+h^{(1)})\n
  &&\qquad=
  L^{(2)}(v;\lambda)L^{(1)}(u;\lambda+h^{(2)})R^{(12)}(u-v;\lambda).
  \label{dRLL}
\ea

\subsection{Corner transfer matrices and vertex operators}\label{sec:4.2}

In the previous subsection the row-to-row transfer matrix is explained.
There is another powerful method, corner trasfer matrix (CTM) method,
which was developed by Baxter \cite{Bax}.
In the following we consider face models in which heights take values 
in a set $I$. 
Vertex models are also treated similarly.

Let us consider the square lattice. We take the central site as the
reference site O and devide a lattice into four quadrants.
\be
\setlength{\unitlength}{1mm}
\begin{picture}(140,50)(-70,-25)
\put(-25,0){\line(1,0){50}}
\put(-21,3){\line(1,0){42}}\put(-21,-3){\line(1,0){42}}
\put(-18,6){\line(1,0){36}}\put(-18,-6){\line(1,0){36}}
\put(-15,9){\line(1,0){30}}\put(-15,-9){\line(1,0){30}}
\put(-12,12){\line(1,0){24}}\put(-12,-12){\line(1,0){24}}
\put(-9,15){\line(1,0){18}}\put(-9,-15){\line(1,0){18}}
\put(-6,18){\line(1,0){12}}\put(-6,-18){\line(1,0){12}}
\put(-3,21){\line(1,0){6}}\put(-3,-21){\line(1,0){6}}
\put(0,-25){\line(0,1){50}}
\put(3,-21){\line(0,1){42}}\put(-3,-21){\line(0,1){42}}
\put(6,-18){\line(0,1){36}}\put(-6,-18){\line(0,1){36}}
\put(9,-15){\line(0,1){30}}\put(-9,-15){\line(0,1){30}}
\put(12,-12){\line(0,1){24}}\put(-12,-12){\line(0,1){24}}
\put(15,-9){\line(0,1){18}}\put(-15,-9){\line(0,1){18}}
\put(18,-6){\line(0,1){12}}\put(-18,-6){\line(0,1){12}}
\put(21,-3){\line(0,1){6}}\put(-21,-3){\line(0,1){6}}
\put(-16,16){\makebox(0,0){$A$}}
\put(-16,-16){\makebox(0,0){$B$}}
\put(16,-16){\makebox(0,0){$C$}}
\put(16,16){\makebox(0,0){$D$}}
\put(0,0){\makebox(0,0){$\bullet$}}
\put(2,2){\makebox(0,0){$O$}}
%
\put(25,0){\oval(15,15)[br]}
\put(25,0){\oval(15,15)[tr]}
\put(26,7.5){\vector(-1,0){1}}
\end{picture}
\ee
For each quadrant we assign a matrix $A,B,C,D$, which are called
corner transfer matrices. 
We take a multiplication of matrices counterclockwise.
Sites on each column (row) are numbered increasingly from south to north
(from east to west),  $\cdots,-2,-1,0,1,2,\cdots$, where the zeroth site
is the reference site $O$.
For example $A(u)=(A_{\underline{a},\underline{b}}(u))$ is
\be
\setlength{\unitlength}{1mm}
\begin{picture}(140,50)(-10,0)
\put(10,25){\makebox(0,0)[l]{
$A_{\underline{a},\underline{b}}(u)=\delta_{a_0,b_0}$}}
\put(40,6){\vector(1,0){25.5}}\put(65.5,6){\vector(1,0){7}}
\put(72.5,6){\vector(1,0){7}}\put(79.5,6){\line(1,0){2.5}}
\put(40,13){\vector(1,0){25.5}}\put(65.5,13){\vector(1,0){7}}
\put(72.5,13){\vector(1,0){7}}\put(79.5,13){\line(1,0){2.5}}
\put(47,20){\vector(1,0){25.5}}
\put(72.5,20){\vector(1,0){7}}\put(79.5,20){\line(1,0){2.5}}
\put(54,27){\vector(1,0){25.5}}\put(79.5,27){\line(1,0){2.5}}
\put(61,34){\line(1,0){21}}
\put(68,41){\line(1,0){14}}
\put(75,48){\line(1,0){7}}
\put(40,13){\line(0,-1){7}}
\put(47,20){\line(0,-1){14}}
\put(54,27){\line(0,-1){21}}
\put(61,34){\vector(0,-1){25.5}}\put(61,8.5){\line(0,-1){2.5}}
\put(68,41){\vector(0,-1){25.5}}
\put(68,15.5){\vector(0,-1){7}}\put(68,8.5){\line(0,-1){2.5}}
\put(75,48){\vector(0,-1){25.5}}\put(75,22.5){\vector(0,-1){7}}
\put(75,15.5){\vector(0,-1){7}}\put(75,8.5){\line(0,-1){2.5}}
\put(82,48){\vector(0,-1){25.5}}\put(82,22.5){\vector(0,-1){7}}
\put(82,15.5){\vector(0,-1){7}}\put(82,8.5){\line(0,-1){2.5}}
\put(64.5,9.5){\makebox(0,0){$u$}}
\put(71.5,9.5){\makebox(0,0){$u$}}
\put(78.5,9.5){\makebox(0,0){$u$}}
\put(71.5,16.5){\makebox(0,0){$u$}}
\put(78.5,16.5){\makebox(0,0){$u$}}
\put(78.5,23.5){\makebox(0,0){$u$}}
\put(61,3){\makebox(0,0){$\cdots$}}
\put(68,3){\makebox(0,0){$a_2$}}
\put(75,3){\makebox(0,0){$a_1$}}
\put(82,3){\makebox(0,0){$a_0$}}
\put(85,6){\makebox(0,0){$b_0$}}
\put(85,13){\makebox(0,0){$b_1$}}
\put(85,20){\makebox(0,0){$b_2$}}
\put(85,27){\makebox(0,0){$\vdots$}}
%
\put(54,13){\makebox(0,0){$\scr\bullet$}}
\put(61,13){\makebox(0,0){$\scr\bullet$}}
\put(68,13){\makebox(0,0){$\scr\bullet$}}
\put(75,13){\makebox(0,0){$\scr\bullet$}}
\put(61,20){\makebox(0,0){$\scr\bullet$}}
\put(68,20){\makebox(0,0){$\scr\bullet$}}
\put(75,20){\makebox(0,0){$\scr\bullet$}}
\put(68,27){\makebox(0,0){$\scr\bullet$}}
\put(75,27){\makebox(0,0){$\scr\bullet$}}
\put(75,34){\makebox(0,0){$\scr\bullet$}}
\put(90,23){,}
\end{picture}
\label{A}
\ee
where $\underline{a}=(a_0,a_1,a_2,\cdots)$, 
$\underline{b}=(b_0,b_1,b_2,\cdots)$, and
summation is taken over inner sites $\bullet$.
Using these CTM's the partition function is expressed as
\be
  Z=\tr\,(DCBA),
\ee
and one-point local height probability (LHP), which is a probability
of finding local height at O to be $k$, is
\be
  P_k=Z^{-1}\tr_{a_0=k}(DCBA).
\ee
To discuss these quantities, however, we have to specify the ground state.

We restrict ourselves to consider the following ground state configuration
\be
\setlength{\unitlength}{1mm}
\begin{picture}(140,50)(-30,5)
\put(10,10){\vector(1,0){32.5}}\put(42.5,10){\vector(1,0){7}}
\put(49.5,10){\vector(1,0){7}}\put(56.5,10){\line(1,0){2.5}}
\put(10,17){\vector(1,0){32.5}}\put(42.5,17){\vector(1,0){7}}
\put(49.5,17){\vector(1,0){7}}\put(56.5,17){\line(1,0){2.5}}
\put(10,24){\vector(1,0){39.5}}
\put(49.5,24){\vector(1,0){7}}\put(56.5,24){\line(1,0){2.5}}
\put(10,31){\line(1,0){49}}
\put(10,38){\line(1,0){49}}
\put(10,45){\line(1,0){49}}
\put(10,45){\line(0,-1){35}}
\put(17,45){\line(0,-1){35}}
\put(24,45){\line(0,-1){35}}
\put(31,45){\line(0,-1){35}}
\put(38,45){\vector(0,-1){32.5}}\put(38,12.5){\line(0,-1){2.5}}
\put(45,45){\vector(0,-1){25.5}}
\put(45,19.5){\vector(0,-1){7}}\put(45,12.5){\line(0,-1){2.5}}
\put(52,45){\vector(0,-1){25.5}}
\put(52,19.5){\vector(0,-1){7}}\put(52,12.5){\line(0,-1){2.5}}
\put(59,45){\vector(0,-1){25.5}}
\put(59,19.5){\vector(0,-1){7}}\put(59,12.5){\line(0,-1){2.5}}
\put(41.5,13.5){\makebox(0,0){$u$}}
\put(48.5,13.5){\makebox(0,0){$u$}}
\put(55.5,13.5){\makebox(0,0){$u$}}
\put(48.5,20.5){\makebox(0,0){$u$}}
\put(55.5,20.5){\makebox(0,0){$u$}}
\multiput(8.5,8.5)(1,1){40}{\makebox(0,0){$\cdot$}}
\multiput(8.5,15.5)(1,1){33}{\makebox(0,0){$\cdot$}}
\multiput(8.5,22.5)(1,1){26}{\makebox(0,0){$\cdot$}}
\multiput(8.5,29.5)(1,1){19}{\makebox(0,0){$\cdot$}}
\multiput(8.5,36.5)(1,1){12}{\makebox(0,0){$\cdot$}}
\multiput(15.5,8.5)(1,1){40}{\makebox(0,0){$\cdot$}}
\multiput(22.5,8.5)(1,1){40}{\makebox(0,0){$\cdot$}}
\multiput(29.5,8.5)(1,1){40}{\makebox(0,0){$\cdot$}}
\put(20.5,50){\makebox(0,0){$i_{\alpha}$}}
\put(27.5,50){\makebox(0,0){$\cdots$}}
\put(34.5,50){\makebox(0,0){$i_2$}}
\put(41.5,50){\makebox(0,0){$i_1$}}
\put(48.5,50){\makebox(0,0){$i_{\alpha}$}}
\put(55.5,50){\makebox(0,0){$\cdots$}}
\put(62.5,50){\makebox(0,0){$i_2$}}
\put(69.5,50){\makebox(0,0){$i_1$}}
\put(70,10){\makebox(0,0){,}}
\end{picture}
\ee
which is invariant under the shift in the NE-SW direction.
Many interesting models have this type of ground states.
We label this ground state configuration (i.e. an ordered set  
$(i_1,i_2,\cdots,i_{\alpha})$ whose cyclic permutations are identified) 
as an integer $l$.
On each column or row the height sequence is 
$\cdots,i_1,i_2,\cdots,i_{\alpha},i_1,i_2,\cdots,i_{\alpha},\cdots$, 
but there are $\alpha$ configurations because the height on the zeroth site
takes one of $\alpha$ values $i_{m+1}$.
We distinguish these configurations by an integer $m\in\Z/\alpha\Z$.
Namely the ground state configuration $(l,m)$, which we denote
$\bar{a}(l,m)$, is
\ba
  &&\bar{a}_{j-m}(l,m)=i_{j+1}\quad(0\leq j\leq\alpha-1),\n
  &&\bar{a}_{j+\alpha}(l,m)=\bar{a}_j(l,m)\quad(j\in\Z).
\ea
When we discuss this face model, we impose the condition that
heights at the boundary (or sites far away from O) 
take one of the ground state configurations.

In the north and west directions the ground state configurations
are $\cdots,i_1,i_2$,$\cdots$,$i_{\alpha},\cdots$.
Let $\cH^{(k)}_{l,m}$ denote the space of states of the half-infinite
lattice where the central height (i.e. the one on the zeroth site) is
fixed to $k$ and the boundary heights are in the ground state $(l,m)$.
Formally it is 
\be
  \cH^{(k)}_{l,m}=\mbox{Span}
  \Bigl\{(a_j)_{j=0}^{\infty}\,\Bigm|\,a_j\in I,\,a_0=k,\,
  a_j=\bar{a}_j(l,m)\;(j\gg 1)\Bigr\}.
\ee
In the south and east directions we relabel the numbering of sites :
$0,-1,-2,\cdots$ $\mapsto$ $0,1,2,\cdots$. 
Then the ground state configurations are in the south and east directions are 
$\cdots,i_{\alpha},\cdots,i_2,i_1,\cdots$.
So we introduce 
\be
  \cH^{\,\prime(k)}_{l,m}=\mbox{Span}
  \Bigl\{(a_j)_{j=0}^{\infty}\,\Bigm|\,a_j\in I,\,a_0=k,\,
  a_j=\bar{a}'_j(l,m)\;(j\gg 1)\Bigr\},
\ee
where $\bar{a}'_j(l,m)=\bar{a}_{-j}(l,m)$. 
We have
\be
\setlength{\unitlength}{1mm}
\begin{picture}(140,50)(-70,-25)
\put(-25,0){\line(1,0){50}}
\put(0,-25){\line(0,1){50}}
\put(-16,16){\makebox(0,0){$A$}}
\put(-16,-16){\makebox(0,0){$B$}}
\put(16,-16){\makebox(0,0){$C$}}
\put(16,16){\makebox(0,0){$D$}}
\put(0,0){\makebox(0,0){$\bullet$}}
\put(2,2){\makebox(0,0){$k$}}
\put(-2,-2){\makebox(0,0){$O$}}
\put(1,25){\makebox(0,0)[lt]{$\bar{a}(l,m)$}}
\put(1,-25){\makebox(0,0)[lb]{$\bar{a}'(l,m)$}}
\put(-25,1){\makebox(0,0)[bl]{$\bar{a}(l,m)$}}
\put(25,1){\makebox(0,0)[br]{$\bar{a}'(l,m)$}}
%
\end{picture}
\ee
For later convenience we define
\be
  \cH^{(k)}_l=\bigoplus_m\cH^{(k)}_{l,m},\quad
  \cH_l=\bigoplus_k\cH^{(k)}_l,\quad
  \cH^{\,\prime(k)}_l=\bigoplus_m\cH^{\,\prime(k)}_{l,m},\quad
  \cH^{\,\prime}_l=\bigoplus_k\cH^{\,\prime(k)}_l.
\ee

Since the factor $\delta_{a_0,b_0}$ in \eq{A}, $A(u)$ has a block structure
\be
  A=\bigoplus_kA^{(k)},
\ee
where $A^{(k)}$ is a matrix in $a_0=k$ sector. $B,C$ and $D$ have also this
block structure. CTM's are linear maps:
\ba
  A^{(k)}&:&\cH^{(k)}_{l,m}\rightarrow\cH^{(k)}_{l,m}\,,\n
  B^{(k)}&:&\cH^{(k)}_{l,m}\rightarrow\cH^{\,\prime(k)}_{l,m},\\
  C^{(k)}&:&\cH^{\,\prime(k)}_{l,m}\rightarrow\cH^{\,\prime(k)}_{l,m},\n
  D^{(k)}&:&\cH^{\,\prime(k)}_{l,m}\rightarrow\cH^{(k)}_{l,m}\,.\nonumber
\ea
One-point LHP $P_k(l,m)$, which is a probability of finding local 
height at O to be $k$ and the boundary heights in the north direction
on the column through $O$ is in the ground state $(l,m)$, is
\be
  P_k(l,m)=Z_{l,m}^{-1}\tr_{\cH^{(k)}_{l,m}}(D^{(k)}C^{(k)}B^{(k)}A^{(k)}),
  \label{Pklm}
\ee
where $Z_{l,m}$ is the partition function for the ground state $(l,m)$ 
\be
  Z_{l,m}=\sum_k\tr_{\cH^{(k)}_{l,m}}(D^{(k)}C^{(k)}B^{(k)}A^{(k)}).
  \label{Zlm}
\ee

Baxter's important observation is the following.
Let us assume that the Boltzmann weight $W$ satisfies the YBE \eq{YBEW},\\
initial condition
\be
  W\BW{a}{b}{c}{d}{0}=\delta_{b,c},
  \label{initial}
\ee
unitarity
\be
  \sum_gW\BW{a}{b}{g}{d}{u}W\BW{a}{g}{c}{d}{-u}=\delta_{b,c},
  \label{unitarity}
\ee
and the second inversion relation
\be
  \sum_gG_gW\BW{a}{b}{c}{g}{\lambda-u}W\BW{d}{c}{b}{g}{\lambda+u}
  =\frac{G_bG_c}{G_a}\delta_{a,d},
  \label{2ndinv}
\ee
where $\lambda$ and $G_a$ are some constants, and
$W$ is a double periodic function in $u$.
Then in the thermodynamic limit (infinite lattice size limit),
$u$-dependence of CTM becomes
\be
  \begin{array}{lcl}
  A^{(a)}(u)=x^{-2uH_C^{(a)}}&:&\cH^{(a)}_{l,m}\rightarrow\cH^{(a)}_{l,m}\,,\\
  B^{(a)}(u)=\sqrt{G_a}{P^{(a)}}^{-1}A^{(a)}(\lambda-u)
    &:&\cH^{(a)}_{l,m}\rightarrow\cH^{\,\prime(a)}_{l,m}\,,\\
  C^{(a)}(u)={P^{(a)}}^{-1}A^{(a)}(u)P^{(a)}
    &:&\cH^{\,\prime(a)}_{l,m}\rightarrow\cH^{\,\prime(a)}_{l,m}\,,\\
  D^{(a)}(u)=\sqrt{G_a}A^{(a)}(\lambda-u)P^{(a)}
    &:&\cH^{\,\prime(a)}_{l,m}\rightarrow\cH^{(a)}_{l,m}\,,
  \end{array}
  \label{ABCD}
\ee
where $H_C^{(a)}$ and $P^{(a)}$ are $u$-independent matrices 
and $x$ is a parameter of the model.
We have neglected multiplicative constant factors (normalization of CTM's)
because they do not contribute to LHP.
$P^{(a)}$ gives an isomorphism of vector spaces
\be
  P^{(a)}\,:\,\cH^{\,\prime(a)}_{l,m}\rightarrow\cH^{(a)}_{l,m}\,.
\ee
The spectrum of the corner Hamiltonian $H_C$ has properties
$$
  \mbox{bounded below},\quad\mbox{discrete},\quad\mbox{equidistance}.
$$
The multiplicity of the spectrum of $H_C$ depends on the model and
it is summarized in a character 
\be
  \chi_{l,m,k}(q)=\tr_{\cH^{(k)}_{l,m}}q^{H_C^{(k)}}.
  \label{chi}
\ee
By using these, one-point LHP \eq{Pklm} becomes
\be
  P_k(l,m)=Z_{l,m}^{-1}G_k\,\tr_{\cH^{(k)}_{l,m}}x^{-4\lambda H_C^{(k)}}
  =Z_{l,m}^{-1}G_k\,\chi_{l,m,k}(x^{-4\lambda}),
  \label{1pt}
\ee
where the partition function \eq{Zlm} is
\be
  Z_{l,m}=\sum_kG_k\,\chi_{l,m,k}(x^{-4\lambda}).
  \label{Zlchi}
\ee
We remark that this $Z_{l,m}$ is independent on $m$ 
because we are considering an infinitely large lattice.

To formulate multi-point LHP let us introduce the vertex operator (VO) of 
type I \cite{Fo94,JM}. Foda et al. \cite{Fo94} presented a formulation of 
VO's in solvable lattice models.
We explain a face model case following refs. \cite{Fo94,JM}.
Here we use intuitive graphical argument.
For precise representation theoretical argument, see subsection \ref{sec:4.4}. 

Graphically the Boltzmann weight is
\be
\setlength{\unitlength}{1mm}
\begin{picture}(140,41)
\put(10,10){\makebox(50,20)[l]{${\displaystyle W\BW{a}{b}{c}{d}{u_1-u_2}\;=}$}}
\put(60,11){\vector(1,0){10}} \put(70,11){\line(1,0){8}}
\put(60,29){\vector(1,0){10}} \put(70,29){\line(1,0){8}}
\put(60,29){\vector(0,-1){10}} \put(60,19){\line(0,-1){8}}
\put(78,29){\vector(0,-1){10}} \put(78,19){\line(0,-1){8}}
\put(55,29){\makebox(5,5){$a$}}
\put(55,6){\makebox(5,5){$c$}}
\put(78,29){\makebox(5,5){$b$}}
\put(78,6){\makebox(5,5){$d$}}
\multiput(84.5,20)(-2,0){3}{\line(-1,0){1}}
\multiput(76.5,20)(-2,0){8}{\line(-1,0){1}}
\multiput(58.5,20)(-2,0){3}{\line(-1,0){1}}
\multiput(69,35.5)(0,-2){3}{\line(0,-1){1}}
\multiput(69,27.5)(0,-2){8}{\line(0,-1){1}}
\multiput(69,9.5)(0,-2){3}{\line(0,-1){1}}
\put(49.5,17.5){\makebox(5,5){$\lhd$}}
\put(66.5,1){\makebox(5,5){${\scriptstyle \bigtriangledown}$}}
\put(66.5,35.5){\makebox(5,5){$u_1$}}
\put(84.5,17.5){\makebox(5,5){$u_2$}}
\put(69,6){\makebox(5,5){$\ve_1$}}
\put(69,29){\makebox(5,5){$\ve'_1$}}
\put(55,15){\makebox(5,5){$\ve_2$}}
\put(78,15){\makebox(5,5){$\ve'_2$}}
\put(89.5,17.5){\makebox(5,5)[b]{,}}
\put(100,23.5){\makebox(20,5)[l]{$b=a+\ve'_1$,}}
\put(100,17.5){\makebox(20,5)[l]{$c=a+\ve_2$,}}
\put(100,11.5){\makebox(20,5)[l]{$d=b+\ve'_2=c+\ve_1$.}}
\end{picture}
\ee
Let $\Phi^{(a,b)}_N(z)$, $\Phi^{(a,b)}_W(z^{-1})$, $\Phi^{(a,b)}_S(z)$ and 
$\Phi^{(a,b)}_E(z^{-1})$ be the half-infinite transfer matrices extending to 
infinity in the north, west, south and east direction respectively
($z=x^{2u}$) :
\be
\setlength{\unitlength}{1mm}
\begin{picture}(140,80)(-10,0)
\put(50,50){
\begin{picture}(15,35)(10,10)
\put(10,38){\vector(0,-1){18.5}}
\put(10,19.5){\vector(0,-1){7}}\put(10,12.5){\line(0,-1){2.5}}
\put(17,38){\vector(0,-1){18.5}}
\put(17,19.5){\vector(0,-1){7}}\put(17,12.5){\line(0,-1){2.5}}
\put(10,10){\vector(1,0){4.5}}\put(14.5,10){\line(1,0){2.5}}
\put(10,17){\vector(1,0){4.5}}\put(14.5,17){\line(1,0){2.5}}
\put(10,24){\vector(1,0){4.5}}\put(14.5,24){\line(1,0){2.5}}
\put(13.5,13.5){\makebox(0,0){$u$}}
\put(13.5,20.5){\makebox(0,0){$u$}}
\put(13.5,34.5){\makebox(0,0){$\vdots$}}
\put(8,8){\makebox(0,0){$a$}}
\put(8,34.5){\makebox(0,0)[r]{$\scr\bar{a}(l,m+1)$}}
\put(19,8){\makebox(0,0){$b$}}
\put(19,34.5){\makebox(0,0)[l]{$\scr\bar{a}(l,m)$}}
\end{picture}
}
\put(65,65){\makebox(0,0)[l]{$=\,\Phi_N^{(a,b)}(z)
$}}
\put(50,0){
\begin{picture}(15,35)(10,10)
\put(10,38){\vector(0,-1){4.5}}
\put(10,33.5){\vector(0,-1){7}}\put(10,26.5){\line(0,-1){16.5}}
\put(17,38){\vector(0,-1){4.5}}
\put(17,33.5){\vector(0,-1){7}}\put(17,26.5){\line(0,-1){16.5}}
\put(10,38){\vector(1,0){4.5}}\put(14.5,38){\line(1,0){2.5}}
\put(10,31){\vector(1,0){4.5}}\put(14.5,31){\line(1,0){2.5}}
\put(10,24){\vector(1,0){4.5}}\put(14.5,24){\line(1,0){2.5}}
\put(13.5,34.5){\makebox(0,0){$u$}}
\put(13.5,27.5){\makebox(0,0){$u$}}
\put(13.5,13.5){\makebox(0,0){$\vdots$}}
\put(8,40){\makebox(0,0){$b$}}
\put(8,13.5){\makebox(0,0)[r]{$\scr\bar{a}'(l,m)$}}
\put(19,40){\makebox(0,0){$a$}}
\put(19,13.5){\makebox(0,0)[l]{$\scr\bar{a}'(l,m-1)$}}
\end{picture}
}
\put(65,10){\makebox(0,0)[l]{$=\,\Phi_S^{(a,b)}(z)
$}}
\put(10,35){
\begin{picture}(35,15)(10,10)
\put(10,10){\vector(1,0){18.5}}
\put(28.5,10){\vector(1,0){7}}\put(35.5,10){\line(1,0){2.5}}
\put(10,17){\vector(1,0){18.5}}
\put(28.5,17){\vector(1,0){7}}\put(35.5,17){\line(1,0){2.5}}
\put(24,17){\vector(0,-1){4.5}}\put(24,12.5){\line(0,-1){2.5}}
\put(31,17){\vector(0,-1){4.5}}\put(31,12.5){\line(0,-1){2.5}}
\put(38,17){\vector(0,-1){4.5}}\put(38,12.5){\line(0,-1){2.5}}
\put(27.5,13.5){\makebox(0,0){$u$}}
\put(34.5,13.5){\makebox(0,0){$u$}}
\put(13.5,13.5){\makebox(0,0){$\cdots$}}
\put(40,8){\makebox(0,0){$a$}}
\put(10,8){\makebox(0,0)[l]{$\scr\bar{a}(l,m-1)$}}
\put(40,19){\makebox(0,0){$b$}}
\put(10,19){\makebox(0,0)[l]{$\scr\bar{a}(l,m)$}}
\end{picture}
}
\put(10,25){\makebox(0,0)[l]{$=\,\Phi_W^{(a,b)}(z^{-1})$}}
%
\put(69,35){
\begin{picture}(35,15)(10,10)
\put(10,10){\vector(1,0){4.5}}
\put(14.5,10){\vector(1,0){7}}\put(21.5,10){\line(1,0){16.5}}
\put(10,17){\vector(1,0){4.5}}
\put(14.5,17){\vector(1,0){7}}\put(21.5,17){\line(1,0){16.5}}
\put(10,17){\vector(0,-1){4.5}}\put(10,12.5){\line(0,-1){2.5}}
\put(17,17){\vector(0,-1){4.5}}\put(17,12.5){\line(0,-1){2.5}}
\put(24,17){\vector(0,-1){4.5}}\put(24,12.5){\line(0,-1){2.5}}
\put(13.5,13.5){\makebox(0,0){$u$}}
\put(20.5,13.5){\makebox(0,0){$u$}}
\put(34.5,13.5){\makebox(0,0){$\cdots$}}
\put(8,8){\makebox(0,0){$b$}}
\put(38,8){\makebox(0,0)[r]{$\scr\bar{a}'(l,m)$}}
\put(8,19){\makebox(0,0){$a$}}
\put(38,19){\makebox(0,0)[r]{$\scr\bar{a}'(l,m+1)$}}
\end{picture}
}
\put(75,25){\makebox(0,0)[l]{$=\,\Phi_E^{(a,b)}(z^{-1})
$}}
\put(100,5){\makebox(0,0){.}}
\end{picture}
\ee
They are linear maps:
\ba
  \Phi_N^{(a,b)}(z)&:&\cH^{(b)}_{l,m}\rightarrow\cH^{(a)}_{l,m+1}\,,\n
  \Phi_S^{(a,b)}(z)&:&\cH^{\,\prime(b)}_{l,m}\rightarrow
  \cH^{\,\prime(a)}_{l,m-1}\,,\\
  \Phi_W^{(a,b)}(z)&:&\cH^{(b)}_{l,m}\rightarrow\cH^{(a)}_{l,m-1}\,,\n
  \Phi_E^{(a,b)}(z)&:&\cH^{\,\prime(b)}_{l,m}\rightarrow
  \cH^{\,\prime(a)}_{l,m+1}\,.\nonumber
\ea
For CTM's and type I VO's we suppress $l$ dependence.

$\Phi_N^{(a,b)}(z)$ satisfies the commutation relation
\be
  \Phi_N^{(a,b)}(z_2)\Phi_N^{(b,c)}(z_1)=\sum_{g}W\BW{a}{g}{b}{c}{u_1-u_2}
  \Phi_N^{(a,g)}(z_1)\Phi_N^{(g,c)}(z_2)\quad (z_j=x^{2u_j}),
  \label{PhiNPhiN}
\ee
which can be shown by repeated use of \eq{YBEW},
\be
\setlength{\unitlength}{1mm}
\begin{picture}(140,60)(-10,5)
\put(0,55){\makebox(0,0){RHS =}}
\put(10,64){\vector(0,-1){23}}
\put(10,41){\vector(0,-1){9}}
\put(10,32){\line(0,-1){4}}
\put(19,64){\vector(0,-1){23}}
\put(19,41){\vector(0,-1){9}}
\put(19,32){\line(0,-1){4}}
\put(28,64){\vector(0,-1){23}}
\put(28,41){\vector(0,-1){9}}
\put(28,32){\line(0,-1){4}}
\put(10,28){\vector(1,0){5}}
\put(15,28){\vector(1,0){9}}
\put(24,28){\line(1,0){4}}
\put(10,37){\vector(1,0){5}}
\put(15,37){\vector(1,0){9}}
\put(24,37){\line(1,0){4}}
\put(10,46){\vector(1,0){5}}
\put(15,46){\vector(1,0){9}}
\put(24,46){\line(1,0){4}}
\put(10,19){\vector(1,1){5}}\put(19,28){\line(-1,-1){4}}
\put(10,19){\vector(1,-1){5}}\put(19,10){\line(-1,1){4}}
\put(19,28){\vector(1,-1){5}}\put(28,19){\line(-1,1){4}}
\put(19,10){\vector(1,1){5}}\put(28,19){\line(-1,-1){4}}
\multiput(10,19)(0,1){9}{\line(0,1){0.5}}
\multiput(28,19)(0,1){9}{\line(0,1){0.5}}
\put(19,28){\makebox(0,0){$\scr\bullet$}}
\put(19,37){\makebox(0,0){$\scr\bullet$}}
\put(19,46){\makebox(0,0){$\scr\bullet$}}
\put(19,19){\makebox(0,0){$u_1-u_2$}}
\put(14.5,32.5){\makebox(0,0){$u_1$}}
\put(14.5,41.5){\makebox(0,0){$u_1$}}
\put(23.5,32.5){\makebox(0,0){$u_2$}}
\put(23.5,41.5){\makebox(0,0){$u_2$}}
\put(7,19){\makebox(0,0){$a$}}
\put(7,28){\makebox(0,0){$a$}}
\put(7,37){\makebox(0,0){$a_1$}}
\put(7,46){\makebox(0,0){$a_2$}}
\put(31,19){\makebox(0,0){$c$}}
\put(31,28){\makebox(0,0){$c$}}
\put(31,37){\makebox(0,0){$c_1$}}
\put(31,46){\makebox(0,0){$c_2$}}
\put(19,7){\makebox(0,0){$b$}}
\put(14.5,55){\makebox(0,0){$\vdots$}}
\put(23.5,55){\makebox(0,0){$\vdots$}}
\put(40,55){\makebox(0,0){=}}
\put(50,64){\vector(0,-1){23}}\put(50,41){\line(0,-1){4}}
\put(59,64){\vector(0,-1){23}}\put(59,41){\line(0,-1){4}}
\put(68,64){\vector(0,-1){23}}\put(68,41){\line(0,-1){4}}
\put(50,37){\vector(1,0){5}}
\put(55,37){\vector(1,0){9}}\put(64,37){\line(1,0){4}}
\put(50,46){\vector(1,0){5}}
\put(55,46){\vector(1,0){9}}\put(64,46){\line(1,0){4}}
\put(50,28){\vector(1,1){5}}\put(59,37){\line(-1,-1){4}}
\put(50,28){\vector(1,-1){5}}\put(59,19){\line(-1,1){4}}
\put(59,37){\vector(1,-1){5}}\put(68,28){\line(-1,1){4}}
\put(59,19){\vector(1,1){5}}\put(68,28){\line(-1,-1){4}}
\put(50,19){\vector(0,-1){5}}\put(50,14){\line(0,-1){4}}
\put(59,19){\vector(0,-1){5}}\put(59,14){\line(0,-1){4}}
\put(68,19){\vector(0,-1){5}}\put(68,14){\line(0,-1){4}}
\put(50,19){\vector(1,0){5}}
\put(55,19){\vector(1,0){9}}\put(64,19){\line(1,0){4}}
\put(50,10){\vector(1,0){5}}
\put(55,10){\vector(1,0){9}}\put(64,10){\line(1,0){4}}
\multiput(50,19)(0,1){18}{\line(0,1){0.5}}
\multiput(68,19)(0,1){18}{\line(0,1){0.5}}
\put(59,19){\makebox(0,0){$\scr\bullet$}}
\put(59,37){\makebox(0,0){$\scr\bullet$}}
\put(59,46){\makebox(0,0){$\scr\bullet$}}
\put(59,28){\makebox(0,0){$u_1-u_2$}}
\put(54.5,41.5){\makebox(0,0){$u_1$}}
\put(54.5,14.5){\makebox(0,0){$u_2$}}
\put(63.5,41.5){\makebox(0,0){$u_2$}}
\put(63.5,14.5){\makebox(0,0){$u_1$}}
\put(47,10){\makebox(0,0){$a$}}
\put(47,19){\makebox(0,0){$a_1$}}
\put(47,28){\makebox(0,0){$a_1$}}
\put(47,37){\makebox(0,0){$a_1$}}
\put(47,46){\makebox(0,0){$a_2$}}
\put(71,10){\makebox(0,0){$c$}}
\put(71,19){\makebox(0,0){$c_1$}}
\put(71,28){\makebox(0,0){$c_1$}}
\put(71,37){\makebox(0,0){$c_1$}}
\put(71,46){\makebox(0,0){$c_2$}}
\put(59,7){\makebox(0,0){$b$}}
\put(54.5,55){\makebox(0,0){$\vdots$}}
\put(63.5,55){\makebox(0,0){$\vdots$}}
\put(80,55){\makebox(0,0){=}}
\put(90,64){\vector(0,-1){32}}
\put(90,32){\vector(0,-1){9}}\put(90,23){\line(0,-1){4}}
\put(99,64){\vector(0,-1){32}}
\put(99,32){\vector(0,-1){9}}\put(99,23){\line(0,-1){4}}
\put(108,64){\vector(0,-1){32}}
\put(108,32){\vector(0,-1){9}}\put(108,23){\line(0,-1){4}}
\put(90,37){\vector(1,0){5}}
\put(95,37){\vector(1,0){9}}\put(104,37){\line(1,0){4}}
\put(90,28){\vector(1,0){5}}
\put(95,28){\vector(1,0){9}}\put(104,28){\line(1,0){4}}
\put(90,19){\vector(1,0){5}}
\put(95,19){\vector(1,0){9}}\put(104,19){\line(1,0){4}}
\put(99,28){\makebox(0,0){$\scr\bullet$}}
\put(99,37){\makebox(0,0){$\scr\bullet$}}
\put(94.5,32.5){\makebox(0,0){$u_2$}}
\put(94.5,23.5){\makebox(0,0){$u_2$}}
\put(103.5,32.5){\makebox(0,0){$u_1$}}
\put(103.5,23.5){\makebox(0,0){$u_1$}}
\put(87,19){\makebox(0,0){$a$}}
\put(87,28){\makebox(0,0){$a_1$}}
\put(87,37){\makebox(0,0){$a_2$}}
\put(111,19){\makebox(0,0){$c$}}
\put(111,28){\makebox(0,0){$c_1$}}
\put(111,37){\makebox(0,0){$c_2$}}
\put(99,16){\makebox(0,0){$b$}}
\put(94.5,55){\makebox(0,0){$\vdots$}}
\put(103.5,55){\makebox(0,0){$\vdots$}}
\put(120,55){\makebox(0,0){= LHS.}}
\end{picture}
\label{WPhiPhi}
\ee
Similarly we have ($z_j=x^{2u_j}$)
\ba
  &&\Phi_W^{(a,b)}(z_2)\Phi_W^{(b,c)}(z_1)=\sum_{g}W\BW{c}{b}{g}{a}{u_1-u_2}
  \Phi_W^{(a,g)}(z_1)\Phi_W^{(g,c)}(z_2),\\
  &&\Phi_N^{(a,b)}(z_1)\Phi_W^{(b,c)}(z_2)=\sum_{g}W\BW{g}{c}{a}{b}{u_1-u_2}
  \Phi_W^{(a,g)}(z_2)\Phi_N^{(g,c)}(z_1),
\ea
and 
\be
  \Phi_S^{(a,b)}(z_2)\Phi_S^{(b,c)}(z_1)=\sum_{g}W\BW{c}{b}{g}{a}{u_1-u_2}
  \Phi_S^{(a,g)}(z_1)\Phi_S^{(g,c)}(z_2),\quad\mbox{etc.}
  \label{PhiSPhiS}
\ee
The initial condition \eq{initial} implies
\ba
  \!\!\!\!\!\!\!\!\!\!&&
  A^{(a)}(u)\Phi_N^{(a,b)}(z)=\Phi_N^{(a,b)}(1)A^{(b)}(u),\quad
  \Phi_W^{(a,b)}(z^{-1})A^{(b)}(u)=A^{(a)}(u)\Phi_W^{(a,b)}(1),\n
  \!\!\!\!\!\!\!\!\!\!&&
  B^{(a)}(u)\Phi_W^{(a,b)}(1)=\Phi_S^{(a,b)}(z)B^{(b)}(u),\quad
  \Phi_S^{(a,b)}(1)B^{(b)}(u)=B^{(a)}(u)\Phi_W^{(a,b)}(z^{-1}),\\
  \!\!\!\!\!\!\!\!\!\!&&
  C^{(a)}(u)\Phi_S^{(a,b)}(z)=\Phi_S^{(a,b)}(1)C^{(b)}(u),\quad
  \Phi_E^{(a,b)}(z^{-1})C^{(b)}(u)=C^{(a)}(u)\Phi_E^{(a,b)}(1),\n
  \!\!\!\!\!\!\!\!\!\!&&
  D^{(a)}(u)\Phi_E^{(a,b)}(1)=\Phi_N^{(a,b)}(z)D^{(b)}(u),\quad
  \Phi_N^{(a,b)}(1)D^{(b)}(u)=D^{(a)}(u)\Phi_E^{(a,b)}(z^{-1}),\nonumber
\ea
and therefore by combining \eq{ABCD} we have 
\ba
  &&\Phi_N^{(a,b)}(z)=x^{2uH_C^{(a)}}\Phi_N^{(a,b)}(1)x^{-2uH_C^{(b)}},
  \label{APhiN}\\
  &&\Phi_W^{(a,b)}(z)=x^{2uH_C^{(a)}}\Phi_W^{(a,b)}(1)x^{-2uH_C^{(b)}},\\
  &&P^{(a)}\Phi_S^{(a,b)}(z){P^{(b)}}^{-1}=
  x^{2uH_C^{(a)}}P^{(a)}\Phi_S^{(a,b)}(1){P^{(b)}}^{-1}x^{-2uH_C^{(b)}},\\
  &&P^{(a)}\Phi_E^{(a,b)}(z){P^{(b)}}^{-1}=
  x^{2uH_C^{(a)}}P^{(a)}\Phi_E^{(a,b)}(1){P^{(b)}}^{-1}x^{-2uH_C^{(b)}},\\
  &&P^{(a)}\Phi_S^{(a,b)}(z){P^{(b)}}^{-1}=
  \sqrt{\frac{G_a}{G_b}}\Phi_W^{(a,b)}(x^{-2\lambda}z),
  \label{PhiSW} \\
  &&P^{(a)}\Phi_E^{(a,b)}(z){P^{(b)}}^{-1}=
  \sqrt{\frac{G_b}{G_a}}\Phi_N^{(a,b)}(x^{2\lambda}z).
\ea
The unitarity \eq{unitarity} gives
\ba
  &&\sum_g\Phi_W^{(a,g)}(z)\Phi_N^{(g,a)}(z)=\id,
  \label{sumPhiWPhiN}\\
  &&\sum_g\Phi_E^{(a,g)}(z)\Phi_S^{(g,a)}(z)=\id.
\ea
So \eq{PhiNPhiN},\eq{initial} and \eq{sumPhiWPhiN} imply
\be
  \Phi_N^{(a,b)}(z)\Phi_W^{(b,a)}(z)=\id.
\ee
The second inversion relation \eq{2ndinv} entails
\be
  G_a^{-1}\sum_gG_g\Phi_N^{(a,g)}(x^{2\lambda}z)\Phi_W^{(g,a)}(x^{-2\lambda}z)
  =\id.
  \label{sumPhiNPhiW}
\ee

Consider neighboring $n+1$ sites in a row, whose most east site is the
reference site $O$.
We divide a lattice into $2n+4$ parts,
\be
\setlength{\unitlength}{1mm}
\begin{picture}(140,60)(-70,-30)
\put(-44,2){\vector(1,0){25}}\put(-19,2){\line(1,0){3}}
\put(-30,9){\vector(1,0){11}}\put(-19,9){\line(1,0){3}}
\put(-23,16){\line(1,0){7}}
\put(-30,9){\line(0,-1){7}}
\put(-23,16){\vector(0,-1){11}}\put(-23,5){\line(0,-1){3}}
\put(-16,30){\vector(0,-1){25}}\put(-16,5){\line(0,-1){3}}
\put(-35,21){\makebox(0,0){$A(u)$}}
\put(-19.5,5.5){\makebox(0,0){$u$}}
\put(-14,2){\vector(1,0){4}}\put(-10,2){\line(1,0){3}}
\put(-14,9){\vector(1,0){4}}\put(-10,9){\line(1,0){3}}
\put(-14,16){\line(1,0){7}}
\put(-14,30){\vector(0,-1){25}}\put(-14,5){\line(0,-1){3}}
\put(-7,30){\vector(0,-1){25}}\put(-7,5){\line(0,-1){3}}
\put(-10.5,5.5){\makebox(0,0){$u$}}
\put(-10.5,26.5){\makebox(0,0){$\Phi_N$}}
\put(0,19.5){\makebox(0,0){$\cdots$}}
\put(7,2){\vector(1,0){4}}\put(11,2){\line(1,0){3}}
\put(7,9){\vector(1,0){4}}\put(11,9){\line(1,0){3}}
\put(7,16){\line(1,0){7}}
\put(7,30){\vector(0,-1){25}}\put(7,5){\line(0,-1){3}}
\put(14,30){\vector(0,-1){25}}\put(14,5){\line(0,-1){3}}
\put(10.5,5.5){\makebox(0,0){$u$}}
\put(10.5,26.5){\makebox(0,0){$\Phi_N$}}
\put(16,2){\vector(1,0){4}}\put(20,2){\line(1,0){24}}
\put(16,9){\vector(1,0){4}}\put(20,9){\line(1,0){10}}
\put(16,16){\line(1,0){7}}
\put(16,30){\vector(0,-1){25}}\put(16,5){\line(0,-1){3}}
\put(23,16){\vector(0,-1){11}}\put(23,5){\line(0,-1){3}}
\put(30,9){\line(0,-1){7}}
\put(35,21){\makebox(0,0){$D(u)$}}
\put(19.5,5.5){\makebox(0,0){$u$}}
\put(-15,0){\makebox(0,0){$\scr\bullet$}}
\put(-6,0){\makebox(0,0){$\scr\bullet$}}
\put(6,0){\makebox(0,0){$\scr\bullet$}}
\put(15,0){\makebox(0,0){$\scr\bullet$}}
\put(-14,0){\makebox(0,0)[l]{$\scr a_n$}}
\put(-5,0){\makebox(0,0)[l]{$\scr a_{n-1}$}}
\put(7,0){\makebox(0,0)[l]{$\scr a_1$}}
\put(16,0){\makebox(0,0)[l]{$\scr a_0$}}
\put(14,0){\makebox(0,0)[r]{$\scr O$}}
\put(-44,-2){\vector(1,0){25}}\put(-19,-2){\line(1,0){3}}
\put(-30,-9){\vector(1,0){11}}\put(-19,-9){\line(1,0){3}}
\put(-23,-16){\line(1,0){7}}
\put(-30,-2){\line(0,-1){7}}
\put(-23,-2){\vector(0,-1){4}}\put(-23,-6){\line(0,-1){10}}
\put(-16,-2){\vector(0,-1){4}}\put(-16,-6){\line(0,-1){24}}
\put(-35,-21){\makebox(0,0){$B(u)$}}
\put(-19.5,-5.5){\makebox(0,0){$u$}}
\put(-14,-2){\vector(1,0){4}}\put(-10,-2){\line(1,0){3}}
\put(-14,-9){\vector(1,0){4}}\put(-10,-9){\line(1,0){3}}
\put(-14,-16){\line(1,0){7}}
\put(-14,-2){\vector(0,-1){4}}\put(-14,-6){\line(0,-1){24}}
\put(-7,-2){\vector(0,-1){4}}\put(-7,-6){\line(0,-1){24}}
\put(-10.5,-5.5){\makebox(0,0){$u$}}
\put(-10.5,-26.5){\makebox(0,0){$\Phi_S$}}
\put(0,-19.5){\makebox(0,0){$\cdots$}}
\put(7,-2){\vector(1,0){4}}\put(11,-2){\line(1,0){3}}
\put(7,-9){\vector(1,0){4}}\put(11,-9){\line(1,0){3}}
\put(7,-16){\line(1,0){7}}
\put(7,-2){\vector(0,-1){4}}\put(7,-6){\line(0,-1){24}}
\put(14,-2){\vector(0,-1){4}}\put(14,-6){\line(0,-1){24}}
\put(10.5,-5.5){\makebox(0,0){$u$}}
\put(10.5,-26.5){\makebox(0,0){$\Phi_S$}}
\put(16,-2){\vector(1,0){4}}\put(20,-2){\line(1,0){24}}
\put(16,-9){\vector(1,0){4}}\put(20,-9){\line(1,0){10}}
\put(16,-16){\line(1,0){7}}
\put(16,-2){\vector(0,-1){4}}\put(16,-6){\line(0,-1){24}}
\put(23,-2){\vector(0,-1){4}}\put(23,-6){\line(0,-1){10}}
\put(30,-2){\line(0,-1){7}}
\put(35,-21){\makebox(0,0){$C(u)$}}
\put(19.5,-5.5){\makebox(0,0){$u$}}
\put(46,-2){\makebox(0,0){$.$}}
\end{picture}
\label{2n+2parts}
\ee
Let $P_{a_n,\cdots,a_1,a_0}(l,m)$ denote the probability 
of finding these local variables to be $(a_n,\cdots,a_0)$
under the condition that the boundary heights in the north direction 
on the column through $O$ are in the ground state $(l,m)$.
$P_{a_n,\cdots,a_0}(l,m)$ satisfies obvious recursion relations
and normalization condition,
\be
  \sum_{a_n}P_{a_n,\cdots,a_0}(l,m)=P_{a_{n-1},\cdots,a_0}(l,m),\quad
  \sum_{a_0}P_{a_n,\cdots,a_0}(l,m)=P_{a_n,\cdots,a_1}(l,m+1),
  \label{recP}
\ee
\be
  \sum_{a_0,\cdots,a_n}P_{a_n,\cdots,a_0}(l,m)=1.
\ee
{}From \eq{2n+2parts} we have 
\ba
  &&P_{a_n,\cdots,a_0}(l,m)=
  Z_{l,m}^{-1}\,\tr_{\cH^{(a_0)}_{l,m}}\Bigl(D^{(a_0)}(u)C^{(a_0)}(u)
  \Phi_S^{(a_0,a_1)}(z)\cdots\Phi_S^{(a_{n-1},a_n)}(z)\n
  &&\qquad\qquad\qquad\qquad\qquad\qquad\times
  B^{(a_n)}(u)A^{(a_n)}(u)
  \Phi_N^{(a_n,a_{n-1})}(z)\cdots\Phi_N^{(a_1,a_0)}(z)\Bigr).
\ea
By using
\be
  \Phi_S^{(a,b)}(w)B^{(b)}(u)A^{(b)}(u)=
  B^{(a)}(u)A^{(a)}(u)\Phi_W^{(a,b)}(w),
\ee
which is derived from \eq{ABCD} and \eq{PhiSW}, it can be written as
\ba
  \!\!\!\!\!\!\!\!&&P_{a_n,\cdots,a_0}(l,m)\n
  \!\!\!\!\!\!\!\!&=\!\!&
  Z_{l,m}^{-1}\,\tr_{\cH^{(a_0)}_{l,m}}\Bigl(
  D^{(a_0)}(u)C^{(a_0)}(u)B^{(a_0)}(u)A^{(a_0)}(u)\n
  \!\!\!\!\!\!\!\!&&\qquad\qquad\qquad\times
  \Phi_W^{(a_0,a_1)}(z)\cdots\Phi_W^{(a_{n-1},a_n)}(z)
  \Phi_N^{(a_n,a_{n-1})}(z)\cdots\Phi_N^{(a_1,a_0)}(z)\Bigr)
  \label{LHP}\\
  \!\!\!\!\!\!\!\!&=\!\!&
  Z_{l,m}^{-1}G_{a_0}\,\tr_{\cH^{(a_0)}_{l,m}}\Bigl(x^{-4\lambda H_C^{(a_0)}}
  \Phi_W^{(a_0,a_1)}(z)\cdots\Phi_W^{(a_{n-1},a_n)}(z)
  \Phi_N^{(a_n,a_{n-1})}(z)\cdots\Phi_N^{(a_1,a_0)}(z)\Bigr),\nonumber
\ea
where the partition function $Z_{l,m}$ is given in \eq{Zlchi}.
Using this expression and \eq{sumPhiWPhiN}, the cyclic property of trace,
\eq{APhiN} and \eq{sumPhiNPhiW}, we can reproduce the 
recursion relations \eq{recP}.
By changing spectral parameters of VO's 
we can formulate more general quantities
\ba
  \!\!\!\!\!\!\!\!&&P_{a_n,\cdots,a_0}
  (l,m\,;\mbox{${z_n,\cdots,z_1\atop z'_n,\cdots,z'_1}$})\\
  \!\!\!\!\!\!\!\!&=\!\!&
  Z_{l,m}^{-1}G_{a_0}\,\tr_{\cH^{(a_0)}_{l,m}}\Bigl(x^{-4\lambda H_C^{(a_0)}}
  \Phi_W^{(a_0,a_1)}(z'_1)\cdots\Phi_W^{(a_{n-1},a_n)}(z'_n)
  \Phi_N^{(a_n,a_{n-1})}(z_n)\cdots\Phi_N^{(a_1,a_0)}(z_1)\Bigr).\nonumber
\ea
We remark that these LHP's are independent of $z=x^{2u}$ because
\ba
  &&\tr\Bigl(x^{-4\lambda H_C}{\cal O}(w_i)\Bigr)=
  \tr\Bigl(w^{H_C}w^{-H_C}x^{-4\lambda H_C}{\cal O}(w_i)\Bigr)\n
  &=\!\!&\tr\Bigl(x^{-4\lambda H_C}w^{-H_C}{\cal O}(w_i)w^{H_C}\Bigr)=
  \tr\Bigl(x^{-4\lambda H_C}{\cal O}(\sfrac{w_i}{w})\Bigr).
\ea
By averaging $P_{a_n,\cdots,a_0}(l,m)$ over $m$ we obtain
\ba
  \!\!\!\!\!\!\!\!&&P_{a_n,\cdots,a_0}(l)
  =\sum_m\frac{Z_{l,m}}{Z_l}P_{a_n,\cdots,a_0}(l,m)
  \label{Pl}\\
  \!\!\!\!\!\!\!\!&=\!\!&
  Z_l^{-1}G_{a_0}\,\tr_{\cH^{(a_0)}_l}\Bigl(x^{-4\lambda H_C^{(a_0)}}
  \Phi_W^{(a_0,a_1)}(z)\cdots\Phi_W^{(a_{n-1},a_n)}(z)
  \Phi_N^{(a_n,a_{n-1})}(z)\cdots\Phi_N^{(a_1,a_0)}(z)\Bigr),\nonumber
\ea
where $Z_l$ is
\be
  Z_l=\sum_mZ_{l,m}=\sum_aG_a\tr_{\cH^{(a)}_l}
  x^{-4\lambda H_C^{(a)}}=\sum_{a,m}G_a\chi_{l,m,a}(x^{-4\lambda}).
  \label{Zl}
\ee

For later use we define the following quantities
\ba
  \!\!\!\!\!\!\!\!&&Q_{a_n,\cdots,a_0}(l,m\,;z|{\cal O})
  =Z_{l,m}^{-1}G_{a_0}\,\tr_{\cH^{(a_0)}_{l,m}}\Bigl(x^{-4\lambda H_C^{(a_0)}}
  {\cal O}_{\Phi}(z){\cal O}\Bigr),
  \label{Qlm}\\
  \!\!\!\!\!\!\!\!&&Q_{a_n,\cdots,a_0}(l\,;z|{\cal O})
  =\sum_m\frac{Z_{l,m}}{Z_l}Q_{a_n,\cdots,a_0}(l,m\,;z|{\cal O})
  =Z_l^{-1}G_{a_0}\,\tr_{\cH^{(a_0)}_l}\Bigl(x^{-4\lambda H_C^{(a_0)}}
  {\cal O}_{\Phi}(z){\cal O}\Bigr),\nonumber
\ea
where 
${\cal O}_{\Phi}(z)=\Phi_W^{(a_0,a_1)}(z)\cdots\Phi_W^{(a_{n-1},a_n)}(z)
\Phi_N^{(a_n,a_{n-1})}(z)\cdots\Phi_N^{(a_1,a_0)}(z)$ and 
${\cal O}$ is a linear map 
${\cal O}:\cH^{(a)}_{l,m}\rightarrow\cH^{(a)}_{l,m}$.
Note that $Q_{a_n,\cdots,a_0}(l,m\,;z|1)=P_{a_n,\cdots,a_0}(l,m)$.
Like as $P_{a_n,\cdots,a_0}(l,m)$, $Q_{a_n,\cdots,a_0}(l,m\,;z|{\cal O})$
satisfies recursion relations. {}From \eq{sumPhiWPhiN} we have
\be
  \sum_{a_n}Q_{a_n,\cdots,a_0}(l,m\,;z|{\cal O})=
  Q_{a_{n-1},\cdots,a_0}(l,m\,;z|{\cal O}).
  \label{recQ1}
\ee
If ${\cal O}$ satisfies 
$\Phi_N^{(a,b)}(z){\cal O}=f(z){\cal O}\Phi_N^{(a,b)}(z)$ where $f(z)$
is a function independent on $a$ and $b$, then we have another recursion
relation from \eq{sumPhiNPhiW}
\be
  \sum_{a_0}Q_{a_n,\cdots,a_0}(l,m\,;z|{\cal O})=
  f(z)Q_{a_{n-1},\cdots,a_1}(l,m+1\,;z|{\cal O}).
  \label{recQ2}
\ee

Type I VO's are half-infinite transfer matrices but a `physical' transfer
matrix is the row-to-row (or column-to-column) transfer matrix discussed
in the previous subsection. 
The column-to-column transfer matrix $\T_{\rm col}$, 
which adds one column from west to east, is a linear map
\be
  \T_{\rm col}(u)\,:\,\widetilde{\cH}_l\rightarrow\widetilde{\cH}_l,
\ee
where $\widetilde{\cH}_l$ is defined by
\be
  \widetilde{\cH}_l=\bigoplus_k\widetilde{\cH}^{(k)}_l,\quad
  \widetilde{\cH}^{(k)}_l=\bigoplus_m
  \cH^{\,\prime(k)}_{l,m}\otimes\cH^{(k)}_{l,m}.
\ee
Here the numbering of sites in the south direction is the original one
(i.e. increasingly from south to north $\cdots,-2,-1,0,1,2,\cdots$) and
the zeroth site is identified. 
Its matrix element is
\be
  \T_{\rm col}(u)_{\cdots a_{-1}a_0a_1\cdots,\cdots b_{-1}b_0b_1\cdots}
  =\Phi_S^{(a_0,b_0)}(z)_{a_0a_{-1}\cdots,b_0b_{-1}\cdots}
  \Phi_N^{(b_0,a_0)}(z)_{b_0b_1\cdots,a_0a_1\cdots},
\ee
and its matrix form is
\be
  \Bigl(\T_{\rm col}(u)\Bigr)^{(a,b)}
  =\Phi_S^{(a,b)}(z)\otimes{}^t\Phi_N^{(b,a)}(z),
\ee
where $a,b$ are heights on the zeroth site.
$\T_{\rm col}$ with different spectral parameters commute each other
\be
  \Bigl[\T_{\rm col}(u),\T_{\rm col}(v)\Bigr]=0,
\ee
because
\ba
  &&\Bigl(\T_{\rm col}(u_2)\T_{\rm col}(u_1)\Bigr)^{(a,c)}
  =\sum_b\Bigl(\T_{\rm col}(u_2)\Bigr)^{(a,b)}
  \Bigl(\T_{\rm col}(u_1)\Bigr)^{(b,c)}\n
  &=\!\!&\sum_b\Phi_S^{(a,b)}(z_2)\Phi_S^{(b,c)}(z_1)\otimes
  {}^t(\Phi_N^{(c,b)}(z_1)\Phi_N^{(b,a)}(z_2))\\
  &=\!\!&\sum_g\Phi_S^{(a,g)}(z_1)\Phi_S^{(g,c)}(z_2)\otimes
  {}^t(\Phi_N^{(c,g)}(z_2)\Phi_N^{(g,a)}(z_1))
  =\Bigl(\T_{\rm col}(u_1)\T_{\rm col}(u_2)\Bigr)^{(a,c)},\nonumber
\ea
where we have used \eq{PhiNPhiN}, \eq{PhiSPhiS} and \eq{unitarity}.

When we consider the action of $\T_{\rm col}$ in this vertex operator approach,
it is convenient to regard a state of 
$\widetilde{\cH}_l\subset\cH^{\,\prime}_l\otimes\cH_l$ 
as a linear map on $\cH_l$.
In general for two vector spaces $V_i$ ($i=1,2$) a vector of 
$V_1\otimes V_2$ can be regarded as a linear map from $V_2$ to $V_1$ by
$f_1\otimes f_2\in V_1\otimes V_2\mapsto f_1{}^tf_2$ where
$f_1$ and $f_2$ are understood as column vectors. When there is an
isomorphism $P:V_1\rightarrow V_2$, a vector of $V_1\otimes V_2$ can 
be regarded as a linear map on $V_2$,
\be
  \ket{f}=f_1\otimes f_2\in V_1\otimes V_2\mapsto 
  Pf_1{}^tf_2\in\mbox{End}(V_2).
\ee
Similarly
\be
  \bra{f}={}^tf_1\otimes {}^tf_2\in (V_1\otimes V_2)^*\mapsto 
  f_2{}^tf_1P^{-1}\in\mbox{End}(V_2).
\ee
Inner product (pairing) of two vectors $\ket{f}=f_1\otimes f_2, 
\ket{g}=g_1\otimes g_2\in V_1\otimes V_2$ can be expressed as a trace
over $V_2$,
\be
  \langle f\ket{g}={}^tf_1g_1\cdot {}^tf_2g_2
  =\tr_{V_2}\Bigl(f_2{}^tf_1P^{-1}\cdot Pg_1{}^tg_2\Bigr),
\ee
For $A=A_1\otimes A_2$ ($A_i\in\mbox{End}(V_i)$ ($i=1,2$)), 
$\ket{f}=f_1\otimes f_2$ and $\bra{f}={}^tf_1\otimes{}^tf_2$,
we have the following correspondence
\ba
  &&A\ket{f}=A_1f_1\otimes A_2f_2\mapsto
  P(A_1f_1){}^t(A_2f_2)=PA_1P^{-1}\cdot Pf_1{}^tf_2\cdot{}^tA_2,
  \label{Af}\\
  &&\bra{f}A={}^tf_1A_1\otimes{}^tf_2A_2\mapsto
  {}^t({}^tf_2A_2)({}^tf_1A_1)P^{-1}
  ={}^tA_2\cdot f_2{}^tf_1P^{-1}\cdot PA_1P^{-1},
  \nonumber
\ea
and 
\be
  \bra{f}A\ket{g}=\langle f\ket{Ag}
  =\tr_{V_2}\Bigl(f_2{}^tf_1P^{-1}\cdot(PA_1P^{-1}\cdot 
  Pg_1{}^tg_2\cdot{}^tA_2)\Bigr).
\ee

Using this correspondence $\T_{\rm col}$ acts on a vector 
$\ket{f}\in\widetilde{\cH}_l$ as
\be
  \Bigl(\T_{\rm col}(u)\ket{f}\Bigr)^{(a)}
  =\sum_bP^{(a)}\Phi_S^{(a,b)}(z){P^{(b)}}^{-1}\cdot f\cdot\Phi_N^{(b,a)}(z),
  \label{Tcolf}
\ee
where in the RHS $f$ is understood as a linear map on $\cH_l$, which does
not change the central height ($f:\cH^{(a)}_l\rightarrow\cH^{(a)}_l$).
The translation invariant vacuum state $\ket{\mbox{vac}}$ is given by
\be
  \ket{\mbox{vac}}^{(a)}=\sqrt{G_a}x^{-2\lambda H_C^{(a)}}.
  \label{vac}
\ee
In fact it is translationally invariant
\ba
  \Bigl(\T_{\rm col}(u)\ket{\mbox{vac}}\Bigr)^{(a)}
  &\!\!=\!\!&\sum_bP^{(a)}\Phi_S^{(a,b)}(z){P^{(b)}}^{-1}\cdot 
  \sqrt{G_b}x^{-2\lambda H_C^{(b)}}\cdot\Phi_N^{(b,a)}(z)\n
  &\!\!=\!\!&\sum_b\sqrt{G_b}x^{-2\lambda H_C^{(a)}}
  P^{(a)}\Phi_S^{(a,b)}(x^{2\lambda}z){P^{(b)}}^{-1}\Phi_N^{(b,a)}(z)\n
  &\!\!=\!\!&\sqrt{G_a}x^{-2\lambda H_C^{(a)}}\sum_b
  \Phi_W^{(a,b)}(z)\Phi_N^{(b,a)}(z)\\
  &\!\!=\!\!&\sqrt{G_a}x^{-2\lambda H_C^{(a)}}=\ket{\mbox{vac}}^{(a)}.
  \nonumber
\ea
For $\widetilde{{\cal O}}=1\otimes{}^t{\cal O}$ where ${\cal O}$ is 
a linear map ${\cal O}\,:\,\cH^{(a)}_l\rightarrow\cH^{(b)}_l$, we have
\be
  \bra{\mbox{vac}}\widetilde{{\cal O}}\ket{\mbox{vac}}=
  \sum_{a,b}\tr_{\cH^{a}_l}\Bigl(\sqrt{G_a}x^{-2\lambda H_C^{(a)}}\cdot
  P^{(a)}{P^{(b)}}^{-1}\cdot\sqrt{G_b}x^{-2\lambda H_C^{(b)}}\cdot 
  {\cal O}^{(b,a)}\Bigr).
\ee
In particular when ${\cal O}$ is a linear map 
${\cal O}\,:\,\cH^{(a)}_l\rightarrow\cH^{(a)}_l$, we have
\ba
  \!\!\!\!\!\!\!\!\!\!&&\bra{\mbox{vac}}\widetilde{{\cal O}}\ket{\mbox{vac}}=
  \sum_a{}^{(a)}\!\bra{\mbox{vac}}\widetilde{{\cal O}}
  \ket{\mbox{vac}}^{(a)},\quad
  {}^{(a)}\!\bra{\mbox{vac}}\widetilde{{\cal O}}\ket{\mbox{vac}}^{(a)}=
  G_a\tr_{\cH^{(a)}_l}\Bigl(x^{-4\lambda H_C^{(a)}}{\cal O}\Bigr),\\
  \!\!\!\!\!\!\!\!\!\!&&\langle\mbox{vac}\ket{\mbox{vac}}=
  \sum_aG_a\tr_{\cH^{(a)}_l}\Bigl(x^{-4\lambda H_C^{(a)}}\Bigr)=Z_l\,.
\ea
Therefore LHP \eq{Pl} can be written as a vacuum expectation value
with the fixed central value
\be
  P_{a_n,\cdots,a_0}(l)=
  \frac{{}^{(a_0)}\bra{\mbox{vac}}\widetilde{{\cal O}}
  \ket{\mbox{vac}}^{(a_0)}}{\langle\mbox{vac}\ket{\mbox{vac}}},
\ee
where $\widetilde{{\cal O}}=1\otimes{}^t{\cal O}$ with 
${\cal O}=\Phi_W^{(a_0,a_1)}(z)\cdots\Phi_W^{(a_{n-1},a_n)}(z)
\Phi_N^{(a_n,a_{n-1})}(z)\cdots\Phi_N^{(a_1,a_0)}(z)$.

Similarly the row-to-row transfer matrix, which adds one row from 
south to north, can be written as
\be
  \Bigl(\T_{\rm row}(u)\Bigr)^{(a,b)}
  =\Phi_E^{(a,b)}(z^{-1})\otimes{}^t\Phi_W^{(b,a)}(z^{-1})
  \;:\;\widetilde{\cH}_l\rightarrow\widetilde{\cH}_l,
\ee
and it acts on a state $\ket{f}\in\widetilde{\cH}_l$ as
\be
  \Bigl(\T_{\rm row}(u)\ket{f}\Bigr)^{(a)}
  =\sum_bP^{(a)}\Phi_E^{(a,b)}(z^{-1}){P^{(b)}}^{-1}\cdot 
  f\cdot\Phi_W^{(b,a)}(z^{-1}).
  \label{Trowf}
\ee
The state \eq{vac} is translationally invariant 
\ba
  \Bigl(\T_{\rm row}(u)\ket{\mbox{vac}}\Bigr)^{(a)}
  &\!\!=\!\!&\sum_bP^{(a)}\Phi_E^{(a,b)}(z^{-1}){P^{(b)}}^{-1}\cdot 
  \sqrt{G_b}x^{-2\lambda H_C^{(b)}}\cdot\Phi_W^{(b,a)}(z^{-1})\n
  &\!\!=\!\!&\sum_b\sqrt{G_b}x^{-2\lambda H_C^{(a)}}
  P^{(a)}\Phi_E^{(a,b)}(x^{2\lambda}z^{-1}){P^{(b)}}^{-1}
  \Phi_W^{(b,a)}(z^{-1})\n
  &\!\!=\!\!&\sqrt{G_a}x^{-2\lambda H_C^{(a)}}\sum_b
  \frac{G_b}{G_a}\Phi_N^{(a,b)}(x^{4\lambda}z^{-1})\Phi_W^{(b,a)}(z^{-1})\\
  &\!\!=\!\!&\sqrt{G_a}x^{-2\lambda H_C^{(a)}}=\ket{\mbox{vac}}^{(a)}.
  \nonumber
\ea

We remark that these $\T_{\rm col}$ and $\T_{\rm row}$ are defined on an 
infinitely large lattice from the beginning in contrast to those 
in subsection \ref{sec:4.1}. 
Their excited states are created by another type of vertex operators,
type II VO's, from the vacuum state 
(Note that we need also the translation non-invariant `vacuum' states
to obtain a complete set of excited states) \cite{JM}.
We will give an example in subsection \ref{sec:5.4}.
Commutation relations for type II VO will be given in subsection 
\ref{sec:4.4}, section \ref{sec:5} and section \ref{sec:6}.

In the rest of this subsection we assume that $W$ has the crossing symmetry,
\be
  W\BW{b}{d}{a}{c}{\lambda-u}=
  \sqrt{\frac{G_aG_d}{G_bG_c}}W\BW{a}{b}{c}{d}{u}.
  \label{crossing}
\ee
Remark that unitarity \eq{unitarity} and crossing symmetry \eq{crossing} 
imply the second inversion relation \eq{2ndinv}.
We also remark that there are models which enjoy the second inversion
relation but do not have crossing symmetry. 
Here we assume the crossing symmetry.
Then graphical argument shows that CTM's in the thermodynamic limit are
(\eq{ABCD} with $P^{(a)}=1$)
\be
  \begin{array}{l}
  A^{(a)}(u)=C^{(a)}(u)=x^{-2uH_C^{(a)}}\\
  B^{(a)}(u)=D^{(a)}(u)=\sqrt{G_a}x^{-2(\lambda-u)H_C^{(a)}}
  \end{array}
  \,:\,\cH^{(a)}_{l,m}\rightarrow\cH^{(a)}_{l,m}\,,
  \label{CTM}
\ee
and type I VO's are related each other,
\ba
  \Phi_W^{(a,b)}(z)&\!\!=\!\!&
  \sqrt{\sfrac{G_b}{G_a}}\Phi_N^{(a,b)}(x^{2\lambda}z),
  \label{PhiWPhiN}\\
  \Phi_S^{(a,b)}(z)&\!\!=\!\!&\Phi_N^{(a,b)}(z),\\
  \Phi_E^{(a,b)}(z)&\!\!=\!\!&
  \sqrt{\sfrac{G_b}{G_a}}\Phi_N^{(a,b)}(x^{2\lambda}z)
  =\Phi_W^{(a,b)}(z).
\ea
We write
\ba
  \Phi^{(a,b)}(z)=\Phi_N^{(a,b)}(z)&:&
  \cH^{(b)}_{l,m}\rightarrow\cH^{(a)}_{l,m+1},\\
  \Phi^{*(a,b)}(z)=\Phi_W^{(a,b)}(z)&:&
  \cH^{(b)}_{l,m}\rightarrow\cH^{(a)}_{l,m-1}.
\ea
Then $\Phi^{(a,b)}(z)$ and $\Phi^{*(a,b)}(z)$ enjoy the following properties,
\ba
  &&\Phi^{*(a,b)}(z)=\sqrt{\sfrac{G_b}{G_a}}\Phi^{(a,b)}(x^{2\lambda}z),
  \label{Phi*}\\
  &&\Phi^{(a,b)}(z_2)\Phi^{(b,c)}(z_1)=\sum_{g}W\BW{a}{g}{b}{c}{u_1-u_2}
  \Phi^{(a,g)}(z_1)\Phi^{(g,c)}(z_2),
  \label{comPhiPhi}\\
  &&w^{H_C}\Phi^{(a,b)}(z)w^{-H_C}=\Phi^{(a,b)}(wz),
  \label{wHCPhi}\\
  &&\sum_{g}\Phi^{*(a,g)}(z)\Phi^{(g,a)}(z)=1,
  \label{sumPhiPhi}\\
  &&\Phi^{(a,b)}(z)\Phi^{*(b,c)}(z)=\delta_{a,c}.
  \label{PhiPhi*}
\ea
The first equation is \eq{PhiWPhiN}, the second one is \eq{PhiNPhiN},
the third one is obtained by \eq{APhiN}, the fourth one is \eq{sumPhiWPhiN},
and the fifth one is derived by using \eq{Phi*}, \eq{comPhiPhi}, 
\eq{crossing}, \eq{initial} and \eq{sumPhiPhi}.
Multi-point LHP \eq{LHP} becomes
\ba
  \!\!\!\!\!\!\!\!&&P_{a_n,\cdots,a_0}(l,m)
  \label{LHP2}\\
  \!\!\!\!\!\!\!\!&=\!\!&
  Z_{l,m}^{-1}G_{a_0}\,\tr_{\cH^{(a_0)}_{l,m}}\Bigl(x^{-4\lambda H_C^{(a_0)}}
  \Phi^{*(a_0,a_1)}(z)\cdots\Phi^{*(a_{n-1},a_n)}(z)
  \Phi^{(a_n,a_{n-1})}(z)\cdots\Phi^{(a_1,a_0)}(z)\Bigr).\nonumber
\ea
Since the Boltzmann weight is invariant under $180^{\circ}$ rotation 
$\ds W\BW{a}{b}{c}{d}{u}=W\BW{d}{c}{b}{a}{u}$,
$P_{a_n,\cdots,a_0}(l)$ satisfies
\be
  P_{a_n,\cdots,a_0}(l)=P_{a_0,\cdots,a_n}(l).
  \label{reverse}  
\ee
Eq. \eq{Tcolf} becomes
\be
  \Bigl(\T_{\rm col}(u)\ket{f}\Bigr)^{(a)}=
  \sum_b\Phi^{(a,b)}(z)\cdot f\cdot\Phi^{(b,a)}(z).
  \label{Tcolf2}
\ee

In section \ref{sec:5} and subsection \ref{sec:6.4}
we will calculate LHP's by using bosonization technique for the VO's.

\subsection{Introduction to quasi-Hopf algebra}\label{sec:4.3}

In this subsection we illustrate an outline of a quasi-Hopf algebra.
For more details we refer the readers to refs.\cite{D,JKOS1}

In quantum mechanics, we know an addition of angular momentums very well.
For two particles system the total angular momentum $\vec{J}$ is obtained
simply by an addition of each angular momentum $\vec{J}^{(1)}$ and 
$\vec{J}^{(2)}$,
\be
  \vec{J}=\vec{J}^{(1)}+\vec{J}^{(2)}.
\ee
In mathematics, this formula is written in the following way;
$\vec{J}^{(1)}$ acts on the representation space $V_1$,
$\vec{J}^{(2)}$ acts on $V_2$, and the total angular momentum
$\vec{J}$ acts on $V_1\otimes V_2$ by,
\be
  \vec{J}=\vec{J}\otimes 1+1\otimes\vec{J}.
\ee
This is called the tensor product representation of Lie algebra $so(3)$.
If a system has rotational symmetry, for example the Heisenberg spin chain
(XXX spin chain), one can apply the representation theory of the
rotational group $SO(3)$ (or its Lie algebra $so(3)$) to it.
But if the system is perturbed and looses the rotational symmetry,
then one can not apply $so(3)$ to it.
Some models, however, have a good property.
For example the XXZ spin chain has the same degeneracy of energy
as the XXX spin chain.
To treat such models we need some deformation of the Lie algebra or
some deformation of the tensor product representation.

\subsubsection*{1. algebra and coalgebra}
Let us begin with the definitions of an algebra and a coalgebra.
For simplicity we take the complex field $\C$ as a base field.
An \underline{algebra} $A$ is a vector space with two operations,
product (multiplication) $m$ and unit $u$, which satisfy
\be
\begin{array}{ccccl}
  A\otimes A\otimes A&\maprightu{15mm}{m\otimes{\rm id}}&
  A\otimes A&\mbox{product}&m:A\otimes A\rightarrow A\\
  \mapdownl{{\rm id}\otimes m}&&\mapdownr{m}&\mbox{unit}&
  u:\C\rightarrow A\\
  A\otimes A&\maprightd{15mm}{m}&A&\mbox{associativity}&
  m\circ(m\otimes{\rm id})=m\circ({\rm id}\otimes m),
\end{array}
\ee
and $m\circ({\rm id}\otimes u)={\rm id}=m\circ(u\otimes{\rm id})$ 
($A\otimes\C$, $A$ and $\C\otimes A$ are identified).
If we write $m(a\otimes b)=ab$, the associativity becomes a usual form 
$(ab)c=a(bc)$.

A coalgebra is defined by reversing the arrows.
A \underline{coalgebra} $A$ is a vector space with two operations,
coproduct $\Delta$ and counit $\varepsilon$, which satisfy
\be
\begin{array}{ccccl}
  A\otimes A\otimes A&\mapleftu{15mm}{\Delta\otimes{\rm id}}&
  A\otimes A&\mbox{coproduct}&\Delta:A\rightarrow A\otimes A\\
  \mapupl{{\rm id}\otimes\Delta}&&\mapupr{\Delta}&\mbox{counit}&
  \varepsilon:A\rightarrow\C\\
  A\otimes A&\mapleftd{15mm}{\Delta}&A&\mbox{coassociativity}&
  (\Delta\otimes{\rm id})\circ\Delta=({\rm id}\otimes\Delta)\circ\Delta,
\end{array}
\label{coalg}
\ee
and $({\rm id}\otimes\varepsilon)\circ\Delta={\rm id}=(\varepsilon\otimes
{\rm id})\circ\Delta$
($A\otimes\C$, $A$ and $\C\otimes A$ are identified).

Let us introduce $\sigma$  
($\sigma:A\otimes A\rightarrow A\otimes A$, 
$\sigma(a\otimes b)=b\otimes a$) 
and define $m'=m\circ\sigma$ and $\Delta'=\sigma\circ\Delta$.
An algebra is called commutative if  $m'=m$,
and a coalgebra is called cocommutative if $\Delta'=\Delta$.

\subsubsection*{2. Hopf algebra}
A \underline{Hopf algebra} is a set $(A,m,u,\Delta,\varepsilon,S)$ 
satisfying the following conditions: $A$ is an algebra and a coalgebra; 
$m$, $u$, $\Delta$, $\varepsilon$ are homomorphism;
antipode $S:A\rightarrow A$ satisfies
$m\circ(S\otimes{\rm id})\circ\Delta=u\circ\varepsilon=
m\circ({\rm id}\otimes S)\circ\Delta$. 
$S$ is an anti-homomorphism.

We give two examples of a Hopf algebra, a group $G$ and 
a Lie algebra $\g$,
exactly speaking a function algebra of group $\mbox{Fun}(G)$ and
an enveloping algebra of Lie algebra $U(\g)$ respectively.
Their Hopf algebra structures are
\be
\begin{tabular}{lll}
&$\mbox{Fun}(G)=\mbox{Map}(G,\C)$&$U(\g)$\\
product&$(f_1f_2)(x)=f_1(x)f_2(x)$&$XY$\\
unit&$(u(a))(x)=a$&$u(a)=a1$\\
coproduct&$(\Delta(f))(x_1,x_2)=f(x_1x_2)$&$\Delta(X)=X\otimes 1+1\otimes X$\\
counit&$(\varepsilon(f))(x)=f(e)$&$\varepsilon(X)=0$\\
antipode&$(S(f))(x)=f(x^{-1})$&$S(X)=-X$,
\end{tabular}
\ee
and $(m(g))(x)=g(x,x)$ for $g(x_1,x_2)\in\mbox{Map}(G\times G,\C)$,
and $\Delta(1)=1\otimes 1$, $\varepsilon(1)=1$, $S(1)=1$ for $U(\g)$. 
Roughly speaking $\Delta$, $\varepsilon$ and $S$ correspond to
\be
\begin{tabular}{lcll}
&&$\mbox{Fun}(G)$&$U(\g)$\\
$\Delta$&$\leftrightarrow$&product of $G$
  &tensor product rep. of $U(\g)$\\
$\varepsilon$&$\leftrightarrow$&unit element of $G$
  &trivial rep. of $U(\g)$\\
$S$&$\leftrightarrow$&inverse element of $G$
  &contragredient rep. of $U(\g)$.
\end{tabular}
\ee

$\mbox{Fun}(G)$ is a commutative (and non-cocommutative) Hopf algebra and
$U(\g)$ is a cocommutative (and non-commutative) Hopf algebra.
Non-commutative and non-cocommutative Hopf algebra may be regarded as
an extension (deformation) of group or Lie algebra in this sense.
This is the idea of quantum group (quantum algebra).

The quantum group (quantum algebra) is a Hopf algebra.
We give an example of the quantum group, $U_q(\slt)$, 
which is a deformation of $U(\slt)$. 
$U_q(\slt)$ is generated by $t=q^h$, $e$ and $f$, which satisfy
\be
\begin{array}{lllllll}
  \lbrack h,e\rbrack=2e&~&\Delta(h)=h\otimes 1+1\otimes h
  &~&\varepsilon(h)=0&~&S(h)=-h\\
  \lbrack h,f\rbrack=-2f&&\Delta(e)=e\otimes 1+t\otimes e
  &&\varepsilon(e)=0&&S(e)=-t^{-1}e\\
  \lbrack e,f\rbrack=\frac{t-t^{-1}}{q-q^{-1}}
  &&\Delta(f)=f\otimes t^{-1}+1\otimes f&&\varepsilon(f)=0&&S(f)=-ft.
\end{array}
\ee
This quantum algebra appears in the XXZ spin chain as a symmetry,
\ba
  &&\lbrack H_{\mbox{\scriptsize XXZ}},U_q(\slt)\rbrack=0,\n
  &&H_{\mbox{\scriptsize XXZ}}=
  J\sum_{i=1}^{N-1}\Bigl(s_i^xs_{i+1}^x+s_i^ys_{i+1}^y
  +\frac{q+q^{-1}}{2}s_i^zs_{i+1}^z\Bigr)
  +J\frac{q-q^{-1}}{4}(s_1^z-s_N^z),\\
  &&h=\sum_i2s_i^z,\quad e=\sum_iq^{\sum_{j<i}2s_j^z}s_i^+,\quad
  f=\sum_is_i^-q^{-\sum_{j>i}2s_j^z},\nonumber
\ea
where $\vec{s}=\frac12\vec{\sigma}$ and $s^{\pm}=s^1\pm is^2$.
In the $q\rightarrow 1$ limit, $U_q(\slt)$ reduces to $U(\slt)$.

\subsubsection*{3. quasi-triangular Hopf algebra}
Using the coproduct, a tensor product representation of two 
representations $(\pi_i,V_i)$ ($i=1,2$) of the Hopf algebra
can be defined in the following way,
\be
  \Bigl((\pi_1\otimes\pi_2)\circ\Delta,V_1\otimes V_2\Bigr).
\ee
Coassociativity implies the isomorphism,
\be
  (V_1\otimes V_2)\otimes V_3\cong V_1\otimes(V_2\otimes V_3)
  \quad \mbox{(as $A$ module)}.
\ee
But cocommutativity does not hold in general, so the following 
isomorphism depends on the detail of the Hopf algebra:
\be
  V_1\otimes V_2\stackrel{?}{\cong}V_2\otimes V_1
  \quad \mbox{(as $A$ module)}.
  \label{VV=VV}
\ee
Of course we have $V_1\otimes V_2\cong V_2\otimes V_1$ as vector space,
by $P_{V_1V_2}:V_1\otimes V_2\stackrel{\cong}{\rightarrow}V_2\otimes V_1$,
$P_{V_1V_2}(v_1\otimes v_2)=v_2\otimes v_1$.
But the problem is the commutativity of the action of $A$ and $P_{V_1V_2}$.

Drinfeld considered the situation that the isomorphism \eq{VV=VV} does hold. 
A \underline{quasi-} \underline{triangular Hopf algebra} 
$(A,m,u,\Delta,\varepsilon,S,\cR)$ (we abbreviate it as $(A,\Delta,\cR)$)
is a Hopf algebra with a universal $R$ matrix $\cR$, 
which satisfies
\ba
  &\cR&\in A\otimes A : \underline{\mbox{universal $R$ matrix}}\n
  &&\Delta'(a)=\cR\Delta(a)\cR^{-1}\quad(\forall a\in A), \n
  &&(\Delta\otimes{\rm id})\cR=\cR^{(13)}\cR^{(23)},\quad
  (\varepsilon\otimes{\rm id})\cR=1,
  \label{univR}\\
  &&({\rm id}\otimes\Delta)\cR=\cR^{(13)}\cR^{(12)},\quad
  ({\rm id}\otimes\varepsilon)\cR=1.\nonumber
\ea
Then we have an intertwiner
\be
  R_{V_1V_2}=P_{V_1V_2}\circ(\pi_1\otimes\pi_2)(\cR):
  V_1\otimes V_2 \stackrel{\cong}{\longrightarrow}V_2\otimes V_1
  \quad \mbox{(as $A$ module)},
\ee
and $\cR$ satisfies the Yang-Baxter equation,
\be
  \cR^{(12)}\cR^{(13)}\cR^{(23)}=\cR^{(23)}\cR^{(13)}\cR^{(12)}.
\ee

\subsubsection*{4. quasi-triangular quasi-Hopf algebra}
As presented in {\bf 2} 
the quantum group is obtained from the Lie algebra by relaxing one 
condition, cocommutativity.
Here we relax one more condition, coassociativity.
Coassociativity \eq{coalg} is modified by a coassociator $\Phi$
in the following way,
\ba
  \!\!\!\!\!\!\!\!\!\!
  &\Phi&\in A\otimes A\otimes A : \underline{\mbox{coassociator}}\n
  \!\!\!\!\!\!\!\!\!\!&&({\rm id}\otimes\Delta)\Delta(a)=
  \Phi(\Delta\otimes{\rm id})\Delta(a)\Phi^{-1}\quad
  (\forall a\in A),\n
  \!\!\!\!\!\!\!\!\!\!&&
  (\id\otimes\id\otimes\Delta)\Phi\cdot(\Delta\otimes\id\otimes\id)\Phi 
  =(1\otimes\Phi)\cdot(\id\otimes\Delta\otimes\id)\Phi\cdot(\Phi\otimes 1),
  \label{coast}\\
  \!\!\!\!\!\!\!\!\!\!&&(\id\otimes\ve\otimes\id)\Phi=1.\nonumber
\ea
A \underline{quasi-triangular quasi-Hopf algebra} 
$(A,m,u,\Delta,\varepsilon,\Phi,S,\alpha,\beta,\cR)$ (we abbreviate it
as $(A,\Delta,\Phi,\cR)$) satisfies \eq{coast} and
\ba
  &\cR&\in A\otimes A\n
  &&\Delta'(a)=\cR\Delta(a)\cR^{-1}\quad(\forall a\in A), \n
  &&(\Delta\otimes{\rm id})\cR
  =\Phi^{(312)}\cR^{(13)}{\Phi^{(132)}}^{-1}\cR^{(23)}\Phi^{(123)},\\
  &&({\rm id}\otimes\Delta)\cR
  ={\Phi^{(231)}}^{-1}\cR^{(13)}\Phi^{(213)}\cR^{(12)}{\Phi^{(123)}}^{-1},
  \nonumber
\ea
where $\Phi^{(312)}$ means $\Phi^{(312)}=\sum_iZ_i\otimes X_i\otimes Y_i$ 
for $\Phi=\sum_iX_i\otimes Y_i\otimes Z_i$.
$\alpha,\beta\in A$ satisfy
\ba
  &&\sum_i X_i \beta S(Y_i)\alpha Z_i=1,\\
  &&\sum_iS(b_i)\alpha c_i=\ve(a)\alpha,\quad 
  \sum_ib_i\beta S(c_i)=\ve(a)\beta\quad (\forall a\in A),
\ea
where $\Delta(a)=\sum_ib_i\otimes c_i$.
Then $\cR$ enjoys the Yang-Baxter type equation,
\be
  \cR^{(12)}\Phi^{(312)}\cR^{(13)}{\Phi^{(132)}}^{-1}\cR^{(23)}\Phi^{(123)}
  =\Phi^{(321)}\cR^{(23)}{\Phi^{(231)}}^{-1}\cR^{(13)}\Phi^{(213)}\cR^{(12)}.
\ee
A quasi-Hopf algebra with $\Phi=1$ is nothing but a Hopf algebra.

\subsubsection*{5. twist}
Quasi-Hopf algebras admit an important operation, \underline{twist}.
For any invertible element $F\in A\otimes A$ 
($(\varepsilon\otimes{\rm id})F=({\rm id}\otimes\varepsilon)F=1$), 
which is called as a twistor,
there is a map from quasi-Hopf algebras to quasi-Hopf algebras:
\ba
  \mbox{quasi-Hopf algebra}&\longrightarrow&\mbox{quasi-Hopf algebra}\n
  (A,\Delta,\Phi,\cR)&\stackrel{F}{\longrightarrow}&
  (A,\widetilde{\Delta},\widetilde{\Phi},\widetilde{\cR}).
\ea
New coproduct, coassociator, R matrix etc. are given by
\ba
  &F&\in A\otimes A : \underline{\mbox{twistor}}\n
  &&\widetilde{\Delta}=F\Delta(a)F^{-1}\quad (\forall a\in A),\n
  &&\widetilde{\Phi}=\Bigl(F^{(23)}({\rm id}\otimes\Delta)F\Bigr)
    \Phi\Bigl(F^{(12)}(\Delta\otimes{\rm id})F\Bigr)^{-1},\\
  &&\widetilde{\cR}=F^{(21)}\cR F^{-1},\n
  &&\widetilde{\ve}=\ve,\quad\widetilde{S}=S,\quad
  \widetilde{\alpha}=\sum_iS(d_i)\alpha e_i,\quad
  \tilde{\beta}=\sum_if_i\beta S(g_i),\nonumber
\ea
where $\sum_id_i\otimes e_i=F^{-1}$ and $\sum_if_i\otimes g_i=F$.
We remark that an algebra $A$ itself is unchanged.
If a twistor $F$ satisfies the cocycle condition, this twist operation
maps a Hopf algebra to a Hopf algebra. For a general twistor $F$, however, 
a Hopf algebra is mapped to a quasi-Hopf algebra:
\ba
  \mbox{Hopf algebra}&\longrightarrow&\mbox{quasi-Hopf algebra}\n
  (A,\Delta,\cR)&\stackrel{F}{\longrightarrow}&
  (A,\widetilde{\Delta},\widetilde{\Phi},\widetilde{\cR}).
\ea

Let $H$ be an Abelian subalgebra of $A$, with the product written
additively. A twistor $F(\lambda)\in A\otimes A$ depending on 
$\lambda\in H$ is a shifted cocycle if it satisfies the relation
(shifted cocycle condition),
\ba
  &&F(\lambda) : \underline{\mbox{shifted cocycle}}\n
  &\Leftrightarrow&
  F^{(12)}(\lambda)(\Delta\otimes{\rm id})F(\lambda)
  =F^{(23)}(\lambda+h^{(1)})({\rm id}\otimes\Delta)F(\lambda),
\ea
for some $h\in H$.

When a twistor $F(\lambda)$ satisfies the shifted cocycle condition,
we obtain a quasi-triangular quasi-Hopf algebra from a quasi-triangular
Hopf algebra by twisting,
\ba
  \mbox{Hopf algebra}&\verylongrightarrow&\mbox{quasi-Hopf algebra}\n
  (A,\Delta,\cR)&\stackrel{F(\lambda)}{\verylongrightarrow}&
  (A,\Delta_{\lambda},\Phi(\lambda),\cR(\lambda)),
\ea
and we have
\ba
  &&\Phi(\lambda)=F^{(23)}(\lambda)F^{(23)}(\lambda+h^{(1)})^{-1},\n
  &&(\Delta_{\lambda}\otimes{\rm id})\cR(\lambda)
  =\Phi^{(312)}(\lambda)\cR^{(13)}(\lambda)\cR^{(23)}(\lambda+h^{(1)}),\\
  &&({\rm id}\otimes\Delta_{\lambda})\cR(\lambda)
  =\cR^{(13)}(\lambda+h^{(2)})\cR^{(12)}(\lambda){\Phi^{(123)}(\lambda)}^{-1},
  \nonumber
\ea
and $R$ matrix satisfies the dynamical Yang-Baxter equation,
\be
  \cR^{(12)}(\lambda+h^{(3)})\cR^{(13)}(\lambda)\cR^{(23)}(\lambda+h^{(1)})
  =\cR^{(23)}(\lambda)\cR^{(13)}(\lambda+h^{(2)})\cR^{(12)}(\lambda).
  \label{dYBE}
\ee

\subsection{Elliptic quantum groups}\label{sec:4.4}

The solutions of YBE ($R$-matrices) are classified into two types, 
vertex-type and face-type. 
Corresponding to these two there are two types of elliptic quantum
groups (algebras). 

The vertex-type elliptic algebras are associated with 
the $R$-matrix $R(u)$ of Baxter \cite{Bax72} and Belavin \cite{Bel}.
The first example of this sort 
is the Sklyanin algebra \cite{Skl82}, designed as an elliptic 
deformation of the Lie algebra $\slt$. 
It is presented by the `$RLL$'-relation 
\be
  R^{(12)}(u_1-u_2)L^{(1)}(u_1)L^{(2)}(u_2)
  =L^{(2)}(u_2)L^{(1)}(u_1)R^{(12)}(u_1-u_2), 
 \label{vRLL}
\ee 
together with a specific choice of the form for $L(u)$. 
$R$ and $L$ depend on an elliptic modulus $r$.
Foda et al.\cite{FIJKMY} proposed its affine version, $\Aqp(\slth)$,
\be
  R^{(12)}(u_1-u_2,r)L^{(1)}(u_1)L^{(2)}(u_2)
  =L^{(2)}(u_2)L^{(1)}(u_1)R^{(12)}(u_1-u_2,r-c), 
  \label{vRLL2}
\ee
whose main point of is the shift of $r$ by a central element $c$.
 
The face-type algebras are based on $R$-matrices 
of Andrews, Baxter, Forrester \cite{ABF} and generalizations 
\cite{DJMO86,JMO1,JMO3}. 
In this case, besides the elliptic modulus, 
$R$ and $L$ depend also on extra parameter(s) $\lambda$. 
As Felder has shown \cite{Fel95}, 
the $RLL$ relation undergoes a `dynamical' shift 
by elements $h$ of the Cartan subalgebra, 
\ba
  &&R^{(12)}(u_1-u_2,\la+h)L^{(1)}(u_1,\la)L^{(2)}(u_2,\la+h^{(1)})\n
  &&\qquad=L^{(2)}(u_2,\la)L^{(1)}(u_1,\la+h^{(2)})R^{(12)}(u_1-u_2,\la), 
  \label{fRLL}
\ea
and the YBE itself is modified to a dynamical one \eq{dYBE}, \eq{dYBER}.
As we shall see, a central extension of this algebra is obtained by  
introducing further a shift of the elliptic modulus 
analogous to \eq{vRLL2} (see \eq{RLL1}-\eq{RLL2}
and the remark following them). 

These two algebras, $RLL$ relations \eq{vRLL2} and \eq{fRLL}, 
seemed to be different but Fr\o nsdal \cite{Fron} 
pointed out that they have a common structure; they are quasi-Hopf 
algebras obtained by twisting quantum affine algebras. 
Namely, there exist two types of twistors which give rise to 
different comultiplications on the quantum affine algebras $U_q(\g)$, 
and the resultant quasi-Hopf algebras are nothing but the two types of 
elliptic quantum groups. 
In \cite{JKOS1} explicit formulas for the twistors satisfying the
shifted cocycle condition were presented and 
two types of elliptic quantum groups, 
$\Aqp(\slnh)$ and $\Bqla(\g)$, were defined: 
\be
\begin{tabular}{ccccc}
  type&twistor&Hopf algebra&&quasi-Hopf algebra\\
  face&$F(\lambda)$&$U_q(\g)$&$\hspace{-3mm}
  \stackrel{F(\lambda)}{\verylongrightarrow}\hspace{-3mm}$&$\Bqla(\g)$\\ 
  vertex&$E(r)$&$U_q(\slnh)$&$\hspace{-3mm}
  \stackrel{E(r)}{\verylongrightarrow}\hspace{-3mm}$&$\Aqp(\slnh)$.
\end{tabular}  
\label{eQG}
\ee

The face-type algebra has also been given an alternative 
formulation in terms of the Drinfeld currents. 
This is the approach adopted by Enriquez and Felder \cite{EF} 
and Konno \cite{Ko97}.
The Drinfeld currents are suited to deal with infinite dimensional 
representations. 

\subsubsection*{1. quantum group}

First let us fix the notation.
Let $\g$ be the Kac-Moody Lie algebra associated with 
a symmetrizable generalized Cartan matrix $A=({a_{ij}})_{i,j\in I}$ 
\cite{Kac}. 
We fix an invariant inner product $(~,~)$ on 
the Cartan subalgebra $\gothh$
and identify $\gothh^*$ with $\gothh$ via $(~,~)$.  
If $\{\alpha_i\}_{i\in I}$ denotes the set of 
simple roots, then $(\alpha_i,\alpha_j)=d_ia_{ij}$, where 
$d_i={1\over 2}(\alpha_i,\alpha_i)$. 

Consider the corresponding quantum group $U=U_q(\g)$. 
Hereafter we fix a complex number $q\neq 0$, $|q|<1$. 
The algebra $U$ has generators $e_i$, $f_i$ ($i\in I$) 
and $h$ ($h\in \gothh$), 
satisfying the standard relations 
\ba
  &&[h,h']=0\qquad (h,h'\in\gothh),
  \label{Uqrel0}\\
  &&[h,e_i]=(h,\alpha_i)e_i,\qquad [h,f_i]=-(h,\alpha_i)f_i
  \qquad (i\in I,h\in\gothh),
  \label{Uqrel1}\\
  &&[e_i,f_j]=\delta_{ij}\frac{t_i-t_i^{-1}}{q_i-q^{-1}_i}
  \qquad (i,j\in I),
  \label{Uqrel2}
\ea
and the Serre relations which we omit. 
In \eq{Uqrel2} we have set $q_i=q^{d_i}$, $t_i=q^{\alpha_i}$. 
We adopt the Hopf algebra structure given as follows. 
\ba
  &&\Delta(h)=h\otimes 1+1\otimes h,\\
  &&\Delta(e_i)=e_i\otimes 1+t_i\otimes e_i,\quad
  \Delta(f_i)=f_i\otimes t_i^{-1}+1\otimes f_i, 
  \label{copro}\\
  &&\ve(e_i)=\ve(f_i)=\ve(h)=0, 
  \label{counit}\\
  &&S(e_i)=-t_i^{-1}e_i,\quad S(f_i)=-f_it_i,\quad S(h)=-h,
  \label{antipode} 
\ea
where $i\in I$ and $h\in\gothh$. 

Let $\cR\in U^{\otimes 2}$ denote the universal $R$ matrix of $U$. 
It has the form 
\ba
  &&\cR=q^{-T}{\cal C} ,\\
  &&{\cal C}=\sum_{\beta\in Q^+} q^{(\beta,\beta)}
  \Bigl(q^{-\beta}\otimes q^{\beta}\Bigr){\cal C}_\beta
  =1-\sum_{i\in I}(q_i-q_i^{-1})e_i t_i^{-1}\otimes t_i f_i+\cdots.
\ea
Here the notation is as follows. 
Take a basis $\{h_l\}$ of $\gothh$, and its dual basis $\{h^l\}$. 
Then 
\be
  T=\sum_l h_l\otimes h^l 
  \label{Thh}
\ee
denotes the canonical element of $\gothh\otimes \gothh$. 
The element 
${\cal C}_\beta=\sum_j u_{\beta,j}\otimes u^j_{-\beta}$ 
is the canonical element of $U^+_\beta\otimes U^-_{-\beta}$ 
with respect to a certain Hopf pairing, 
where $U^+$ (resp. $U^-$) denotes the subalgebra of $U$ 
generated by the $e_i$ (resp. $f_i$), and 
$U^\pm_{\pm \beta}$ ($\beta\in Q^+$)
signifies the homogeneous components 
with respect to the natural gradation by $Q^+=\sum_i\Z_{\ge 0}\alpha_i$.
(For the details the reader is referred e.g. to \cite{D,Tani}.)
Basic properties of the universal $R$ matrix is given in \eq{univR}.

\subsubsection*{2. face type algebra}

Let $\rho\in\gothh$ be an element such that 
$(\rho,\alpha_i)=d_i$ for all $i\in I$. 
Let $\phi$ be an automorphism of $U$ given by 
\be
  \phi=\mbox{Ad}(q^{\frac{1}{2}\sum_l h_lh^l-\rho}), 
  \label{phi}
\ee
where $\{h_l\}$, $\{h^l\}$ are as in \eq{Thh}. 
In other words, 
\be
  \phi(e_i)=e_i t_i,\quad \phi(f_i)=t_i^{- 1}f_i,\quad \phi(q^{h})=q^{h}.
\ee
Since 
\be
  \mbox{Ad}(q^T)\circ(\phi\otimes\phi)=
  \mbox{Ad}(q^{\frac{1}{2}\sum_l \Delta(h_lh^l)-\Delta(\rho)}),
  \label{phiphi}
\ee
we have 
\be
  \mbox{Ad}(q^{T})\circ(\phi\otimes\phi)\circ\Delta=\Delta\circ\phi. 
  \label{delphi}
\ee
For $\la\in\gothh$, introduce an automorphism 
\be
  \varphi_\la=\mbox{Ad}(q^{\sum_l h_lh^l+2(\la-\rho)})
  =\phi^2\circ\mbox{Ad}(q^{2\la}).
  \label{phila}
\ee
Then the expression $(\varphi_\la\otimes\id)(q^T\cR)$
is a formal power series in the variables 
$x_i=q^{2(\la,\alpha_i)}$ ($i\in I$) of the form 
$\ds
  1-\sum_i(q_i-q_i^{-1})x_i e_it_i\otimes t_i f_i+\cdots
$.
We define the twistor $F(\la)$ as follows.\\
{\bf [Face type twistor]}
\be
  F(\la)=
%
  \mathop{\prod_{k\geq 1}}^{\kuru}
  \Bigl(\varphi_\la^k\otimes\id\Bigr)\Bigl(q^T\cR\Bigr)^{-1}.
  \label{facetwistor}
\ee
Here and after, we use the ordered product symbol 
$\displaystyle\mathop{\prod_{k\geq 1}}^{\kuru}A_k=\cdots A_3A_2A_1$. 
Note that the $k$-th factor in the product \eq{facetwistor}
is a formal power series in the $x_i^k$ with leading term $1$, 
and hence the infinite product makes sense. 
Then the face type twistor \eq{facetwistor} satisfies the
shifted cocycle condition 
\be
  F^{(12)}(\la)(\Delta\otimes \id)F(\la)
  =F^{(23)}(\la+h^{(1)})(\id \otimes \Delta)F(\la),
  \label{facecocy}
\ee
and 
\be
  \left(\varepsilon\otimes\id\right)F(\la)
  =\left(\id\otimes\varepsilon\right)F(\la)=1.
  \label{epF}
\ee
A proof of this property and examples are given in \cite{JKOS1}.
In \eq{facecocy}, if $\la=\sum_l\la_lh^l$, 
then $\la+h^{(1)}$ means $\sum_l(\la_l+h_l^{(1)})h^l$. 
Hence we have, for example, 
\ban
  &&\mbox{Ad}(q^{2l T^{(12)}}) F^{(23)}_k(\la)=F^{(23)}_k(\la+ l h^{(1)}),\\
  &&\mbox{Ad}(q^{2l T^{(13)}}) F^{(23)}_k(\la)=F^{(23)}_k(\la- l h^{(1)}).
\ean

For convenience, let us give a name to the quasi-Hopf algebra
associated with the twistor \eq{facetwistor}.\\ 
{\bf [Face type algebra]} {\it
We define the quasi-Hopf algebra $\Bqla(\g)$ of face type
to be the set 
$(U_q(\g),\Delta_\la,\Phi(\la),\cR(\la))$
together with 
$\alpha_\la=\sum_i S(d_i)e_i$, 
$\beta_\la=\sum_i f_i S(g_i)$, 
the antiautomorphism $S$ defined by (\ref{antipode}) and
$\ve$ defined by \eq{counit},
\ba
  \Delta_{\la}(a)&\!\!=\!\!&F^{(12)}(\la)\,\Delta(a)\,F^{(12)}(\la)^{-1},
  \label{facecopro}\\
  \cR(\la)&\!\!=\!\!&F^{(21)}(\la)\,\cR\,F^{(12)}(\la)^{-1},
  \label{faceR}\\
  \Phi(\la)&\!\!=\!\!&F^{(23)}(\la)F^{(23)}(\la+h^{(1)})^{-1},
  \label{facephi}
\ea
and $\sum_id_i\otimes e_i=F(\la)^{-1}$, $\sum_if_i\otimes g_i=F(\la)$.
}

Let us consider the case where $\g$ is of affine type, 
in which we are mainly interested. 
Let $c$ be the canonical central element and $d$ the scaling element. 
We set 
\be
  \la-\rho=rd+s'c+\bar{\la}-\bar{\rho}\quad (r,s'\in \C),
\ee
where $\bar{\la}$ stands for the classical part of $\la\in\gothh$. 
Denote by $\{\bar{h}_j\}$, $\{\bar{h}^j\}$ the 
classical part of the dual basis of $\gothh$. 
Since $c$ is central, $\varphi_\la$ is independent of $s'$. 
Writing $p=q^{2r}$, we have 
\be
  \varphi_\la=\mbox{Ad}(p^dq^{2cd})\circ\bar{\varphi}_\la,\quad 
  \bar{\varphi}_\la
  =\mbox{Ad}(q^{\sum \bar{h}_j\bar{h}^j+2(\bar{\la}-\bar{\rho})}).
\ee
Set further
\ba
  \cR(z)&\!\!=\!\!&\mbox{Ad}(z^d\otimes 1)(\cR),
  \label{Rg}\\
  F(z,\la)&\!\!=\!\!&\mbox{Ad}(z^d\otimes 1)(F(\la)),
  \label{Fg}\\
  \cR(z,\la)&\!\!=\!\!&\mbox{Ad}(z^d\otimes 1)(\cR(\la))
  =\sigma\Bigl(F(z^{-1},\la)\Bigr)\cR(z)F(z,\la)^{-1}.
  \label{Rg2}
\ea
\eq{Rg} and \eq{Fg} are formal power series in $z$, whereas 
\eq{Rg2} contains both positive and negative powers of $z$. 
Note that $q^{c\otimes d+d\otimes c}\cR(z)\bigl|_{z=0}$ reduces to the 
universal $R$ matrix of $U_q(\bar{\g})$ 
corresponding to the underlying finite dimensional Lie algebra 
$\bar{\g}$. {}From the definition \eq{facetwistor} of $F(\la)$ we 
have the difference equation 
\be
  F(pq^{2c^{(1)}}z,\la)=(\bar{\varphi}_\la\otimes\id)^{-1}
  \Bigl(F(z,\la)\Bigr)\cdot q^{T}\cR(pq^{2c^{(1)}}z),
\ee
with the initial condition $F(0,\la)=F_{\bar{\g}}(\bar{\la})$,
where $F_{\bar{\g}}(\bar{\la})$ signifies the twistor 
corresponding to $\bar{\g}$. 

\subsubsection*{3. vertex type algebra}

When $\g=\slnh$, it is possible to construct a different 
type of twistor. We call it {\it vertex type}. 

Let us write $h_i=\alpha_i$ ($i=0,\ldots,n-1$). 
A basis of $\gothh$ is $\{h_0,\ldots,h_{n-1},d\}$. 
The element $d$ gives the homogeneous grading, 
\be
  [d,e_i]=\delta_{i0}e_i,\quad [d,f_i]=-\delta_{i0}f_i,
\ee
for all $i=0,\ldots,n-1$. 
Let the dual basis be $\{\La_0,\ldots,\La_{n-1},c\}$. 
The $\La_i$ are the fundamental weights and $c$ is the canonical 
central element. Let $\tau$ be the automorphism of $U_q(\slnh)$ such that 
\be
  \tau(e_i)=e_{i+1\,\bmod\,n},\quad\tau(f_i)=f_{i+1\,\bmod\,n},\quad
  \tau(h_i)=h_{i+1\,\bmod\,n}
\ee
and $\tau^n=\id$. Then we have 
\be
  \tau(\La_i)=\La_{i+1\,\bmod\,n}-\frac{n-1-2i}{2n}c.
\ee
The element $\rho=\sum_{i=0}^{n-1}\La_i$ is invariant under $\tau$. 
It gives the principal grading 
\be
  [\rho,e_i]=e_i,\quad [\rho,f_i]=-f_i,
\ee
for all $i=0,\ldots,n-1$. 
Note also that 
\be
  (\tau\otimes\tau)\circ \Delta=\Delta\circ\tau,\quad
  (\tau\otimes\tau) ({\cal C}_\beta)={\cal C}_{\tau(\beta)}.
\ee

For $r\in \C$, we introduce an automorphism 
\be
  \widetilde{\varphi}_r
  =\tau\circ\mbox{Ad}\Bigl(q^{\frac{2(r+c)}{n}\rho}\Bigr),
  \label{philatilde}
\ee
and set 
\be
  \widetilde{T}=\frac{1}{n}\Bigl(\rho\otimes c+c\otimes \rho
  -\frac{n^2-1}{12}c\otimes c\Bigr).
\ee
Then $\Bigl(\widetilde{\varphi}_r\otimes\id\Bigr)
\Bigl(q^{\widetilde{T}}\cR\Bigr)^{-1}$
is a formal power series in $p^{\frac{1}{n}}$ where $p=q^{2r}$. 
Unlike the face case, this is a formal series with a non-trivial 
leading term $q^{T-\widetilde{T}}\left(1+\cdots\right)$. 
Nevertheless, the $n$-fold product 
\be
  \mathop{\prod_{n\ge k\ge 1}}^{\kuru}
  \Bigl(\widetilde{\varphi}_r^k\otimes\id\Bigr)
  \Bigl(q^{\widetilde{T}}\cR\Bigr)^{-1}
\ee
takes the form $1+\cdots$, because of the relation
\be
  \sum_{k=1}^n\left(\tau^k\otimes\id\right)\left(T-\widetilde{T}\right)=0.
\ee
We now define the vertex type twistor $E(r)$ as follows.\\
{\bf [Vertex type twistor]}
\be
  E(r)=\mathop{\prod_{k\geq 1}}^{\kuru}
  \left(\widetilde{\varphi}_r^k\otimes\id\right)
  \Bigl(q^{\widetilde{T}}\cR\Bigr)^{-1}.
  \label{vertextwistor}
\ee
The infinite product $\displaystyle\mathop{\prod_{k\geq 1}}^{\kuru}{}$
is to be understood as $\displaystyle\lim_{N\rightarrow\infty}
\mathop{\prod_{nN\ge k\ge 1}}^{\kuru}{}$. 
In view of the remark made above, $E(r)$ is 
a well defined formal series in $p^{\frac{1}{n}}$. 
Then the vertex type twistor \eq{vertextwistor} satisfies the
shifted cocycle condition 
\be
  E^{(12)}(r)(\Delta\otimes \id)E(r)
  =E^{(23)}(r+c^{(1)})(\id \otimes \Delta)E(r),
  \label{vertexcocy}
\ee
and 
\be
  \left(\varepsilon\otimes\id\right)E(r)
  =\left(\id\otimes\varepsilon\right)E(r)=1.
  \label{epE}
\ee
A proof of this property and an example are given in \cite{JKOS1}.\\
{\bf [Vertex type algebra]}{\it 
We define the quasi-Hopf algebra $\Aqp{}(\slnh)$ 
($p=q^{2r}$) of vertex type to be the set 
$(U_q(\slnh),\Delta_r,\Phi(r),\cR(r))$ together with 
$\alpha_r=\sum_i S(d_i)e_i$, $\beta_r=\sum_i f_i S(g_i)$, 
the antiautomorphism $S$ defined by \eq{antipode} and 
$\ve$ defined by \eq{counit},
\ba
  \Delta_{r}(a)&\!\!=\!\!&E^{(12)}(r)\,\Delta(a)\,E^{(12)}(r)^{-1},
  \label{vertexcopro}\\
  \cR(r)&\!\!=\!\!&E^{(21)}(r)\,\cR\,E^{(12)}(r)^{-1},
  \label{vertexR}\\
  \Phi(r)&\!\!=\!\!&E^{(23)}(r)E^{(23)}(r+c^{(1)})^{-1},
  \label{vertexphi}
\ea
and $\sum_i d_i\otimes e_i=E(r)^{-1}$, $\sum_i f_i \otimes g_i=E(r)$.
}

Let us set 
\ba
  \widetilde{\cR}'(\zeta)&\!\!=\!\!&\Bigl(\mbox{Ad}(\zeta^\rho)\otimes\id\Bigr)
  (q^{\widetilde{T}}\cR),\\
  E(\zeta,r)&\!\!=\!\!&\Bigl(\mbox{Ad}(\zeta^\rho)\otimes\id\Bigr)E(r).
\ea
In just the same way as in the face type case, 
the definition \eq{vertextwistor} can be alternatively described as the 
unique solution of the difference equation 
\be
  E(p^{\frac{1}{n}}q^{\frac{2}{n}c^{(1)}}\zeta,r)=
  (\tau\otimes\id)^{-1}\bigl(E(\zeta,r)\bigr)\cdot 
  \widetilde{\cR}'(p^{\frac{1}{n}}q^{\frac{2}{n}c^{(1)}}\zeta),
\ee
with the initial condition $E(0,r)=1$, where $p=q^{2r}$.

\subsubsection*{4. dynamical $RLL$-relations and vertex operators}

The $L$-operators and vertex operators for the elliptic algebras 
can be constructed from those of $U_q(\g)$ 
by `dressing' the latter with the twistors. 
In this subsection, we examine various commutation relations among these 
operators. 
We shall mainly discuss 
the case of the face type algebra $\Bqla(\g)$
where $\g$ is of affine type. 

Hereafter we write $U=U_q(\g)$, $\B=\Bqla(\g)$.  
By a representation of the quasi-Hopf algebra $\B$ we mean  
that of the underlying associative algebra $U$. 
Let $(\pi_V,V)$ be a finite dimensional module over $U$, 
and $(\pi_{V,z},V_z)$ be the evaluation representation associated with it
where $\pi_{V,z}=\pi_V\circ\mbox{Ad}(z^d)$. 

We define $L$-operators for $\B$ by 
\ba
  L_V^\pm(z,\la)&\!\!=\!\!&\left(\pi_{V,z}\otimes \id \right)\cR^{'\pm}(\la),
  \label{Lpm}\\
  \cR^{'+}(\la)&\!\!=\!\!&q^{c\otimes d+d\otimes c}\cR(\la),
  \label{Rp}\\
  \cR^{'-}(\la)&\!\!=\!\!&\cR^{(21)}(\la)^{-1}q^{-c\otimes d-d\otimes c}.
  \label{Rm}
\ea
Likewise we set
\be
  R^{\pm}_{VW}(\sfrac{z_1}{z_2},\la)
  =(\pi_{V,z_1}\!\otimes\pi_{W,z_2})\cR^{'\pm}(\la).
\ee
Setting further 
\be
  \cR^{'\pm}(z,\la)=\mbox{Ad}(z^d\otimes 1)\cR^{'\pm}(\la), 
  \label{Rpmz}
\ee
we find from the dynamical YBE \eq{dYBE} that 
\ba
  &&\cR^{'\pm(12)}(\sfrac{z_1}{z_2},\la+h^{(3)})
  \cR^{'\pm(13)}(q^{\mp c^{(2)}}\sfrac{z_1}{z_3},\la)
  \cR^{'\pm(23)}(\sfrac{z_2}{z_3},\la+h^{(1)})\n
  &&\qquad=
  \cR^{'\pm(23)}(\sfrac{z_2}{z_3},\la)
  \cR^{'\pm(13)}(q^{\pm c^{(2)}}\sfrac{z_1}{z_3},\la+h^{(2)})
  \cR^{'\pm(12)}(\sfrac{z_1}{z_2},\la),\\
  &&\cR^{'+(12)}(q^{c^{(3)}}\sfrac{z_1}{z_2},\la+h^{(3)})
  \cR^{'+(13)}(\sfrac{z_1}{z_3},\la)
  \cR^{'-(23)}(\sfrac{z_2}{z_3},\la+h^{(1)})\n
  &&\qquad=
  \cR^{'-(23)}(\sfrac{z_2}{z_3},\la)
  \cR^{'+(13)}(\sfrac{z_1}{z_3},\la+h^{(2)})
  \cR^{'+(12)}(q^{-c^{(3)}}\sfrac{z_1}{z_2},\la).
\ea
Applying $\pi_V\otimes\pi_W\otimes\id$, we obtain 
the dynamical $RLL$ relation,
\ba
  &&R^{\pm(12)}_{VW}(\sfrac{z_1}{z_2},\la+ h)
  L_V^{\pm(1)}(z_1,\la)L_W^{\pm(2)}(z_2,\la+ h^{(1)})\n
  &&\qquad=
  L_W^{\pm(2)}(z_2,\la)L_V^{\pm(1)}(z_1,\la+ h^{(2)})
  R^{\pm(12)}_{VW}(\sfrac{z_1}{z_2},\la),
  \label{RLL1}\\
  &&R^{+(12)}_{VW}(q^c\sfrac{z_1}{z_2},\la+ h)
  L_V^{+(1)}(z_1,\la)L_W^{-(2)}(z_2,\la+ h^{(1)})\n
  &&\qquad=
  L_W^{-(2)}(z_2,\la)L_V^{+(1)}(z_1,\la+ h^{(2)})
  R^{+(12)}_{VW}(q^{-c}\sfrac{z_1}{z_2},\la).
  \label{RLL2}
\ea
Here the index ($1$) (resp. ($2$)) refers to $V$ (resp. $W$), and 
$h$, $c$ (without superfix) are elements of $\gothh\subset \B$. 
If we write 
\be
  \la-\rho=rd+s'c+\bar{\la}-\bar{\rho}\quad 
  (r,s'\in \C, \bar{\la}\in\bar{\gothh}), 
\ee
then 
\be
  \la+h^{(1)}=(r+h^{\vee}+c^{(1)})d+(s'+d^{(1)})c+(\bar{\la}+\bar{h^{(1)}}),
  \label{shift}
\ee
where $h^\vee$ is the dual Coxeter number. 
The parameter $r$ plays the role of the elliptic modulus.
Note that, in \eq{RLL1}-\eq{RLL2},  
$r$ also undergoes a shift depending on the central element $c$. 

Actually the two $L$-operators \eq{Lpm} are not independent. 
We have 
\be
  L_V^+(pq^cz,\la)=q^{-2\bar{T}_{V,\bullet}}
  (\mbox{Ad}(\bar{X}_\la)\otimes\id)^{-1}L_V^-(z,\la),
  \label{LpLm}
\ee
where 
\be
  \bar{T}_{V,\bullet}=\sum_j\pi(\bar{h}_j)\otimes \bar{h}^j,\quad
  \bar{X}_\la=\pi(q^{\sum \bar{h}_j\bar{h}^j+2(\bar{\la}-\bar{\rho})}).
\ee

Next let us consider vertex operators.
Let $(\pi_{V,z},V_z)$ be as before, and let 
$V(\mu)$ be a highest weight module with highest weight $\mu$. 
Consider intertwiners of $U$-modules of the form 
\ba
  \Phi_V^{(\nu,\mu)}(z)&:&V(\mu)\longrightarrow V(\nu)\otimes V_{z},\\
  \Psi_V^{*(\nu,\mu)}(z)&:& V_{z}\otimes V(\mu)\longrightarrow V(\nu),
\ea
which are called vertex operators of type I and type II respectively.
Define the corresponding VO's for $\B$ as follows \cite{Fron2}: 
\ba
  \Phi_V^{(\nu,\mu)}(z,\la)&\!\!=\!\!&
  (\id\otimes \pi_z)F(\la)\circ\Phi_V^{(\nu,\mu)}(z),
  \label{twVO1}\\
  \Psi_V^{*(\nu,\mu)}(z,\la)&\!\!=\!\!&
  \Psi_V^{*(\nu,\mu)}(z)\circ (\pi_z\otimes \id)F(\la)^{-1}.
  \label{twVO2}
\ea
When there is no fear of confusion, we often drop the sub(super)scripts 
$V$ or $(\nu,\mu)$. 
It is clear that \eq{twVO1} and \eq{twVO2} satisfy the intertwining relations 
relative to the coproduct $\Delta_\la$ \eq{facecopro}, 
\ba
  &&\Delta_\la(a)\Phi(z,\la)=\Phi(z,\la)a\qquad (\forall a\in \B),\\
  &&a\Psi^*(z,\la)=\Psi^*(z,\la)\Delta_\la(a)\qquad (\forall a\in \B).
\ea
These intertwining relations can be encapsulated to  
commutation relations with the $L$-operators. 

The `dressed' VO's \eq{twVO1}, \eq{twVO2} 
satisfy the following dynamical intertwining relations
(see the diagram below):
\ba
  \!\!\!\!\!&&\Phi_W(z_2,\la)L^+_V(z_1,\la)
  =R^+_{VW}(q^c\sfrac{z_1}{z_2},\la+h)L^+_V(z_1,\la)\Phi_W(z_2,\la+h^{(1)}),
  \label{dint1}\\
  \!\!\!\!\!&&\Phi_W(z_2,\la)L^-_V(z_1,\la)
  =R^-_{VW}(\sfrac{z_1}{z_2},\la+h)L^-_V(z_1,\la)\Phi_W(z_2,\la+h^{(1)}),
  \label{dint1m}\\
  \!\!\!\!\!&&L^+_V(z_1,\la)\Psi_W^*(z_2,\la+h^{(1)})
  =\Psi_W^*(z_2,\la)L^+_V(z_1,\la+h^{(2)})R^+_{VW}(\sfrac{z_1}{z_2},\la),
  \label{dint2}\\
  \!\!\!\!\!&&L^-_V(z_1,\la)\Psi_W^*(z_2,\la+h^{(1)})
  =\Psi_W^*(z_2,\la)L^-_V(z_1,\la+h^{(2)})R^-_{VW}(q^c\sfrac{z_1}{z_2},\la).
  \label{dint2m}
\ea
\be
\begin{array}{ccc}
  V_{z_1}\otimes V(\mu)&\maprightu{1cm}{\Phi_W}\;\;
  V_{z_1}\otimes V(\nu)\otimes W_{z_2}\;\;\maprightu{1cm}{L^\pm_V}&
  V_{z_1}\otimes V(\nu)\otimes W_{z_2} \\
  \hspace{-0.4cm}\mapdownl{\scriptstyle L^\pm_V}&
  \makebox{\rule[-0.7cm]{0cm}{1.6cm}}&
  \mapdownr{\scriptstyle R^\pm_{VW}} \\
  V_{z_1}\otimes V(\mu)&\maprightd{5.5cm}{\Phi_W}&
  V_{z_1}\otimes V(\nu)\otimes W_{z_2}
\end{array}
\ee

\medskip\medskip

\be
\begin{array}{ccc}
  V_{z_1}\otimes W_{z_2}\otimes V(\mu)\hspace{-0.7cm}&
  \maprightu{1cm}{R^\pm_{VW}}\;\;
  V_{z_1}\otimes W_{z_2}\otimes V(\mu)\;\;\maprightu{1cm}{L^\pm_V}&
  \hspace{-0.7cm}V_{z_1}\otimes W_{z_2}\otimes V(\mu) \\
  \hspace{0.1cm}\mapdownl{\scriptstyle \Psi^*_W}&
  \makebox{\rule[-0.7cm]{0cm}{1.6cm}}&
  \hspace{-1.1cm}\mapdownr{\scriptstyle \Psi^{*}_W} \\
  \hspace{0.4cm}V_{z_1}\otimes V(\nu)&\maprightd{7cm}{L^\pm_V}&
  \hspace{-0.8cm}V_{z_1}\otimes V(\nu)
\end{array}
\ee
\medskip\medskip

{}From the theory of $q$-KZ-equation\cite{FR92}, 
we know the VO's for $U$ satisfy the commutation relations of the form 
\ba
  \!\!\!\!\!\!\!\!\!\!&&\check{R}_{VV}(\sfrac{z_1}{z_2})
  \Phi_V^{(\nu,\mu)}(z_1)\Phi_V^{(\mu,\kappa)}(z_2)
  =\sum_{\mu'}
  \Phi_V^{(\nu,\mu')}(z_2)\Phi_V^{(\mu',\kappa)}(z_1)
  W_{I}\BW{\kappa}{\mu}{\mu'}{\nu}{\sfrac{z_1}{z_2}},
  \label{VVt1}\\
  \!\!\!\!\!\!\!\!\!\!&&\Psi_V^{*(\nu,\mu)}(z_1)\Psi_V^{*(\mu,\kappa)}(z_2)
  {\check{R}_{VV}(\sfrac{z_1}{z_2})}^{-1}
  =\sum_{\mu'}
  W_{II}\BW{\kappa}{\mu}{\mu'}{\nu}{\sfrac{z_1}{z_2}}
  \Psi_V^{*(\nu,\mu')}(z_2)\Psi_V^{*(\mu',\kappa)}(z_1),
  \label{VVt2}\\
  \!\!\!\!\!\!\!\!\!\!&&\Phi_V^{(\nu,\mu)}(z_1)\Psi_V^{*(\mu,\kappa)}(z_2)
  =\sum_{\mu'}
  W_{I,II}\BW{\kappa}{\mu}{\mu'}{\nu}{\sfrac{z_1}{z_2}}
  \Psi_V^{*(\nu,\mu')}(z_2)\Phi_V^{(\mu',\kappa)}(z_1).
  \label{VVt3}
\ea
Here 
\ba
  &&\check{R}_{VV}(z)=PR_{VV}(z),
  \qquad P(v\otimes v')=v'\otimes v,\\
  &&R_{VV}(\sfrac{z_1}{z_2})=(\pi_{V,z_1}\!\otimes\pi_{V,z_2})\cR
\ea
is the `trigonometric' $R$ matrix. 
In \eq{VVt1}-\eq{VVt3} we used a slightly abbreviated notation. 
For example, the left hand side of \eq{VVt1} means the composition 
$$
  V(\kappa)\maprightu{1cm}{\Phi(z_2)}V(\mu)\otimes V_{z_2}
  \maprightu{2cm}{\Phi(z_1)\otimes \mbox{\footnotesize id}}
  V(\nu)\otimes V_{z_1}\otimes V_{z_2}
  \maprightu{2.5cm}{\mbox{\footnotesize id}\otimes\check{R}(\sfrac{z_1}{z_2})}
  V(\nu)\otimes V_{z_2}\otimes V_{z_1}.
$$
Similarly \eq{VVt2}, \eq{VVt3} are maps 
\be
  V_{z_2}\otimes V_{z_1}\otimes V(\kappa)\longrightarrow V(\nu),\quad
  V_{z_2}\otimes V(\kappa) \longrightarrow V(\nu)\otimes V_{z_1},
\ee
respectively. For $U_q(\slth)$, the formulas for the $W$-factors 
in the simplest case can be found e.g. in \cite{IIJMNT}. 

The `dressed' VO's satisfy similar relations with appropriate dynamical shift. 
Setting $\check{R}_{VV}(z,\la)=PR_{VV}(z,\la)$, we have 
\ba
  &&\check{R}_{VV}(\sfrac{z_1}{z_2},\la+h^{(1)})
  \Phi_V^{(\nu,\mu)}(z_1,\la)\Phi_V^{(\mu,\kappa)}(z_2,\la)\n
  &&\qquad\qquad =\sum_{\mu'}
  \Phi_V^{(\nu,\mu')}(z_2,\la)\Phi_V^{(\mu',\kappa)}(z_1,\la)
  W_{I}\BW{\kappa}{\mu}{\mu'}{\nu}{\sfrac{z_1}{z_2}},\\
  &&\Psi_V^{*(\nu,\mu)}(z_1,\la)\Psi_V^{*(\mu,\kappa)}(z_2,\la+h^{(1)})
  \check{R}_{VV}(\sfrac{z_1}{z_2},\la)^{-1}\n
  &&\qquad\qquad =\sum_{\mu'}
  W_{II}\BW{\kappa}{\mu}{\mu'}{\nu}{\sfrac{z_1}{z_2}}
  \Psi_V^{*(\nu,\mu')}(z_2,\la)\Psi_V^{*(\mu',\kappa)}(z_1,\la+h^{(1)}),\\
  &&\Phi_V^{(\nu,\mu)}(z_1,\la)\Psi_V^{*(\mu,\kappa)}(z_2,\la)\n
  &&\qquad\qquad =\sum_{\mu'}
  W_{I,II}\BW{\kappa}{\mu}{\mu'}{\nu}{\sfrac{z_1}{z_2}}
  \Psi_V^{*(\nu,\mu')}(z_2,\la)\Phi_V^{(\mu',\kappa)}(z_1,\la+h^{(1)}).
\ea
Notice that the $W$-factors stay the same with the trigonometric case, 
and are not affected by a dynamical shift. 

The case of vertex type algebras can be treated in a parallel way. 
See \cite{JKOS1}.

\subsubsection*{5. Drinfeld currents}

The Drinfeld currents are suited to deal with infinite dimensional 
representations. We explain it by taking $U_q(\slth)$ as an example.
(See \cite{JKOS2} for general case.)

First let us recall the Drinfeld currents of $U_q(\slth)$ \cite{Dri88}. 
Let $x^\pm_n$ ($n\in\Z$), $a_n$ ($n\in\Z_{\neq 0}$), $h$, $c$, $d$ 
denote the Drinfeld generators of $U_q(\slth)$. 
In terms of the generating functions 
\ba
  x^\pm(z)&\!\!=\!\!&\sum_{n\in \Z}x^\pm_n z^{-n},
  \label{Dcur1}\\
  \psi(q^{\frac{c}{2}}z)&\!\!=\!\!&q^h
  \exp\Bigl((q-q^{-1})\sum_{n>0}a_{n}z^{-n}\Bigr),
  \label{Dcur2}\\
  \varphi(q^{-\frac{c}{2}}z)&\!\!=\!\!&q^{-h}
  \exp\Bigl(-(q-q^{-1})\sum_{n>0}a_{-n}z^{n}\Bigr), 
  \label{Dcur3}
\ea
the defining relations read as follows: 
\ba
  &&c :\hbox{ central },\\
  &&[h,d]=0,\quad [d,a_n]=na_n,\quad [d,x^{\pm}_n]=n x^{\pm}_n, \\
  &&[h,a_n]=0,\quad [h,x^\pm(z)]=\pm 2 x^{\pm}(z),\\
  &&[a_n,a_m]=\frac{[2n]_q[cn]_q}{n}q^{-c|n|}\delta_{n+m,0},
  \label{Uqsl2an}\\
  &&[a_n,x^+(z)]=\frac{[2n]_q}{n}q^{-c|n|}z^n x^+(z),\quad
  [a_n,x^-(z)]=-\frac{[2n]_q}{n} z^n x^-(z),\\
  &&(z-q^{\pm 2}w)x^\pm(z)x^\pm(w)=(q^{\pm 2}z-w)x^\pm(w)x^\pm(z),\\
  &&[x^+(z),x^-(w)]=\frac{1}{q-q^{-1}}
  \Bigl(\delta(q^c\sfrac{w}{z})\psi(q^{\frac{c}{2}}w)
  -\delta(q^{-c}\sfrac{w}{z})\varphi(q^{-\frac{c}{2}}w)
  \Bigr),
\ea
and the Serre relation which we omit.

For the level $1$ case ($c=1$), $x^{\pm}(z)$ has the following free boson
realization \cite{FJ},
\ba
  x^+(z)&\!\!=\!\!&:\exp\Bigl(-\sum_{n\neq 0}\frac{a_n}{[n]_q}z^{-n}\Bigr):
  e^Qz^{a_0+1},\\
  x^-(z)&\!\!=\!\!&:\exp\Bigl(\sum_{n\neq 0}\frac{a_n}{[n]_q}
  z^{-n}q^{|n|}\Bigr):
  e^{-Q}z^{-a_0+1},
\ea
where $a_n$ ($n\in\Z_{\neq 0}$) is given in \eq{Uqsl2an},
$a_0$ and $Q$ are given in \eq{an}, and $h=a_0$.
For general level $c$, $U_q(\slth)$ has the Wakimoto realization 
(see e.g. \cite{AOS93} and references therein).

We now introduce a new parameter $p$ and 
 modify \eq{Dcur1}-\eq{Dcur3} to define another set of currents. 
For notational convenience,  we will frequently write 
\be
  p=q^{2r},\quad p^*=p q^{-2c}=q^{2\rs},
\ee
where $\rs$ is
\be
  \rs=r-c.
\ee
Let us introduce two currents $u^\pm(z,p)\in U_q(\slth)$ depending on $p$ by 
\ba
  u^+(z,p)&\!\!=\!\!&
  \exp\Bigl(\sum_{n>0}\frac{1}{[r^*n]_q}a_{-n}(q^rz)^n\Bigr),
  \label{dressup}\\
  u^-(z,p)&\!\!=\!\!&
  \exp\Bigl(-\sum_{n>0}\frac{1}{[rn]_q}a_{n}(q^{-r}z)^{-n}\Bigr).
  \label{dressum}
\ea
We define the `dressed' currents $x_+(z,p)$, $x_-(z,p)$, $\psi^\pm(z,p)$  by
\ba
  x_+(z,p)&\!\!=\!\!&u^+(z,p)x^+(z), 
  \label{dress1}\\
  x_-(z,p)&\!\!=\!\!&x^-(z)u^-(z,p),
  \label{dress2}\\
  \psi^+(z,p)&\!\!=\!\!&
  u^+(q^{\frac{c}{2}} z,p)\psi(z)u^-(q^{-\frac{c}{2}}z,p),
  \label{dress3}\\
  \psi^{-}(z,p)&\!\!=\!\!&
  u^+(q^{-\frac{c}{2}}z,p)\varphi(z)u^-(q^{\frac{c}{2}}z,p). 
  \label{dress4}
\ea
We call these as elliptic currents. They are `Drinfeld currents' of
$U_q(\slth)$ or $\Bqla(\slth)$ which is nothing but $U_q(\slth)$ 
equipped with a different coproduct.
We will often drop $p$, and write $x_-(z,p)$ as $x_-(z)$ and so forth. 

The merit of these currents is that they obey the following `elliptic' 
commutation relations:
\ba
  &&\psi^{\pm}(z)\psi^{\pm}(w)
  =\frac{\Theta_{p}(q^{-2}\sfrac{z}{w})}{\Theta_{p}(q^2\sfrac{z}{w})}
  \frac{\Theta_{p^*}(q^{2}\sfrac{z}{w})}{\Theta_{p^*}(q^{-2}\sfrac{z}{w})}
  \psi^{\pm}(w)\psi^{\pm}(z),
  \label{Drcom1}\\
  &&\psi^{+}(z)\psi^{-}(w)
  =\frac{\Theta_{p}(pq^{-c-2}\sfrac{z}{w})}{\Theta_{p}(pq^{-c+2}\sfrac{z}{w})}
  \frac{\Theta_{p^*}(p^*q^{c+2}\sfrac{z}{w})}
  {\Theta_{p^*}(p^*q^{c-2}\sfrac{z}{w})}
  \psi^{-}(w)\psi^{+}(z),
  \label{Drcom2}\\
  &&\psi^{\pm}(z)x_+(w)\psi^{\pm}(z)^{-1}=
  q^{-2}\frac{\Theta_{p^*}(q^{\pm\frac{c}{2}+2}\sfrac{z}{w})}
  {\Theta_{p^*}(q^{\pm\frac{c}{2}-2}\sfrac{z}{w})}
  x_+(w),
  \label{Drcom3}\\
  &&\psi^{\pm}(z)x_-(w)\psi^{\pm}(z)^{-1}=
  q^2\frac{\Theta_{p}(q^{\mp\frac{c}{2}-2}\sfrac{z}{w})}
  {\Theta_{p}(q^{\mp\frac{c}{2}+2}\sfrac{z}{w})}
  x_-(w),
  \label{Drcom4}\\
  &&[x_+(z),x_-(w)]=\frac{1}{q-q^{-1}}
  \Bigl(\delta(q^c\sfrac{w}{z})\psi^+(q^{\frac{c}{2}}w)
  -\delta(q^{-c}\sfrac{w}{z})\psi^-(q^{-\frac{c}{2}}w)\Bigr),
  \label{Drcom5}
\ea
where $\Theta_p(z)$ is given in \eq{Theta}.
It is convenient to consider also the current 
\be
  k(z)=\exp\Bigl(\sum_{n>0}\frac{[n]_q}{[2n]_q[r^*n]_q}a_{-n}(q^cz)^n\Bigr)
  \exp\Bigl(-\sum_{n>0}\frac{[n]_q}{[2n]_q[rn]_q}a_nz^{-n}\Bigr). 
  \label{dress5}
\ee
The $\psi^\pm(z)$ are related to $k(z)$ by the formula 
\ba
  &&\psi^\pm(p^{\mp(r-\frac{c}{2})}z)=\kappa q^{\pm h}k(q z)k(q^{-1}z),  
  \label{eq:psipm}\\
  &&\kappa=\frac{\xi(z;p^*,q)}{\xi(z;p,q)} \Biggl|_{z=q^{-2}},
  \label{kappac}
\ea
where the function 
\be
  \xi(z;p,q)=\frac{(q^2z,pq^2z;p,q^4)_\infty}{(q^4z,pz;p,q^4)_\infty}
  \label{xi}
\ee
is a solution of the difference equation 
\be
  \xi(z;p,q)\xi(q^2z;p,q)=\frac{(q^2z;p)_\infty}{(pz;p)_\infty}.
\ee
We have the commutation relations supplementing \eq{Drcom1}-\eq{Drcom5},
\ba
  &&k(z)k(w)=\frac{\xi(\sfrac{w}{z};p,q)}{\xi(\sfrac{w}{z};p^*,q)}
  \frac{\xi(\sfrac{z}{w};p^*,q)}{\xi(\sfrac{z}{w};p,q)}k(w)k(z),
  \label{Drcom6}\\
  &&k(z)x_+(w)k(z)^{-1}
  =\frac{\Theta_{p^*}(p^{*\frac12}q\sfrac{z}{w})}
  {\Theta_{p^*}(p^{*\frac12}q^{-1}\sfrac{z}{w})}x_+(w),
  \label{Drcom7}\\
  &&k(z)x_-(w)k(z)^{-1}
  =\frac{\Theta_{p}(p^{\frac12}q^{-1}\sfrac{z}{w})}
  {\Theta_{p}(p^{\frac12}q \sfrac{z}{w})}x_-(w).
  \label{Drcom8}
\ea

Elliptic algebra $U_{q,p}(\slth)$ is obtained as a tensor product
of $U_q(\slth)$ ($\Bqla(\slth)$) and a Heisenberg algebra $[Q,P]=1$.
For details see \cite{Ko97,JKOS2}. (See also \cite{HY97}.)

\setcounter{section}{4}
\setcounter{equation}{0}
\section{Free Field Approach to ABF Model}\label{sec:5}

Lukyanov and Pugai studied the Andrews-Baxter-Forrester (ABF) model 
and calculated LHS's by bosonizing VO's in subsection \ref{sec:4.2} 
\cite{LP96}. Here we explain their work.

$r$, $\beta$, $x_{\pm}(z)$, $T(z)$, $\alpha_n$, $\alpha'_n$, $h_n$, 
$C_{l,k}$, etc. 
in this section are same as those in section \ref{sec:3}.
In this section $r$ is a positive integer,
\be
  r\in\Z,\quad r\geq 4,
\ee
and $\rs$ is
\be
  \rs=r-1.
\ee
So we have
\be
  \beta=\frac{r}{r-1},\quad \az=\frac{1}{\sqrt{r(r-1)}},\quad 
  p'=r-1,\quad p''=r.
\ee

\subsection{ABF model}\label{sec:5.1}

The ABF model \cite{ABF} is a RSOS model associated to
the vector representation of $A_1^{(1)}$ algebra \cite{JMO3}.
The height variables $a,b,\cdots$ take one of the $r-1$ states 
$1,2,\cdots r-1$ and those on neighboring sites are subject to the
condition $|a-b|=1$.
We write the Boltzmann weight in the form
\be
  W\BW{a}{b}{c}{d}{u}= \rho(u)\Wb\BW{a}{b}{c}{d}{u}, 
\ee
where an overall scalar factor $\rho(u)$ is given in \eq{rho} and
chosen so that the partition function per site equals to $1$.
Non-zero components of $\Wb$ are given by
\ba
  &&\Wb\BW{a}{a\pm 1}{a\pm 1}{a\pm 2}{u}=1,\n
  &&\Wb\BW{a}{a\pm 1}{a\pm 1}{a}{u}=\frac{[a\mp u]}{[a]}\frac{[1]}{[1+u]},
  \label{Wb}\\
  &&\Wb\BW{a}{a\pm 1}{a\mp 1}{a}{u}=\frac{\sqrt{[a+1][a-1]}}{[a]}
  \frac{[-u]}{[1+u]}.\nonumber
\ea
Here $[u]$ is given in \eq{[u]} and we use a parameter $x$ which is obtained
from the original nome $p$ by the modular transformation 
($p=e^{2\pi i\tau}\mapsto x^{2r}=e^{2\pi i\frac{-1}{\tau}}$).
Low temperature limit corresponds to $x\sim 0$ ($p\sim 1$).

This Boltzmann weight $W$ enjoys
YBE \eq{YBEW}, initial condition \eq{initial}, unitarity \eq{unitarity}
and crossing symmetry \eq{crossing} ($\lambda=-1$, $G_a=[a]$),
\be
  W\BW{b}{d}{a}{c}{-1-u}=
  \sqrt{\frac{[a][d]}{[b][c]}}W\BW{a}{b}{c}{d}{u}.
\ee
In this gauge $W$ enjoys also a reflection symmetry
\be
  W\BW{a}{c}{b}{d}{u}=W\BW{a}{b}{c}{d}{u}.
  \label{refl}
\ee
Along the additive variable $u$, we often use the multiplicative variable 
$z=x^{2u}$.

We will restrict ourselves to the `regime III' region defined by
\be
  0<x<1,\quad -1<u<0.
\ee
In this region $W$ is positive.
On the critical point ($x\rightarrow 1$ ($p\rightarrow 0$)), a correlation 
length becomes infinite and this model becomes scale invariant, i.e.
it has conformal symmetry. So it is described by CFT, i.e. Virasoro algebra. 
In this case it corresponds to the minimal unitary series 
with $\beta=\frac{r}{r-1}$.
On the off-critical point ($x<1$), which corresponds to the 
$(1,3)$-perturbation of the minimal unitary CFT, the Virasoro symmetry 
is lost but the DVA symmetry remains.
In the low temperature limit ($x\rightarrow 0$ with $z=x^{2u}(\geq 1)$ 
fixed kept),
$W$ behaves as
\ba
  &&\rho(u)\rightarrow z^{-\frac{\rs}{2r}},\n
  &&\Wb\BW{a}{a\pm 1}{a\pm 1}{a\pm 2}{u}\rightarrow 1,\quad
  \Wb\BW{a}{a\pm 1}{a\mp 1}{a}{u}\rightarrow 0,\\
  &&\Wb\BW{a}{a+1}{a+1}{a}{u}\rightarrow z^{\frac{r-a-1}{r}}(\geq 1),\quad
  \Wb\BW{a}{a-1}{a-1}{a}{u}\rightarrow z^{\frac{a-1}{r}}(\geq 1).\nonumber
\ea
Therefore ground state configuration is
\be
\setlength{\unitlength}{1mm}
\begin{picture}(140,45)(-30,5)
\put(5,10){\line(1,0){40}}
\put(5,20){\line(1,0){40}}
\put(5,30){\line(1,0){40}}
\put(5,40){\line(1,0){40}}
\put(10,5){\line(0,1){40}}
\put(20,5){\line(0,1){40}}
\put(30,5){\line(0,1){40}}
\put(40,5){\line(0,1){40}}
\put(11,12){\makebox(0,0)[l]{$\scr l+1$}}
\put(11,22){\makebox(0,0)[l]{$\scr l$}}
\put(11,32){\makebox(0,0)[l]{$\scr l+1$}}
\put(11,42){\makebox(0,0)[l]{$\scr l$}}
\put(21,12){\makebox(0,0)[l]{$\scr l$}}
\put(21,22){\makebox(0,0)[l]{$\scr l+1$}}
\put(21,32){\makebox(0,0)[l]{$\scr l$}}
\put(21,42){\makebox(0,0)[l]{$\scr l+1$}}
\put(31,12){\makebox(0,0)[l]{$\scr l+1$}}
\put(31,22){\makebox(0,0)[l]{$\scr l$}}
\put(31,32){\makebox(0,0)[l]{$\scr l+1$}}
\put(31,42){\makebox(0,0)[l]{$\scr l$}}
\put(41,12){\makebox(0,0)[l]{$\scr l$}}
\put(41,22){\makebox(0,0)[l]{$\scr l+1$}}
\put(41,32){\makebox(0,0)[l]{$\scr l$}}
\put(41,42){\makebox(0,0)[l]{$\scr l+1$}}
\put(50,10){.}
\end{picture}
\ee
In the notation of subsection \ref{sec:4.2} this corresponds to 
$(i_1,i_2)=(l,l+1)$. 
We label this ground state by an integer $l$ ($l=1,2,\cdots,r-2$)
and $m$ takes two values $0,1\in\Z/2\Z$.
Since heights on neighboring sites differ $\pm 1$, we have
\be
  \cH^{(k)}_{l,m}=0\quad(m\not\equiv l-k\!\!\!\pmod{2}).
\ee
Therefore we can identify
\be
  \cH^{(k)}_l=\bigoplus_{m=0}^1\cH^{(k)}_{l,m}\cong\cH^{(k)}_{l,l-k}.
\ee
This space of states $\cH^{(k)}_l$, 
on which the corner Hamiltonian $H_C^{(k)}$ acts,
has two labels $l$ and $k$ with range $1\leq l\leq r-2$, $1\leq k\leq r-1$.
This range is same as \eq{lkrange}. 

In the thermodynamic limit CTM's \eq{CTM} become
\be
  A^{(k)}(u)=C^{(k)}(u)=x^{-2uH_C^{(k)}},\quad 
  B^{(k)}(u)=D^{(k)}(u)=\sqrt{[k]}x^{2(u+1)H_C^{(k)}}.
\ee
Careful study of the corner Hamiltonian shows that the character \eq{chi} 
agrees with the Virasoro minimal unitary character \cite{ABF,DJKMO},
\be
  \chi_{l,l-k,k}(q)=\chi_{l,k}^{{\rm Vir}}(q),
\ee
where $\chi_{l,k}^{{\rm Vir}}(q)$ is given in \eq{Virch}.
This character $\chi_{l,k}^{{\rm Vir}}(q)$, which agrees with DVA one,
should be understood as a character of the representation of DVA.
Comparing the free filed realization given in subsection \ref{sec:3.4},
we make an identification \cite{LP96}
\ba
  \cH^{(k)}_l&\!\!=\!\!&\cL_{l,k}\,,
  \label{H=L}\\
  H_C^{(k)}&\!\!=\!\!&d\Bigm|_{\cL_{l,k}},
  \label{H_C=d}
\ea
where $\cL_{l,k}$ is an irreducible DVA module given in \eq{cohomologyDVA} 
as a cohomology of the Felder complex and $d$ is realized in \eq{d}.
We remark that $Z_{l,m}$ \eq{Zlchi} is in fact $m$-independent, 
$Z_{l,0}=Z_{l,1}$.

For later use we define $W^*$,
\be
  W^*\BW{a}{b}{c}{d}{u}=
  \Wb\BW{a}{b}{c}{d}{u}\Biggl|_{r\rightarrow\rs}\times\rho^*(u),
  \label{W*}
\ee
where $\rho^*(u)$ is given in \eq{rho*}.

\subsection{Vertex operators}\label{sec:5.2}

Although bosons are already introduced in subsection \ref{sec:3.4},
we present their definitions again.

Let us introduce free boson oscillator $\alpha_n$ ($n\in\Z_{\neq 0}$),
\be
  [\alpha_n,\alpha_m]=\frac{[n]_x[2n]_x}{n}
  \frac{[rn]_x}{[\rs n]_x}\delta_{n+m,0},
\ee
and use zero mode $a_0$ and $Q$ defined in \eq{an} (or $a'_0$ in \eq{a'0}).
The Fock space $\F_{l,k}$ is defined by
\be
  \F_{l,k}=\bigoplus_{m\geq 0}\bigoplus_{n_1\geq\cdots\geq n_m>0}\C
  \alpha_{-n_1}\cdots \alpha_{-n_m}\ketb{\alpha_{l,k}},
\ee
where $\ketb{\alpha}$ is given by \eq{ketalpha} with replacing $a_n$ by $h_n$,
\be
  a'_0\ketb{\alpha_{l,k}}=(\alpha_{l,k}-\az)\ketb{\alpha_{l,k}},\quad
  \alpha_{l,k}-\az=-\sqrt{\frac{r}{r-1}}l+\sqrt{\frac{r-1}{r}}k.
\ee
We use also free boson oscillator $\alpha'_n$ ($n\in\Z_{\neq 0}$),
\be
  \alpha'_n=(-1)^n\frac{[\rs n]_x}{[rn]_x}\alpha_n,\quad
  [\alpha'_n,\alpha'_m]=\frac{[n]_x[2n]_x}{n}
  \frac{[\rs n]_x}{[rn]_x}\delta_{n+m,0}.
\ee
Operators $\hat{l},\hat{k}:\F_{l,k}\rightarrow\F_{l,k}$ are defined by 
\be
  \hat{l}\,|_{\F_{l,k}}=l\times\id_{\F_{l,k}},\quad
  \hat{k}\,|_{\F_{l,k}}=k\times\id_{\F_{l,k}}.
  \label{hatlk}
\ee
If $\hat{l}$ and $\hat{k}$ appear in arguments of $[u]^*$ and $[u]$ 
respectively, they can be realized by $a'_0$,
\ba
  &&\bigl[u+\hat{l}\,\bigr]^*=(-1)^{l-k}\Bigl[u-\sqrt{r\rs}a'_0\Bigr]^*
  \quad\mbox{on $\F_{l,k}$},\\
  &&\bigl[u+\hat{k}\bigr]=(-1)^{l-k}\Bigl[u-\sqrt{r\rs}a'_0\Bigr]
  \quad\mbox{on $\F_{l,k}$}.
\ea 

Elliptic currents $x_{\pm}(z)$ for $U_x(\slth)$ (or $\B_{x,\lambda}(\slth)$)
of level $1$ ($c=1$) 
are obtained by a `dressing' procedure described in \ref{sec:4.4} {\bf 5}
\ba
  &x_+(z)&:\,\F_{l,k}\rightarrow \F_{l-2,k}\n
  &&x_+(z)=\,
  :\exp\Bigl(-\sum_{n\neq 0}\frac{\alpha_n}{[n]_x}z^{-n}\Bigr):\times 
  e^{\sqrt{\frac{r}{\rs}}Q}z^{\sqrt{\frac{r}{\rs}}a'_0+\frac{r}{\rs}},
  \label{x+}\\
  &x_-(z)&:\,\F_{l,k}\rightarrow \F_{l,k-2}\n
  &&x_-(z)=\,
  :\exp\Bigl(\sum_{n\neq 0}\frac{\alpha'_n}{[n]_x}z^{-n}\Bigr):\times
  e^{-\sqrt{\frac{\rs}{r}}Q}z^{-\sqrt{\frac{\rs}{r}}a'_0+\frac{\rs}{r}},
  \label{x-}
\ea
and they are interpreted as screening currents in subsection \ref{sec:3.4}.

The elliptic version of VO's (of type I and type II) are defined  
in terms of their trigonometric ones and a `twistor' given by 
an infinite product of the universal $R$ matrix 
described in subsection \ref{sec:4.4} {\bf 4}.  
They satisfy the commutation relations of the type 
\eq{com1}-\eq{com3} below.
As we do not know how to evaluate the twistor in the bosonic realization,   
we have solved the relations \eq{com1}-\eq{com3} directly 
for $\Phi_\ve(z),\Psi^*_\ve(z)$ ($\ve=\pm 1$). 
We write $\Phi_{\pm}(z)=\Phi_{\pm 1}(z)$ and 
$\Psi^*_{\pm}(z)=\Psi^*_{\pm 1}(z)$.
We obtain the following : \cite{LP96,MW96} 
\ba
  \!\!\!\!\!&\mbox{type I}&
  \Phi_\ve(z)\,:\,\F_{l,k}\rightarrow \F_{l,k-\ve}\n
  \!\!\!\!\!&&\Phi_-(z)=\sqrt{g}\,
  :\exp\Biggl(-\sum_{n\neq 0}\frac{\alpha'_n}{[2n]_x}z^{-n}\Biggr):\times
  e^{\frac12\sqrt{\frac{\rs}{r}}Q}
  z^{\frac12\sqrt{\frac{\rs}{r}}a'_0+\frac{\rs}{4r}},
  \label{Phi-}\\
  \!\!\!\!\!&&\Phi_+(z)=
  \oint_{C_{\Phi}(z)}\dz'\,\Phi_-(z)x_-(z')
  \frac{[u-u'-\frac12+\hat{k}]}{[u-u'+\frac12]}
  \frac{1}{\sqrt{[\hat{k}][\hat{k}-1]}},
  \label{Phi+}
\\
  \!\!\!\!\!&\mbox{type II}&
  \Psi^*_\ve(z)\,:\,\F_{l,k}\rightarrow \F_{l-\ve,k}\n
  \!\!\!\!\!&&\Psi^*_-(z)=\frac{1}{\sqrt{g^*}}\,
  :\exp\Biggl(\sum_{n\neq 0}\frac{\alpha_n}{[2n]_x}z^{-n}\Biggr):\times
  e^{-\frac12\sqrt{\frac{r}{\rs}}Q}
  z^{-\frac12\sqrt{\frac{r}{\rs}}a'_0+\frac{r}{4\rs}},
  \label{Psi-}\\
  \!\!\!\!\!&&\Psi^*_+(z)=
  \oint_{C_{\Psi^*}(z)}\dz'\,\Psi^*_-(z)x_+(z')
  \frac{[u-u'+\frac12-\hat{l}\,]^*}{[u-u'-\frac12]^*}
  \frac{1}{\sqrt{[\hat{l}\,]^*[\hat{l}-1]^*}}.
  \label{Psi+}
\ea
Here $z=x^{2u}$, $z'=x^{2u'}$, $\dz'=\frac{dz'}{2\pi iz'}$,
and normalization constants $g$ and $g^*$ will be given in \eq{g} and \eq{g^*}.
For $k=1$ ($l=1$), we adopt the following prescription; set $k=1+\epsilon$ 
($l=1+\epsilon$) and take a limit $\epsilon\rightarrow 0$ after all the
calculation.
(Another method of getting rid of this factor is a gauge transformation,
see subsection \ref{sec:5.4}.)
The poles of the integrand of \eq{Phi-}-\eq{Psi+}
and the integration contours are listed in the following table. 
For example, $C_{\Phi}(z)$ is a simple closed contour 
that encircles $x^{1+2rn}z$ ($n\ge 0$) but not $x^{-1-2rn}z$ ($n\ge 0$).  
\be
\begin{tabular}{|c|c|c|}
\hline
&inside&outside\\
\hline
$C_{\Phi}(z)$&$z'=x^{1+2rn}z$&$z'=x^{-1-2rn}z$\\
\hline
$C_{\Psi^*}(z)$&$z'=x^{-1+2\rs n}z$&$z'=x^{1-2\rs n}z$\\
\hline
\end{tabular}\quad(n=0,1,2,\cdots).
\label{tableCVO}
\ee
OPE formulas are given in subsection \ref{sec:5.a}.
In the conformal limit \eq{conflim}, $g^{-\frac12}\Phi_-(z)$ reduces to 
a primary field $V_{\alpha_{1,2}}(z)$ and 
$g^{*\frac12}\Psi^*_-(z)$ reduces to 
$V_{\alpha_{2,1}}(-z)$ (up to phase factor) where $V_{\alpha}(z)$ is 
given in \eq{Valpha}. $g^{-\frac12}\Phi_+(z)$ and $g^{*\frac12}\Psi^*_+(z)$ 
reduce to primary fields 
with screening current (screened vertex operator \cite{Fel89}).
We can show that type I VO $\Phi_{\ve}(z)$ commutes with BRST 
operator \eq{Xj}
\be
  [X_j,\Phi_{\ve}(z)]=0.
  \label{XjPhi}
\ee
Type II VO $\Psi^*_{\ve}(z)$ commutes with another BRST operator obtained
by using\eq{X'}, see subsection \ref{sec:5.b}.

The VO's given above satisfy the following commutation relations
($\ve_1,\ve_2=\pm 1$),
\ba
  \!\!\!\!\!\!\!\!\!\!\!\!\!\!\!
  &&\Phi_{\ve_2}(z_2)\Phi_{\ve_1}(z_1)
  =\!\!\!\!\!\sum_{\ve_1',\ve_2'=\pm 1\atop \ve_1'+\ve_2'=\ve_1+\ve_2}
  \!\!\!\!\!
  W\BW{\hat{k}}{\hat{k}+\ve_1'}{\hat{k}+\ve_2}{\hat{k}+\ve_1+\ve_2}{u_1-u_2}
  \Phi_{\ve'_1}(z_1)\Phi_{\ve'_2}(z_2),
  \label{com1}\\
  \!\!\!\!\!\!\!\!\!\!\!\!\!\!\!
  &&\Psi^*_{\ve_1}(z_1)\Psi^*_{\ve_2}(z_2)
  =\!\!\!\!\!\sum_{\ve_1',\ve_2'=\pm 1\atop \ve_1'+\ve_2'=\ve_1+\ve_2}
  \!\!\!\!\!
  W^*\BW{\hat{l}}{\hat{l}+\ve_1}{\hat{l}+\ve_2'}{\hat{l}+\ve_1+\ve_2}{u_1-u_2}
  \Psi^*_{\ve'_2}(z_2)\Psi^*_{\ve'_1}(z_1),
  \label{com2}\\
  \!\!\!\!\!\!\!\!\!\!\!\!\!\!\!
  &&\Phi_{\ve_2}(z_2)\Psi^*_{\ve_1}(z_1)=
  \tau(u_1-u_2)\Psi^*_{\ve_1}(z_1)\Phi_{\ve_2}(z_2),
  \label{com3}
\ea
where $z_i=x^{2u_i}$ and $\tau(u)$ is given in \eq{tau}.\\
{\it Proof.}\quad
First let us show \eq{com1}. 
For $(\ve_1,\ve_2)=(-1,-1)$, \eq{com1} is
$$
  \Phi_-(z_2)\Phi_-(z_1)=\rho(u_1-u_2)\Phi_-(z_1)\Phi_-(z_2).
$$
This is an OPE rule \eq{PhiPhirho} itself.
For $(\ve_1,\ve_2)=(-1,1)$, \eq{com1} is
\ban
  \Phi_+(z_2)\Phi_-(z_1)&\!\!=\!\!&
  W\BW{\hat{k}}{\hat{k}+1}{\hat{k}+1}{\hat{k}}{u_1-u_2}\Phi_+(z_1)\Phi_-(z_2)
  \\
  &&
  +W\BW{\hat{k}}{\hat{k}-1}{\hat{k}+1}{\hat{k}}{u_1-u_2}\Phi_-(z_1)\Phi_+(z_2).
\ean
We set (see subsection \ref{sec:5.a})
\ba
  &&\Phi_+(z)=\oint\dz'\,\Phi_-(z)x_-(z')\varphi_+(u,u';\hat{k}),\n
  &&\qquad\quad
  \varphi_+(u,u';\hat{k})=\frac{[u-u'-\frac12+\hat{k}]}{[u-u'+\frac12]}
  \frac{1}{\sqrt{[\hat{k}][\hat{k}-1]}},\\
  &&x_-(z_1)\Phi_-(z_2)=f(u_1-u_2)\Phi_-(z_2)x_-(z_1),\quad
  f(u)=\frac{[u+\frac12]}{[-u+\frac12]},\\
  &&x_-(z_1)x_-(z_2)=h(u_1-u_2)x_-(z_2)x_-(z_1),\quad
  h(u)=\frac{[u-1]}{[u+1]}.
\ea
By using OPE rules in subsection \ref{sec:5.a}, the above equation becomes
\ban
  &&
  \oint\dz'\,\Phi_-(z_2)\Phi_-(z_1)x_-(z')
  \varphi_+(u_2,u';\hat{k}+1)f(u'-u_1)\\
  &=\!\!&
  \oint\dz'\,\Phi_-(z_2)\Phi_-(z_1)x_-(z')
  \Wb\BW{\hat{k}}{\hat{k}+1}{\hat{k}+1}{\hat{k}}{u_1-u_2}
  \varphi_+(u_1,u';\hat{k}+1)f(u'-u_2)\\
  &&+\oint\dz'\,\Phi_-(z_2)\Phi_-(z_1)x_-(z')
  \Wb\BW{\hat{k}}{\hat{k}-1}{\hat{k}+1}{\hat{k}}{u_1-u_2}
  \varphi_+(u_2,u';\hat{k}).
\ean
Careful analysis of the location of poles shows that we can take a common 
integration contour. Therefore it is sufficient to compare the integrands,
\ban
  \varphi_+(u_2,u';\hat{k}+1)f(u'-u_1)&\!\!=\!\!&
  \Wb\BW{\hat{k}}{\hat{k}+1}{\hat{k}+1}{\hat{k}}{u_1-u_2}
  \varphi_+(u_1,u';\hat{k}+1)f(u'-u_2)\\
  &&+\Wb\BW{\hat{k}}{\hat{k}-1}{\hat{k}+1}{\hat{k}}{u_1-u_2}
  \varphi_+(u_2,u';\hat{k}).
\ean
This equation does hold by the Riemann identity \eq{Rid}.
$(\ve_1,\ve_2)=(1,-1)$ case is similar.
For $(\ve_1,\ve_2)=(1,1)$, \eq{com1} is
$$
  \Phi_+(z_2)\Phi_+(z_1)=\rho(u_1-u_2)\Phi_+(z_1)\Phi_+(z_2).
$$
By using OPE rules, this becomes
\ban
  &&
  \oint\dz'_2\oint\dz'_1\,\Phi_-(z_2)\Phi_-(z_1)x_-(z'_2)x_-(z'_1)
  \varphi_+(u_2,u'_2;\hat{k}-1)\varphi_+(u_1,u'_1;\hat{k})f(u'_2-u_1)\\
  &=\!\!&
  \oint\dz'_1\oint\dz'_2\,\Phi_-(z_2)\Phi_-(z_1)x_-(z'_2)x_-(z'_1)\\
  &&\qquad\qquad\qquad\qquad\qquad\times
  \varphi_+(u_1,u'_1;\hat{k}-1)\varphi_+(u_2,u'_2;\hat{k})
  f(u'_1-u_2)h(u'_1-u'_2).
\ean
Detailed analysis of the location of poles shows that we can take a common 
integration contour which is symmetric in $z'_1$ and $z'_2$. 
Consequently it is sufficient to compare the integrands after
symmetrization in $z'_1$ and $z'_2$.
Here symmetrization means
\ba
  &&\oint\dz'_1\oint\dz'_2\,x_-(z'_1)x_-(z'_2)F(u'_1,u'_2)\n
  &=\!\!&\oint\dz'_1\oint\dz'_2\,x_-(z'_1)x_-(z'_2)
  \sfrac12\Bigl(F(u'_1,u'_2)+F(u'_2,u'_1)h(u'_1-u'_2)\Bigr).
\ea
We remark that $h(-u)=h(u)^{-1}$.
Therefore we are enough to show
\ban
  &&\varphi_+(u_2,u'_2;\hat{k}-1)\varphi_+(u_1,u'_1;\hat{k})f(u'_2-u_1)
  +(u'_1\leftrightarrow u'_2)\times h(u'_1-u'_2)\\
  &=\!\!&\varphi_+(u_1,u'_1;\hat{k}-1)\varphi_+(u_2,u'_2;\hat{k})
  f(u'_1-u_2)h(u'_1-u'_2)+(u'_1\leftrightarrow u'_2)\times h(u'_1-u'_2), 
\ean
and this is correct due to the Riemann identity \eq{Rid}.

Next let us show \eq{com2}. 
We set (see subsection \ref{sec:5.a})
\ba
  &&\Psi^*_+(z)=\oint\dz'\,\Psi^*_-(z)x_+(z')\varphi^*_+(u,u';\hat{l}\,),\n
  &&\qquad\quad
  \varphi^*_+(u,u';\hat{l}\,)
  =\frac{[u-u'+\frac12-\hat{l}\,]^*}{[u-u'-\frac12]^*}
  \frac{1}{\sqrt{[\hat{l}\,]^*[\hat{l}-1]^*}},\\
  &&x_+(z_1)\Psi^*_-(z_2)=f^*(u_1-u_2)\Psi^*_-(z_2)x_+(z_1),\quad
  f^*(u)=\frac{[u-\frac12]^*}{[-u-\frac12]^*},\\
  &&x_+(z_1)x_+(z_2)=h^*(u_1-u_2)x_+(z_2)x_+(z_1),\quad
  h^*(u)=\frac{[u+1]^*}{[u-1]^*}.
\ea
We remark that
\ba
  &&\varphi^*_+(u,u';a)=\varphi_+(-u,-u';a)\Bigl|_{r\rightarrow\rs},\\
  &&f^*(u)=f(-u)\Bigl|_{r\rightarrow\rs},\\
  &&h^*(u)=h(-u)\Bigl|_{r\rightarrow\rs},
\ea
and \eq{W*} and \eq{refl}. 
Integrands of \eq{com2} are obtained from those of \eq{com1} by replacement
$r\rightarrow\rs$, $u\rightarrow -u$ and $\hat{k}\rightarrow\hat{l}$.
Careful analysis of the location of poles shows that we can take a common 
(symmetric) integration contour for each case. 
Therefore \eq{com2} holds.

\eq{com3} is easily shown, although we have to take care of integration 
contours.
\qed

The dual VO's are realized in the following way,
\ba
  \Phi^*_\ve(z)&\!\!=\!\!&
  \sqrt{[\hat{k}]}^{-1}\Phi_{-\ve}(x^{-2}z)\sqrt{[\hat{k}]},
  \label{Phi*=Phi}\\
  \Psi_\ve(z)&\!\!=\!\!&
  \sqrt{[\hat{l}\,]^*}\Psi^*_{-\ve}(x^{-2}z)\sqrt{[\hat{l}\,]^*}^{-1},
\ea
and normalization constants $g$ and $g^*$ in \eq{Phi-} and \eq{Psi-} are
\ba
  g^{-1}&\!\!=\!\!&x^{\frac{\rs}{2r}}
  \frac{1}{(x^2,x^{2r},x^{2r};x^{2r})_\infty}
  \frac{(x^4,x^{2r};x^4,x^{2r})_\infty}{(x^2,x^{2r+2};x^4,x^{2r})_\infty},
  \label{g}\\
  g^*&\!\!=\!\!&x^{-\frac{r}{2\rs}}
  \frac{1}{(x^{-2},x^{2\rs},x^{2\rs};x^{2\rs})_\infty}
  \frac{(x^2,x^{2\rs+2};x^4,x^{2\rs})_\infty}
       {(x^4,x^{2\rs};x^4,x^{2\rs})_\infty}.
  \label{g^*}
\ea
Then we have 
\ignore{
\ba
  \!\!\!\!\!\!\!\!\!\!
  &&\Phi_{\ve_2}(z)\Phi^*_{\ve_1}(z)=\delta_{\ve_1,\ve_2}\times\id,\quad
  \Psi_{\ve_1}(z_1)\Psi^*_{\ve_2}(z_2)
  =\frac{\delta_{\ve_1,\ve_2}}{1-\frac{z_1}{z_2}}\times\id+\cdots,\quad
  (z_1\rightarrow z_2),
  \label{inv1}\\
  \!\!\!\!\!\!\!\!\!\!
  &&\sum_{\ve=\pm 1}\Phi^*_{\ve}(z)\Phi_{\ve}(z)=\id,\qquad
  \sum_{\ve=\pm 1}\Psi^*_{\ve}(z_2)\Psi_{\ve}(z_1)
  =\frac{1}{1-\frac{z_1}{z_2}}\times\id+\cdots,\quad 
  (z_1\rightarrow z_2).
  \label{inv2}
\ea
}
\ba
  &&\sum_{\ve=\pm 1}\Phi^*_{\ve}(z)\Phi_{\ve}(z)=\id,
  \label{inv2I}\\
  &&\Phi_{\ve_2}(z)\Phi^*_{\ve_1}(z)=\delta_{\ve_1,\ve_2}\times\id,
  \label{inv1I}\\
  &&\sum_{\ve=\pm 1}\Psi^*_{\ve}(z_2)\Psi_{\ve}(z_1)
  =\frac{1}{1-\frac{z_1}{z_2}}\times\id+\cdots\quad 
  (z_1\rightarrow z_2),
  \label{inv2II}\\
  &&\Psi_{\ve_1}(z_1)\Psi^*_{\ve_2}(z_2)
  =\frac{\delta_{\ve_1,\ve_2}}{1-\frac{z_1}{z_2}}\times\id+\cdots\quad
  (z_1\rightarrow z_2),
  \label{inv1II}
\ea
and for $d$ in \eq{d}
\be
  w^d{\cal O}(z)w^{-d}={\cal O}(wz),\quad\mbox{for }
  {\cal O}=\Phi_{\ve},\Phi^*_{\ve},\Psi^*_{\ve},\Psi_{\ve},x_{\pm}.
  \label{wdO}
\ee

We identify type I VO's in subsection \ref{sec:4.2} and those here 
in the following way:
\ba
  \Phi^{(a-\ve,a)}(z)&\!\!=\!\!&\Phi_{\ve}(z)\Bigm|_{\cL_{l,a}},
  \label{Phi=Phi}\\
  \Phi^{*(a+\ve,a)}(z)&\!\!=\!\!&\Phi^*_{\ve}(z)\Bigm|_{\cL_{l,a}}.
  \label{Phi*=Phi*}
\ea
Then eqs. 
\eq{Phi*}-\eq{PhiPhi*} 
correspond to 
\eq{Phi*=Phi},\eq{com1},\eq{wdO},\eq{inv2I} and \eq{inv1I} respectively.

Next let us see how the DVA current is obtained from VO's.
Let introduce free boson oscillator $h_n$ ($n\in\Z_{\neq 0}$),
\ba
  &&h_n=(x-x^{-1})(-1)^n\frac{[\rs n]_x}{[2n]_x}\alpha_n
  =(x-x^{-1})\frac{[rn]_x}{[2n]_x}\alpha'_n,\\
  &&[h_n,h_m]=(x-x^{-1})^2\frac{1}{n}
  \frac{[n]_x}{[2n]_x}[rn]_x[\rs n]_x\delta_{n+m,0}.
\ea
As explained in subsection \ref{sec:3.4}, the DVA current $T(z)$ is 
realized as
\ba
  &&T(z)=\Lambda_+(z)+\Lambda_-(z),\n
  &&\Lambda_{\pm}(z)=\,
  :\exp\Bigl(\pm\sum_{n\neq 0}h_n(x^{\pm 1}z)^{-n}\Bigr):
  \times x^{\pm\sqrt{r\rs}a'_0}.
\ea
This $T(z)$ is obtained from type I VO's by fusing them \cite{JS97},
\ba
  &&\Phi_{\ve_2}(x^{1+r}z')\Phi^*_{\ve_1}(x^{1-r}z)\n
  &=\!\!&
  \Bigl(1-\frac{z}{z'}\Bigr)
  \delta_{\ve_1,\ve_2}T(z)\cdot x^{-\frac{\rs}{2}}
  \frac{(x^4,x^{4-2r};x^4)_\infty}{(x^2,x^{2-2r};x^4)_\infty}
  +\cdots\quad(z'\rightarrow z),
\ea
or from type II VO's
\ba
  &&\Psi_{\ve_1}(x^{1+\rs}z')\Psi^*_{\ve_2}(x^{1-\rs}z)\n
  &=\!\!&
  \frac{1}{1-\frac{z'}{z}}\delta_{\ve_1,\ve_2}
  \Bigl(-T(-z)\Bigr)\cdot(-x^{-\frac{r}{2}})
  \frac{(x^2,x^{2-2\rs};x^4)_\infty}{(x^4,x^{-2\rs};x^4)_\infty}
  +\cdots\quad(z'\rightarrow z).
\ea
Higher DVA currents \eq{T_{(n)}} in subsection \ref{sec:3.5} are 
also obtained by fusion,
\be
  \Phi_{\ve_2}(x^{1+rj}z')\Phi^*_{\ve_1}(x^{1-rj}z)
  =\delta_{\ve_1,\ve_2}\left\{\begin{array}{ll}
  \id&(j=0)\\
  (1-\frac{z}{z'})T_{(j)}(z)A_j&(j\geq 1)
  \end{array}\right.
  \;\;+\cdots\quad(z'\rightarrow z),
\ee
where $A_j$ is
\be
  A_j=x^{-\frac12r^*j}
  \frac{(x^4,x^{4-2rj};x^4)_{\infty}}{(x^2,x^{2-2rj};x^4)_{\infty}}
  \prod_{i=1}^{j-1}\frac{(x^{4-2ri},x^{-2ri};x^4)_{\infty}}
  {(x^{2-2ri},x^{2-2ri};x^4)_{\infty}},
\ee
and we have assumed that $r$ is generic.

\subsection{Local height probability}\label{sec:5.3}

Let us consider local height probabilities $P_{a_n,\cdots,a_0}(l)$. 
As remarked in subsection \ref{sec:5.1}, we have
\be
  P_{a_n,\cdots,a_0}(l)=\sum_{m=0}^1\frac{Z_{l,m}}{Z_l}
  P_{a_n,\cdots,a_0}(l,m)=\frac12P_{a_n,\cdots,a_0}(l,l-a_0).
\ee

One-point LHP is already obtained in \eq{1pt},
\be
  P_k(l)=Z_{l}^{-1}\,[k]\,\chi_{l,k}(x^4).
\ee
Here the partition function $Z_l$ \eq{Zl} is
\be
  Z_l=\sum_{k=1}^{r-1}\,[k]\,\chi_{l,k}(x^4),
\ee
where $\chi_{l,k}(q)$ is given in \eq{Virch}.
Two-point LHP's satisfy recursion relations \eq{recP}
\be
  \sum_aP_{a,b}(l)=P_b(l),\quad\sum_bP_{a,b}(l)=P_a(l).
  \label{2ptrec}
\ee
Since $P_{a,b}(l)$ vanishes unless $|a-b|=1$ and $1\leq a,b\leq r-1$, 
two-point LHP can be determined uniquely by these recursion relations 
and expressed in terms of one-point LHP.
Starting from $P_{0,1}(l)=P_{1,0}(l)=0$, we obtain
\ba
  P_{k+1,k}(l)=P_{k,k+1}(l)=\sum_{a=1}^k(-1)^{k-a}P_a(l)
  =Z_l^{-1}\sum_{a=1}^k(-1)^{k-a}\,[a]\,\chi_{l,a}(x^4),
  \label{2ptans}
\ea
and this satisfies $P_{r,r-1}(l)=P_{r-1,r}(l)=0$ 
(which is equivalent to $Z_{l,0}=Z_{l,1}$).
$P_{a,b}(l)=P_{b,a}(l)$ agrees with physical requirement \eq{reverse}.
For higher-point LHP's, however, the recursion relations can not 
determine them. So we will use vertex operator approach.

As explained in subsection \ref{sec:4.2}, local height probabilities
can be expressed in terms of corner transfer matrices and type I VO's.
In this case \eq{Pl} with \eq{LHP2} becomes
\ba
  \!\!\!\!\!&&P_{a_n,\cdots,a_0}(l)\\
  \!\!\!\!\!&=\!\!&
  Z_{l}^{-1}\,[a_0]\,\tr_{\cH^{(a_0)}_l}\Bigl(x^{4H_C^{(a_0)}}
  \Phi^{*(a_0,a_1)}(z)\cdots\Phi^{*(a_{n-1},a_n)}(z)
  \Phi^{(a_n,a_{n-1})}(z)\cdots\Phi^{(a_1,a_0)}(z)\Bigr).\nonumber
\ea
We evaluate this LHP using a free field realization of VO.
Our identification of the space of state and operators are 
\eq{H=L},\eq{H_C=d},\eq{Phi=Phi} and \eq{Phi*=Phi*}.
Then we have
\be
  P_{a_n,\cdots,a_0}(l)=
  Z_{l}^{-1}\,[a_0]\,\tr_{\cL_{l,a_0}}\Bigl(x^{4d}
  \Phi^*_{\ve_1}(z)\cdots\Phi^*_{\ve_n}(z)
  \Phi_{\ve_n}(z)\cdots\Phi_{\ve_1}(z)\Bigr),
  \label{LHPABF}
\ee
where $\ve_i=a_{i-1}-a_i$.
Since the type I VO has the property \eq{XjPhi} 
(see also subsection \ref{sec:5.b}), 
we can apply the formula \eq{trO} to the trace in the above LHP.

We illustrate this free field calculation 
by taking two-point LHP as an example.
Two-point LHP is ($\ve=\pm 1$)
\ba
  P_{k-\ve,k}(l)&\!\!=\!\!&
  Z_{l}^{-1}\,[k]\,\tr_{\cL_{l,k}}\Bigl(x^{4d}
  \Phi^*_{\ve}(z)\Phi_{\ve}(z)\Bigr)\n
  &\!\!=\!\!&Z_{l}^{-1}\sqrt{[k][k-\ve]}\,\tr_{\cL_{l,k}}\Bigl(x^{4d}
  \Phi_{-\ve}(x^{-2}z)\Phi_{\ve}(z)\Bigr).
  \label{2pt}
\ea
{}From \eq{2pt}, \eq{g} and OPE in subsection \ref{sec:5.a}, we have
\ba
  \!\!\!\!\!\!\!\!\!\!&&P_{k-1,k}(l)\n
  \!\!\!\!\!\!\!\!\!\!&=\!\!&
  Z_l^{-1}(x^2,x^2,x^{2r};x^{2r})_{\infty}\oint_C\dz'
  \frac{(x^{2r-1}\zeta,x^{2r+1}\zeta;x^{2r})_{\infty}}
  {(x\zeta,x^3\zeta;x^{2r})_{\infty}}\\
  \!\!\!\!\!\!\!\!\!\!&&\times\tr_{\cL_{l,k}}\Bigl(x^{4d}g^{-1}
  :\Phi_-(x^{-2}z)\Phi_-(z)x_-(z'):
  z^{-\frac32\frac{\rs}{r}}x^{\frac{\rs}{2r}}
  \sqrt{\frac{[k][k-1]}{[\hat{k}][\hat{k}-1]}}
  \frac{[u-u'-\frac12+\hat{k}]}{[u-u'+\frac12]}\Bigr),\nonumber
\ea
where $z=x^{2u}$, $z'=x^{2u'}$ and $\zeta=\frac{z'}{z}$.
Calculation of trace is separated into two parts, oscillator part and
zero mode part.
The oscillator parts are common for all Fock spaces
\ba
  &&\tr_{\F_{l',k}}^{\rm osc}\Biggl(
  x^{4d^{\rm osc}}:\exp\Bigl(\sum_{n\neq 0}\frac{\alpha'_n}{[n]_x}
  ({z'}^{-n}-x^nz^{-n})\Bigr):\Biggr)\n
  &=\!\!&\frac{1}{(x^4;x^4)_{\infty}}
  \frac{(x^4,x^4,x^{2r+3}\zeta,x^{2r+1}\zeta^{-1};x^2,x^{2r})_{\infty}}
  {(x^{2r+2},x^{2r+2},x^5\zeta,x^3\zeta^{-1};x^2,x^{2r})_{\infty}},
\ea
where $\zeta=\frac{z'}{z}$ and we have used \eq{traceCS1}
(Or use the formula in subsection \ref{sec:5.a}).
On the other hand, the zero mode part,
\be
  \tr_{\cL_{l,k}}^{\rm zero}\Bigl(x^{4d^{\rm zero}}
  (\sfrac{z'}{z})^{-\sqrt{\frac{\rs}{r}}a'_0+\frac{\rs}{r}}
  x^{-\sqrt{\frac{\rs}{r}}a'_0}
  \sqrt{\frac{[k][k-1]}{[\hat{k}][\hat{k}-1]}}
  \frac{[u-u'-\frac12+\hat{k}]}{[u-u'+\frac12]}\Bigr),
  \label{trzero}
\ee
can be calculated by \eq{trO} with the Felder complex \eq{Cj}
\be
  C_{2j}=\F_{l-2\rs j,k},\quad C_{2j+1}=\F_{-l-2\rs j,k}.
\ee
\eq{trzero} becomes
\be
  \frac{[u-u'-\frac12+k]}{[u-u'+\frac12]}{\cal O}_{l,k}(\sfrac{z'}{z}),
\ee
where ${\cal O}_{l,k}(\zeta)$ is
(this definition is different from \cite{LP96},
${\cal O}_{l,k}^{LP}(\zeta^{-2})=\frac{\zeta^{-l+\frac{\rs}{r}(k-1)}
x^{\frac{\rs}{r}}}{(x^4;x^4)_{\infty}}{\cal O}_{l,k}(x^{-1}\zeta)$)
\ba
  {\cal O}_{l,k}(\zeta)&\!\!=\!\!&
  \sum_{j\in\Z}(-1)^j\tr_{C_j}^{{\rm zero}}\Bigl(x^{4d}
  \zeta^{-\sqrt{\frac{\rs}{r}}a'_0+\frac{\rs}{r}}
  x^{-\sqrt{\frac{\rs}{r}}a'_0}\Bigr)\n
  &=\!\!&\sum_{j\in\Z}\Bigl(
  x^{4(h_{l-2\rs j,k}-\frac{c}{24})}
  (x\zeta)^{l-2\rs j-\frac{\rs}{r}k}
  -x^{4(h_{-l-2\rs j,k}-\frac{c}{24})}
  (x\zeta)^{-l-2\rs j-\frac{\rs}{r}k}\Bigr)
  \zeta^{\frac{\rs}{r}}\\
  &=\!\!&x^{4(h_{l,k}-\frac{c}{24})}
  (x\zeta)^{l-\frac{\rs}{r}(k-1)}x^{-\frac{\rs}{r}}
  \times\Bigl(\Theta_{8r\rs}(-x^{4(\rs k-rl+r\rs)}(x\zeta)^{-2\rs})\n
  &&\qquad\qquad\qquad\qquad\qquad\qquad\qquad
  -\Theta_{8r\rs}(-x^{4(\rs k+rl+r\rs)}(x\zeta)^{-2\rs})x^{4lk}(x\zeta)^{-2l}
  \Bigr).\nonumber
\ea
So we have obtained an integral representation of the two-point LHP,
\be
  P_{k-1,k}(l)=Z_l^{-1}\frac{(x^2;x^2)_{\infty}^3}{(x^4;x^4)_{\infty}}
  I_{l,k},
\ee
where $I_{l,k}$ is
\ba
  &&I_{l,k}=\oint_{|z'|=|z|}\dz'F_{l,k}(\zeta)
  =\oint_{|\zeta|=1}\underline{d\zeta}F_{l,k}(\zeta),\\
  &&F_{l,k}(\zeta)=
  \frac{\Theta_{x^{2r}}(x^{-1+2k}\zeta^{-1})}{\Theta_{x^{2r}}(x\zeta)}
  {\cal O}_{l,k}(\zeta)\zeta^{-\frac{1}{r}(k-1)}x^{\frac{1}{r}k(k-1)+1-k}.
\ea

Next let us evaluate this integral.
$F_{l,k}(\zeta)$ has simple poles at $\zeta=x^{1+2n}$ ($n\in\Z$).
Since ${\cal O}_{l,k}(\zeta)$ satisfies
\be
  {\cal O}_{l,k}(x^{4rm}\zeta)={\cal O}_{l,k}(\zeta)
  \zeta^{-2\rs m}x^{-4r\rs m^2+2\rs m}\quad(m\in\Z),
\ee
$F_{l,k}(\zeta)$ has the property
\be
  F_{l,k}(x^{4r}\zeta)=F_{l,k}(\zeta).
  \label{Flaprop}
\ee
Consequently $I_{l,k}$ requires some regularization.
We regularize $I_{l,k}$ in the following way,
\be
  I_{l,k}=\lim_{\epsilon\rightarrow 0}I_{l,k}^{\epsilon},\quad
  I_{l,k}^{\epsilon}=\oint_{|\zeta|=1}\underline{d\zeta}
  \,\zeta^{\epsilon}F_{l,k}(\zeta).
\ee
Deforming the contour
\be
  I_{l,k}^{\epsilon}=
  \oint_{\zeta=x,x^3,\cdots,x^{4r-1}}
  \underline{d\zeta}\,\zeta^{\epsilon}F_{l,k}(\zeta)
  +\oint_{|\zeta|=x^{4r}}
  \underline{d\zeta}\,\zeta^{\epsilon}F_{l,k}(\zeta),
\ee
and using \eq{Flaprop}, we obtain
\be
  I_{l,k}^{\epsilon}=\frac{1}{1-x^{4r\epsilon}}
  \oint_{\zeta=x,x^3,\cdots,x^{4r-1}}
  \underline{d\zeta}\,\zeta^{\epsilon}F_{l,k}(\zeta).
\ee
We remark that
\be
  \oint_{\zeta=x,x^3,\cdots,x^{4r-1}}
  \underline{d\zeta}F_{l,k}(\zeta)=0,
\ee
namely
\be
  \sum_{a=0}^{2r-1}(-1)^a[k-1-a]x^{\frac{\rs}{r}a^2}{\cal O}_{l,k}(x^{1+2a})
  =0.
  \label{sum=0}
\ee
Using these and picking up residues at $\zeta=x,x^3,\cdots,x^{4r-1}$, 
we obtain
\be
  P_{k-1,k}(l)=\frac{1}{Z_l(x^4;x^4)_{\infty}}
  \sum_{a=0}^{2r-1}\frac{-a}{2r}(-1)^a[k-1-a]
  x^{\frac{\rs}{r}a^2}{\cal O}_{l,k}(x^{1+2a}).
  \label{2ptboson}
\ee
We can check that this answer (and $P_{k,k-1}(l)=P_{k-1,k}(l)$)
satisfies \eq{2ptrec} and $P_l(0,1)=0$ (use \eq{sum=0}). 
Therefore this answer obtained by free field approach agrees with \eq{2ptans}.

For multi-point LHP \eq{LHPABF}, we can write down its integral representation
in a similar way. (Evaluation of the integral is another problem.)
See \cite{LP96}.
\ignore{
Here we give only the integral formula for the traces over the Fock space
in a general situation
\be
  \tr_{\F_{l,k}}\Bigl(x^{4d}\Phi_{\ve_1}(z_1)\cdots\Phi_{\ve_N}(z_N)\Bigr),
\ee
with $\sum_{i=1}^N\ve_i=0$.
}

\subsection{Form factor}\label{sec:5.4}

Let us recall the translation invariant vacuum state \eq{vac} 
\be
  \ket{\mbox{vac}}=\sqrt{[\hat{k}]}\,x^{2d},
  \label{vacABF}
\ee
and the action of the column-to-column transfermatrix \eq{Tcolf2}
\be
  \Bigl(\T_{\rm col}(u)\ket{f}\Bigr)^{(k)}
  =\sum_{\ve=\pm1}\Phi_{\ve}(z)\cdot f\cdot\Phi_{-\ve}(z)\Bigm|_{\cL_{l,k}},
\ee
where $f$ is a linear map $f:\cL_{l,a}\rightarrow\cL_{l,a}$.

For the time being, we consider operators on the Fock spaces $\F_{l,k}$ 
not on the cohomology $\cL_{l,k}$.
Action of $\T_{\rm col}$ is $\T_{\rm col}(u)\ket{f}
=\sum_{\ve=\pm1}\Phi_{\ve}(z)\cdot f\cdot\Phi_{-\ve}(z)$.
Let us define an `excited' state
\be
  \ket{w_m,\cdots,w_1}_{\ve_m,\cdots,\ve_1}
  =\Psi^*_{\ve_m}(w_m)\cdots\Psi^*_{\ve_1}(w_1)\ket{\mbox{vac}}
  \,:\,\F_{l,k}\rightarrow\F_{l-\sum_i\ve_i,k}.
  \label{excited}
\ee
(We remark that the original description of this $\Psi^*$ is 
$\Psi^*\otimes 1$, see \eq{Af}.)
This state is an eigenstate of $\T_{\rm col}$
\ba
  \T_{\rm col}(u)\ket{w_m,\cdots,w_1}_{\ve_m,\cdots,\ve_1}
  &\!\!=\!\!&\sum_{\ve=\pm1}\Phi_{\ve}(z)\cdot
  \Psi^*_{\ve_m}(w_m)\cdots\Psi^*_{\ve_1}(w_1)\sqrt{[\hat{k}]}\,x^{2d}\cdot
  \Phi_{-\ve}(z)\n
  &\!\!=\!\!&\prod_{j=1}^m\tau(v_j-u)\cdot
  \ket{w_m,\cdots,w_1}_{\ve_m,\cdots,\ve_1},
\ea
where $z=x^{2u}$, $w_j=x^{2v_j}$ and we have used \eq{com3}.
Therefore type II VO creates (or annihilates) `particles'. 
But we remark that this state is a `true' eigenstate of $\T_{\rm col}$ 
because it is not a linear map on $\cL_{l,k}$ in general. 
We will show when this state becomes a true eigenstate.

To avoid $\frac{1}{\sqrt{[\hat{k}][\hat{k}-1]}}$ and 
$\frac{1}{\sqrt{[\hat{l}\,]^*[\hat{l}-1]^*}}$ factors
we perform a gauge transformation (see \eq{Wgauge})
\ba
  &&\widetilde{W}\BW{a}{b}{c}{d}{u}
  =\frac{F(a,b)F(b,d)}{F(a,c)F(c,d)}W\BW{a}{b}{c}{d}{u},\quad
  F(a,b)=([a][b])^{-\frac14},\\
  &&\widetilde{W}^*\BW{a}{b}{c}{d}{u}
  =\frac{F^*(a,b)F^*(b,d)}{F^*(a,c)F^*(c,d)}W^*\BW{a}{b}{c}{d}{u},\quad
  F^*(a,b)=([a]^*[b]^*)^{\frac14},\\
  &&\widetilde{\Phi}_-(z)=\Phi_-(z),\quad
  \widetilde{\Phi}_+(z)=\Phi_+(z)\sqrt{[\hat{k}][\hat{k}-1]},
  \label{tPhi}\\
  &&\widetilde{\Psi}^*_-(z)=\Psi^*_-(z),\quad
  \widetilde{\Psi}^*_+(z)=\Psi^*_+(z)\sqrt{[\hat{l}\,]^*[\hat{l}-1]^*}.
  \label{tPsi}
\ea
$\widetilde{W}$ differs from $W$ only in the following component
\be
  \widetilde{W}\BW{a}{a\pm 1}{a\mp 1}{a}{u}=\frac{[a\mp 1]}{[a]}
  \frac{[-u]}{[1+u]}\rho(u).
\ee
In this gauge the reflection symmetry \eq{refl} is lost.
$\widetilde{W}^*$ is
\be
  \widetilde{W}^*\BW{a}{b}{c}{d}{u}=
  \left(\rho(u)^{-1}\widetilde{W}\BW{a}{c}{b}{d}{u}\right)
  \Biggl|_{r\rightarrow\rs}\times\rho^*(u).
\ee
$\widetilde{\Phi}_{\ve}(z)$ and $\widetilde{\Psi}^*_{\ve}(z)$ satisfy
\eq{com1}-\eq{com3} with $(\Phi,\Psi^*,W,W^*)$ replaced by 
$(\widetilde{\Phi},\widetilde{\Psi}^*,\widetilde{W},\widetilde{W}^*)$ 
respectively. \eq{com3} is also hold by replacing $\Psi^*$ with 
$\widetilde{\Psi}^*$ and $\Phi$ unchanged.
Therefore the state
\be
  \ket{w_m,\cdots,w_1}_{\ve_m,\cdots,\ve_1}^{\sim}
  =\widetilde{\Psi}^*_{\ve_m}(w_m)\cdots\widetilde{\Psi}^*_{\ve_1}(w_1)
  \ket{\mbox{vac}}
  \,:\,\F_{l,k}\rightarrow\F_{l-\sum_i\ve_i,k},
  \label{excitedt}
\ee
which differs from \eq{excited} by a multiplicative constant factor
dependent on $\hat{l}$, is also an eigenstate of $\T_{\rm col}$.

How to realize the type II VO on the cohomology is discussed in 
subsection \ref{sec:5.b}.
In order to interpret the state \eq{excitedt} as a linear map on $\cL_{l,k}$,
the first requirement is $\ds\sum_{i=1}^m\ve_i=0$. 
Hence $m$ is an even integer, $m=2n$. 
Then $\ds\prod_{i=1}^{\kuru\atop 2n}\widetilde{\Psi}^*_{\ve_i}(w_i)$
and the BRST charge satisfy the intertwining property
\be
  Q_m\cdot\prod_{i=1}^{\kuru\atop 2n}\widetilde{\Psi}^*_{\ve_i}(w_i)
  =(-1)^n\prod_{i=1}^{\kuru\atop 2n}\widetilde{\Psi}^*_{-\ve_i}(w_i)\cdot 
  Q_m\quad(m=l,\rs-l).
\ee
Therefore this product of type II VO's is a well-defined operator on the
cohomology $\cL_{l,k}$. Explicitly it is
\be
  \prod_{i=1}^{\kuru\atop 2n}\widetilde{\Psi}^*_{\ve_i}(w_i)
  \;\;\mbox{on $C_{2j}$},\quad
  (-1)^n\prod_{i=1}^{\kuru\atop 2n}\widetilde{\Psi}^*_{-\ve_i}(w_i)
  \;\;\mbox{on $C_{2j+1}$}.
\ee
Consequently the state 
$\ket{w_{2n},\cdots,w_1}_{\ve_{2n},\cdots,\ve_1}^{\sim}$ with 
$\sum_{i=1}^{2n}\ve_i=0$ is a true eigenstate of $\T_{\rm col}$.
(If $m$ is an odd integer or $\sum_i\ve_i$ does not vanish, then
$\ket{w_m,\cdots,w_1}_{\ve_m,\cdots,\ve_1}$ is a map from $\F_{l,k}$ to 
$\F_{l',k}$ ($l'=l-\sum_i\ve_i\neq l$), namely in the original description
it corresponds to a state whose half is in the ground state $l$ and 
the other half in $l'$.
As remarked in subsection \ref{sec:4.2}, to obtain a complete set of 
excited states, we need not only the translation invariant vacuum
state \eq{vacABF} but also the translation non-invariant vacuum states.)
The form factors of local operator $\widetilde{{\cal O}}$ are defined by
\be
  \bra{\mbox{vac}}\widetilde{{\cal O}}
  \ket{w_m,\cdots,w_1}_{\ve_m,\cdots,\ve_1}.
\ee

Next let us study \eq{Qlm} with the following ${\cal O}={\cal O}_{\Psi^*}$
\be
  {\cal O}_{\Psi^*}=\prod_{i=1}^{\kuru\atop 2n}\widetilde{\Psi}^*_{\ve_i}(w_i)
  \qquad(w_j=x^{2v_j},\;\sum_{i=1}^{2n}\ve_i=0).
\ee
In this model we have
\be
  Q_{a_n,\cdots,a_0}(l\,;z|{\cal O}_{\Psi^*})
  =\sum_{m=0}^1\frac{Z_{l,m}}{Z_l}Q_{a_n,\cdots,a_0}(l,m\,;z|{\cal O}_{\Psi^*})
  =\frac12Q_{a_n,\cdots,a_0}(l,l-a_0\,;z|{\cal O}_{\Psi^*}).
\ee
This is the form factor of 
$\Phi^*_{a_0-a_1}(z)\cdots\Phi^*_{a_{n-1}-a_n}(z)
\Phi_{a_{n-1}-a_n}(z)\cdots\Phi_{a_0-a_1}(z)$.
\eq{recQ1} and \eq{recQ2} with $n=1$ become
\ba
  \sum_{\ve=\pm1}Q_{a+\ve,a}(l\,;z|{\cal O}_{\Psi^*})
  &\!\!=\!\!&Q_a(l|{\cal O}_{\Psi^*}),\\
  \sum_{\ve=\pm1}Q_{a,a+\ve}(l\,;z|{\cal O}_{\Psi^*})
  &\!\!=\!\!&G(z)Q_a(l|{\cal O}_{\Psi^*}),
\ea
where $Q_a(l|{\cal O}_{\Psi^*})=Q_a(l\,;z|{\cal O}_{\Psi^*})$ and $G(z)$ is 
\be
  G(z)=\prod_{j=1}^{2n}\tau(w_j-u).
\ee
This $Q_{a,b}(l\,;z|{\cal O}_{\Psi^*})$ is uniquely determined by these 
recursion relations and it is expressed in terms of $Q_a(l|{\cal O}_{\Psi^*})$.
Starting from 
$Q_{0,1}(l\,;z|{\cal O}_{\Psi^*})=Q_{1,0}(l\,;z|{\cal O}_{\Psi^*})=0$, 
we obtain
\ba
  Q_{a+1,a}(l\,;z|{\cal O}_{\Psi^*})&\!\!=\!\!&
  \sum_{b=1\atop b\equiv a\!\!\!\!\!\pmod{2}}^aQ_b(l|{\cal O}_{\Psi^*})
  -G(z)\sum_{b=1\atop b\not\equiv a\!\!\!\!\!\pmod{2}}^a
  Q_b(l|{\cal O}_{\Psi^*}),\\
  Q_{a,a+1}(l\,;z|{\cal O}_{\Psi^*})&\!\!=\!\!&
  -\sum_{b=1\atop b\not\equiv a\!\!\!\!\!\pmod{2}}^aQ_b(l|{\cal O}_{\Psi^*})
  +G(z)\sum_{b=1\atop b\equiv a\!\!\!\!\!\pmod{2}}^aQ_b(l|{\cal O}_{\Psi^*}).
  \label{Qaa+1}
\ea
This answer should satisfy 
$Q_{r,r-1}(l\,;z|{\cal O}_{\Psi^*})=Q_{r-1,r}(l\,;z|{\cal O}_{\Psi^*})=0$.

As an example let us calculate 
$Q_k(l|\widetilde{\Psi}^*_+(w_2)\widetilde{\Psi}^*_-(w_1))$ 
by free field realization. By using \eq{trO} it becomes
\ba
  \!\!\!\!\!\!\!\!\!\!&&
  Q_k(l|\widetilde{\Psi}^*_+(w_2)\widetilde{\Psi}^*_-(w_1))
  =Z_l^{-1}[k]\,\tr_{\cL_{l,k}}
  \Bigl(x^{4d}\widetilde{\Psi}^*_+(w_2)\widetilde{\Psi}^*_-(w_1)\Bigr)\n
  \!\!\!\!\!\!\!\!\!\!&=\!\!&
  Z_l^{-1}[k]\sum_{j\in\Z}\Biggl(\tr_{C_{2j}}
  \Bigl(x^{4d}\widetilde{\Psi}^*_+(w_2)\widetilde{\Psi}^*_-(w_1)\Bigr)
  +\tr_{C_{2j+1}}
  \Bigl(x^{4d}\widetilde{\Psi}^*_-(w_2)\widetilde{\Psi}^*_+(w_1)\Bigr)
  \Biggr).
\ea
By using OPE and trace rule listed in subsection \ref{sec:5.a}, we obtain
\ba
  &&Q_k(l|\widetilde{\Psi}^*_+(w_2)\widetilde{\Psi}^*_-(w_1))\n
  &=\!\!&Z_l^{-1}[k]{g^*}^{-1}\Biggl(
  \oint_{C_{\Psi^*}(w_2)}\dw'(w_1w_2)^{\frac{r}{2\rs}l}{w'}^{-\frac{r}{\rs}l}
  \dbr{x_+(w')\Psi^*_-(w_1)}\frac{[v_2-v'-\frac12-l]^*}{[v_2-v'-\frac12]^*}\n
  &&\qquad\qquad\qquad\qquad\times
  \Theta_{x^{8r\rs}}(-(\sfrac{{w'}^2}{w_1w_2})^rx^{4(-rl+\rs k+r\rs)})\n
  &&\qquad\qquad
  +\oint_{C_{\Psi^*}(w_1)}\dw'
  (w_1w_2)^{-\frac{r}{2\rs}l}{w'}^{\frac{r}{\rs}l}
  \dbr{\Psi^*_-(w_1)x_+(w')}\frac{[v_1-v'+\frac12+l]^*}{[v_1-v'-\frac12]^*}\n
  &&\qquad\qquad\qquad\qquad\times
  \Theta_{x^{8r\rs}}(-(\sfrac{{w'}^2}{w_1w_2})^rx^{4(rl+\rs k+r\rs)})
  x^{4lk}\Biggr)\n
  &&\qquad\times
  (w_1w_2)^{-\frac12 k+\frac{r}{4\rs}}{w'}^{k+\frac{r}{\rs}}
  x^{4h_{l,k}-\frac{c}{24}}
  \dbr{\Psi^*_-(w_2)\Psi^*_-(w_1)}\dbr{\Psi^*_-(w_2)x_+(w')}\n
  &&\qquad\times
  (x^4;x^4)_{\infty}^{-1}
  \prod_{i,j=1}^2F_{\Psi^*_-,\Psi^*_-}(\sfrac{w_i}{w_j})\cdot
  \prod_{i=1}^2F_{\Psi^*_-,x_+}(\sfrac{w_i}{w'})
  F_{x_+,\Psi^*_-}(\sfrac{w'}{w_i}),
\ea
where $F_{A,B}(z)$ is given in subsection \ref{sec:5.a}.
$Q_{k-1,k}(l\,;z|\widetilde{\Psi}^*_+(w_2)\widetilde{\Psi}^*_-(w_1))$ can
be obtained by using this result and \eq{Qaa+1}.
If you want to calculate it directly, you will follow
\ba
  \!\!\!\!\!\!\!\!\!\!&&
  Q_{k-1,k}(l\,;z|\widetilde{\Psi}^*_+(w_2)\widetilde{\Psi}^*_-(w_1))
  =Z_l^{-1}[k]\,\tr_{\cL_{l,k}}
  \Bigl(x^{4d}\Phi^*_+(z)\Phi_+(z)
  \widetilde{\Psi}^*_+(w_2)\widetilde{\Psi}^*_-(w_1)\Bigr)\n
  \!\!\!\!\!\!\!\!\!\!&=\!\!&
  Z_l^{-1}[k]\sum_{j\in\Z}\Biggl(
  \tr_{C_{2j}}\Bigl(x^{4d}\Phi^*_+(z)\Phi_+(z)
  \widetilde{\Psi}^*_+(w_2)\widetilde{\Psi}^*_-(w_1)\Bigr)\n
  \!\!\!\!\!\!\!\!\!\!&&\qquad\qquad\qquad
  +\tr_{C_{2j+1}}\Bigl(x^{4d}\Phi^*_+(z)\Phi_+(z)
  \widetilde{\Psi}^*_-(w_2)\widetilde{\Psi}^*_+(w_1)\Bigr)
  \Biggr)\n
  \!\!\!\!\!\!\!\!\!\!&=\!\!&
  \cdots
\ea

\renewcommand{\thesubsection}{\thesection.\alph{subsection}}
\setcounter{subsection}{0}
\subsection{OPE and trace}\label{sec:5.a}

{\bf OPE}\\
We list the normal ordering relations used in section \ref{sec:5}. 
Recall
$$
  \rs=r-1.
$$
For operators $A(z),B(w)$ that have the form 
$:\exp(\mbox{linear in boson}):$, we use the notation
\be
  A(z)B(w)=\dbr{A(z)B(w)}:A(z)B(w):,
  \label{dbrAB}
\ee
and write down only the part $\dbr{A(z)B(w)}$ : 
\ba
  \dbr{x_+(z_1)x_+(z_2)}&\!\!=\!\!&z_1^{\frac{2r}{\rs}}(1-\zeta)
  \frac{(x^{-2}\zeta;x^{2\rs})_\infty}{(x^{2\rs+2}\zeta;x^{2\rs})_\infty},\\
  \dbr{x_-(z_1)x_-(z_2)}&\!\!=\!\!&z_1^{\frac{2\rs}{r}}(1-\zeta)
  \frac{(x^{2}\zeta;x^{2r})_\infty}{(x^{2r-2}\zeta;x^{2r})_\infty},\\
  \dbr{x_\pm(z_1)x_\mp(z_2)}&\!\!=\!\!&z_1^{-2}
  \frac{1}{(1+x\zeta)(1+x^{-1}\zeta)},\\
  \dbr{\Phi_-(z_1)x_+(z_2)}&\!\!=\!\!&\dbr{x_+(z_1)\Phi_-(z_2)}=z_1+z_2,\\
  \dbr{\Phi_-(z_1)x_-(z_2)}&\!\!=\!\!&\dbr{x_-(z_1)\Phi_-(z_2)}
  =z_1^{-\frac{\rs}{r}}
  \frac{(x^{2r-1}\zeta;x^{2r})_\infty}{(x\zeta;x^{2r})_\infty},\\
  \dbr{\Psi^*_-(z_1)x_+(z_2)}&\!\!=\!\!&\dbr{x_+(z_1)\Psi_-^*(z_2)}
  =z_1^{-\frac{r}{\rs}}
  \frac{(x^{2\rs+1}\zeta;x^{2\rs})_\infty}{(x^{-1}\zeta;x^{2\rs})_\infty},\\
  \dbr{\Psi^*_-(z_1)x_-(z_2)}&\!\!=\!\!&\dbr{x_-(z_1)\Psi^*_-(z_2)}=z_1+z_2,\\
  \dbr{\Phi_-(z_1)\Phi_-(z_2)}&\!\!=\!\!&z_1^{\frac{\rs}{2r}}
  \frac{(x^2\zeta,x^{2r+2}\zeta;x^4,x^{2r})_\infty}
       {(x^4\zeta,x^{2r}\zeta;x^4,x^{2r})_\infty},\\
  \dbr{\Psi^*_-(z_1)\Psi^*_-(z_2)}&\!\!=\!\!&z_1^{\frac{r}{2\rs}}
  \frac{(\zeta,x^{2\rs+4}\zeta;x^4,x^{2\rs})_\infty}
       {(x^2\zeta,x^{2\rs+2}\zeta;x^4,x^{2\rs})_\infty},\\
  \dbr{\Phi_-(z_1)\Psi^*_-(z_2)}&\!\!=\!\!&\dbr{\Psi^*_-(z_1)\Phi_-(z_2)}
  =z_1^{-\frac12}\frac{(-x^3\zeta;x^4)_\infty}{(-x\zeta;x^4)_\infty},
\ea
where $\zeta=\frac{z_2}{z_1}$ and we have used \eq{Hausdorff}.

As meromorphic functions we have ($z_i=x^{2u_i}$)
\ba
  x_+(z_1)x_+(z_2)&\!\!=\!\!&x_+(z_2)x_+(z_1)
  \frac{[u_1-u_2+1]^*}{[u_1-u_2-1]^*},\\
  x_-(z_1)x_-(z_2)&\!\!=\!\!&x_-(z_2)x_-(z_1)
  \frac{[u_1-u_2-1]}{[u_1-u_2+1]},\\
  x_\pm(z_1)x_\mp(z_2)&\!\!=\!\!&x_\mp(z_2)x_\pm(z_1),\\
  \Phi_-(z_1)x_+(z_2)&\!\!=\!\!&x_+(z_2)\Phi_-(z_1),\\
  \Phi_-(z_1)x_-(z_2)&\!\!=\!\!&x_-(z_2)\Phi_-(z_1)
  \frac{[u_1-u_2+\frac12]}{[-u_1+u_2+\frac12]},\\
  \Psi^*_-(z_1)x_+(z_2)&\!\!=\!\!&x_+(z_2)\Psi^*_-(z_1)
  \frac{[u_1-u_2-\frac12]^*}{[-u_1+u_2-\frac12]^*},\\
  \Psi^*_-(z_1)x_-(z_2)&\!\!=\!\!&x_-(z_2)\Psi^*_-(z_1),\\
  \Phi_-(z_1)\Phi_-(z_2)&\!\!=\!\!&\Phi_-(z_2)\Phi_-(z_1)\rho(u_2-u_1),
  \label{PhiPhirho}\\
  \Psi^*_-(z_1)\Psi^*_-(z_2)&\!\!=\!\!&\Psi^*_-(z_2)\Psi^*_-(z_1)
  \rho^*(u_1-u_2),\\
  \Phi_-(z_1)\Psi^*_-(z_2)&\!\!=\!\!&\Psi^*_-(z_2)\Phi_-(z_1)\tau(u_2-u_1).
\ea
Here $\rho(u)$, $\rho^*(u)$ and $\tau(u)$ 
are given by 
\ba
  &&z^{\frac{\rs}{2r}}\rho(u)=\frac{\rho_+(u)}{\rho_+(-u)},\quad
  \rho_+(u)=\frac{(x^2z,x^{2r+2}z;x^4,x^{2r})_{\infty}}
                 {(x^4z,x^{2r}z;x^4,x^{2r})_{\infty}},
  \label{rho}\\
  &&z^{-\frac{r}{2\rs}}\rho^*(u)=\frac{\rho^*_+(u)}{\rho^*_+(-u)},\quad
  \rho^*_+(u)=\frac{(x^2z,x^{2\rs+2}z;x^4,x^{2\rs})_{\infty}}
                 {(z,x^{2\rs+4}z;x^4,x^{2\rs})_{\infty}},
  \label{rho*}\\
  &&\tau(u)=z^{\frac12}\frac{\Theta_{x^4}(-xz^{-1})}{\Theta_{x^4}(-xz)}.
  \label{tau}
\ea
Note that
\be
  \rho^*(u)=-\rho(u)\Bigl|_{r\rightarrow\rs}.
\ee

{}~

\noindent{\bf Trace}\\
For an operator $A(z)$ that have the form 
$:\exp(\mbox{linear in boson}):$, we write
\be
  A(z)=A^{\rm osc}(z)A^{\rm zero}(z),\quad
  A^{\rm osc}(z)=\exp\Bigl(\sum_{n>0}f^A_{-n}\alpha_{-n}z^n\Bigr)
  \exp\Bigl(\sum_{n>0}f^A_n\alpha_nz^{-n}\Bigr),
\ee
where $f^A_n$ is a coefficient.
In this model $A$ is $x_{\pm},\Phi_-,\Psi^*_-$.
For example, 
$$
  \Phi_-^{\rm osc}(z)=
  \exp\Biggl(\sum_{n>0}\frac{\alpha'_n}{[2n]_x}z^n\Biggr)
  \exp\Biggl(-\sum_{n>0}\frac{\alpha'_n}{[2n]_x}z^{-n}\Biggr),\quad
  \Phi_-^{\rm zero}(z)=
  \sqrt{g}e^{\frac12\sqrt{\frac{\rs}{r}}Q}
  z^{\frac12\sqrt{\frac{\rs}{r}}a'_0+\frac{\rs}{4r}}.
$$
For such operators Wick's theorem tells us
\be
  A_1(z_1)\cdots A_n(z_n)=\prod_{i<j}\dbr{A_i(z_i)A_j(z_j)}\times
  :\prod_iA_i(z_i):.
\ee
The trace of oscillator parts over the Fock space $\F=\F_{l,k}$ 
can be calculated by using the trace technique \eq{traceCS1}.
We have
\be
  \tr_{\F}\Bigl(x^{4d^{\rm osc}}:\prod_iA_i^{\rm osc}(z_i):\Bigr)
  =\frac{1}{(x^4;x^4)_{\infty}}
  \prod_{i,j}F_{A_i,A_j}(\sfrac{z_i}{z_j}),
\ee
where $F_{A,B}(z)$ is given by
\be
  F_{A,B}(z)=\exp\Biggl(\sum_{n>0}\frac{1}{n}[n]_x[2n]_x
  \frac{[rn]_x}{[\rs n]_x}\frac{x^{4n}}{1-x^{4n}}f^A_{-n}f^B_nz^n\Biggr).
\ee
We write down $F_{A,B}(z)$ (Remark $F_{A,B}(z)=F_{B,A}(z)$) :
\ba
  F_{x_+,x_+}(z)&\!\!=\!\!&
  \frac{(x^2z;x^2,x^{2\rs})_{\infty}}{(x^{2\rs+4}z;x^2,x^{2\rs})_{\infty}},\\
  F_{x_-,x_-}(z)&\!\!=\!\!&
  \frac{(x^4z;x^2,x^{2r})_{\infty}}{(x^{2r+2}z;x^2,x^{2r})_{\infty}},\\  
  F_{x_+,x_-}(z)&\!\!=\!\!&
  \frac{1}{(-x^3z,-x^5z;x^4)_{\infty}},\\
  F_{\Phi_-,x_+}(z)&\!\!=\!\!&
  (-x^4z;x^4)_{\infty},\\
  F_{\Phi_-,x_-}(z)&\!\!=\!\!&
  \frac{(x^{2r+3}z;x^4,x^{2r})_{\infty}}{(x^5z;x^4,x^{2r})_{\infty}},\\
  F_{\Psi^*_-,x_+}(z)&\!\!=\!\!&
  \frac{(x^{2\rs+5}z;x^4,x^{2\rs})_{\infty}}{(x^3z;x^4,x^{2\rs})_{\infty}},\\
  F_{\Psi^*_-,x_-}(z)&\!\!=\!\!&
  (-x^4z;x^4)_{\infty},\\
  F_{\Phi_-,\Phi_-}(z)&\!\!=\!\!&
  \frac{(x^6z,x^{2r+6}z;x^4,x^4,x^{2r})_{\infty}}
       {(x^8z,x^{2r+4}z;x^4,x^4,x^{2r})_{\infty}},\\
  F_{\Psi^*_-,\Psi^*_-}(z)&\!\!=\!\!&
  \frac{(x^4z,x^{2\rs+8}z;x^4,x^4,x^{2\rs})_{\infty}}
       {(x^6z,x^{2\rs+6}z;x^4,x^4,x^{2\rs})_{\infty}},\\
  F_{\Phi_-,\Psi^*_-}(z)&\!\!=\!\!&
  \frac{(-x^7z;x^4,x^4)_{\infty}}{(-x^5z;x^4,x^4)_{\infty}}.
\ea

\subsection{Screening operators and vertex operators}\label{sec:5.b}

All the formulas in this subsection are understood as meromorphic functions.

Let us recall the screening operators \eq{X} and \eq{X'}
\ba
  X(z)&\!\!=\!\!&\oint_{C_X(z)}\dz'\,x_+(z')
  \frac{[u-u'+\frac12-\hat{l}\,]^*}{[u-u'-\frac12]^*}
  \;:\,\F_{l,k}\rightarrow\F_{l-2,k},\\
  X'(z)&\!\!=\!\!&\oint_{C_{X'}(z)}\dz'\,x_-(z')
  \frac{[u-u'-\frac12+\hat{k}]}{[u-u'+\frac12]}
  \;:\,\F_{l,k}\rightarrow\F_{l,k-2},
\ea
where integration contours are 
\be
\begin{tabular}{|c|c|c|}
\hline
&inside&outside\\
\hline
$C_X(z)$&$z'=x^{-1+2\rs n}z$&$z'=x^{-1-2\rs(n+1)}z$\\
\hline
$C_{X'}(z)$&$z'=x^{1+2rn}z$&$z'=x^{1-2r(n+1)}z$\\
\hline
\end{tabular}\;\;(n=0,1,2,\cdots).
\ee
These screening operators satisfy the following exchange relations \cite{JKOS2}
\ba
  \!\!\!\!\!\!\!\!\!\!&&X(z_1)X(z_2)\frac{[u_1-u_2-1]^*}{[u_1-u_2]^*}
  -X(z_2)^2\frac{[u_1-u_2-\hat{l}+2]^*}{[u_1-u_2]^*}\frac{[1]^*}{[\hat{l}-2]^*}
  =(u_1\leftrightarrow u_2),\\
  \!\!\!\!\!\!\!\!\!\!&&X'(z_1)X'(z_2)\frac{[u_1-u_2+1]}{[u_1-u_2]}
  -X'(z_2)^2\frac{[u_1-u_2+\hat{k}-2]}{[u_1-u_2]}\frac{[1]}{[\hat{k}-2]}
  =(u_1\leftrightarrow u_2),\\
  \!\!\!\!\!\!\!\!\!\!&&X(z_1)X'(z_2)=X'(z_2)X(z_1).
\ea
$\widetilde{\Phi}_+$ and $\widetilde{\Psi}^*_+$ can be expressed as
(see \eq{Phi+},\eq{Psi+},\eq{tPhi},\eq{tPsi})
\be
  \widetilde{\Phi}_+(z)=\widetilde{\Phi}_-(z)X'(z),\quad
  \widetilde{\Psi}^*_+(z)=\widetilde{\Psi}^*_-(z)X(z).
\ee
Here RHS's are understood as the analytic continuation
of $A(z)B(z')$ from the region $|z|\gg|z'|$ and 
hence integration contours become \eq{tableCVO}.

Since screening operators and VO's satisfy
\ba
  X(z)\widetilde{\Phi}_{\ve}(w)=\widetilde{\Phi}_{\ve}(w)X(z),\\
  X'(z)\widetilde{\Psi}^*_{\ve}(w)=\widetilde{\Psi}^*_{\ve}(w)X'(z),
\ea
$\prod_i\widetilde{\Phi}_{\ve_i}(w_i)$ and
$\prod_i\widetilde{\Psi}^*_{\ve_i}(w_i)$ ($\sum_i\ve_i=0$) are
well-defined operators on the cohomology $H^0(C_{l,k})$ and $H^0(C'_{l,k})$
respectively where the BRST charges are $Q_m=X(1)^m$ and $Q'_m=X'(1)^m$ 
(see subsection \ref{sec:3.4.3}).
On the other hand 
$\widetilde{\Phi}_{\ve}(w)$ and $\widetilde{\Psi}^*_{\ve}(w)$ do not
commute with $X'(z)$ and $X(z)$ respectively, so they are not well-defined
operators on the cohomology $H^0(C'_{l,k})$ and $H^0(C_{l,k})$ in general.
However we can find the following intertwining properties.

We can show that
\ba
  \!\!\!\!\!\!\!\!\!\!\!\!\!\!\!&&X'(z)\widetilde{\Phi}_-(w)=
  -\widetilde{\Phi}_-(w)X'(z)\frac{[u-v+1]}{[u-v]}\frac{[\hat{k}]}{[\hat{k}-1]}
  +\widetilde{\Phi}_+(w)\frac{[u-v+\hat{k}]}{[u-v]}\frac{[1]}{[\hat{k}-1]},\\
  \!\!\!\!\!\!\!\!\!\!\!\!\!\!\!&&X(z)\widetilde{\Psi}^*_-(w)=
  -\widetilde{\Psi}^*_-(w)X(z)
  \frac{[u-v-1]^*}{[u-v]^*}\frac{[\hat{l}\,]^*}{[\hat{l}-1]^*}
  +\widetilde{\Psi}^*_+(w)
  \frac{[u-v-\hat{l}\,]^*}{[u-v]^*}\frac{[1]^*}{[\hat{l}-1]^*},
\ea
where $z=x^{2u}$ and $w=x^{2v}$. By induction we obtain 
\ba
  \!\!\!\!\!\!\!\!\!\!\!\!\!\!\!\!\!\!\!\!\!\!\!\!\!&&
  X'(z)^n\widetilde{\Phi}_-(w)=
  \widetilde{\Phi}_-(w)X'(z)^nA_n(u-v,\hat{k})
  +\widetilde{\Phi}_+(w)X'(z)^{n-1}B_n(u-v,\hat{k}),\\
  \!\!\!\!\!\!\!\!\!\!\!\!\!\!\!\!\!\!\!\!\!\!\!\!\!&&
  X'(z)^n\widetilde{\Phi}_+(w)=
  \widetilde{\Phi}_-(w)X'(z)^{n+1}B_n(v-u,\hat{k}-2)
  +\widetilde{\Phi}_+(w)X'(z)^nA_n(v-u,\hat{k}-2),\\
  \!\!\!\!\!\!\!\!\!\!\!\!\!\!\!\!\!\!\!\!\!\!\!\!\!&&
  X(z)^n\widetilde{\Psi}^*_-(w)=
  \widetilde{\Psi}^*_-(w)X(z)^nA^*_n(u-v,\hat{l})
  +\widetilde{\Psi}^*_+(w)X(z)^{n-1}B^*_n(u-v,\hat{l}),\\
  \!\!\!\!\!\!\!\!\!\!\!\!\!\!\!\!\!\!\!\!\!\!\!\!\!&&
  X(z)^n\widetilde{\Psi}^*_+(w)=
  \widetilde{\Psi}^*_-(w)X(z)^{n+1}B^*_n(v-u,\hat{l}-2)
  +\widetilde{\Psi}^*_+(w)X(z)^nA^*_n(v-u,\hat{l}-2),
\ea
where coefficients are
\ba
  A_n(u,k)&\!\!=\!\!&(-1)^n\frac{[u+n]}{[u]}\frac{[k-n+1]}{[k-2n+1]},\\
  B_n(u,k)&\!\!=\!\!&(-1)^{n-1}\frac{[u+k-n+1]}{[u]}\frac{[n]}{[k-2n+1]},\\
  A^*_n(u,l)&\!\!=\!\!&
  (-1)^n\frac{[u-n]^*}{[u]^*}\frac{[l-n+1]^*}{[l-2n+1]^*},\\
  B^*_n(u,l)&\!\!=\!\!&
  (-1)^{n-1}\frac{[u-k+n-1]^*}{[u]^*}\frac{[n]^*}{[l-2n+1]^*}.
\ea
Noticing the following properties of these coefficients
\ba
  &&A_{k+1}(u,k')=0,\quad B_{k+1}(u,k')=(-1)^{k+1}\quad
  \mbox{for $k'\equiv k\!\!\pmod{r}$},\\
  &&A^*_{l+1}(u,l')=0,\quad B^*_{l+1}(u,l')=(-1)^{l+1}\quad
  \mbox{for $l'\equiv l\!\!\pmod{\rs}$},
\ea
we obtain the intertwining properties 
($1\leq l\leq\rs-1$, $1\leq k\leq r-1$)
\ba
  \!\!\!\!\!\!\!\!\!\!&&X'(z)^{k-\ve}\widetilde{\Phi}_{\ve}(w)
  =(-1)^{k-\ve}\widetilde{\Phi}_{-\ve}(w)X'(z)^k\quad
  \qquad\mbox{on $\F_{l',k'}$ $k'\equiv k\!\!\pmod{r}$},\\
  \!\!\!\!\!\!\!\!\!\!&&X'(z)^{r-k-\ve}\widetilde{\Phi}_{\ve}(w)
  =(-1)^{r-k-\ve}\widetilde{\Phi}_{-\ve}(w)X'(z)^{r-k}\quad
  \mbox{on $\F_{l',k'}$ $k'\equiv-k\!\!\pmod{r}$},\\
  \!\!\!\!\!\!\!\!\!\!&&X(z)^{l-\ve}\widetilde{\Psi}^*_{\ve}(w)
  =(-1)^{l-\ve}\widetilde{\Psi}^*_{-\ve}(w)X(z)^l\quad
  \qquad\mbox{on $\F_{l',k'}$ $l'\equiv l\!\!\pmod{\rs}$},\\
  \!\!\!\!\!\!\!\!\!\!&&X(z)^{\rs-l-\ve}\widetilde{\Psi}^*_{\ve}(w)
  =(-1)^{\rs-l-\ve}\widetilde{\Psi}^*_{-\ve}(w)X(z)^l\quad
  \mbox{on $\F_{l',k'}$ $l'\equiv-l\!\!\pmod{\rs}$}.
\ea
Therefore $\prod_i\widetilde{\Phi}_{\ve_i}(w_i)$ and
$\prod_i\widetilde{\Psi}^*_{\ve_i}(w_i)$ ($\sum_i\ve_i=0$) are
well-defined operators on the cohomology $H^0(C'_{l,k})$ and $H^0(C_{l,k})$ 
respectively. However we remark that on $C'_{2j+1}$ and $C_{2j+1}$ 
those operators should be replaced
by $\pm\prod_i\widetilde{\Phi}_{-\ve_i}(w_i)$ and
$\pm\prod_i\widetilde{\Psi}^*_{-\ve_i}(w_i)$ respectively.

\renewcommand{\thesubsection}{\thesection.\arabic{subsection}}
\setcounter{section}{5}
\setcounter{equation}{0}
\section{DVA ($A_2^{(2)}$ type) and Dilute $A_L$ Models}\label{sec:6}

In this section we introduce another deformed Virasoro algebra
and study dilute $A_L$ models by free field approach \cite{HJKOS99}. 
Both of them are associated to $A_2^{(2)}$ algebra.

$r$, $\beta$, $x_{\pm}(z)$, $T(z)$, $\alpha_n$, $\alpha'_n$, $h_n$, 
$C_{l,k}$, etc. 
in this section are different form those in section \ref{sec:5}.

\subsection{DVA ($A_2^{(2)}$)}

Brazhnikov and Lukyanov \cite{BL97} pointed out that 
one can associate to the algebra $A^{(2)}_2$ 
a deformed Virasoro algebra which is different from the 
one discussed in section \ref{sec:3}. 
We denote it DVA($A_2^{(2)}$).

DVA($A_2^{(2)}$) is an associative algebra 
over $\C$ generated by $T_n$ ($n\in\Z$) with two parameters $x$ and $r$, 
and their relation is \cite{BL97}
\ba
  [T_n,T_m]&\!\!=\!\!&-\sum_{\ell=1}^{\infty}f_{\ell}
  (T_{n-\ell}T_{m+\ell}-T_{m-\ell}T_{n+\ell})\n
  &&+(x-x^{-1})^2\frac{[r+\frac12]_x[r]_x[r-1]_x[r-\frac32]_x}
  {[\frac12]_x[\frac32]_x}[3n]_x\delta_{n+m,0}
  \label{DVA2}\\
  &&+(x-x^{-1})^2\frac{[r]_x[r-\frac12]_x[r-1]_x}{[\frac12]_x}
  [n-m]_xT_{n+m},\nonumber
\ea
where the structure constants $f_{\ell}$ is given by
\ba
  f(z)&\!\!=\!\!&\sum_{\ell=0}^{\infty}f_{\ell}z^{\ell}
  =\exp\Biggl(-\sum_{n>0}(x-x^{-1})^2
  \frac{1}{n}\frac{[n]_x[rn]_x[(r-1)n]_x}{[2n]_x-[n]_x}z^n\Biggr)
  \label{f2}\\
  &\!\!=\!\!&
  \frac{1}{1-z}
  \frac{(x^{2-2r}z,x^{3-2r}z,x^4z,x^5z,x^{2r}z,x^{2r+1}z;x^6)_{\infty}}
       {(x^{5-2r}z,x^{6-2r}z,xz,x^2z,x^{2r+3}z,x^{2r+4}z;x^6)_{\infty}}.
  \nonumber
\ea
By introducing DVA current $\ds T(z)=\sum_{n\in\Z}T_nz^{-n}$, 
the above relation can be written as a formal power series,
\ba
  &&f(\sfrac{w}{z})T(z)T(w)
  -T(w)T(z)f(\sfrac{z}{w})\n
  &=\!\!&(x-x^{-1})\frac{[r+\frac12]_x[r]_x[r-1]_x[r-\frac32]_x}
  {[\frac12]_x[\frac32]_x}\Bigl(\delta(x^3\sfrac{w}{z})
  -\delta(x^{-3}\sfrac{w}{z})\Bigr)
  \label{DVA2fTT}\\
  &&+(x-x^{-1})\frac{[r]_x[r-\frac12]_x[r-1]_x}{[\frac12]_x}
  \Bigl(\delta(x^2\sfrac{w}{z})T(xw)
  -\delta(x^{-2}\sfrac{w}{z})T(x^{-1}w)\Bigr).\nonumber
\ea
The notation of \cite{BL97} is related to ours by 
$x_{BL}=x^{\frac32}$, $\frac{b}{Q}=r$, $\frac{1}{Qb}=1-r$, 
$g(z)=f(z)$, ${\bf V}(z)=T(z)$.
For later use we add a grading operator $d$,
\be
  [d,T_n]=-nT_n.
\ee
The above relation \eq{DVA2} is invariant under 
\ba
  \mbox{(i)}&&x\mapsto x^{-1},\quad r\mapsto r,\quad T(z)\mapsto T(z),\\
  \mbox{(ii)}&&x\mapsto x,\quad r\mapsto 1-r,\quad T(z)\mapsto -T(z). 
  \label{rto1-r}
\ea
In the case of (i) $f(z)$ is understood as the first line of \eq{f2}. 
Let us introduce $\beta$ as
\be
  \beta=\frac{r}{2(r-1)},
  \label{betar2}
\ee
then \eq{az} becomes 
\be 
  \az=\frac{2-r}{\sqrt{2r(r-1)}}.
  \label{azr2}
\ee

In the conformal limit ($x=e^{\hbar}\rightarrow 1$, $r$ : fixed), 
\eq{DVA2} admits two limits \cite{BL97} related by \eq{rto1-r},
\ba
  \!\!\!\!\!\!\!\!\!\!\!\!\!\!\!\!\!\!\!\!&&
  T(z)=3-2r+8r(r-1)\hbar^2\Bigl(z^2L(z)+\sfrac{1}{48}(1-2r)
  +\sfrac{(2-r)^2}{8r(r-1)}\Bigr)
  +O(\hbar^4),
  \label{3-2r}\\
  \!\!\!\!\!\!\!\!\!\!\!\!\!\!\!\!\!\!\!\!&&
  T(z)=-1-2r-8r(r-1)\hbar^2\Bigl(z^2\widetilde{L}(z)
  -\sfrac{1}{48}(1-2r)+\sfrac{(1+r)^2}{8r(r-1)}\Bigr)+O(\hbar^4),
  \label{-1-2r}
\ea
where $L(z), \widetilde{L}(z)$ are the Virasoro currents with the central 
charges $c,\widetilde{c}$ respectively,
\be
  c=1-\frac{3(2-r)^2}{r(r-1)}=1-6\az^2,\quad
  \widetilde{c}=1-\frac{3(1+r)^2}{r(r-1)}.
\ee

The highest weight representation is defined by the same manner 
presented in subsection \ref{sec:3.3}.
The Kac determinant at level $N$ is \cite{HJKOS99}
\ba
  &&\det\Bigl(\langle\lambda;N,i|\lambda;N,j\rangle\Bigr)_{1\leq i,j\leq p(N)}
  \n
  &=\!\!&\prod_{l,k\geq 1 \atop lk\leq N}\left(
  \frac{(x^{rl}-x^{-rl})(x^{(r-1)l}-x^{-(r-1)l})}{x^l-1+x^{-l}}
  (\lambda-\lambda_{l,k})(\lambda-\widetilde{\lambda}_{l,k})
  \right)^{p(N-lk)},
  \label{DVAKacdet2}
\ea
where $\lambda_{l,k}$ and $\widetilde{\lambda}_{l,k}$ are given by
\ba
  \lambda_{l,k}&\!\!=\!\!&
  x^{-lr+2k(r-1)}+x^{lr-2k(r-1)}-\frac{[r-\frac12]_x}{[\frac12]_x},\\
  \widetilde{\lambda}_{l,k}&\!\!=\!\!&
  -x^{l(r-1)-2kr}-x^{-l(r-1)+2kr}-\frac{[r-\frac12]_x}{[\frac12]_x},
\ea
and they are related by \eq{rto1-r}.
In the conformal limit this Kac determinant reduces to the Virasoro one
with the proportional constant expected from \eq{3-2r} or \eq{-1-2r}.

\subsection{Free field realization}\label{sec:6.2}

\subsubsection{free field realization}

Let us introduce free boson oscillator $h_n$ ($n\in\Z_{\neq 0}$),
\be
  [h_n,h_m]=(x-x^{-1})^2\frac{1}{n}
  \frac{[n]_x[rn]_x[(r-1)n]_x}{[2n]_x-[n]_x}\delta_{n+m,0},
  \label{hn2}
\ee
and use zero mode $a_0$ and $Q$ defined in \eq{an} (or $a'_0$ in \eq{a'0}).
Notice that $[2n]_x-[n]_x=[3n]_x[\frac{n}{2}]_x/[\frac{3n}{2}]_x$.
The Fock space $\F_{\alpha}$ is defined as before.

The DVA($A_2^{(2)}$) current $T(z)$ is realized as follows:
\ba
  &&T(z)=\Lambda_+(z)+\Lambda_0(z)+\Lambda_-(z),\n
  &&\Lambda_{\pm}(z)=\,:\exp\Bigl(\pm\sum_{n\neq0}
  h_n(x^{\pm\frac32}z)^{-n}\Bigr):\times x^{\pm\sqrt{2r(r-1)}a'_0},
  \label{Tfr2}\\
  &&\Lambda_{0}(z)=-\frac{[r-\frac12]_x}{[\frac12]_x}
  :\exp\Bigl(\sum_{n\neq0}h_n(x^{-n/2}-x^{n/2})z^{-n}\Bigr):.\nonumber
\ea
To prove this we need \eq{deltaformula} and the OPE formula,
\ba
  f(\sfrac{w}{z})\Lambda_{\pm}(z)\Lambda_{\pm}(w)&\!\!=\!\!&
  :\Lambda_{\pm}(z)\Lambda_{\mp}(w):,\n
  f(\sfrac{w}{z})\Lambda_{\pm}(z)\Lambda_{\mp}(w)&\!\!=\!\!&
  :\Lambda_{\pm}(z)\Lambda_{\mp}(w):
  \gamma(x^{\mp 1}\sfrac{w}{z})\gamma(x^{\mp 2}\sfrac{w}{z}),\n
  f(\sfrac{w}{z})\Lambda_0(z)\Lambda_0(w)&\!\!=\!\!&
  :\Lambda_0(z)\Lambda_0(w):
  \gamma(\sfrac{w}{z}),\\
  f(\sfrac{w}{z})\Lambda_{\pm}(z)\Lambda_0(w)&\!\!=\!\!&
  :\Lambda_{\pm}(z)\Lambda_0(w):
  \gamma(x^{\mp 1}\sfrac{w}{z}),\n
  f(\sfrac{w}{z})\Lambda_0(z)\Lambda_{\pm}(w)&\!\!=\!\!&
  :\Lambda_0(z)\Lambda_{\pm}(w):
  \gamma(x^{\pm 1}\sfrac{w}{z}),\nonumber
\ea
where $\gamma(z)$ is given in \eq{gamma}.
The grading operator $d$ is realized by
\ba
  &&d=d^{\rm osc}+d^{\rm zero},\n
  &&d^{\rm osc}=\sum_{n>0}
  \frac{n^2([2n]_x-[n]_x)}{(x-x^{-1})^2[n]_x[rn]_x[(r-1)n]_x}
  h_{-n}h_n,\quad d^{\rm zero}=\frac14a_0^{\prime\,2}-\frac{1}{24},
  \label{d2}
\ea
which satisfies 
\be
  [d,h_n]=-nh_n,\quad [d,Q]=a'_0,\quad
  d\ketb{\alpha_{l,k}}=(h_{l,k}-\frac{c}{24})\ketb{\alpha_{l,k}},
\ee 
where $c$ and $h_{l,k}$ are
given by \eq{cbeta} and \eq{hlkbeta} respectively.
We remark that $\widetilde{T}(z)=-\Lambda_+(z)+\Lambda_0(z)-\Lambda_-(z)$ 
also satisfies (\ref{DVA2}).

$\ketb{\alpha}$ is the highest weight state of DVA with 
$\lambda=\lambda(\alpha)$,
\ba
  &&\ketb{\alpha}=\ket{\lambda(\alpha)},\\
  &&\lambda(\alpha)
  =x^{\sqrt{2r(r-1)}(\alpha-\az)}+x^{-\sqrt{2r(r-1)}(\alpha-\az)}
  -\frac{[r-\frac12]_x}{[\frac12]_x}.
  \label{lambdaalpha2}
\ea
The dual space $\F_{\alpha}^*$ becomes a DVA module by \eq{pairing} with
\ba
  {}^tT_n&\!\!=\!\!&T_{-n},\\
  {}^th_n&\!\!=\!\!&-h_{-n},\quad  \mbox{eq.}\eq{ta0},
  \label{thn2}
\ea
and \eq{<|>=1}.
By \eq{thn2}, $\F_{\alpha}^*$ is isomorphic to $\F_{2\az-\alpha}$ as 
DVA module,
\ba
  \F_{\alpha}^*&\!\!\cong\!\!&\F_{2\az-\alpha}\quad(\mbox{DVA module}).\\
  \ketbs{\alpha}&\!\!\leftrightarrow\!\!&\ketb{2\az-\alpha}.
\ea
Note that $\lambda(\alpha)=\lambda(2\az-\alpha)$ by \eq{lambdaalpha2}.

In the conformal limit ($x=e^{\hbar}\rightarrow 1$, $r$ : fixed), 
oscillator $h_n$ is expressed by $a_n$ in \eq{an} as follows:
\be
  h_n=\hbar\sqrt{2r(r-1)}\sqrt{\frac{1}{x^n-1+x^{-n}}
  \frac{x^{rn}-x^{-rn}}{2rn\hbar}  
  \frac{x^{(r-1)n}-x^{-(r-1)n}}{2(r-1)n\hbar}}a_n.
\ee
Substituting this expression into \eq{Tfr2} and expanding in $\hbar$, 
we get \eq{3-2r} with $L(z)$ in \eq{L-phi} (or \eq{L-phi2}).

The representation with $\lambda=\lambda_{l,k}$ is realized on 
$\F_{l,k}=\F_{\alpha_{l,k}}$,
\be
  \lambda_{l,k}=\lambda(\alpha_{l,k}),
  \label{lambdalk2}
\ee
where $\alpha_{l,k}$ is given in \eq{alphalk}.

\subsubsection{Kac determinant}

In the free boson realization the singular vectors can be expressed by
screening currents.
A screening current $x_+(z)$ is defined by
\be
  x_+(z)=
  :\exp\Bigl(-\sum_{n\neq 0}\frac{\alpha_n}{[n]_x}z^{-n}\Bigr):\times 
  e^{\sqrt{\frac{r}{2(r-1)}}Q}z^{\sqrt{\frac{r}{2(r-1)}}a'_0+\frac{r}{2(r-1)}},
  \label{x+2first}
\ee
where oscillators $\alpha_n$ ($n\in\Z_{\neq 0}$) are related
to $h_n$ as
\be
  h_n=(x-x^{-1})(-1)^n\frac{[(r-1)n]_x}{[2n]_x-[n]_x}\alpha_n.
\ee
Like as \eq{DVAsv} singular vector is obtained by the BRST operator.
Screening charges and BRST charges will be discussed in the
next subsection.

Like as in subsection \ref{sec:3.4.2} let us introduce matrices 
$C(N,\alpha)$ and $C'(N,\alpha)$ by \eq{CDVA} and \eq{C'DVA}.
Then their determinants are given by
\ba
  \det C(N,\alpha)_{I,J}&\!\!=\!\!&\prod_{l,k\geq 1\atop lk\leq N}
  \Biggl(\Bigl(z^{\frac12}x^{\frac12lr-k(r-1)}
  -z^{-\frac12}x^{-\frac12lr+k(r-1)}\Bigr)\n
  &&\qquad\times
  \Bigl(z^{\frac12}x^{-\frac12l(r-1)+kr}
  +z^{-\frac12}x^{\frac12l(r-1)-kr)}\Bigr)\Biggr)^{p(N-lk)},\\
  \det C'(N,\alpha)_{I,J}&\!\!=\!\!&\det C(N,2\az-\alpha)\cdot\det D\n
  &=\!\!&\prod_{l,k\geq 1\atop lk\leq N}
  \Biggl(\Bigl(z^{\frac12}x^{-\frac12lr+k(r-1)}
  -z^{-\frac12}x^{\frac12lr-k(r-1)}\Bigr)\n
  &&\qquad\times
  \Bigl(z^{\frac12}x^{\frac12l(r-1)-kr}
  +z^{-\frac12}x^{-\frac12l(r-1)+kr)}\Bigr)\Biggr)^{p(N-lk)},
\ea
where $z=x^{\sqrt{2r(r-1)}(\alpha-\az)}$.
The inner product of two states in the Verma module is given by \eq{hTTh}
where $G_{K,L}$ is
\ba
  G_{K,L}&\!\!=\!\!&<({}^th)_K\ketbs{\alpha},h_{-L}\ketb{\alpha}>\n
  &=\!\!&\delta_{K,L}\prod_i\Biggl(
  \frac{1}{i}\frac{(x^{ri}-x^{-ri})(x^{(r-1)i}-x^{-(r-1)i})}{x^i-1+x^{-i}}
  \Biggr)^{k_i}k_i!,
\ea
and its determinant is (use \eq{k_i!} and \eq{Fk}) 
\ba
  \det G_{K,L}&\!\!=\!\!&
  \prod_{\{k_i\}\atop{\scriptstyle\Sigma}_iik_i=N}
  \prod_i\Biggl(
  \frac{1}{i}\frac{(x^{ri}-x^{-ri})(x^{(r-1)i}-x^{-(r-1)i})}{x^i-1+x^{-i}}
  \Biggr)^{k_i}k_i!\n
  &=\!\!&\prod_{l,k\geq 1\atop lk\leq N}\Biggl(
  \frac{(x^{rl}-x^{-rl})(x^{(r-1)l}-x^{-(r-1)l})}{x^l-1+x^{-l}}
  \Biggr)^{p(N-lk)}.
\ea
Therefore we obtain the Kac determinant \eq{DVAKacdet2},
\ba
  &&\det\bra{h}T_IT_{J}\ket{h}=
  \det C'(N,\alpha)\cdot\det G\cdot\det{}^tC(N,\alpha)\n
  &=\!\!&\prod_{l,k\geq 1 \atop lk\leq N}\left(
  \frac{(x^{rl}-x^{-rl})(x^{(r-1)l}-x^{-(r-1)l})}{x^l-1+x^{-l}}
  (\lambda-\lambda_{l,k})(\lambda-\widetilde{\lambda}_{l,k})
  \right)^{p(N-lk)}.
\ea

\subsubsection{Felder complex}

We consider the representation of $\lambda=\lambda_{l,k}$ in \eq{lambdalk2} 
with \eq{minimalbeta} and \eq{lkrange}, i.e.
\be
  r=\frac{2p''}{2p''-p'}. 
  \label{r2}
\ee

Let us consider the Felder complex $C_{l,k}$,
\be
  \cdots ~{\buildrel X_{-3} \over \longrightarrow }~
  C_{-2} ~{\buildrel X_{-2} \over \longrightarrow }~
  C_{-1} ~{\buildrel X_{-1} \over \longrightarrow }~
  C_0 ~{\buildrel X_0 \over \longrightarrow }~
  C_1 ~{\buildrel X_1 \over \longrightarrow }~
  C_2 ~{\buildrel X_2 \over \longrightarrow }~ \cdots, 
\ee
where $C_j$ and $X_j\;:\;C_j\rightarrow C_{j+1}$ ($j\in\Z$) are
\ba
  &&C_{2j}=\F_{l-2p'j,k},\quad C_{2j+1}=\F_{-l-2p'j,k},\\
  &&X_{2j}=Q_l,\quad X_{2j+1}=Q_{p'-l}.
  \label{Xj2}
\ea
We assume that $p'$ is odd 
(we have not obtained the result for general even $p'$). 
BRST charge $Q_m$ ($1\leq m\leq p'-1$) is defined by 
\ba
  Q_{2m+1}&\!\!=\!\!&Q_1Q_2^{(1)}Q_2^{(2)}\cdots Q_2^{(m)},\\
  Q_{2m}&\!\!=\!\!&Q_2^{(\frac{p'+1}{2}-m)}\cdots 
  Q_2^{(\frac{p'-3}{2})}Q_2^{(\frac{p'-1}{2})}.
\ea
Here $Q_1$ and $Q_2^{(a)}$ are
\ba
  Q_1&\!\!=\!\!&\oint_{|z|=1}\dz\,x_+(z)
  \frac{[u+\frac12\hat{l}\,]^*}{[u+\frac12]^*},\\
  Q_2^{(a)}&\!\!=\!\!&\oint\!\!\oint_{|z_1|=|z_2|=1}
  \dz_1\dz_2\,x_+(z_1)x_+(z_2)
  \frac{1}{[u_1+\frac12]^*[u_2+\frac12]^*}\n
  &&\times \frac{[u_1-u_2]^*}{[u_1-u_2+1]^*[u_1-u_2-\frac12]^*}
  f_2^{(a)}(u_1+\sfrac12\hat{l},u_2+\sfrac12\hat{l}),
\ea
where $[u]^*$ is given in \eq{[u]^*} with $\rs=r-1$ and 
$z=x^{2u}$, $z_i=x^{2u_i}$ and
\ba
  f^{(a)}_2(u_1,u_2)&\!\!=\!\!&
  [2a+1]^*[a-\sfrac12]^*[u_1-a]^*[u_2+a-1]^*[u_1-u_2+a-\sfrac12]^*\n
  &&\!\!\!\!
  -[2a-1]^*[a+\sfrac12]^*[u_1+a]^*[u_2-a-1]^*[u_1-u_2-a-\sfrac12]^*.  
\ea
$X$ satisfies the BRST property,
\be
  X_jX_{j-1}=0.
\ee
We assume that this Felder complex has the same structure as the Virasoro
case because it formally tends to Virasoro one in the conformal limit
($x\rightarrow 1$ and $r$ and $z=x^{2u}$ fixed kept).
Then the cohomology groups of the complex $C_{l,k}$ are
\be
  H^j(C_{l,k})=\mbox{Ker }X_j/\mbox{Im }X_{j-1}=
  \left\{\begin{array}{ll}
  0&j\neq 0,\\
  \cL_{l,k}&j=0,
  \end{array}\right.
  \label{cohomologyDVA2}
\ee
where $\cL_{l,k}$ is the irreducible DVA module of 
$\lambda=\lambda_{l,k}$.
The trace of operator ${\cal O}$ over $\cL_{l,k}$ can be written as
\eq{trO}.

The BRST charge commutes with DVA \cite{HJKOS99}
\be
  [T_n,Q_l]=0 \quad\mbox{on $\F_{l',k}$}\quad l'\equiv l\;(\mbox{mod }p').
\ee
Singular vector can be obtained by $Q_l$ like as \eq{DVAsv}.

\subsection{Dilute $A_L$ models}

In the following we fix a positive integer $L\ge 3$. 
The dilute $A_L$ model \cite{WNS92,WPSN94} is an integrable 
RSOS model obtained by restricting the face model of 
type $A^{(2)}_2$ \cite{Ku91}. 
In  the dilute $A_L$ model, the local fluctuation variables 
$a,b,\cdots$ take one of the $L$ states $1,2,\cdots, L$, and 
those on neighboring lattice sites are subject to the condition $a-b=0,\pm 1$. 
The Boltzmann weights can be found in \cite{WPSN94}, eq.(3.1).
For our purpose it is convenient to use the 
parametrization given in Appendix A of \cite{WPSN94},  
which is suitable in the `low-temperature' regime. 
With some change of notation we recall the formula below \cite{HJKOS99}.

Let $x=e^{-2\pi \lambda/\varepsilon}$, $r=\pi/(2\lambda)$ 
and $u=-u_{orig}/(2\lambda)$,  
where $\lambda$, $\varepsilon$ are the variables used in \cite{WPSN94} and 
$u_{orig}$ stands for `$u$' there. 
We shall restrict ourselves to the `regime $2^+$' defined by 
\be
  0<x<1, \quad r=2\,\frac{L+1}{L+2},\quad-\frac{3}{2}<u<0.
\ee
We have taken $p'=L$, $p''=L+1$ in \eq{r2}, which corresponds to minimal
unitary series.
\eq{betar2} and \eq{azr2} become
\be
  \beta=\frac{L+1}{L},\quad \az=\frac{1}{\sqrt{L(L+1)}},\quad
  p'=L,\quad p''=L+1,
\ee
and we set
\be
  \rs=r-1.
\ee
On the critical point ($x\rightarrow 1$), this model is described by CFT, 
i.e. the minimal unitary series with $\beta=\frac{L+1}{L}$.
On the off-critical point ($x<1$), which corresponds to the 
$(1,2)$-perturbation of the minimal unitary CFT, the Virasoro symmetry 
is lost but the DVA($A_2^{(2)}$) symmetry remains.
Changing an overall scalar factor we put the Boltzmann weights in the form 
\be
  W\BW{a}{b}{c}{d}{u}= \rho(u)\Wb\BW{a}{b}{c}{d}{u}, 
\ee
where $\rho(u)$ is given in \eq{rho2} and chosen
so that the partition function per site of the model equals to $1$. 
$\Wb$ is
\ba
  &&\Wb\BW{a\pm1}{a}{a}{a\mp1}{u}=1,\n
  &&\Wb\BW{a}{a\pm1}{a}{a\pm1}{u}=\Wb\BW{a\pm1}{a\pm1}{a}{a}{u}=
  -\left(\frac{[\pm a+\frac32]_+[\pm a-\frac12]_+}
  {[\pm a+\frac12]_+^2}\right)^{\frac12}
  \frac{[u]}{[1+u]},\n
  &&\Wb\BW{a\pm1}{a}{a}{a}{u}=\Wb\BW{a}{a}{a}{a\pm1}{u}=
  \frac{[\pm a+\frac12+u]_+}{[\pm a+\frac12]_+}\frac{[1]}{[1+u]},\n
  &&\Wb\BW{a}{a\mp1}{a\pm1}{a}{u}=
  \left(G_a^+G_a^-\right)^{\frac12}\frac{[\frac12+u]}{[\frac32+u]}
  \frac{[u]}{[1+u]},\\
  &&\Wb\BW{a}{a}{a\pm1}{a}{u}=\Wb\BW{a}{a\pm1}{a}{a}{u}=
  -\left(G_a^{\pm}\right)^{\frac12}
  \frac{[\pm a-1-u]_+}{[\pm a+\frac12]_+}\frac{[1]}{[1+u]}
  \frac{[u]}{[\frac32+u]},\n
  &&\Wb\BW{a}{a\pm1}{a\pm1}{a}{u}=
  \frac{[\pm 2a+1-u]}{[\pm 2a+1]}\frac{[1]}{[1+u]}
  -G_a^{\pm}\frac{[\pm 2a-\frac12-u]}{[\pm 2a+1]}
  \frac{[u]}{[\frac32+u]}\frac{[1]}{[1+u]},\n
  &&\Wb\BW{a}{a}{a}{a}{u}=
  \frac{[3+u]}{[3]}\frac{[1]}{[1+u]}\frac{[\frac32-u]}{[\frac32+u]}
  +H_a\frac{[1]}{[3]}\frac{[u]}{[1+u]}.\nonumber
\ea
Here
\be
  G_a^{\pm}=\frac{S(a\pm1)}{S(a)},\quad
  S(a)=(-1)^{a}\frac{[2a]}{[a]_+},\quad
  H_a=G_a^+\frac{[a-\frac52]_+}{[a+\frac12]_+}
  +G_a^-\frac{[a+\frac52]_+}{[a-\frac12]_+},
\ee
where $[u]$ and $[u]_+$ are given in \eq{[u]} and \eq{[u]_+}.

This Boltzmann weight $W$ enjoys
YBE \eq{YBEW}, initial condition \eq{initial}, unitarity \eq{unitarity}
and crossing symmetry \eq{crossing} ($\lambda=-\frac32$, $G_a=S(a)$),
\be
  W\BW{b}{d}{a}{c}{-\frac{3}{2}-u}=
  \sqrt{\frac{S(a)S(d)}{S(b)S(c)}}W\BW{a}{b}{c}{d}{u}.
\ee
In this gauge $W$ enjoys also a reflection symmetry
\be
  W\BW{a}{c}{b}{d}{u}=W\BW{a}{b}{c}{d}{u}.
\ee
Along with the additive variable $u$, we often use the multiplicative 
variable $z=x^{2u}$.
For later use we define $W^*$,
\be
  W^*\BW{a}{b}{c}{d}{u}=
  \Wb\BW{a}{b}{c}{d}{u}\Biggl|_{r\rightarrow\rs}\times\rho^*(u),
  \label{W*2}
\ee
where $\rho^*(u)$ is given in \eq{rho*2}.

Hereafter we assume that $L$ is odd. 
The model has ground states labeled by odd 
integers $l=1,3,\cdots,L-2$ \cite{WPSN94}.
They are characterized as configurations 
in which all heights take the same value $b$. 
If $L=4n\pm 1$, then the possible values are 
$b=l$ ($1\le l\le 2n-1$, $l$ : odd) or $b=l+1$ ($2n+1\le l\le L-2$, $l$ : odd).
Therefore in the notation of subsection \ref{sec:4.2} we have
$(i_1)=(b)$, $m=0\in\Z/\Z$ and $\cH^{(k)}_l=\cH^{(k)}_{l,0}$.
In the thermodynamic limit CTM's \eq{CTM} become
\be
  A^{(k)}(u)=C^{(k)}(u)=x^{-2uH_C^{(k)}},\quad 
  B^{(k)}(u)=D^{(k)}(u)=\sqrt{S(k)}x^{2(u+\frac32)H_C^{(k)}}.
\ee
Careful study of the corner Hamiltonian shows that the character \eq{chi} 
coincides with the Virasoro minimal unitary character \cite{WPSN94},
\be
  \chi_{l,0,k}(q)=\chi_{l,k}^{{\rm Vir}}(q),
\ee
where $\chi_{l,k}^{{\rm Vir}}(q)$ is given in \eq{Virch}.
Comparing the free filed realization given in subsection \ref{sec:6.2},
we make an identification
\ba
  \cH^{(k)}_l&\!\!=\!\!&\cL_{l,k}\,,
  \label{H=L2}\\
  H_C^{(k)}&\!\!=\!\!&d\Bigm|_{\cL_{l,k}},
  \label{H_C=d2}
\ea
where $\cL_{l,k}$ is given in \eq{cohomologyDVA2} as 
a cohomology of the Felder complex and $d$ is realized in \eq{d2}.

\subsection{Free field approach}\label{sec:6.4}

Although bosons are already introduced in subsection \ref{sec:6.2},
we present their definitions again.
Recall that
$$
  r=2\,\frac{L+1}{L+2},\quad \rs=r-1.
$$

Let us introduce free boson oscillator $\alpha_n$ ($n\in\Z_{\neq 0}$)
\be
  [\alpha_n,\alpha_m]=\frac{[n]_x([2n]_x-[n]_x)}{n}
  \frac{[rn]_x}{[\rs n]_x}\delta_{n+m,0},
\ee
and use zero mode $a_0$ and $Q$ defined in \eq{an} (or $a'_0$ in \eq{a'0}).
The Fock space $\F_{l,k}$ is defined by
\be
  \F_{l,k}=\bigoplus_{m\geq 0}\bigoplus_{n_1\geq\cdots\geq n_m>0}\C
  \alpha_{-n_1}\cdots \alpha_{-n_m}\ketb{\alpha_{l,k}},
\ee
where $\ketb{\alpha}$ is given by \eq{ketalpha} with replacing $a_n$ by $h_n$,
\be
  a'_0\ketb{\alpha_{l,k}}=(\alpha_{l,k}-\az)\ketb{\alpha_{l,k}},\quad
  \alpha_{l,k}-\az=-\sqrt{\frac{L+1}{L}}l+\sqrt{\frac{L}{L+1}}k.
\ee
We use also free boson oscillator $\alpha'_n$ ($n\in\Z_{\neq 0}$)
\be
  \alpha'_n=(-1)^n\frac{[\rs n]_x}{[rn]_x}\alpha_n,\quad
  [\alpha'_n,\alpha'_m]=\frac{[n]_x([2n]_x-[n]_x)}{n}
  \frac{[\rs n]_x}{[rn]_x}\delta_{n+m,0}.
\ee
Operators $\hat{l},\hat{k}:\F_{l,k}\rightarrow\F_{l,k}$ are defined by 
\eq{hatlk}.

Elliptic currents $x_{\pm}(z)$ for $U_x(A_2^{(2)})$ 
(or $\B_{x,\lambda}(A_2^{(2)})$) of level $1$ ($c=1$) 
are obtained by a `dressing' procedure described in subsection 
\ref{sec:4.4} {\bf 5}
\ba
  &x_+(z)&:\,\F_{l,k}\rightarrow \F_{l-2,k}\n
  &&x_+(z)=\,
  :\exp\Bigl(-\sum_{n\neq 0}\frac{\alpha_n}{[n]_x}z^{-n}\Bigr):\times 
  e^{\sqrt{\frac{r}{2\rs}}Q}z^{\sqrt{\frac{r}{2\rs}}a'_0+\frac{r}{2\rs}},
  \label{x+2}\\
  &x_-(z)&:\,\F_{l,k}\rightarrow \F_{l,k-1}\n
  &&x_-(z)=\,
  :\exp\Bigl(\sum_{n\neq 0}\frac{\alpha'_n}{[n]_x}z^{-n}\Bigr):\times
  e^{-\sqrt{\frac{\rs}{2r}}Q}z^{-\sqrt{\frac{\rs}{2r}}a'_0+\frac{\rs}{2r}}.
  \label{x-2}
\ea
$x_+(z)$ is interpreted as a screening current in subsection \ref{sec:6.2}.
Note that $x_-(z)$ is not a screening current.

VO's are obtained by solving the relations \eq{com12}-\eq{com32} below
directly for $\Phi_\ve(z),\Psi^*_\ve(z)$ ($\ve=0,\pm 1$). 
We write $\Phi_{\pm}(z)=\Phi_{\pm 1}(z)$ 
and $\Psi^*_{\pm}(z)=\Psi^*_{\pm 1}(z)$.
Results are

\medskip
\noindent type I \quad
$\ds\Phi_\ve(z):\F_{l,k}\rightarrow \F_{l,k-\ve}$
\ba
%
  \Phi_-(z)&\!\!=\!\!&\sqrt{g}
  :\exp\Bigl(-\sum_{n\neq 0}\frac{\alpha'_n}{[2n]_x-[n]_x}z^{-n}\Bigr):\times
  e^{\sqrt{\frac{\rs}{2r}}Q}z^{\sqrt{\frac{\rs}{2r}}a'_0+\frac{\rs}{2r}},
  \label{Phi-2}\\
  \Phi_0(z)&\!\!=\!\!&
  x^{-\frac{\rs}{2r}}\oint_{C_0(z)}\dz_1\Phi_-(z)x_-(z_1)
  \frac{1}{\sqrt{[\hat{k}+\frac12]_+[\hat{k}-\frac12]_+}}
  \frac{[u-u_1+\hat{k}]_+}{[u-u_1+\frac12]},
  \label{Phi02}\\
  \Phi_+(z)&\!\!=\!\!&
  x^{-\frac{\rs}{r}}\oint\!\!\oint_{C_+(z)}
  \dz_1\dz_2\Phi_-(z)x_-(z_1)x_-(z_2)\n
  &&\times\sqrt{\frac{S(\hat{k}-1)}{S(\hat{k})}}
  \frac{1}{[\hat{k}-\frac12]_+[2\hat{k}-2]}
  \frac{[u-u_1+2\hat{k}-\frac32]}{[u-u_1+\frac12]}
  \frac{[u_1-u_2+\hat{k}]_+}{[u_1-u_2+\frac12]},
  \label{Phi+2}
\ea

\noindent type II \quad 
$\ds\Psi^*_\ve(z):\F_{l,k}\rightarrow \F_{l-2\ve,k}$
\ba
%
  \!\!\!\!\!\!\!
  \Psi^*_-(z)&\!\!=\!\!&\frac{1}{\sqrt{g^*}}
  :\exp\Bigl(\sum_{n\neq 0}\frac{\alpha_n}{[2n]_x-[n]_x}z^{-n}\Bigr):\times
  e^{-\sqrt{\frac{r}{2\rs}}Q}z^{-\sqrt{\frac{r}{2\rs}}a'_0+\frac{r}{2\rs}},
  \label{Psi-2}\\
  \!\!\!\!\!\!\!
  \Psi^*_0(z)&\!\!=\!\!&
  ix^{\frac{r}{2\rs}}\oint_{C^*_0(z)}\dz_1\Psi^*_-(z)x_+(z_1)
  \frac{1}{\sqrt{[\frac12(\hat{l}+1)]^*_+[\frac12(\hat{l}-1)]^*_+}}
  \frac{[u-u_1-\frac12\hat{l}\,]^*_+}{[u-u_1-\frac12]^*},
  \label{Psi02}\\
  \!\!\!\!\!\!\!
  \Psi^*_+(z)&\!\!=\!\!&
  x^{\frac{r}{\rs}}\oint\!\!\oint_{C^*_+(z)}\dz_1\dz_2
  \Psi^*_-(z)x_+(z_1)x_+(z_2)\n
  \!\!\!\!\!\!\!
  &&\!\!\!\!\!\times\sqrt{\frac{S^*(\frac12\hat{l}-1)}{S^*(\frac12\hat{l}\,)}}
  \frac{1}{[\frac12(\hat{l}-1)]^*_+[\hat{l}-2]^*}
  \frac{[u-u_1-\hat{l}+\frac32]^*}{[u-u_1-\frac12]^*}
  \frac{[u_1-u_2-\frac12\hat{l}\,]^*_+}{[u_1-u_2-\frac12]^*}.
  \label{Psi+2}
\ea
Here $z=x^{2u}$, $z_j=x^{2u_j}$, $\dz_j=\frac{dz_j}{2\pi iz_j}$,
and normalization constants $g$ and $g^*$ will be given in \eq{g2} and
\eq{g^*2}.
The poles of the integrand of \eq{Phi02}-\eq{Psi+2}
and the integration contours are listed in the following table.
For example, $C_0(z)$ is a simple closed contour 
that encircles $x^{1+2rn}z$ ($n\ge 0$) but not $x^{-1-2rn}z$ ($n\ge 0$).  
\be
\begin{tabular}{|c|c|c|}
\hline
&inside&outside\\
\hline
$C_0(z)$&$z_1=x^{1+2rn}z$&$z_1=x^{-1-2rn}z$\\
\hline
$C_+(z)$&$z_1=x^{1+2rn}z$&$z_1=x^{-1-2rn}z$\\
&$z_2=x^{1+2rn}z_1$&$z_2=x^{-1-2rn}z,x^{-1-2rn}z_1,x^{2-2r(n+1)}z_1$\\
\hline
$C^*_0(z)$&$z_1=x^{-1+2\rs n}z$&$z_1=x^{1-2\rs n}z$\\
\hline
$C^*_+(z)$&$z_1=x^{-1+2\rs n}z$&$z_1=x^{1-2\rs n}z$\\
&$z_2=x^{-1+2\rs n}z_1$&
$z_2=x^{1-2\rs n}z,x^{1-2\rs n}z_1,x^{-2-2\rs (n+1)}z_1$\\
\hline
\end{tabular}
\quad (n=0,1,2,\cdots).
\ee
OPE formulas are given in subsection \ref{sec:6.a}.
In the conformal limit ($x=e^{\hbar}\rightarrow 1$, $r$ : fixed), 
$x_+(z)\rightarrow zV_{\alpha_{-1,1}}(-z)$ (up to phase), 
$x_-(z)\rightarrow V_{\alpha_{1,0}}(z)$ (up to power of $z$),
$g^{-\frac12}\Phi_-(z)\rightarrow V_{\alpha_{1,2}}(z)$ (up to power of $z$), 
and $g^{*\frac12}\Psi^*_-(z)\rightarrow V_{\alpha_{3,1}}(-z)$ (up to phase and 
power of $z$), where $V_{\alpha}(z)$ is given in \eq{Valpha}. 
Type I VO $\Phi_{\ve}(z)$ commutes with BRST 
operator \eq{Xj2}
\be
  [X_j,\Phi_{\ve}(z)]=0.
\ee

The VO's given above satisfy the following commutation relations 
($\ve_1,\ve_2=0,\pm 1$),
\ba
  \!\!\!\!\!\!\!\!\!\!\!\!\!\!\!
  &&\Phi_{\ve_2}(z_2)\Phi_{\ve_1}(z_1)
  =\!\!\!\!\!\sum_{\ve_1',\ve_2'=0,\pm 1\atop \ve_1'+\ve_2'=\ve_1+\ve_2}
  \!\!\!\!\!
  W\BW{\hat{k}}{\hat{k}+\ve_1'}{\hat{k}+\ve_2}{\hat{k}+\ve_1+\ve_2}{u_1-u_2}
  \Phi_{\ve'_1}(z_1)\Phi_{\ve'_2}(z_2),
  \label{com12}\\
  \!\!\!\!\!\!\!\!\!\!\!\!\!\!\!
  &&\Psi^*_{\ve_1}(z_1)\Psi^*_{\ve_2}(z_2)
  =\!\!\!\!\!\sum_{\ve_1',\ve_2'=0,\pm 1\atop \ve_1'+\ve_2'=\ve_1+\ve_2}
  \!\!\!\!\!
  W^*\BW{\frac12\hat{l}}{\frac12\hat{l}+\ve_1}{\frac12\hat{l}+\ve_2'}
  {\frac12\hat{l}+\ve_1+\ve_2}{u_1-u_2}
  \Psi^*_{\ve'_2}(z_2)\Psi^*_{\ve'_1}(z_1),
  \label{com22}\\
  \!\!\!\!\!\!\!\!\!\!\!\!\!\!\!
  &&\Phi_{\ve_2}(z_2)\Psi^*_{\ve_1}(z_1)=
  \tau(u_1-u_2)\Psi^*_{\ve_1}(z_1)\Phi_{\ve_2}(z_2),
  \label{com32}
\ea
where $z_i=x^{2u_i}$ and $\tau(u)$ is given in \eq{tau2}.
We do not present the tedious but straightforward 
verification of \eq{com12}-\eq{com32}.

For the description of correlation functions we need also the `dual' VO's. 
Define
\ba
  \Phi^*_\ve(z)&\!\!=\!\!&\sqrt{S(\hat{k})}^{-1}\Phi_{-\ve}(x^{-3}z)
  \sqrt{S(\hat{k})},
  \label{Phi*=Phi2}\\
  \Psi_\ve(z)&\!\!=\!\!&\sqrt{S^*(\sfrac12\hat{l}\,)}
  \Psi^*_{-\ve}(x^{-3}z)\sqrt{S^*(\sfrac12\hat{l}\,)}^{-1},
\ea
and normalization constants $g$ and $g^*$ in \eq{Phi-2} and \eq{Psi-2} are
\ba
  \!\!\!\!\!g^{-1}&\!\!=\!\!&\frac{(x;x^{2r})_\infty}
  {(x^2;x^{2r})^2_\infty(x^{2r-1};x^{2r})_\infty(x^{2r};x^{2r})^4_\infty}
  \frac{(x^5,x^6,x^{2r},x^{2r+1};x^6,x^{2r})_\infty}
  {(x^2,x^3,x^{2r+3},x^{2r+4};x^6,x^{2r})_\infty},
  \label{g2}\\
  \!\!\!\!\!g^*&\!\!=\!\!&\frac{(x^{-1};x^{2\rs})_\infty}
  {(x^{-2};x^{2\rs})^2_\infty(x^{2\rs+1};x^{2\rs})_\infty
  (x^{2\rs};x^{2\rs})^5_\infty}
  \frac{(x^3,x^4,x^{2\rs+2},x^{2\rs+3};x^6,x^{2\rs})_\infty}
  {(x,x^6,x^{2\rs+5},x^{2\rs+6};x^6,x^{2\rs})_\infty}.
  \label{g^*2}
\ea
Then we have 
\ignore{
\ba
  \!\!\!\!\!\!\!\!\!\!
  &&\Phi_{\ve_2}(z)\Phi^*_{\ve_1}(z)=\delta_{\ve_1,\ve_2}\times\id,\quad
  \Psi_{\ve_1}(z_1)\Psi^*_{\ve_2}(z_2)
  =\frac{\delta_{\ve_1,\ve_2}}{1-\frac{z_1}{z_2}}+\cdots\quad
  (z_1\rightarrow z_2),
  \label{inv12}\\
  \!\!\!\!\!\!\!\!\!\!
  &&\sum_{\ve=0,\pm 1}\Phi^*_{\ve}(z)\Phi_{\ve}(z)=\id,\qquad
  \sum_{\ve=0,\pm 1}\Psi^*_{\ve}(z_2)\Psi_{\ve}(z_1)
  =\frac{1}{1-\frac{z_1}{z_2}}+\cdots\quad 
  (z_1\rightarrow z_2).
  \label{inv22}
\ea
}
\ba
  &&\sum_{\ve=0,\pm 1}\Phi^*_{\ve}(z)\Phi_{\ve}(z)=\id,
  \label{inv2I2}\\
  &&\Phi_{\ve_2}(z)\Phi^*_{\ve_1}(z)=\delta_{\ve_1,\ve_2}\times\id,
  \label{inv1I2}\\
  &&\sum_{\ve=0,\pm 1}\Psi^*_{\ve}(z_2)\Psi_{\ve}(z_1)
  =\frac{1}{1-\frac{z_1}{z_2}}+\cdots\quad 
  (z_1\rightarrow z_2),\\
  &&\Psi_{\ve_1}(z_1)\Psi^*_{\ve_2}(z_2)
  =\frac{\delta_{\ve_1,\ve_2}}{1-\frac{z_1}{z_2}}+\cdots\quad
  (z_1\rightarrow z_2),
\ea
and for $d$ in \eq{d2}
\be
  w^d{\cal O}(z)w^{-d}={\cal O}(wz),\quad\mbox{for }
  {\cal O}=\Phi_{\ve},\Phi^*_{\ve},\Psi^*_{\ve},\Psi_{\ve},x_{\pm}.
  \label{wdO2}
\ee

We identify type I VO's in subsection \ref{sec:4.2} and those here 
in the following way:
\ba
  \Phi^{(a-\ve,a)}(z)&\!\!=\!\!&\Phi_{\ve}(z)\Bigm|_{\cL_{l,a}},
  \label{Phi=Phi2}\\
  \Phi^{*(a+\ve,a)}(z)&\!\!=\!\!&\Phi^*_{\ve}(z)\Bigm|_{\cL_{l,a}}.
  \label{Phi*=Phi*2}
\ea
Then eqs. 
\eq{Phi*}-\eq{PhiPhi*}
correspond to 
\eq{Phi*=Phi2},\eq{com12},\eq{wdO2},\eq{inv2I2} and \eq{inv1I2} respectively.

Next let us see how the DVA($A_2^{(2)}$) current is obtained from VO's.
Let introduce free boson oscillator $h_n$ ($n\in\Z_{\neq 0}$),
\ba
  &&\lambda_n=(-1)^n(x-x^{-1})\frac{[\rs n]_x}{[2n]_x-[n]_x}\alpha_n
  =(x-x^{-1})\frac{[rn]_x}{[2n]_x-[n]_x}\alpha'_n,\\
  &&[\lambda_n,\lambda_m]=(x-x^{-1})^2
  \frac{1}{n}\frac{[n]_x[rn]_x[\rs n]_x}{[2n]_x-[n]_x}\delta_{m+n,0}.
\ea
As explained in subsection \ref{sec:6.2}, the DVA($A_2^{(2)}$) current 
$T(z)$ is realized as
\ba
  &&T(z)=\Lambda_+(z)+\Lambda_0(z)+\Lambda_-(z),\n
  &&\Lambda_{\pm}(z)=\,:\exp\Bigl(\pm\sum_{n\neq0}
  h_n(x^{\pm\frac32}z)^{-n}\Bigr):\times x^{\pm\sqrt{2r\rs}a'_0},\\
  &&\Lambda_{0}(z)=-\frac{[r-\frac12]_x}{[\frac12]_x}
  :\exp\Bigl(\sum_{n\neq0}h_n(x^{-n/2}-x^{n/2})z^{-n}\Bigr):.\nonumber
\ea
This $T(z)$ is obtained from type I VO's by fusing them,
\ba
  \!\!\!\!\!\!\!\!\!\!
  &&\Phi_{\ve_2}(x^{\frac32+r}z')\Phi^*_{\ve_1}(x^{\frac32-r}z)\\
  \!\!\!\!\!\!\!\!\!\!&=\!\!&
  \left(1-\frac{z}{z'}\right)(-1)^{\ve_1+1}
  \delta_{\ve_1,\ve_2}T(z)\cdot x^{1-r}
  \frac{(x,x^6,x^{5-2r},x^{6-2r};x^6)_\infty}
       {(x^3,x^4,x^{2-2r},x^{3-2r};x^6)_\infty}
  +\cdots\quad(z'\rightarrow z).\nonumber
\ea
$\widetilde{T}(z)=-\Lambda_+(z)+\Lambda_0(z)-\Lambda_-(z)$ 
also satisfies \eq{DVA2}, and it is obtained from type II VO's,
\ba
  &&\Psi_{\ve_1}(x^{\frac32+\rs}z')\Psi^*_{\ve_2}(x^{\frac32-\rs}z)\\
  &=\!\!&
  \frac{1}{1-\frac{z'}{z}}(-1)^{\ve_1+1}\delta_{\ve_1,\ve_2}
  \widetilde{T}(-z)\cdot(-x^{-r})
  \frac{(x^2,x^3,x^{3-2\rs},x^{4-2\rs};x^6)_\infty}
       {(x^5,x^6,x^{-2\rs},x^{1-2\rs};x^6)_\infty}
  +\cdots\quad(z'\rightarrow z).\nonumber
\ea

Using these bosonized VO's let us calculate LHP \eq{Pl} with \eq{LHP2},
\ba
  \!\!\!\!\!&&P_{a_n,\cdots,a_0}(l)\n
  \!\!\!\!\!&=\!\!&
  Z_{l}^{-1}\,S(a_0)\,\tr_{\cH^{(a_0)}_l}\Bigl(x^{6H_C^{(a_0)}}
  \Phi^{*(a_0,a_1)}(z)\cdots\Phi^{*(a_{n-1},a_n)}(z)
  \Phi^{(a_n,a_{n-1})}(z)\cdots\Phi^{(a_1,a_0)}(z)\Bigr).\n
  \!\!\!\!\!&=\!\!&
  Z_{l}^{-1}\,S(a_0)\,\tr_{\cL_{l,a_0}}\Bigl(x^{6d}
  \Phi^*_{\ve_1}(z)\cdots\Phi^*_{\ve_n}(z)
  \Phi_{\ve_n}(z)\cdots\Phi_{\ve_1}(z)\Bigr),
\ea
where $\ve_i=a_{i-1}-a_i$.
Here we have identified the space of state and operators as
\eq{H=L2},\eq{H_C=d2},\eq{Phi=Phi2} and \eq{Phi*=Phi*2}.

One-point LHP is already obtained in \eq{1pt},
\be
  P_{k}(l)=Z_{l}^{-1}\,S(k)\,\chi_{l,k}(x^6),
\ee
where $\chi_{l,k}(q)$ is given in \eq{Virch} and $Z_l$ \eq{Zl} is
\be
  Z_l=\sum_{k=1}^L\,S(k)\,\chi_{l,k}(x^6),
\ee
which can be expressed in product of theta functions with conjugate 
modulus \cite{WPSN94}.
Two-point LHP satisfies \eq{reverse} and \eq{recP}
\be
  P_{a,b}(l)=P_{b,a}(l),\quad \sum_{b}P_{a,b}(l)=P_a(l).    
\ee
In contrast to subsection \ref{sec:5.3}, however, this recursion relation
does not determine $P_{a,b}(l)$ uniquely.
So we will use free field realization of vertex operator approach.
Two-point LHP $P_{k-\ve,k}(l)$ is ($\ve=0,\pm1)$
\ba
  P_{k-\ve,k}(l)&\!\!=\!\!&Z_l^{-1}\,S(k)\, 
  \tr_{\cL_{l,k}}\Bigl(x^{6d}\Phi^{*}_\ve(z)\Phi_\ve(z)\Bigr)\n
  &\!\!=\!\!&Z_l^{-1}\sqrt{S(k)S(k-\ve)}\,\tr_{\cL_{l,k}}
  \Bigl(x^{6d}\Phi_{-\ve}(x^{-3}z)\Phi_\ve(z)\Bigr)
  \label{tpf}.
\ea
Since this is independent on $z$, we take $z=1$.
The evaluation of the trace yields the following expressions:
\ba
  P_{k-1,k}(l)&\!\!=\!\!&-\frac{g}{Z_l}
  \frac{S(k-1)\,x^{-\frac{\rs}{r}}}{[k-\frac12]_+[2k-2]}
  \oint\oint_{C_+(1)}{\dw}_1\,\dw_2\,{\cal I}(w_1,w_2)\n
  &&\qquad\qquad\qquad\qquad\qquad\times 
  \frac{[v_1-2k+\frac32][v_1-v_2+k]_+}{[v_1+\frac12][v_1-v_2+\frac12]},\\
  P_{k,k}(l)&\!\!=\!\!&\frac{g}{Z_l}
  \frac{S(k)\,x^{-\frac{\rs}{r}}}{[k+\frac12]_+[k-\frac12]_+}
  \oint_{C_0(x^{-3})}\dw_1\oint_{C_0(1)}\dw_2\,{\cal I}(w_1,w_2)\n
  &&\qquad\qquad\qquad\qquad\qquad\times 
  \frac{[v_1-k+\frac32]_+[v_2-k]_+}{[v_1+1][v_2-\frac12]},\\
  P_{k+1,k}(l)&\!\!=\!\!&-\frac{g}{Z_l}
  \frac{S(k)\,x^{\frac{-\rs}{r}}}{[k+\frac12]_+[2k]}
  \oint\oint_{C_+(x^{-3})}\dw_1\,\dw_2\,{\cal I}(w_1,w_2)\n
  &&\qquad\qquad\qquad\qquad\times 
  \frac{[v_1-2k+1][v_1-v_2+k+1]_+[v_2+\frac12]}
  {[v_1+1][v_1-v_2+\frac12][v_2-\frac12]},
\ea
where $w_i=x^{2v_i}$ ($i=1,2$) and ${\cal I}(w_1,w_2)$ is 
\ba
  {\cal I}(w_1,w_2)&\!\!=\!\!&
  g^{-1}\tr_{\cL_{l,k}}\Bigl(
  x^{6d}\Phi_-(x^{-3})x_-(w_1)\Phi_-(1)x_-(w_2)\Bigr)\n
  &=\!\!&{\cal I}^{\rm OPE}(w_1,w_2){\cal I}^{\rm osc}(w_1,w_2)
  {\cal I}^{\rm zero}(w_1,w_2).
\ea
Here ${\cal I}^{\rm OPE}(w_1,w_2)$ is the OPE contribution
\ba
  {\cal I}^{\rm OPE}(w_1,w_2)&\!\!=\!\!&
  \dbr{\Phi_-(x^{-3})x_-(w_1)}\dbr{\Phi_-(x^{-3})\Phi_-(1)}
  \dbr{\Phi_-(x^{-3})x_-(w_2)}\n
  &&\times\dbr{x_-(w_1)\Phi_-(1)}\dbr{x_-(w_1)x_-(w_2)}
  \dbr{\Phi_-(1)x_-(w_2)},
\ea
and ${\cal I}^{\rm osc}(w_1,w_2)$ is the oscillator contribution
\ba
  {\cal I}^{\rm osc}(w_1,w_2)&\!\!=\!\!&
  \tr_{\F}\Bigl(x^{6d^{\rm osc}}:\Phi_-^{\rm osc}(x^{-3})x_-^{\rm osc}(w_1)
  \Phi_-^{\rm osc}(1)x_-^{\rm osc}(w_2):\Bigr)\n
  &=\!\!&\frac{1}{(x^6;x^6)_{\infty}}
  F_{\Phi_-,\Phi_-}(1)^2 F_{\Phi_-,\Phi_-}(x^3) F_{\Phi_-,\Phi_-}(x^{-3})
  F_{x_-,x_-}(1)^2\n
  &&\times\prod_{i=1,2} F_{\Phi_-,x_-}(w_i) F_{\Phi_-,x_-}(w_i^{-1})
  F_{\Phi_-,x_-}(x^3w_i) F_{\Phi_-,x_-}(x^{-3}w_i^{-1})\n
  &&\times F_{x_-,x_-}(\sfrac{w_2}{w_1}) F_{x_-,x_-}(\sfrac{w_1}{w_2}),
\ea
and ${\cal I}^{\rm zero}(w_1,w_2)$ is the zero mode contribution
\ba
  {\cal I}^{\rm zero}(w_1,w_2)&\!\!=\!\!&
  g^{-1}\tr_{\cL_{l,k}}^{\rm zero}\Bigl(x^{6d^{\rm zero}}
  :\Phi_-^{\rm zero}(x^{-3})
  x_-^{\rm zero}(w_1)\Phi_-^{\rm zero}(1)x_-^{\rm zero}(w_2):\Bigr)\n
  &=\!\!&\tr_{\cL_{l,k}}^{\rm zero}\Bigl(x^{6d^{\rm zero}}
  (x^3w_1w_2)^{-\sqrt{\frac{\rs}{2r}}a'_0}(x^{-3}w_1w_2)^{\frac{\rs}{2r}}
  \Bigr)\n
  &=\!\!& \sum_{j\in\Z}\Bigl(x^{6(h_{l-2Lj,k}-\frac{c}{24})}
  (x^3w_1w_2)^{\frac12(l-2Lj)-\frac{\rs}{r}k}\n
  &&\qquad -x^{6(h_{-l-2Lj,k}-\frac{c}{24})}
  (x^3w_1w_2)^{\frac12(-l-2Lj)-\frac{\rs}{r}k}\Bigr)
  (x^{-3}w_1w_2)^{\frac{\rs}{2r}}\n
  &=\!\!&x^{6(h_{l,k}-\frac{c}{24})}x^{-3\frac{\rs}{r}}
  (x^3w_1w_2)^{\frac12l-\frac{\rs}{r}k+\frac{\rs}{2r}}\\
  &&\times\Bigl(
  \Theta_{x^{12L(L+1)}}(-x^{6(-(L+1)l+Lk+L(L+1))}(x^3w_1w_2)^{-L})\n
  &&\quad 
  -\Theta_{x^{12L(L+1)}}(-x^{6((L+1)l+Lk+L(L+1))}(x^3w_1w_2)^{-L})
  x^{6lk}(x^3w_1w_2)^{-l}\Bigr).\nonumber
\ea
Using formulas in subsection \ref{sec:6.a} we have
\ba
  &&{\cal I}^{\rm OPE}(w_1,w_2){\cal I}^{\rm osc}(w_1,w_2)\n
  &=\!\!&x^{3\frac{\rs}{r}}
  \frac{(x^5,x^5,x^6,x^6,x^8,x^8,x^{2r+5},x^{2r+5};x^6,x^{2r})_{\infty}}
       {(x^7,x^7,x^{2r+3},x^{2r+3},x^{2r+4},x^{2r+4},x^{2r+4},x^{2r+4};
         x^6,x^{2r})_{\infty}}\n
  &&\times
  \frac{(x^{2r-1}w_1^{-1},x^{2r+2}w_1;x^3,x^{2r})_{\infty}}
       {(xw_1^{-1},x^4w_1;x^3,x^{2r})_{\infty}}
  \frac{(x^{2r-1}w_2,x^{2r+2}w_2^{-1};x^3,x^{2r})_{\infty}}
       {(xw_2,x^4w_2^{-1};x^3,x^{2r})_{\infty}}\\
  &&\times\Theta_{x^6}(z)
  \frac{(x^2z,x^8z^{-1},x^{2r-1}z,x^{2r+5}z^{-1};x^6,x^{2r})_{\infty}}
       {(xz,x^7z^{-1},x^{2r-2}z,x^{2r+4}z^{-1};x^6,x^{2r})_{\infty}}
  \Biggm|_{z=\frac{w_2}{w_1}}.\nonumber
\ea
The contours $C_+(1)$, $C_0(x^{-3})\cup C_0(1)$, $C_+(x^{-3})$ are chosen
as follows $(n,m\geq 0)$;
For all the contours,
the poles $w_1=x^{4+3m+2rn}$, $w_2=x^{4+3m+2rn}$, $x^{4+6m+2r(n+1)}w_1$,
$x^{1+6(m+1)+2rn}w_1$ are inside and the poles $w_1=x^{-4-3m-2rn}$, 
$w_2=x^{-1-3m-2rn}$, $x^{2-6m-2r(n+1)}w_1$, $x^{-1-6m-2rn}w_1$ are outside.
In addition,  
$$ 
\begin{tabular}{|c|c|c|}
\hline
&inside&outside\\
\hline
$C_+(1)$&$w_1=x^{-1+2r(n+1)}$&$w_1=x^{-1-2rn}$\\
&$w_2=x^{1+2rn}w_1$&$w_2=x^{-1-2rn},x^{-1-2rn}w_1,x^{2-2r(n+1)}w_1$\\
\hline
$C_0(x^{-3})\cup C_0(1)$&$w_1=x^{-2+2rn}$&$w_1=x^{-4-2rn}$\\
&$w_2=x^{1+2rn}$&$w_2=x^{-1-2rn}$\\
\hline
$C_+(x^{-3})$&$w_1=x^{-2+2rn}$&$w_1=x^{-4-2rn}$\\
&$w_2=x^{1+2rn},x^{1+2rn}w_1$&$w_2=x^{1-2r(n+1)},x^{-1-2rn}w_1,
x^{2-2r(n+1)}w_1$\\
\hline
\end{tabular}\ .
$$

For integral representations of general LHP, see \cite{HJKOS99}.

Excited states are obtained by using type II VO and
traces of type I and type II VO's are calculated similarly like as 
subsection \ref{sec:5.4}.

\renewcommand{\thesubsection}{\thesection.\alph{subsection}}
\setcounter{subsection}{0}
\subsection{OPE and trace}\label{sec:6.a}

{\bf OPE}\\
We list the normal ordering relations used in section \ref{sec:6}.
$\rs$ is 
$$
  \rs=r-1.
$$ 
Notation $\dbr{A(z)B(w)}$ is given in \eq{dbrAB} :
\ba
  \dbr{x_+(z_1)x_+(z_2)}&\!\!=\!\!&z_1^{\frac{r}{\rs}}(1-\zeta)
  \frac{(x^{-2}\zeta,x^{2\rs+1}\zeta;x^{2\rs})_\infty}
  {(x^{-1}\zeta,x^{2\rs+2}\zeta;x^{2\rs})_\infty},\\
  \dbr{x_-(z_1)x_-(z_2)}&\!\!=\!\!&z_1^{\frac{\rs}{r}}(1-\zeta)
  \frac{(x^2\zeta,x^{2r-1}\zeta;x^{2r})_\infty}
  {(x\zeta,x^{2r-2}\zeta;x^{2r})_\infty},\\
  \dbr{x_\pm(z_1)x_\mp(z_2)}&\!\!=\!\!&z_1^{-1}
  \frac{1+\zeta}{(1+x\zeta)(1+x^{-1}\zeta)},\\
  \dbr{\Phi_-(z_1)x_+(z_2)}&\!\!=\!\!&\dbr{x_+(z_1)\Phi_-(z_2)}=z_1+z_2,\\
  \dbr{\Phi_-(z_1)x_-(z_2)}&\!\!=\!\!&\dbr{x_-(z_1)\Phi_-(z_2)}
  =z_1^{-\frac{\rs}{r}}
  \frac{(x^{2r-1}\zeta;x^{2r})_\infty}{(x\zeta;x^{2r})_\infty},\\
  \dbr{\Psi^*_-(z_1)x_+(z_2)}&\!\!=\!\!&\dbr{x_+(z_1)\Psi_-^*(z_2)}
  =z_1^{-\frac{r}{\rs}}
  \frac{(x^{2\rs+1}\zeta;x^{2\rs})_\infty}{(x^{-1}\zeta;x^{2\rs})_\infty},\\
  \dbr{\Psi^*_-(z_1)x_-(z_2)}&\!\!=\!\!&\dbr{x_-(z_1)\Psi^*_-(z_2)}
  =z_1+z_2,\\
  \dbr{\Phi_-(z_1)\Phi_-(z_2)}&\!\!=\!\!&z_1^{\frac{\rs}{r}}
  \frac{(x^2\zeta,x^3\zeta,x^{2r+3}\zeta,x^{2r+4}\zeta;x^6,x^{2r})_\infty}
  {(x^5\zeta,x^6\zeta,x^{2r}\zeta,x^{2r+1}\zeta;x^6,x^{2r})_\infty},\\
  \dbr{\Psi^*_-(z_1)\Psi^*_-(z_2)}&\!\!=\!\!&z_1^{\frac{r}{\rs}}
  \frac{(\zeta,x\zeta,x^{2\rs+5}\zeta,x^{2\rs+6}\zeta;x^6,x^{2\rs})_\infty}
  {(x^3\zeta,x^4\zeta,x^{2\rs+2}\zeta,x^{2\rs+3}\zeta;x^6,x^{2\rs})_\infty},\\
  \dbr{\Phi_-(z_1)\Psi^*_-(z_2)}&\!\!=\!\!&\dbr{\Psi^*_-(z_1)\Phi_-(z_2)}
  =z_1^{-1}\frac{(-x^4\zeta,-x^5\zeta;x^6)_\infty}
  {(-x\zeta,-x^2\zeta;x^6)_\infty},
\ea
where $\zeta=\frac{z_2}{z_1}$ and we have used \eq{Hausdorff}.

As meromorphic functions we have ($z_i=x^{2u_i}$)
\ba
  x_+(z_1)x_+(z_2)&\!\!=\!\!&x_+(z_2)x_+(z_1)
  \frac{[u_1-u_2+1]^*}{[u_1-u_2-1]^*}
  \frac{[u_1-u_2-\frac12]^*}{[-u_1+u_2-\frac12]^*},\\
  x_-(z_1)x_-(z_2)&\!\!=\!\!&x_-(z_2)x_-(z_1)
  \frac{[u_1-u_2-1]}{[u_1-u_2+1]}
  \frac{[u_1-u_2+\frac12]}{[-u_1+u_2+\frac12]},\\
  x_\pm(z_1)x_\mp(z_2)&\!\!=\!\!&x_\mp(z_2)x_\pm(z_1),\\
  \Phi_-(z_1)x_+(z_2)&\!\!=\!\!&x_+(z_2)\Phi_-(z_1),\\
  \Phi_-(z_1)x_-(z_2)&\!\!=\!\!&x_-(z_2)\Phi_-(z_1)
  \frac{[u_1-u_2+\frac12]}{[-u_1+u_2+\frac12]},\\
  \Psi^*_-(z_1)x_+(z_2)&\!\!=\!\!&x_+(z_2)\Psi^*_-(z_1)
  \frac{[u_1-u_2-\frac12]^*}{[-u_1+u_2-\frac12]^*},\\
  \Psi^*_-(z_1)x_-(z_2)&\!\!=\!\!&x_-(z_2)\Psi^*_-(z_1),\\
  \Phi_-(z_1)\Phi_-(z_2)&\!\!=\!\!&\Phi_-(z_2)\Phi_-(z_1)\rho(u_2-u_1),\\
  \Psi^*_-(z_1)\Psi^*_-(z_2)&\!\!=\!\!&\Psi^*_-(z_2)\Psi^*_-(z_1)
  \rho^*(u_1-u_2),\\
  \Phi_-(z_1)\Psi^*_-(z_2)&\!\!=\!\!&\Psi^*_-(z_2)\Phi_-(z_1)\tau(u_2-u_1).
\ea
Here $\rho(u)$, $\rho^*(u)$ and $\tau(u)$ 
are given by 
\ba
  &&z^{\frac{\rs}{r}}\rho(u)=\frac{\rho_+(u)}{\rho_+(-u)},\quad
  \rho_+(u)=\frac{(x^2z,x^3z, x^{2r+3}z, x^{2r+4}z;x^6,x^{2r})_\infty}
  {(x^5z,x^6z,x^{2r}z,x^{2r+1}z;x^6,x^{2r})_\infty}, 
  \label{rho2}\\
  &&z^{-\frac{r}{\rs}}\rho^*(u)=\frac{\rho^*_+(u)}{\rho^*_+(-u)},\quad
  \rho^*_+(u)=\frac{(x^3z,x^4z,x^{2\rs+2}z,x^{2\rs+3}z;x^6,x^{2\rs})_\infty}
  {(z,xz,x^{2\rs+5}z,x^{2\rs+6}z;x^6,x^{2\rs})_\infty},
  \label{rho*2}\\
  &&\tau(u)=z\frac{\Theta_{x^6}(-xz^{-1})\Theta_{x^6}(-x^2z^{-1})}
  {\Theta_{x^6}(-xz)\Theta_{x^6}(-x^2z)}.
  \label{tau2}
\ea
Note that
\be
  \rho^*(u)=-\rho(u)\Bigl|_{r\rightarrow \rs}\times 
  z\frac{\Theta_{x^6}(xz^{-1})\Theta_{x^6}(x^2z^{-1})}
        {\Theta_{x^6}(xz)\Theta_{x^6}(x^2z)}.
\ee

{}~

\noindent{\bf Trace}\\
We use the same notation as the second part of subsection \ref{sec:5.a}.
The trace of oscillator parts over the Fock space $\F=\F_{l,k}$ is
\be
  \tr_{\F}\Bigl(x^{6d^{\rm osc}}:\prod_iA_i^{\rm osc}(z_i):\Bigr)
  =\frac{1}{(x^6;x^6)_{\infty}}
  \prod_{i,j}F_{A_i,A_j}(\sfrac{z_i}{z_j}),
\ee
where $F_{A,B}(z)$ is given by
\be
  F_{A,B}(z)=\exp\Biggl(\sum_{n>0}\frac{1}{n}[n]_x([2n]_x-[n]_x)
  \frac{[rn]_x}{[\rs n]_x}\frac{x^{6n}}{1-x^{6n}}f^A_{-n}f^B_nz^n\Biggr).
\ee
We write down $F_{A,B}(z)$ (Remark $F_{A,B}(z)=F_{B,A}(z)$) :
\ba
  F_{x_+,x_+}(z)&\!\!=\!\!&
  (x^6z;x^6)_{\infty}
  \frac{(x^4z,x^{2\rs+7}z;x^6,x^{2\rs})_{\infty}}
       {(x^5z,,x^{2\rs+8};x^6,x^{2\rs})_{\infty}},\\
  F_{x_-,x_-}(z)&\!\!=\!\!&
  (x^6z;x^6)_{\infty}
  \frac{(x^8z,x^{2r+5}z;x^6,x^{2r})_{\infty}}
       {(x^7z,x^{2r+4}z;x^6,x^{2r})_{\infty}},\\
  F_{x_+,x_-}(z)&\!\!=\!\!&
  \frac{(-x^6z;x^6)_{\infty}}{(-x^5z,-x^7z;x^6)_{\infty}},\\
  F_{\Phi_-,x_+}(z)&\!\!=\!\!&
  (-x^6z;x^6)_{\infty},\\
  F_{\Phi_-,x_-}(z)&\!\!=\!\!&
  \frac{(x^{2r+5}z;x^6,x^{2r})_{\infty}}{(x^7z;x^6,x^{2r})_{\infty}},\\
  F_{\Psi^*_-,x_+}(z)&\!\!=\!\!&
  \frac{(x^{2\rs+7}z;x^6,x^{2\rs})_{\infty}}{(x^5z;x^6,x^{2\rs})_{\infty}},\\
  F_{\Psi^*_-,x_-}(z)&\!\!=\!\!&
  (-x^6z;x^6)_{\infty},\\
  F_{\Phi_-,\Phi_-}(z)&\!\!=\!\!&
  \frac{(x^8z,x^9z,x^{2r+9}z,x^{2r+10}z;x^6,x^6,x^{2r})_{\infty}}
       {(x^{11}z,x^{12}z,x^{2r+6}z,x^{2r+7}z;x^6,x^6,x^{2r})_{\infty}},\\
  F_{\Psi^*_-,\Psi^*_-}(z)&\!\!=\!\!&
  \frac{(x^6z,x^7z,x^{2\rs+11}z,x^{2\rs+12}z;x^6,x^6,x^{2\rs})_{\infty}}
       {(x^9z,x^{10}z,x^{2\rs+8}z,x^{2\rs+9}z;x^6,x^6,x^{2\rs})_{\infty}},\\
  F_{\Phi_-,\Psi^*_-}(z)&\!\!=\!\!&
  \frac{(-x^{10}z,-x^{11}z;x^6,x^6)_{\infty}}
       {(-x^7z,-x^8z;x^6,x^6)_{\infty}}.
\ea

\renewcommand{\thesubsection}{\thesection.\arabic{subsection}}
\setcounter{section}{6}
\setcounter{equation}{0}
\section{Conclusion}

In this lecture we have explained deformed Virasoro algebras ($A_1^{(1)}$ 
type and $A_2^{(2)}$ type) and elliptic quantum groups (face type algebra 
$\Bqla(\g)$ and vertex type algebra $\Aqp(\slnh)$) and studied solvable 
lattice models (ABF model in regime III and dilute $A_L$ model in regime 
$2^+$) by using vertex operators and free field realizations.

\medskip
We close this lecture by mentioning some related topics.

\medskip\noindent{\it deformed $W$algebras (DWA's):}\quad
In CFT there are several extensions of the Virasoro algebra, e.g.
superconformal algebras, current algebras (affine Lie algebras), $W$ algebras,
parafermions, which contain the Virasoro algebra as a subalgebra.
For $W$ algebras, see review \cite{BS93}. $W_N$ algebra is a $W$ algebra
associated to $A_{N-1}$ algebra. Deformation of $W_N$ algebra, which we
denote DWA($A_{N-1}$), was obtained in \cite{AKOS95}(see also \cite{AKOS2}) 
and \cite{FF95} by
using correspondence between singular vectors and Macdonald symmetric
polynomials or quantization of the deformed $W_N$ Poisson algebra
respectively. The deformed $W_n$ Poisson algebra was obtained 
from the Wakimoto realization of $U_q(\slNh)$ at the critical level by
E.~Frenkel and Reshetikhin \cite{FR95} and they pointed out that deformed 
$W$ currents in a free field realization have the same forms of transfer 
matrices in analytic Bethe ansatz, dressed vacuum form. 
(Bethe ansatz is also a powerful method to study solvable models 
\cite{Bax,KBI})
Based on this observation, DWA's for arbitrary simple Lie algebras were 
constructed \cite{FR97}. 
See \cite{FJMOP} for screening currents,
\cite{AFRS98} for relation to $\Aqp(\slNh)$, 
\cite{FRS97} for connection to $q$-difference version of the 
Drinfeld-Sokorov reduction, and
see \cite{BP98,FR98,HJKOS99} for higher currents.
DWA($A_{N-1}$) appears in the $A_{N-1}^{(1)}$ face model 
\cite{AJMP96,FJMOP} and also in the ABF model in regime II \cite{JKOPS99}.
Since $A_1^{(1)}$ and $A_2^{(2)}$ are the only affine Lie algebras of 
rank $1$, those DWA closes for one current $T(z)$, i.e. DVA.
In CFT, $W_N$ algebra does not contain $W_n$ ($n\leq N$) algebra as a 
subalgebra explicitly except for $n=2$ case which corresponds to the 
Virasoro algebra.
In deformed case, DWA($A_{N-1}$) does not contain even DVA($A_1^{(1)}$)
explicitly.
As shortly explained in subsection \ref{sec:2.2.3}, singular vectors of 
Virasoro and $W$ algebras imply that the correlation functions containing 
corresponding primary fields satisfy the differential equations.
For singular vectors of DVA and DWA, do the correlation functions
satisfy some difference equations?

\medskip\noindent{\it KZ and $q$-KZ equations :}\quad
In CFT Wess-Zumino-Novikov-Witten (WZNW) model has gauge symmetries, i.e.
affine Lie algebra symmetries \cite{CFT}. Virasoro current is realized
as a quadratic form of affine Lie algebra currents (Sugawara construction).
Consequently conformal Ward identity has rich structure, which is known as
the Knizhnik-Zamolodchikov (KZ) equation \cite{KZ84,TK88}. 
Since the Virasoro algebra is a Lie algebra, we know a rule for its tensor
product representation. Moreover we obtain new realizations and character
formulas of the Virasoro algebra by coset construction \cite{GKO86}.
On the other hand DVA is not a Lie algebra and its tensor product 
representation is unknown. (We remark that tensor product representations
of $W_N$ algebras in CFT are also unknown.) If there exists some deformation
of the Sugawara construction, it could give tensor product representations
of DVA but we do not know it at present.
Rather we derive DVA or DWA from (elliptic) quantum groups by fusion of
VO's \cite{JS97}.
$q$-deformation of KZ equation were presented by I.~Frenkel and Reshetikhin
\cite{FR92}. $q$-KZ equations are holonomic $q$-difference equations for
the matrix coefficients of the products of intertwining operators for 
representations of quantum affine algebra. Connection matrix of their 
solutions gives a face type elliptic solution of YBE.
See \cite{FR92,AKM92,Smi92,DFJMN,IIJMNT}.

\medskip\noindent{\it massive integrable models :}\quad
Integrable perturbations of CFT were studied in \cite{Zam87,EY,Zam89,Smi91}
and $(1,3)$-perturbed CFT is described by the sine-Gordon model.
Sine-Gordon model is a typical massive integrable model.
$S$-matrix was obtained by Zamlodochikov's bootstrap approach \cite{Zam79}.
For form factors see Smirnov's bootstrap approach \cite{Smi} 
(see also \cite{L95}).
Lukyanov pointed out that DVA($A_1^{(1)}$) current $T(z)$ in certain 
scaling limit gives the Zamolodchikov-Faddeev (ZF) algebra of sine-Gordon 
model (before taking a scaling limit, $T(z)$ is interpreted as the ZF algebra
for basic scalar excitation of XYZ spin chain) \cite{L96}.
In section \ref{sec:3}, $z$ of $T(z)$ is introduced as a formal parameter,
but here $z$ of $T(z)$ is related to the spectral parameter of the particle,
like as lattice models in section \ref{sec:4}, \ref{sec:5} ($z=x^{2u}$). 
This is contrasted with the CFT case, where 
$z$ of $L(z)$ is interpreted as a complex coordinate of the Riemann surface.
We can obtain integrable massive field theory models from solvable lattice 
models by taking appropriate scaling limit. 
Particles in a field theory are created by type II VO's.
For XXZ model, sine-Gordon model, Bullough-Dodd model and affine Toda
model see \cite{JKM96,MW96,BL97,L97,O96}. 
Field theory analog of transfer matrix and Baxter's $Q$-operator is 
studied in \cite{BLZ}.

\medskip\noindent{\it eight vertex model :}\quad
ABF model was studied by Lukyanov and Pugai by bosonization of type I VO's.
An algebraic approach to the fusion ABF models was presented in 
\cite{Ko97,JKOS2} on the basis of the quasi-Hopf algebra $\Bqla(\slth)$ 
and the elliptic algebra $U_{q,p}(\slth)$. Bosonization of VO's for 
the $A_{N-1}^{(1)}$ face model was given in \cite{AJMP96}.
Another interesting direction is to study Baxter's eight vertex model 
and Belavin's generalization. 
Lashkevich and Pugai proposed a remarkable bosonization formula 
of the type I VO for the eight vertex model \cite{LP97}. 
They succeeded in reducing the problem to the already known bosonization 
for the ABF model through the use of intertwining vectors and Lukyanov's 
screening operators. To understand their bosonization scheme, it seems 
necessary to clarify the relationship between the intertwining vectors 
and the two twistors $F(\lambda)$ and $E(r)$, which define $\Bqla(\slnh)$ 
and $\Aqp(\slnh)$ respectively.  
It is also interesting to seek a more direct bosonization, 
which is intrinsically connected with the quasi-Hopf structure  
of $\Aqp(\slth)$ and does not rely on the bosonization of the ABF model. 

\medskip\noindent{\it supersymmetry :}\quad
In string theory supersymmetries are essential to cancelation of divergence
and consistency of theory, 
and $N=2$ superconformal algebra is related to many interesting topics,
e.g. chiral ring, mirror symmetry, topological field theory.
Are there `good' deformations of superconformal algebras?
Super version ($\Z_2$ graded algebra) of elliptic quantum group was 
formulated in \cite{ZG98} along the line of \cite{JKOS1} 
(See also \cite{ABRR97}).
Can we obtain deformed superconformal currents by fusion of VO's of this
elliptic superalgebra or higher level VO's of $U_q(\slth)$ \cite{H98} ?
See also \cite{DF98}.

\medskip
As explained in the introduction our motivation is to find the symmetry
of massive integrable models, but present status is far from satisfactory.
We hope that this lecture can help the study in this (and also other) field.

%
\section*{Acknowledgments}
\addcontentsline{toc}{section}{Acknowledgments}
This lecture is based on our collaboration and
I would like to thank 
Hidetoshi Awata, Yuji Hara, Michio Jimbo, Harunobu Kubo, Hitoshi Konno, 
Yaroslav Pugai and Jun'ichi Shiraishi 
for valuable discussions and comments.
I thank also the organizers and participants of this summer school 
for their kind hospitality, and CRM for financial support.

\appendix
\setcounter{equation}{0}
\section{Some Formulas}

In this appendix we give a summary of  notations and formulas 
used throughout this lecture.

\subsection{Some functions}

Let us fix $x,r,\rs$. Following functions are used in this lecture:
\ba
  &&[n]_x=\frac{x^n-x^{-n}}{x-x^{-1}},
  \label{[n]_x}\\
  &&(z;p_1,\cdots,p_k)_\infty=\prod_{n_1,\cdots,n_k=0}^\infty
    (1-p_1^{n_1}\cdots p_k^{n_k}z),\\
  &&(z_1,\cdots,z_n;p_1,\cdots,p_k)_\infty=\prod_{j=1}^n
    (z_j;p_1,\cdots,p_k)_\infty,
  \label{pinf}\\
%
  &&\Theta_p(z)=(p,z,pz^{-1};p)_\infty
  =\sum_{n\in\Z}(-1)^nz^np^{\frac12n(n-1)},
  \label{Theta}\\
  &&[u]=x^{\frac{u^2}{r}-u}\Theta_{x^{2r}}(x^{2u}),
  \label{[u]}\\
  &&[u]^*=x^{\frac{u^2}{\rs}-u}\Theta_{x^{2\rs}}(x^{2u}).
  \label{[u]^*}\\
  &&[u]_+=x^{\frac{u^2}{r}-u}\Theta_{x^{2r}}(-x^{2u}),
  \label{[u]_+}\\
  &&[u]_+^*=x^{\frac{u^2}{\rs}-u}\Theta_{x^{2\rs}}(-x^{2u}).
  \label{[u]_+^*}
\ea
$[u]$ and $[u]^*$, $[u]_+$ and $[u]_+^*$, are related by
\be
  [u]^*=[u]\Bigl|_{r\rightarrow\rs},\quad 
  [u]_+^*=[u]_+\Bigl|_{r\rightarrow\rs}.
\ee
$[u]$ satisfies 
\be
  [-u]=-[u],\quad [u+r]=-[u],\quad
  [u+\tau]=-[u]e^{\frac{2\pi i}{r}(u+\frac{\tau}{2})}\quad
  (x=e^{\frac{\pi i}{\tau}}),
  \label{prop[u]}
\ee
and the Riemann identity
\ba
  &&[2u_1][2u_2][2u_3][2u_4]
  \label{Rid}\\
  &=\!\!&
  [u_1+u_2+u_3+u_4][u_1-u_2-u_3+u_4][u_1+u_2-u_3-u_4][u_1-u_2+u_3-u_4]\n
  &&+[-u_1+u_2+u_3+u_4][u_1-u_2+u_3+u_4][u_1+u_2-u_3+u_4][u_1+u_2+u_3-u_4].
  \nonumber
\ea
Lemma 4 in \cite{JLMP} is
\ba
  &&\frac{1}{m!}\sum_{\sigma\in S_m}\prod_{i=1}^m[v_{\sigma(i)}-2i+2]
  \cdot\prod_{1\leq i<j\leq m\atop\sigma(i)>\sigma(j)}
  \frac{[v_{\sigma(i)}-v_{\sigma(j)}-1]}{[v_{\sigma(i)}-v_{\sigma(j)}+1]}\n
  &=\!\!&
  \frac{1}{m!}\prod_{i=1}^m\frac{[i]}{[1]}\cdot
  \prod_{1\leq i<j\leq m}\frac{[v_i-v_j]}{[v_i-v_j-1]}\cdot
  \prod_{i=1}^m[v_i-m+1].
  \label{JLMPlem4}
\ea

Along with the additive variable $u$, we often use the multiplicative variable 
\be
  z=x^{2u},\quad z_j=x^{2u_j},
\ee
and the following abbreviation for an integration measure
\be
  \dz=\frac{dz}{2\pi iz},\quad \dz_j=\frac{dz_j}{2\pi iz_j}.
  \label{dz}
\ee
$(z;p)_{\infty}$, $\Theta_p(z)$ and $[u]$ have no poles and have
simple zeros,
\ba
  (z;p)_{\infty}&:&z=p^{-n}\quad(n\in\Z_{\geq 0}),\\
  \Theta_p(z)&:&z=p^m\quad(m\in\Z),\\
  \lbrack u\rbrack&:&u=rm\quad(m\in\Z)\quad\mbox{(i.e. $z=x^{2rm}$)}.
\ea
For $m\in\Z$ we have
\be
  \oint_{u=rm+a}\dz\,\frac{1}{[u-a]}f(u)
  =\frac{(-1)^{m-1}}{(x^{2r};x^{2r})_{\infty}^3}f(rm+a),
\ee
where $f(u)$ is regular at $u=rm+a$, and $\frac{1}{[u-a]}f(u)$
does not contain a fractional power of $z$.

\subsection{Delta function}\label{app:a.2}

Delta function $\delta(z)$ is a formal power series
\be
  \delta(z)=\sum_{n\in\Z}z^n,
\ee
and has the property
\be
  \oint_0\frac{dz}{2\pi iz}f(z)\delta(\sfrac{z}{a})=f(a).
\ee

Let $F(z;q)$ be the following Taylor series in $z$,
\be
  F(z;q)=\prod_{i=1}^n\frac{1-q^{\beta_i}z}{1-q^{\alpha_i}z}.
\ee
Here $\frac{1}{1-z}$ means $\sum_{m\geq 0}z^m$ and parameters $\alpha_i$, 
$\beta_i$ satisfy
\be
  \sum_{i=1}^n\alpha_i=\sum_{i=1}^n\beta_i, \quad
  \mbox{$\alpha_i$ are all distinct.}
\ee
Then we have a formula,
\be
  F(z;q)-F(z^{-1};q^{-1})=\sum_{i=1}^na_i\delta(q^{\alpha_i}z),\qquad
  a_i=\frac{\ds\prod_{j=1}^n(1-q^{\beta_j-\alpha_i})}
  {\ds\prod_{j=1\atop j\neq i}^n(1-q^{\alpha_j-\alpha_i})}.
  \label{deltaformula}
\ee
{\it Example}:\quad For $a\neq a'$, we have
\ba
  &&\frac{(1-q^{2b}z)(1-q^{2(a+a'-b)}z)}{(1-q^{2a}z)(1-q^{2a'}z)}
  -\frac{(1-q^{-2b}z^{-1})(1-q^{-2(a+a'-b)}z^{-1})}
  {(1-q^{-2a}z^{-1})(1-q^{-2a'}z^{-1})}\n
  &=\!\!&-(q-q^{-1})\frac{[b-a]_q[b-a']_q}{[a-a']_q}
  \Bigl(\delta(q^{2a}z)-\delta(q^{2a'}z)\Bigr).
  \label{deltaformulaex}
\ea
By taking a limit $a'\rightarrow a$, we get
\ba
  &&\frac{(1-q^{2b}z)(1-q^{2(2a-b)}z)}{(1-q^{2a}z)^2}
  -\frac{(1-q^{-2b}z^{-1})(1-q^{-2(2a-b)}z^{-1})}{(1-q^{-2a}z^{-1})^2}\n
  &=\!\!&-(q-q^{-1})^2[b-a]_q^2\,q^{2a}z\delta'(q^{2a}z).
\ea

\subsection{Some summations}

As it is well known in statistical mechanics, 
in order to calculate the following summation
\be
  \sum_{\{k_i\geq 0\}\atop
  {\ds\mathop{\scriptstyle\Sigma}_{\scriptscriptstyle i=1}^{
  \scriptscriptstyle N}}ik_i=N}
  f(k_1,k_2,\cdots),
\ee
it is convenient to introduce its generating function
\be
  \sum_{N=0}^{\infty}\sum_{\{k_i\}\atop{\scriptstyle\Sigma}_iik_i=N}
  f(k_1,k_2,\cdots)y^N=
  \sum_{k_1=0}^{\infty}\sum_{k_2=0}^{\infty}\cdots 
  f(k_1,k_2,\cdots)y^{k_1+2k_2+\cdots}.
\ee

We have two formulas:
\ba
  \prod_{\{k_i\}\atop{\scriptstyle\Sigma}_iik_i=N}\prod_ik_i!&=\!\!&
  \prod_{\{k_i\}\atop{\scriptstyle\Sigma}_iik_i=N}\prod_ii^{k_i},
  \label{k_i!}\\
  \sum_{\{k_i\}\atop{\scriptstyle\Sigma}_iik_i=N}k_l&=\!\!&
  \sum_{k\geq 1\atop lk\leq N}p(N-lk)\qquad(1\leq l\leq N),
  \label{k_i}
\ea
where $p(N)$ is the number of partition (see \eq{p(N)}).
Generating function of $\log$\eq{k_i!} is 
$\ds\prod_{n>0}\frac{1}{1-y^n}\times\sum_{i>0}\frac{y^i}{1-y^i}\log i$,
and that of \eq{k_i} is $\ds\prod_{n>0}\frac{1}{1-y^n}\times\frac{y^l}{1-y^l}$.
{}From \eq{k_i} we obtain
\be
  \prod_{\{k_i\}\atop{\scriptstyle\Sigma}_iik_i=N}\prod_iF(i)^{k_i}=
  \prod_{l,k\geq 1\atop lk\leq N}F(l)^{p(N-lk)},
  \label{Fk}
\ee
where $F(z)$ is any function.

\subsection{Some integrals}\label{app:a.4}

We summarize the relations among the following integral
with various contours \cite{KM,Fel89,BMP,DF},
\ba
  I&\!\!=\!\!&\int dz_1\cdots dz_mF,\\
  F&\!\!=\!\!&\prod_{1\leq i<j\leq m}(z_i-z_j)^{2\alpha}
  \cdot\prod_{i=1}^mz_i^{\alpha'}\cdot f(z_1,\cdots,z_m).\nonumber
\ea
Here $f(z_1,\cdots,z_m)$ is a symmetric function and has no pole at $z_i=z_j$. 
Parameters $\alpha$ and $\alpha'$ are assumed to satisfy the condition
(i) (and (ii) for \eq{IBIF}),
\ba
  &\mbox{(i)}&m(m-1)\alpha+m\alpha'\in\Z,
  \label{condi}\\
  &\mbox{(ii)}&m(m-1)\alpha+m\alpha'\in m\Z.
  \label{condii}
\ea
Let us consider the following contours:

{~}\vspace{3mm}
\input epsf.tex

\centerline{\epsfbox{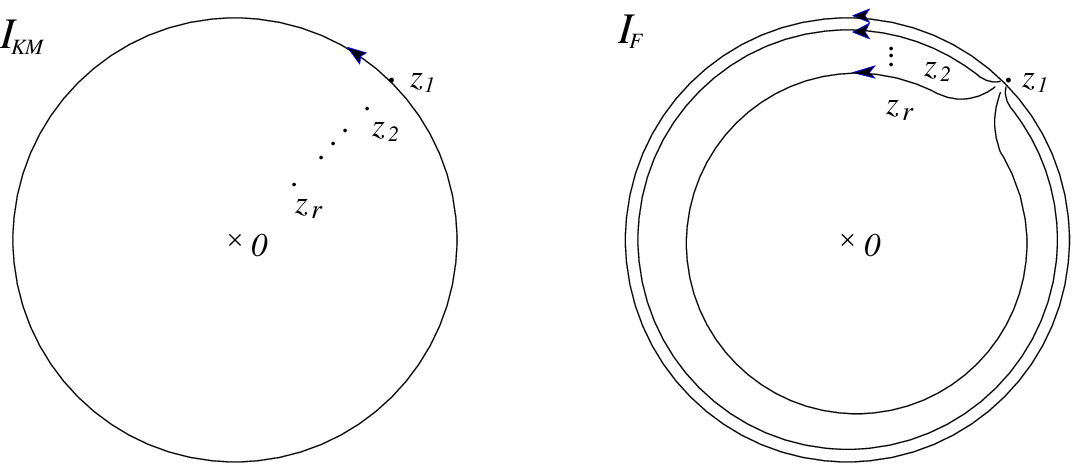}}

{~}

\centerline{\epsfbox{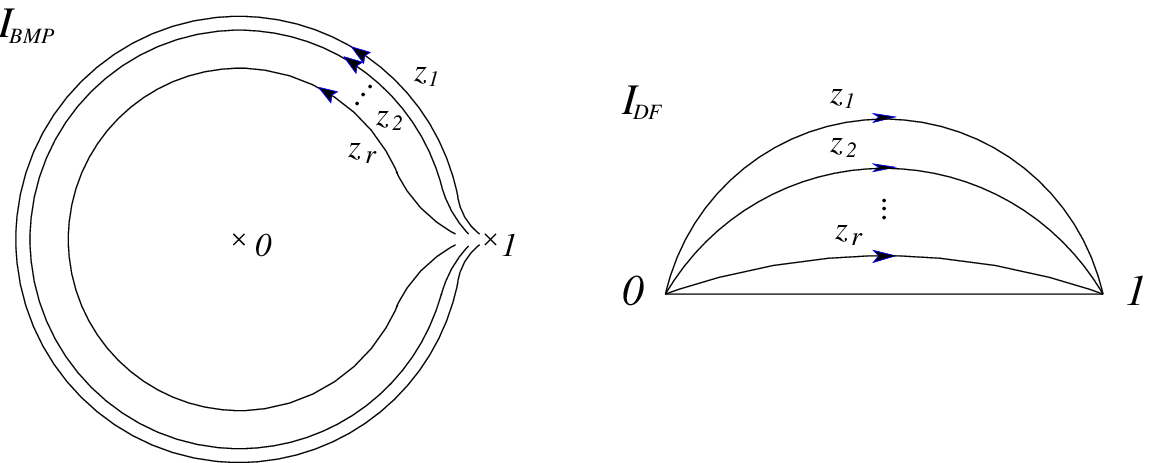}}
\ba
  I_{KM}&\!\!=\!\!&
  \oint dz_1\int_0^{z_1}dz_2\int_0^{z_2}dz_3\cdots\int_0^{z_{m-1}}dz_mF,\\
  I_{F}&\!\!=\!\!&
  \oint dz_1\int dz_2\cdots dz_mF,\\
  I_{BMP}&\!\!=\!\!&
  \int dz_1\cdots dz_mF,\\
  I'_{BMP}&\!\!=\!\!&
  \int dz_1\cdots dz_mF, \quad (z_j=r_je^{i\theta_j},\quad
  0<\theta_1<\theta_2<\cdots<\theta_m<2\pi),\\
  I_{DF}&\!\!=\!\!&
  \int dz_1\cdots dz_mF,\\
  I'_{DF}&\!\!=\!\!&
  \int_0^1dz_1\int_0^{z_1}dz_2\int_0^{z_2}dz_3\cdots\int_0^{z_{m-1}}dz_mF.
\ea
For $I_{KM}$ and $I_{F}$ the contour for $z_1$ closes.

By analytic continuation these integrals are related as follows: 
\ba
  I_F&\!\!=\!\!&(-1)^{m-1}\prod_{j=0}^{m-2}(1-a'a^j)\cdot
  \prod_{j=1}^{m-1}(1+a+a^2+\cdots+a^{j-1})\cdot I_{KM},\n
  &=\!\!&\prod_{j=0}^{m-2}(a'a^j-1)\cdot\prod_{j=1}^{m-1}
  \frac{a^j-1}{a-1}\cdot I_{KM},\\
  I_{BMP}&\!\!=\!\!&\prod_{j=1}^m(1+a+a^2+\cdots+a^{j-1})\cdot I'_{BMP}
  =\prod_{j=1}^m\frac{a^j-1}{a-1}\cdot I'_{BMP},
  \label{IBI'B}\\
  I_{BMP}&\!\!=\!\!&(-1)^m\prod_{j=0}^{m-1}(1-a'a^j)\cdot I_{DF}
  =\prod_{j=0}^{m-1}(a'a^j-1)\cdot I_{DF},\\
  I_{DF}&\!\!=\!\!&\prod_{j=1}^m(1+a+a^2+\cdots+a^{j-1})\cdot I'_{DF}
  =\prod_{j=1}^m\frac{a^j-1}{a-1}\cdot I'_{DF},\\
  I_F&\!\!=\!\!&\prod_{j=1}^{m-1}(1+a+a^2+\cdots+a^{j-1})\n
  &&\quad\times\Bigl(1+a'a^{m-1}+(a'a^{m-1})^2+\cdots
  +(a'a^{m-1})^{m-1}\Bigr)\cdot I'_{BMP},
  \label{IFI'B}
\ea
where $a$ and $a'$ are
\be
  a=e^{2\pi i\alpha},\quad a'=e^{2\pi i\alpha'},
\ee
and we remark that
\be
  a'a^{m-1}=1\mbox{ for (ii)},\quad
  a'a^{m-1}\neq 1\mbox{ for (i) but not (ii)}.
\ee

Under the condition (ii), from \eq{IBI'B} and \eq{IFI'B}, we have
\be
  I_{BMP}=\frac{1}{m}\frac{a^m-1}{a-1}I_{F}.
  \label{IBIF}
\ee

We give an example of this kind of integral:
\ba
  &&f(z_1,\cdots,z_m)=\prod_{i=1}^m(1-z_i)^{\alpha''},\n
  &&I'_{DF}=\prod_{j=1}^m\frac{\Gamma(j\alpha)}{\Gamma(\alpha)}
  \frac{\Gamma((j-1)\alpha+\alpha'+1)\Gamma((j-1)\alpha+\alpha''+1)}
  {\Gamma((m-2+j)\alpha+\alpha'+\alpha''+2)}.
  \label{exI'DF}
\ea

\subsection{Hausdorff formula}

For two operators $A$ and $B$, we have a formula
\be
  e^ABe^{-A}=e^{{\rm ad}A}B,\quad ({\rm ad}A)B=[A,B],
\ee
because two functions of $t$, $e^{tA}Be^{-tA}$ and $e^{t{\rm ad}\,A}B$,
satisfy the same differential equation $f'(t)=Af(t)-f(t)A$ and the
same initial condition $f(0)=B$.
When $A$ and $[A,B]$ commute each other, we have $e^ABe^{-A}=B+[A,B]$ and 
so $e^Ae^Be^{-A}=e^{B+[A,B]}$. Therefore we have a formula
\be
  [A,[A,B]]=[B,[A,B]]=0\;\Longrightarrow\; e^Ae^B=e^{[A,B]}e^Be^A,
  \label{Hausdorff}
\ee
which is a special case of the Campbell-Baker-Hausdorff formula
$e^Ae^B=e^{A+B+\frac12[A,B]+\cdots}$.
A harmonic oscillator $[a,a^{\dagger}]=1$ and this formula are basic tools
of free field realization.

\subsection{trace technique}

Let us consider one free boson oscillator, $[a,a^{\dagger}]=1$.
The Fock space $\F$ is generated by $\ket{0}$ ($a\ket{0}=0$),
and its orthonormal basis is 
$\ket{n}=\frac{1}{\sqrt{n!}}a^{\dagger\,n}\ket{0}$ ($n\geq 0$).
A trace over the Fock space ($\tr_{\F}{\cal O}=\sum_{n\geq 0}
\bra{n}{\cal O}\ket{n}$) can be expressed as a vacuum to vacuum amplitude.
Let us introduce another oscillator $b$, $[b,b^{\dagger}]=1$, which commute 
with $a$ and satisfy $b\ket{0}=0$.
Then we have the following Clavelli-Shapiro's trace formula \cite{CS73},
\be
  \tr_{\F}\Bigl(y^{a^{\dagger}a}{\cal O}(a,a^{\dagger})\Bigr)
  =\frac{1}{1-y}\bra{0}{\cal O}(d,\bar{d})\ket{0}.
\ee
Here $d,\bar{d}$ are
\be
  d=\frac{1}{1-y}a+b^{\dagger},\quad \bar{d}=a^{\dagger}+\frac{y}{1-y}b,
  \quad [d,\bar{d}\,]=1.
\ee
Especially we have
\be
  \tr_{\F}\Bigl(y^{a^{\dagger}a}e^{Aa^{\dagger}}e^{Ba}\Bigr)
  =\frac{1}{1-y}e^{AB\frac{y}{1-y}},
  \label{traceCS1}
\ee
where $A,B$ are constants.

For reader's convenience, we give three direct proofs of \eq{traceCS1}.\\
(i) The first method uses a cyclic property of trace. We have
\ba
   f(A,B)&\!\!=\!\!&
   \tr_{\F}\Bigl(y^{a^{\dagger}a}e^{Aa^{\dagger}}e^{Ba}\Bigr)\quad(|y|<1)\n
  &=\!\!&
  \tr_{\F}\Bigl(e^{Aya^{\dagger}}y^{a^{\dagger}a}e^{Ba}\Bigr)
  =\tr_{\F}\Bigl(y^{a^{\dagger}a}e^{Ba}e^{Aya^{\dagger}}\Bigr)\n
  &=\!\!&
  e^{ABy}\,\tr_{\F}\Bigl(y^{a^{\dagger}a}e^{Aya^{\dagger}}e^{Ba}\Bigr)
  =e^{ABy}f(Ay,B),
  \label{f(Ay,B)}
\ea 
and
\ba
   f(A,B)&\!\!=\!\!&
  e^{-AB}\,\tr_{\F}\Bigl(y^{a^{\dagger}a}e^{Ba}e^{Aa^{\dagger}}\Bigr)
  =e^{-AB}\,\tr_{\F}\Bigl(e^{By^{-1}a}y^{a^{\dagger}a}e^{Aa^{\dagger}}\Bigr)\n
  &=\!\!&
  e^{-AB}\,\tr_{\F}\Bigl(y^{a^{\dagger}a}e^{Aa^{\dagger}}e^{By^{-1}a}\Bigr)
  =e^{-AB}f(A,By^{-1}).
  \label{f(A,By^{-1})}
\ea
\eq{f(Ay,B)} implies
$$
  f(A,B)=e^{AB(y+y^2+\cdots)}f(Ay^{\infty},B)=e^{AB\frac{y}{1-y}}f(0,B),
$$
and \eq{f(A,By^{-1})} implies
$$
  f(0,B)=f(0,By)=f(0,By^{\infty})=f(0,0).
$$
Since $f(0,0)=\frac{1}{1-y}$, we obtain \eq{traceCS1}.\\
(ii) The second method uses a coherent state 
$|\alpha)=e^{\alpha a^{\dagger}}\ket{0}$ ($\alpha\in\C$),
which satisfies $a|\alpha)=\alpha|\alpha)$, 
$(\alpha|\alpha')=e^{\bar{\alpha}\alpha'}$ and the completeness condition
$1=\sum_{n\geq 0}\ket{n}\bra{n}=\int\frac{d^2\alpha}{\pi}e^{-|\alpha|^2}
|\alpha)(\alpha|$,
where $d^2\alpha=d\alpha_1d\alpha_2$ with $\alpha=\alpha_1+i\alpha_2$.
The trace becomes
$$
  \tr_{\F}\Bigl(y^{a^{\dagger}a}e^{Aa^{\dagger}}e^{Ba}\Bigr)  
  =\int\frac{d^2\alpha}{\pi}e^{-|\alpha|^2}
  (\alpha|y^{a^{\dagger}a}e^{Aa^{\dagger}}e^{Ba}|\alpha).
$$
Since
$
  (\alpha|y^{a^{\dagger}a}e^{Aa^{\dagger}}e^{Ba}|\alpha)
  =e^{B\alpha}(\alpha|y^{a^{\dagger}a}e^{Aa^{\dagger}}|\alpha)
  =e^{B\alpha}(\alpha|y^{a^{\dagger}a}|\alpha+A)
  =e^{B\alpha}(\alpha|y(\alpha+A))
  =e^{B\alpha}e^{\bar{\alpha}y(\alpha+A)}
$,
by completing the square and performing the Gauss integral,
we obtain the result.\\
(iii) The third method is a direct calculation.
\ban
  &&\tr_{\F}\Bigl(y^{a^{\dagger}a}e^{Aa^{\dagger}}e^{Ba}\Bigr)
  =\sum_{n=0}^{\infty}\frac{1}{n!}\bra{0}a^n
  y^{a^{\dagger}a}e^{Aa^{\dagger}}e^{Ba}a^{\dagger\,n}\ket{0}\\
  &=\!\!&\sum_{n=0}^{\infty}\frac{1}{n!}y^n\bra{0}a^n
  e^{Aa^{\dagger}}(a^{\dagger}+B)^n\ket{0}
  =\sum_{n=0}^{\infty}\frac{1}{n!}y^n\bra{0}a^n
  \sum_{l=0}^n\frac{A^{n-l}}{(n-l)!}{n\choose l}B^{n-l}a^{\dagger\,n}\ket{0}\\
  &=\!\!&\sum_{n=0}^{\infty}y^n
  \sum_{l=0}^n\frac{1}{(n-l)!}{n\choose l}(AB)^{n-l}.
\ean
By interchanging the order of summations $\sum_{n=0}^{\infty}\sum_{l=0}^n=
\sum_{l=0}^{\infty}\sum_{n=l}^{\infty}$ and shifting $n=m+l$, it becomes
$$
  \sum_{m=0}^{\infty}\frac{1}{m!}(ABy)^m
  \sum_{l=0}^{\infty}{m+l\choose l}y^l.
$$
Since $\sum_{l=0}^{\infty}{m+l\choose l}y^l$ equals to 
$\frac{1}{m!}(y\partial_y+m)\cdots(y\partial_y+1)\sum_{l=0}^{\infty}y^l
=\frac{1}{(1-y)^{m+1}}$,
we obtain the result.

\addcontentsline{toc}{section}{References}

\newpage
\pagestyle{empty}
\centerline{{\Large\bf List of Revised Points} (misprints etc.)}

\noindent{\bf hep-th/9910226}
\begin{itemize}
\item  page 11 : $\cup \Rightarrow +$ (5 places)
\item  below eq.(2.153) : add references
\item  $d \Rightarrow d^{\rm osc}+d^{\rm zero}$\\
eqs.(3.43),(5.73),(5.74),(5.140),(6.18),(6.103),(6.104),(6.130).
\end{itemize}

\noindent
These points are revised in hep-th/9910226v2.

\medskip
\noindent{\bf hep-th/9910226v2}

I will update this list (misprints etc.). See the following page:
\begin{verbatim}
          http://azusa.shinshu-u.ac.jp/~odake/paper.html
\end{verbatim}

\end{document}